\documentclass[12pt]{article}
\usepackage{authblk}
\usepackage{geometry}
\usepackage{mathtools}
\usepackage{amsmath}
\usepackage{amssymb}
\usepackage{amsthm}
\usepackage{mathrsfs}
\usepackage{subfigure}
\usepackage{customcommands}
\usepackage{bm}
\usepackage[normalem]{ulem}
\usepackage{Nettastyle}
\usepackage{comment}

\definecolor{SHCcolor}{rgb}{0.,0.75,0.}

\title{A Tale of Two Saddles}
\author[1]{Venkatesa Chandrasekaran,}
\author[2]{Netta Engelhardt,}
\author[3,4]{Sebastian Fischetti,}
\author[5]{and Sergio Hern\'andez-Cuenca}

\affiliation[1]{School of Natural Sciences, Institute for Advanced Study,\\ 1 Einstein Drive, Princeton, NJ 08540 USA}
\affiliation[2]{Center for Theoretical Physics,\\
Massachusetts Institute of Technology, Cambridge, MA 02139, USA}
\affiliation[3]{Department of Physics, McGill University, Montr\'eal, QC, H3A 2T8, Canada}
\affiliation[4]{Department of Physics and Astronomy, Weber State University, Ogden, UT 84408, USA}
\affiliation[5]{Department of Physics, University of California, Santa Barbara, CA 93106, USA}

\emailAdd{venchandrasekaran@ias.edu}
\emailAdd{engeln@mit.edu}
\emailAdd{sfischetti@weber.edu}
\emailAdd{sergiohc@ucsb.edu}

\abstract{We find a new on-shell replica wormhole in a computation of the generating functional of JT gravity coupled to matter.  We show that this saddle has lower action than the disconnected one, and that it is stable under restriction to real Lorentzian sections, but can be unstable otherwise.  The behavior of the classical generating functional thus may be strongly dependent on the signature of allowed perturbations.  As part of our analysis, we give an LM-style construction for computing the on-shell action of replicated manifolds even as the number of boundaries approaches zero, including a type of one-step replica symmetry breaking that is necessary to capture the contribution of the new saddle.  Our results are robust against quantum corrections; in fact, we find evidence that such corrections  may sometimes stabilize this new saddle.
}

\begin{document}

\maketitle

\section{Introduction}
\label{sec:intro}
The contributions of Euclidean wormhole saddles to gravitational systems have been as puzzling as they have been illuminating (see e.g.~\cite{Col88,GidStr88,MalMao04,ArkOrg07,HarJaf18,SSS,AlmHar19,PenShe19,StaWit19,Ili19,KapMah19,Saa19,MarMax20,MarMax20b,GidTur20}). On the one hand, in the gravitational replica trick for the von Neumann entropy~\cite{LewMal13, FauLew13, AlmHar19,PenShe19} these wormholes provide an explanation for the quantum extremal surface (QES) formula~\cite{EngWal14}, and consequently for the consistency of semiclassical black hole evaporation with unitarity~\cite{AEMM,Pen19}.  On the other hand, the presence of wormholes gives rise to an apparent lack of factorization that raises questions about whether a low-energy description of gravity can be contained in a single theory or must be emergent from an ensemble~\cite{Col88,GidStr88,MalMao04,HarJaf18,SSS,StaWit19,Saa19,KapMah19,MarMax20,AreVol19,Ili19,BouWil20,Wit20b,AfkCoh20,BeldeB20b,Van20,Blo20,MalWit20,CotJen20,PerTro20,JanMir21,TurUsa21,MaxTur21,BenKel21,PenTia22}. 

Given the crucial role these wormholes play in ensuring unitarity, are there other quantities aside from measures of entropy to which wormholes contribute in the semiclassical regime?  This seems likely: we might expect that the imprint of such an important aspect of gravity would be detectable with more general and simpler observables than entropies. The most fundamental such object in an investigation into the basic import of non-factorizing Euclidean geometries is the generating functional. 

How, then, does the factorization problem manifest at the level of the generating functional? An answer to this question would go significantly further in resolving the puzzles raised by the inclusion of replica wormholes in the gravitational path integral than the study of any one individual quantity, be it entropy or any other observable. 

This query was initially raised in the context of gravity in~\cite{EngFis20}, which found significant modifications to the behavior of the generating functional of Jackiw-Teitelboim (JT) gravity (at nonperturbatively low temperatures) depending on whether connected topologies were permitted to contribute. In fact, in the absence of replica wormholes the generating functional gives rise to a negative entropy at low temperature, while the inclusion of wormholes in a nonperturbative completion of JT does in fact give rise to a positive thermodynamic entropy all the way to zero temperature~\cite{Joh19,Joh20a,Joh20b,Joh20c,Joh21a,Joh21b,Joh21c,Joh22a,Joh22b}.

However, these advances offer limited insight in the quest towards an understanding of the role of non-factorization in the emergence of semiclassical gravity.  First, the results of~\cite{Joh19,Joh20a,Joh20b,Joh20c,Joh21a,Joh21b,Joh21c,Joh22a,Joh22b} are highly nonperturbative and thus do not shed light on the nature and behavior of observables in the \textit{semiclassical} regime.  Second, in working with pure JT gravity, we restrict to a theory which is known to feature ensemble averaging, rendering the factorization problem less mysterious than in theories where there is no obvious ensemble.  In particular, we would like to understand replica wormholes in higher-dimensional theories of gravity. 

In the semiclassical regime, the contribution of replica wormholes to the generating functional can be tractably investigated using a replica trick. This replica trick differs substantially from the more familiar one for the von Neumann entropy, so we now briefly review it. In a theory of gravity with Euclidean action~$I$ defined by a boundary geometry or conformal geometry~$(B,h)$ (e.g.~this may be an asymptotically AdS boundary), we schematically denote the gravitational path integral by
\be
\Pcal(B) = \sum_M \int \Dcal g \, \Dcal \psi \, e^{-I[g,\psi]},
\ee
where the sum is over manifolds~$M$ of different topologies with boundary~$B$, the integral is over metrics $g$ on $M$ inducing~$(B,h)$ on their  boundary, and~$\psi$ represents any additional matter fields\footnote{If matter fields are present, their boundary values should also be fixed.}.  The aforementioned factorization puzzle arises from noting that the contribution of different topologies to~$\Pcal(B)$ implies that~$\Pcal(B^n) \neq \Pcal(B)^n$, and hence~$\Pcal(B)$ cannot be interpreted as the partition function $Z(B)$ of a standard quantum system on $B$,  which would factorize on $B^n$. Instead, this observation suggests an interpretation of~$\Pcal(B)$ as some coarse-graining (possibly an averaging) over quantum mechanical partition functions defined by $B$:\footnote{We will be agnostic on the precise protocol that produces $\overline{Z(B)}$ from $Z(B)$, though in certain special cases (such as pure JT gravity) it can be understood precisely; see e.g.~\cite{BouWil20,MalWit20,PerTro20,CotJen20,BeldeB20,BeldeB20b,PolRoz20,Ebe21,SaaShe21,Joh22a,GarGot21,BloIli21,PenTia22} for a number of possible interpretations in more general contexts.}
\be
\label{eq:GPIoverline}
\Pcal(B) = \overline{Z(B)}.
\ee
It prima facie appears that we are simply out of luck in any attempt to directly infer the implications of replica wormholes on the generating functional: a na\"ive semiclassical computation of the generating functional would just correspond to taking $\ln {\cal P}(B)=\ln \overline{Z(B)}$; the gravitational path integral here involves only a single boundary and thus no Euclidean wormholes can be included.

However, $\ln Z(B)$ admits an alternative calculation via a replica trick:
\be
\label{eq:lnZreplica}
\ln Z(B)=  \lim_{m \to 0} \frac{1}{m} \left(Z(B^{m})- 1\right),
\ee
where~$B^m$ is the union of~$m$ copies of the boundary~$B$.  The advantage of this rewriting is that it immediately reveals an alternative semiclassical calculation in which replica wormholes could contribute to the generating functional:
using~\eqref{eq:GPIoverline}, an ``overline'' of~\eqref{eq:lnZreplica} gives
\be
\label{eq:replicatrick}
\overline{\ln Z(B)} = \lim_{m \to 0} \frac{1}{m} \left(\Pcal(B^m) - 1\right).
\ee
While at the fine-grained quantum mechanical level the replica trick~\eqref{eq:lnZreplica} for $\ln Z(B)$ is no different from the direct calculation thereof, the calculations of $\ln \overline{Z(B)}$ and $\overline{\ln Z(B)}$ from the gravitational path integral do not necessarily agree. This discrepancy is easiest to see in the context where $\Pcal(B)$ actually computes an ensemble average: in that case, $\ln \overline{Z(B)}$ -- the so-called \textit{annealed} generating functional, which we shall denote by $\Gamma_{A}$ -- allows the parameters of the ensemble to equilibrate with the dynamical fields, whereas $\overline{\ln Z(B)}$ -- the \textit{quenched} generating functional $\Gamma_{Q}$ -- freezes the ensemble in each computation of the generating functional before averaging.  Since the parameters in such an ensemble interpretation (to which we do not necessarily subscribe) are distinguished by the fact that they are not dynamical, it is clearly the latter that is of physical relevance, and hence the question of whether Euclidean wormholes contribute to the semiclassical generating functional amounts to whether saddles with connected topologies exist and dominate over their disconnected counterparts in the~$m \to 0$ limit above.  

This is what motivates our work in this paper: we look for replica wormholes that dominate over the disconnected topology in a (strictly classical) saddle-point approximation.  The natural starting point is the gravitational replica trick technique of Lewkowycz and Maldacena (LM)~\cite{LewMal13}, which can be applied to explore the analytic continuation to small~$m$ in the replica trick~\eqref{eq:replicatrick} for the generating functional.  Recall that in a saddle-point approximation for $\mathcal{P}(B^m)$, this technique considers on-shell geometries~$(M_m,g_m)$ with a~$\mathbb{Z}_m$ symmetry. Quotienting by this symmetry produces a single-boundary ``quotient'' geometry~$(\widehat{M}_m,\widehat{g}_m)$ containing a (not necessarily connected) codimension-two conical defect with opening angle~$2\pi/m$.  This is analytically continued in $m$ away from the integers (while still imposing the equations of motion), giving an on-shell action~$\widehat{I}_m$ which can be used to approximate the path integral:~$\Pcal(B^m) \approx e^{-m\widehat{I}_m}$.  To compute, say, the von Neumann entropy, the limit~$m \to 1$ is then taken in the corresponding replica trick; working perturbatively about the~$m = 1$ geometry then recovers the QES formula.

In the present context, the appropriate limit is~$m \to 0$, which introduces a number of subtleties not present in the computation of the von Neumann or R\'enyi entropies~\cite{LewMal13,Don16} (in which case $m$ is a positive integer). First, as a consequence of the divergent conical surplus in the $m\to 0$ limit, there appears to be no~$m = 0$ geometry about which we can work perturbatively. Second, in contrast with the entropy replica trick, standard examples of the quenched generating functional (e.g. spin glasses) often require replica symmetry breaking (RSB).  We may therefore expect the need for incorporating RSB when evaluating the replica trick~\eqref{eq:replicatrick} in semiclassical gravity. We will anticipate this and include an algorithmic way of breaking replica symmetry into our protocol for computing $\Gamma_Q$; indeed, this will prove to be necessary in the examples discussed below. 

\begin{figure}[t]
\centering
\includegraphics[width=0.2\textwidth,page=4]{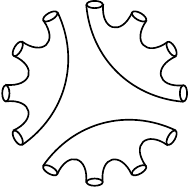}
\caption{An illustration of the kinds of replica symmetry-breaking wormholes we allow in our analysis of the replica trick~\eqref{eq:replicatrick}.  Here~$m = 12$ and~$m_1 = 4$.}
\label{fig:RSB}
\end{figure}

Our prescription for computing $\Gamma_Q$ from the gravitational replica trick is presented in Section~\ref{sec:replicatrick}.  In short, the procedure consists of two steps:
\begin{enumerate} 
    \item First, we break replica symmetry by allowing for partially connected wormholes where the $m$ boundaries cluster into $m_1$-boundary wormholes, with $m_1$ some divisor of $m$, as shown in Figure~\ref{fig:RSB}.  These wormholes have symmetry group~$(\mathbb{Z}_{m_1})^{m/m_1} \times \mathbb{S}_{m/m_1}$, which naturally interpolates between the $\mathbb{S}_{m}$ symmetry of the fully-disconnected phase and the $\mathbb{Z}_m$ one of the fully-connected wormhole.  Quotienting by this symmetry group gives a single-boundary geometry with conical defects of opening angle~$2\pi/m_1$.
    \item We then analytically continue~$m$ and~$m_1$ away from the integers and use~\eqref{eq:replicatrick} to obtain a formula for the quenched generating functional:
    \be
    \Gamma_Q \equiv  \overline{\ln Z} \approx -\min_{m_1 \in [0,1]} \widehat{I}_{m_1}.
    \ee
\end{enumerate}

In practice, evaluating the action~$\widehat{I}_{m_1}$ in order to compute $\Gamma_Q$ requires solving for the quotient metric~$\widehat{g}_m$ explicitly.  In general theories of gravity this is a difficult task, but it is simplified substantially in two-dimensional models.  Consequently, in exploring the contributions of replica wormholes to $\Gamma_Q$ we focus our attention on JT gravity and JT gravity coupled to matter, in which~$\widehat{g}_m$ describes a two-dimensional geometry of constant negative curvature and hence the geometric degrees of freedom are simply the moduli in the space of such geometries.  In the quotient geometry, these moduli reduce to a single degree of freedom that determines the proper distance~$D$ between the conical defects.  This simplification makes an explicit construction of the quotient geometry tractable, and in Section~\ref{sec:quotient} we discuss two methods of constructing it: we obtain~$\widehat{g}_m$ either by patching together appropriate regions of the Poincar\'e disk, or by solving the Liouville equation in the presence of defects.  The former method is very explicit, while the latter is more akin to how we would need to proceed to obtain~$\widehat{g}_m$ in higher dimensions.

With the quotient geometry constructed, we can then compute the quotient action~$\widehat{I}_m$ in our specific models, which involves solving the equation of motion for the ``boundary degree of freedom'' (i.e.~the Schwarzian mode).  This procedure is most straightforward in pure JT, but as is well-known, Euclidean wormholes do not exist as saddles in pure JT due to the tendency of the modulus~$D$ to ``pinch off'' the wormhole throats.  Nevertheless, it is possible to investigate the behavior of off-shell ``constrained'' wormholes in which~$D$ is fixed by hand; we perform this investigation in Section~\ref{sec:pureJT} as an illustrative example that highlights the rather complicated structure of~$\widehat{I}_m$ at~$m < 1$.  In particular, the continuation of~$\widehat{I}_m$ to complex~$m$ is an infinitely-sheeted Riemann surface, and the equations of motion are crucial for determining which of these sheets gives the correct answer for~$\widehat{I}_m$.

In order to stabilize the modulus~$D$ to get genuinely classical wormholes, however, we need to support them by coupling JT to some matter.  In Section~\ref{sec:JTCFT} we therefore couple JT to a massless scalar field and construct the resulting action~$\widehat{I}_m$.  In order for the matter field to exhibit a nontrivial stress tensor -- as is necessary to stabilize the wormholes -- we turn on boundary sources for the matter field\footnote{In higher dimensions, the importance of turning on matter sources to stabilize Euclidean wormholes has been recently discussed by~\cite{MarSan21}.}; these sources break the~$U(1)$ Euclidean time-translation symmetry, so the states that we consider are not thermal states.  Nevertheless, the parameter~$\beta$ that sets the length of the Euclidean time circle (i.e.~the ``inverse temperature'') is a tunable boundary condition, and we find that at sufficiently small $\beta^{-1}$ relative to the strength of our sources the matter is able to support classical wormholes.  In particular, these wormholes exist for~$m < 1$.  Moreover, these new saddles have smaller action than the disconnected saddle that contributes to $\Gamma_A$, which immediately suggests that the quenched generating functional of JT gravity coupled to classical matter has a phase in which it is dominated by replica wormholes.  The inclusion of these wormholes results in quantitatively and qualitatively different behavior of the quenched generating functional $\Gamma_Q$ from its annealed counterpart $\Gamma_A$.

Of course, whether or not these new saddles genuinely contribute to the quenched generating functional depends on their stability properties under a given definition of the path integral.  We find that while the disconnected saddle is stable against all Euclidean perturbations, the new connected saddles at~$m < 1$ (with lower action) are not: they are stable when restricted to perturbations that admit a particular analytic continuation to (real) Lorentzian time, but unstable to arbitrary Euclidean perturbations.  Hence in a purely Euclidean treatment that forgets about the Lorentzian origins of the theory, the quenched generating functional appears to reproduce its annealed version; but in a treatment that imposes a real Lorentzian section (or perhaps that rotates the contour of integration in the path integral to an appropriately ``Lorentzian'' one), the quenched and annealed generating functionals may differ at low temperatures.  In fact, these statements appear to be robust under quantum corrections: in Section~\ref{sec:quantum}, we compute quantum corrections to the matter action perturbatively around~$m=1$ and find that for~$m < 1$ these quantum corrections exhibit a stabilizing effect on the wormholes.  This is to be contrasted with the situation for~$m > 1$, where a Casimir effect actually has a destabilizing influence~\cite{GarGot20,MoiSak21,MoiSak22}.

These observations naturally prompt questions of how to determine what the ``right'' saddles to include in the quenched generating functional are; while we do not answer this question, we discuss some interesting facets and avenues of exploration in Section~\ref{sec:disc}.

\section{The Replica Trick for the Generating Functional}
\label{sec:replicatrick}

Under the interpretation~\eqref{eq:GPIoverline} of the gravitational path integral~$\Pcal(B^m)$, the replica trick~\eqref{eq:replicatrick} is a trivial identity; its nontrivial content arises from how~$\Pcal(B^m)$, which is only well-defined for integer~$m$, is to be continued in~$m$ to a neighborhood of~$m = 0$.  In general this analytic continuation is not unique and must be treated carefully.  For instance, if~$\Pcal(B^m)$ exhibits appropriate behavior in the right-half complex~$m$ plane, Carlson's theorem can be used to ensure a unique analytic continuation; or if the ``ensemble average'' genuinely corresponds to an average over an appropriate distribution of partition functions, one could try to use Carleman's condition in a similar way.  However, these approaches are often insufficient to provide a unique continuation to~$m = 0$; see e.g.~\cite{JanMir21} for a discussion of some of the difficulties involved.

Instead, a fruitful approach is to work in a saddle-point approximation wherein the path integral~$\Pcal(B^m)$ is dominated by a single geometry obeying the saddle-point equations: that is, the classical equations of motion.  By continuing the equations of motion to non-integer~$m$ in a controlled way, one obtains a unique analytic continuation of~$\Pcal(B^m)$ to non-integer~$m$ which typically captures the correct physics.  In condensed matter contexts, the replica trick has been used in this way for decades; see e.g.~\cite{SpinGlassBook}.  In the gravitational context, this analytic continuation of the equations of motion amounts to the Lewkowycz-Maldacena (LM) construction for computing the von Neumann entropy gravitationally~\cite{LewMal13} with some important differences that we now describe.

\subsection{The Saddle-Point Approximation}
\label{subsec:saddlepoint}

Let us review the gravitational replica trick of LM, but adapted to~\eqref{eq:replicatrick} rather than to the von Neumann entropy. The calculation is done in a saddle-point approximation, where the path integral~$\Pcal(B)$ is approximated by the on-shell action of a classical solution:
\be
\Pcal(B) \approx e^{-I[g^{\mathrm{clas}}]},
\ee
where~$g_{\mathrm{clas}}$ solves the classical equations of motion (and we have suppressed any matter fields~$\psi$ -- they are treated classically in the same way as gravity, or semiclassically by including the one-loop effective matter action to the above).  Now, when~$B$ consists of several disconnected regions, as in the replica trick~\eqref{eq:replicatrick}, we should consider all possible topologies of bulk manifold~$M$ and find the configuration with smallest action.  In the case of $\Pcal(B^m)$, the~$m$ boundaries exhibit an~$\mathbb{S}_m$ permutation symmetry.  A common assumption is that the on-shell bulk solution that approximates~$\Pcal(B^m)$ is highly symmetric as well. We will review the construction of the gravitational replica with this in mind and leave all discussion of further replica symmetry breaking -- which is crucial for our construction -- to the subsequent section. 

Let us first consider the maximally symmetric saddles, which are the disconnected solutions. In these saddles, the bulk manifold~$M$ consists of~$m$ disconnected pieces that ``fill in'' each boundary~$B$.  In this case the on-shell action is just~$m I[g_{m = 1}]$, where~$g_{m = 1}$ is the on-shell metric defined by a single boundary.  The analytic continuation in~$m$ is then trivial, and gives a contribution to the quenched generating functional of
\be
\label{eq:lnZdisconnectedcontribution}
\Gamma_Q \equiv \overline{\ln Z} \supset -I[g_{m =1}] = \ln \overline{Z} \equiv \Gamma_A,
\ee
where we identify the average of the partition function as~$\overline{Z} \approx e^{-I[g_{m = 1}]}$.  This contribution is just the annealed generating functional, and is the standard way of computing e.g.~free energies in gravitational theories.

\begin{figure}[t]
\centering
\includegraphics[width=0.5\textwidth,page=1]{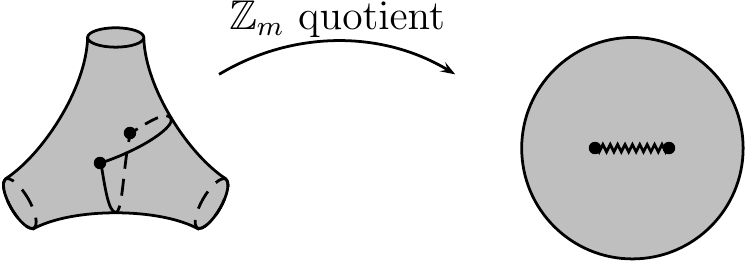}
\caption{The LM construction used to obtain the analytic continuation in~$m$ for the replica trick~\eqref{eq:replicatrick}.  Upon quotienting by~$\mathbb{Z}_m$, the two black lines in the left diagram are stitched together at the branch cut in the right diagram.}
\label{fig:LMquotient}
\end{figure}

The second type of saddle that is typically considered in this context is the wormhole solution that connects various copies of~$B$ in an arrangement exhibiting a~$\mathbb{Z}_m$ symmetry, as shown in the left diagram of Figure~\ref{fig:LMquotient}.  Denoting the metric on the fully-connected manifold as~$g_m$, the on-shell action is simply~$I[g_m]$.  In order to continue this on-shell action away from integer~$m$, the wormhole geometry is quotiented by the~$\mathbb{Z}_m$ symmetry, yielding a quotient manifold~$\widehat{M}_m$ with metric~$\widehat{g}_m$; the codimension-two surfaces of fixed points of the~$\mathbb{Z}_m$ isometry in the full geometry become conical defects in~$(\widehat{M}_m,\widehat{g}_m)$ with opening angle~$2\pi/m$.  All~$m$-dependence appears only in the opening angle about these defects, and so~$m$ can be sensibly continued away from the integers while still imposing the equations of motion.  Note that~$\widehat{g}_m$ does not solve the equations of motion obtained from the action~$I[g]$ at the defects; one can either impose the equations of motion everywhere away from the defects, or modify the action by the contribution of a cosmic brane with tension proportional to~$1-1/m$ sourcing the defects, as in~\cite{Don16}. Either way, the resulting contribution to the quenched generating functional from this replica-symmetric saddle would be
\be
\label{eq:lnZwormholecontribution}
\Gamma_Q \supset \lim_{m \to 0} \frac{1}{m} \left(e^{-m I[\widehat{g}_m]} - 1\right) = -\lim_{m \to 0} \widehat{I}_m,
\ee
where we have introduced the shorthand notation~$\widehat{I}_m \equiv I[\widehat{g}_m]$. Note that we have left the~$m \to 0$ limit explicit because it is not clear whether~$\widehat{g}_0$ is a well-defined geometry.  In particular, while for~$m > 1$ the conical defect is an angular deficit, for~$m < 1$ the defect is an excess, and in fact as~$m \to 0$ the excess angle becomes arbitrarily large\footnote{From the cosmological brane point of view, for~$m > 1$ the brane tension~$T_m \propto (1-1/m)$ is positive, making it gravitationally attractive; for~$m < 1$ the brane tension is negative, making it gravitationally repulsive.}.

The contribution~\eqref{eq:lnZwormholecontribution} to $\Gamma_Q$ captures the effect of replica wormholes in the path integral.  If this contribution is subdominant to the conventional one~\eqref{eq:lnZdisconnectedcontribution} from the disconnected topology, then $\Gamma_Q$ simply reproduces the annealed calculation$\Gamma_A$.  But if wormholes dominate over the disconnected contribution,$\Gamma_Q$ and~$\Gamma_A$ may differ substantially.

It is worth remarking on the differences between this procedure and the analogous one for the von Neumann entropy.  The first difference is clearly the number of replicas: in the derivation of the RT/HRT formula for von Neumann entropy~\cite{RyuTak06,HubRan07}, one ultimately takes the~$m \to 1$ limit of R\'enyi entropies.  This limit allows us to work perturbatively around the ``original geometry''~$m = 1$ to derive a formula for the von Neumann entropy that does not require an explicit construction of the quotient geometry~$\widehat{g}_m$.  On the other hand, our~$m \to 0$ limit for $\Gamma_Q$ genuinely requires a computation~$\widehat{g}_m$ well away from~$m = 1$.  This is more akin to the gravitational computation of R\'enyi entropies~\cite{Don16}, which requires one to compute~$\widehat{I}_m$ at integer~$m > 1$. The second difference is that in computations of von Neumann (or R\'enyi) entropies in gravitational theories, correlations in the boundary conditions of the gravitational path integral render the symmetry group of the boundary to be~$\mathbb{Z}_m$ (e.g.~in the derivation of the RT formula,~$B$ consists of an~$m$-sheeted branched cover whose branch points correspond to the entangling surface).  Hence it is quite natural to take the infilling bulk solution to share this~$\mathbb{Z}_m$ symmetry as well.  On the other hand, in the replica trick~\eqref{eq:replicatrick}, the boundary~$B^m$ consists of~$m$ \textit{completely disconnected} pieces with symmetry group~$\mathbb{S}_m$.  This symmetry is broken to~$\mathbb{Z}_m$ by the wormhole.  In general we expect that connected geometries that preserve the full~$\mathbb{S}_m$ symmetry of the boundary do not exist, so in this sense we may think of the wormhole as a kind of mild form of replica symmetry breaking (RSB).  But if some amount of breaking the~$\mathbb{S}_m$ of the boundaries is inherent to the wormholes, it is natural to wonder whether we can proceed further by generalizing this breaking in a controlled way.  Indeed it can, and incorporating such RSB will be crucial to our later analysis.  We now briefly discuss a form of ``one-step gravitational RSB'' motivated by the structure of the Parisi ansatz for spin glasses~\cite{SpinGlassBook}.

\subsection{Replica Symmetry Breaking}
\label{subsec:RSB}

\begin{figure}[t]
\centering
\includegraphics[width=0.2\textwidth,page=4]{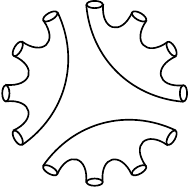}
\caption{An illustration of our replica symmetry-breaking wormholes; here~$m = 12$ and~$m_1 = 4$, giving a wormhole with~$m/m_1 = 3$ connected components.}
\label{fig:LMrsb}
\end{figure}

The one-step RSB procedure that we define incorporates wormholes that are not maximally connected\footnote{ The contributions of various topologies of off-shell wormholes to gravitational computations of von Neumann entropy were considered in e.g.~\cite{PenShe19}.}.  For a given integer~$m$, we take~$m_1$ to be a positive integer that divides~$m$.  We then consider wormholes that connect the~$m$ boundaries into groups of~$m_1$, as illustrated in Figure~\ref{fig:LMrsb} for the case~$m = 12$,~$m_1 = 4$.  Hence~$m_1$ is a parameter that encodes various possible wormhole topologies, and so we should ultimately minimize the action with respect to it to find the dominant contribution.

The bulk geometry consists of~$m/m_1$ disconnected pieces.  Assuming that each of these pieces has the same~$\mathbb{Z}_{m_1}$ symmetry discussed above, the symmetry group of the bulk geometry is~$(\mathbb{Z}_{m_1})^{m/m_1} \times \mathbb{S}_{m/m_1}$, with the~$\mathbb{S}_{m/m_1}$ factor corresponding to the freedom to permute the disconnected pieces amongst each other.  We may quotient the geometry by this symmetry to conclude that the total on-shell action is~$m \widehat{I}_{m_1}$.  We may now analytically continue~$m$ away from the integers.  Because~$m_1$ was constrained to be a divisor of~$m$, continuing in~$m$ naturally leads us to continue in~$m_1$ as well.  However, since~$m_1$ was constrained to range between one and~$m$, we preserve this constraint even after the analytic continuation: that is, we take~$m_1$ to be an arbitrary real number in the range~$m_1 \in [1,m]$.  Hence for arbitrary~$m$, the path integral is approximated by
\be
\Pcal(B^m) \approx \max_{m_1 \in [1,m]} e^{-m \widehat{I}_{m_1}},
\ee
and hence the replica trick~\eqref{eq:replicatrick} gives
\be
\label{eq:lnZRSB}
\Gamma_Q \approx -\min_{m_1 \in [0,1]} \widehat{I}_{m_1}.
\ee
Note that in taking the limit~$m \to 0$, we have maintained the constraint that~$m_1 \in [1,m] \to [0,1]$. If the minimization is achieved by~$m_1 = 1$, then $\Gamma_Q$ coincides with the annealed result $\Gamma_A$ in~\eqref{eq:lnZdisconnectedcontribution}; if the minimization is achieved by~$m_1 = 0$ (or perhaps more carefully, in the limit~$m_1 \to 0$), then we recover the quenched generating functional~\eqref{eq:lnZwormholecontribution} obtained from the completely connected wormholes.  We interpret any other value of~$m_1$ as breaking replica symmetry.

Some comments are in order.  First, though this procedure seems ad hoc (in particular, the analytic continuation of~$m_1$ and its restriction to lie in the interval~$[1,m]$ even after continuing~$m$ to zero), it is precisely analogous to one-step RSB in spin glasses.  As with most instances of the replica trick, ultimately we should interpret~\eqref{eq:lnZRSB} as a \textit{prescription} for computing the quenched generating functional, rather than as a derivation.  Its validity can only be determined on physical grounds.  Second, values of~$m_1$ different from~$m$ actually correspond to a \textit{larger} symmetry group than that of the fully-connected wormhole~$m_1 = m$, so what we are calling ``replica symmetry breaking'' can really be thought of as a ``symmetry enhancement'' relative to the fully-connected wormhole. Regardless, any value of~$m_1$ different from~1 still corresponds to a breaking of the~$\mathbb{S}_m$ symmetry of the boundaries.  Finally, there is a potential subtlety: the action functional could formally change signs for sufficiently small $m$. In that situation, presumably we would need to \textit{maximize} the action instead of minimizing it. This type of behavior is characteristic of spin glasses, where the number of degrees of freedom formally becomes negative in the $m\rightarrow 0$ limit\footnote{ More precisely, in a spin glass the degrees of freedom the action is to be extremized over are the off-diagonal components of an~$m \times m$ correlation matrix~$q_{\alpha\beta}$ encoding correlations between replicas.  The action takes the form of a sum over the components of~$q_{\alpha\beta}$, and the number of off-diagonal components is~$m(m-1)/2$.  When analytically continuing in~$m$, one sets the off-diagonal components of~$q_{\alpha\beta}$ to all be equal to some~$q$, giving the action an explicit overall factor of~$m(m-1)/2$, which becomes negative when~$m < 1$.  Due to this change in the overall sign of the action, the physical minimization of the action with respect to the individual components of~$q_{\alpha\beta}$  corresponds to the \textit{maximization} of the action with respect to the degree of freedom~$q$ in the replica ansatz once~$m$ is taken less than one.}.  Whether or not an analogous change of sign happens in the gravitational contexts we are considering will presumably depend on the details of the gravitational theory.  At least for the theories we will consider later in this paper, this phenomenon does not happen and we expect to need to minimize over~$m_1$.

As a final note, in many cases of interest one takes~$B$ to have topology~$\Sigma \times S^1$, where the circle~$S^1$ has length~$\beta$.  If all boundary sources exhibit a~$U(1)$ symmetry corresponding to translations around the~$S^1$,~$Z(B)$ can be interpreted as a canonical partition function at inverse temperature~$\beta$, and the quenched generating functional is related to the quenched free energy:
\be
F_Q \equiv -\beta^{-1} \Gamma_Q.
\ee
However, in later sections we will consider boundary sources that break the~$U(1)$ isometry of the boundary thermal circle.  In such a case~$F_Q$ no longer has an interpretation as the free energy of a thermal state, and for this reason we restrict our investigation to the generating functional~$\Gamma_Q$.

\section{The Quotient Geometry in Two Dimensions}
\label{sec:quotient}

In certain two-dimensional asymptotically (nearly) AdS models of gravity like JT, the geometry has constant negative curvature.  Moreover, in a classical (or semiclasscal) limit, contributions from higher genera are suppressed.  These simplfications render the construction of the quotient geometry~$(\widehat{M}_m,\widehat{g}_m)$ quite tractable, and we now describe it in this context.  In later sections we will rely on this construction to investigate the behavior of the~$m \to 0$ limit invoked in the computation of the quenched generating functional in pure JT gravity and in JT gravity coupled to matter.

It will be useful to describe two different ways of obtaining the quotient geometry.  The first exploits the fact that since~$\widehat{g}_m$ has constant negative curvature, it is locally AdS$_2$; hence~$\widehat{g}_m$ can be constructed by stitching together appropriate regions of exact AdS$_2$, i.e.~patches of the Poincar\'e disk.  This approach has the advantage of giving an exact and explicit form of~$\widehat{g}_m$, but at the cost of requiring more than one coordinate chart to cover the entire quotient manifold.  On the other hand, the second approach is analogous to the procedure in higher dimensions: we solve the equations of motion directly.  In this case we are only able to obtain an approximate solution for~$\widehat{g}_m$, but we are able to cover the entire quotient manifold with just one coordinate chart.  As we will see, the two constructions are useful in different contexts.

\subsection{Patchwise Construction}
\label{ssec:patchwork}

\begin{figure}[t]
\centering
\includegraphics[width=0.9\textwidth,page=2]{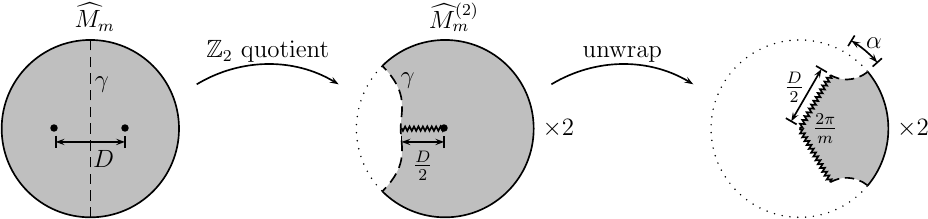}
\caption{In two dimensions, the quotient space~$(\widehat{M}_m, \widehat{g}_m)$, shown left, can be mapped to the unit disk in the complex plane with two conical defects on the real axis separated by a proper distance~$D$.  The imaginary axis is a geodesic~$\gamma$ about which the geometry exhibits a reflection symmetry.  Quotienting by this symmetry gives the geometry~$\widehat{M}_m^{(2)}$ which contains only a single defect; it corresponds to a portion of the Poincar\'e disk with a single defect bounded by a geodesic~$\gamma$, shown center.  Unwrapping the defect by cutting along the jagged line turns~$\widehat{M}^{(2)}_m$ into a subregion of the Poincar\'e disk (with no defect), shown right.  The geodesic~$\gamma$ consists of two segments traversing an angle $\alpha$, related to~$D$ by~\eqref{eq:alphaDrelationship}.}
\label{fig:fundomainquotientgeometry}
\end{figure}

In two dimensions, the quotient manifold has the topology of a disk with two conical defects.  Per the replica trick, the locations of these defects in the geometry should be fixed dynamically.  However, all but one degree of freedom in the locations of the defects can be gauge-fixed: if we think of~$\widehat{M}_m$ as a subset of the complex plane with the disk topology, the automorphisms of~$\widehat{M}_m$ can be used to fix three (real) degrees of freedom in the locations of the defects.  It is convenient to thus take~$\widehat{M}_m$ to be the unit disk in the complex plane and to place the defects on the real axis symmetrically about the origin, leaving the proper distance~$D$ between them as the single dynamical degree of freedom, as shown on the left of Figure~\ref{fig:fundomainquotientgeometry}.  Note that in this construction, the imaginary axis corresponds to a geodesic~$\gamma$ about which the quotient geometry exhibits a~$\mathbb{Z}_2$ reflection symmetry.  There is an additional~$\mathbb{Z}_2$ reflection symmetry about the real axis, so we expect the quotient geometry to exhibit a~$\mathbb{Z}_2 \times \mathbb{Z}_2$ symmetry.

Due to the reflection symmetry across~$\gamma$, we may construct the quotient geometry by stitching together the left and right halves of the disk along~$\gamma$; each of these halves corresponds to a geometry~$(\widehat{M}_m^{(2)}, \widehat{g}_m^{(2)})$ that contains only a \textit{single} conical defect and is bounded by a geodesic, as shown in the middle of Figure~\ref{fig:fundomainquotientgeometry}.  This geometry is simply a portion of the Poincar\'e disk with a defect.  We can then ``unwrap'' the defect by cutting~$\widehat{M}_m^{(2)}$ along the indicated jagged line, ending up with the wedge-shaped region of the Poincar\'e disk (with \textit{no} defect) shown at right in Figure~\ref{fig:fundomainquotientgeometry}\footnote{An alternative way of deriving this construction is to switch the order of the quotients: start with the original~$m$-boundary geometry~$(M_m,g_m)$ for integer~$m$ and first quotient by the~$\mathbb{Z}_2$ symmetry that maps the two fixed points of the replica symmetry to one another.  The resulting geometry is the Poincar\'e disk with $m$ geodesic boundaries corresponding to the fixed points of the~$\mathbb{Z}_2$, identical to the left diagram of Figure~\ref{fig:JTbranewormhole} in the Appendix.  Then quotienting by~$\mathbb{Z}_m$ immediately gives~$\widehat{M}_m^{(2)}$ in the form of the wedge shown at right of Figure~\ref{fig:fundomainquotientgeometry}.}.  Concretely, in complex coordinates in which the metric on the Poincar\'e disk takes the form
\be
\label{eq:Poincaredisk}
ds^2 = \frac{4\, dz \, d\bar{z}}{(1-z \bar{z})^2},\ee
the two jagged lines are the rays~$\Arg(z) = \pm \pi/m$, which are to be identified.  Likewise, the geodesic~$\gamma$ consists of two segments of the circles defined by
\be
\label{eq:bite}
\left|z- e^{\pm i\pi/m}\sec\alpha\right| = \tan\alpha,
\ee
with~$\alpha$ a parameter that sets the size of these circles, related to the proper distance~$D$ between the defects as
\be
\label{eq:alphaDrelationship}
\sin\alpha = \sech (D/2).
\ee
Note that for~$m > 2$, this construction implies that $D$ is bounded from below: in order for the geodesics that define~$\gamma$ to neither intersect one another nor exclude the origin, we must have
\be
\label{eq:alphabound}
0 < \alpha < \min\left\{\frac{\pi}{2},\frac{\pi}{m}\right\}.
\ee
So the minimum proper distance between the defects is~$D_\mathrm{min} = 2 \arcsech(\sin(\pi/m))$ when~$m > 2$.  More intuitively, for~$m > 2$ the mass of the defect in the middle diagram of Figure~\ref{fig:fundomainquotientgeometry} becomes sufficiently large that even when the geodesic~$\gamma$  orbits once around it, it is only able to reach a closest approach distance of $D_\mathrm{min}/2$; achieving~$D < D_\mathrm{min}$ would require~$\gamma$ to self-intersect.

\begin{figure}[t]
\centering
\includegraphics[width=0.25\textwidth,page=5]{Figures/Figures-pics}
\hspace{1cm}
\includegraphics[width=0.3\textwidth]{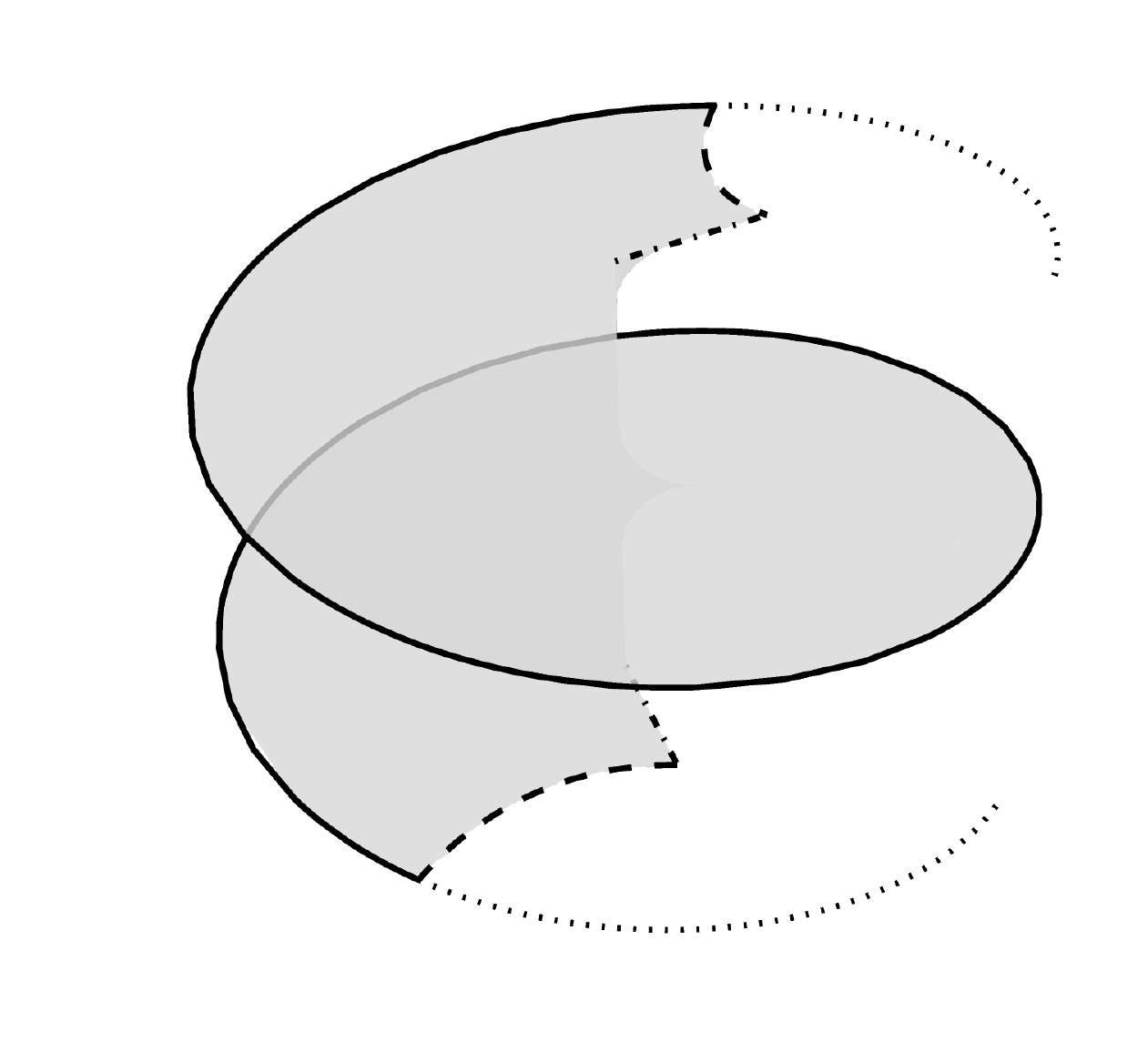}
\caption{For~$m \geq 1$, the opening angle about the conical defects is smaller than~$2\pi$, so the manifold~$\widehat{M}_m^{(2)}$ used in the construction of the quotion geometry consists of the wedge-shaped subregion of the Poincar\'e disk shown at left (with the dot-dashed lines identified).  But for~$m < 1$, the opening angle about the defect is greater than~$2\pi$, so to construct~$\widehat{M}_m^{(2)}$ we must ``unwrap'' the disk into an infinitely-sheeted Riemann surface.  $\widehat{M}_m^{(2)}$ is then a subregion of this Riemann surface, shown at right.}
\label{fig:M2}
\end{figure}

It is worth noting that there is a qualitative difference in this construction between~$m \geq 1$ and~$m < 1$: for~$m \geq 1$, the opening angle~$2\pi/m$ about the conical defects is less than (or equal to)~$2\pi$, so~$\widehat{M}_m^{(2)}$ really is a subregion of the Poincar\'e disk, as shown in the left diagram of Figure~\ref{fig:M2}.  On the other hand, for~$m < 1$ the opening angle~$2\pi/m$ is greater than~$2\pi$, so we must instead interpret~$\widehat{M}_m^{(2)}$ as being a subregion of a \textit{covering} of the Poincar\'e disk -- in terms of the standard angular coordinate~$\theta$ on the disk defined by~$z = r e^{i\theta}$, we no longer identify~$\theta$ with~$\theta + 2\pi$.  This turns the disk into a Riemann surface with infinitely many sheets, as shown in the right diagram of Figure~\ref{fig:M2}.  In particular,~$\widehat{M}_m^{(2)}$ includes arbitrarily many sheets of this Riemann surface as~$m \to 0$.

The upshot is that this patchwise construction of~$(\widehat{M}_m,\widehat{g}_m)$ is very convenient in models that can be reduced to local boundary dynamics.  In such cases we can use the near-boundary behavior of~$\widehat{g}_m^{(2)}$ to construct a local boundary action, and then simply impose appropriate boundary conditions at the (boundaries of the) geodesic~$\gamma$ at which the two copies of~$\widehat{M}_m^{(2)}$ are stitched together.  We will use such a construction in Section~\ref{sec:pureJT} to study pure JT, as well as in Appendix~\ref{app:WCmodel} to study a model of JT coupled to end-of-the-world branes.

However, in models that cannot be reduced to local boundary dynamics, as in the case of JT coupled to a massless scalar that we study in Section~\ref{sec:JTCFT}, the need to impose nontrivial boundary conditions \textit{everywhere} along the stitching geodesic~$\gamma$ makes this patchwise construction less useful.  For such cases, it is instead desirable to construct the quotient geometry~$\widehat{M}_m$ in a \textit{single} coordinate chart.  We turn to such a construction next.

\subsection{Liouville Construction}
\label{subsec:liouville}

\begin{figure}[t]
\centering
\includegraphics[width=0.7\textwidth,page=6]{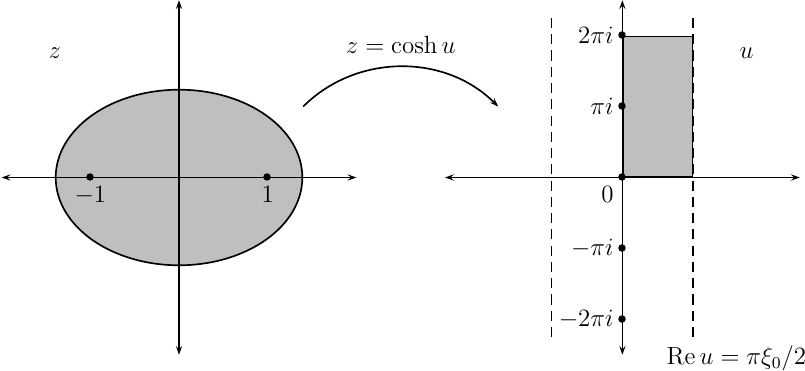}
\caption{The elliptical domain used in constructing the quotient geometry (left) and the rectangle to which we map it (right); the marked points are the locations of the defects, placed at the foci~$z = \pm 1$ of the ellipse.  The rectangle can be extended to an infinite strip (dashed lines) containing infinitely many images of these defects.}
\label{fig:ellipse}
\end{figure}

To construct the quotient geometry in a single coordinate chart, we take~$\widehat{M}_m$ to be a subregion~$\Omega$ of the complex plane with disk topology, with two points~$z_1$,~$z_2$ chosen to be the locations of the conical defects.  As remarked above, there is ample gauge freedom in these choices which we must fix.  In particular, we are free to conformally map~$\Omega$ to any other region of the complex plane with disk topology, as well as to change the locations of~$z_1$ and~$z_2$ using automorphisms of~$\Omega$.  Since such automorphisms contain three real degrees of freedom, the choice of~$\Omega$,~$z_1$, and~$z_2$ contains only a single real physical degree of freedom.  For instance, as in the left diagram of Figure~\ref{fig:fundomainquotientgeometry} we could take~$\Omega$ to be the unit disk and set~$z_1 = -a$,~$z_2 = a$ with~$a \in (0,1)$ the physical modulus that controls the proper distance between the defects.  For our purposes, it is instead convenient to take~$\Omega$ to be the interior of an ellipse with eccentricity~$\eps$ with foci at~$z = \pm 1$ and to place~$z_1$ and~$z_2$ at these foci, as shown in the left diagram of Figure~\ref{fig:ellipse}.  In this case the physical degree of freedom is~$\eps \in (0,1)$, with~$\eps$ near~0 and~1 corresponding to the defects being close together and far apart, respectively.  Also note that this choice of~$\Omega$ naturally manifests the expected~$\mathbb{Z}_2 \times \mathbb{Z}_2$ symmetry of the quotient geometry through the reflection symmetries across the principal axes of the ellipse.

The advantage of this choice of~$\Omega$ is that it can easily be mapped to a coordinate rectangle by converting to elliptic coordinates~$(\xi,\phi)$ through
\be
\label{eq:ellipticcoords}
z = \cosh u, \qquad u \equiv \xi + i\phi.
\ee
In the original~$z$ plane, curves of constant~$\xi$ correspond to confocal ellipses of eccentricity~$\sech \xi$ with foci at~$z = \pm 1$, while curves of constant~$\phi$ correspond to confocal hyperbolae with foci at~$z = \pm 1$.  Hence the ellipse~$\Omega$ is mapped to the rectangular region~$\phi \in [0,2\pi)$,~$\xi \in [0, \arcsech \eps)$ and the defects are mapped to~$u = 0$ and~$i\pi$, as shown in the right diagram of Figure~\ref{fig:ellipse}.  In what follows we will define~$\xi_0 \equiv (2/\pi)\arcsech \eps$, so that the conformal boundary is at~$\xi = \pi \xi_0/2$.  Note that the map to the rectangle in the~$u$ plane doubles the angle around the defects, so if the total angle around them was~$2\pi/m$ in the original ellipse, it is~$4\pi/m$ in the rectangle.

With the quotient manifold thus fixed, we can solve for the quotient metric~$\widehat{g}_m$ on it.  We work in conformal gauge in which the metric on the rectangle takes the form
\be
\label{eq:gmcflat}
ds^2 = e^{2\sigma(u,\bar{u})} \,du\,d\bar{u}.
\ee
Since~$\widehat{g}_m$ has constant curvature~$R = -2$ everywhere except at the defects, the equation of motion for~$\sigma$ is the Liouville equation~\cite{AlmHar19}
\be
\label{eq:preliouville}
- 4 \partial_u\partial_{\bar{u}} \sigma + e^{2\sigma} = 2\pi \left(1 - \frac{2}{m} \right) \left(\delta_{0}(u,\bar{u}) + \delta_{i\pi}(u,\bar{u})\right),
\ee
where we define the complex delta function as~$\delta_{u_*}(u,\bar{u}) \equiv \delta(\xi - \xi_*) \delta(\phi - \phi_*)$ (with~$u \equiv \xi + i\phi$ and~$u_* \equiv \xi_* + i\phi_*$ as above).  The delta functions on the right can be thought of as contributions to the curvature localized to the defects; note that their strength is proportional to~$1-2/m$ due to the doubling of the angle in going to the rectangular domain.  As shown in Appendix~\ref{app:Liouville}, extending the problem to the entire strip~$\xi \in (-\pi \xi_0/2, \pi\xi_0/2)$,~$\phi\in (-\infty,\infty)$ and defining~$\tilde{\sigma}$ via
\bea
\label{subeq:conformalshift}
\sigma &\equiv \tilde{\sigma} + \ln\left(\frac{1}{\xi_0} \sec \left(\frac{u + \bar{u}}{2\xi_0} \right) \right) + \frac{H}{\nu}, \\
\label{subeq:Hdef}
    \mbox{where } H &\equiv \sum_{k = -\infty}^\infty \ln \left|\tan\left(\frac{u - ik\pi}{2\xi_0}\right)\right| \mbox{ and } \nu \equiv \frac{m}{2-m},
\eea
the Liouville equation becomes
\bea
\label{subeq:shiftedliouville}
&-\left(\partial_\xi^2 + \partial_\phi^2\right) \tilde{\sigma} + \frac{1}{\xi_0^2} \sec^2\left(\frac{\xi}{\xi_0}\right) \left(e^{2H/\nu} e^{2\tilde{\sigma}} - 1\right) = 0, \\
\label{subeq:sigmatbndry}
&\tilde{\sigma}\left(\xi = \pm \frac{\pi \xi_0}{2}, \phi\right) = 0, \qquad \tilde{\sigma}(\xi,\phi + 2\pi) = \tilde{\sigma}(\xi,\phi).
\eea
Although we are unable to solve for~$\tilde{\sigma}$ analytically, in practice it is straightforward to obtain~$\tilde{\sigma}$ numerically as discussed in Appendix~\ref{subapp:Liouville}, and indeed we will make use of these numerical solutions later.  However, it \textit{is} possible to construct an analytic approximation for~$\tilde{\sigma}$ which we will use extensively in Section~\ref{sec:JTCFT}.

To obtain this approximate solution, we will work at small~$\xi_0$ and small~$\nu$ -- in fact, it will be convenient to take the relative scaling of~$\nu$ and~$\xi_0$ to be such that~$e^{-\pi/2\xi_0} \ll \nu$.  If~$\xi_0$ is small, the sum defining~$H$ is rapidly convergent since the~$k^\mathrm{th}$ term is sharply peaked around~$\phi = k\pi$.  Although~\eqref{subeq:shiftedliouville} is not linear in~$H$ or~$\tilde{\sigma}$, we consequently expect it to approximately linearize.  This motivates us to define
\be
\label{eq:sigmaredef}
\tilde{\sigma}(\xi,\phi) \equiv \sum_{k = -\infty}^\infty f(\xi,\phi-k\pi) + \delta \tilde{\sigma}(\xi,\phi),
\ee
where~$f$ is a solution to
\begin{subequations}
\label{eqs:fdef}
\begin{align}
&-\left(\partial_\xi^2 + \partial_\phi^2\right) f + \frac{1}{\xi_0^2} \sec^2\left(\frac{\xi}{\xi_0}\right) \left(\left|\tan\left(\frac{\xi + i\phi}{2\xi_0}\right) \right|^{2/\nu} e^{2f} - 1\right) = 0, \\
&f(\pm \pi\xi_0/2,\phi) = 0, \quad f(\xi,\phi \to \pm \infty) = 0
\end{align}
\end{subequations}
and~$\delta \tilde{\sigma}$ is a correction term which must vanish at~$\xi = \pm \pi \xi_0/2$.  Just as with each term in~$H$, we expect~$f$ to be sharply localized around~$\phi = 0$ at small~$\xi_0$.

If~$f$ localizes around~$\phi = 0$, then~$\delta \tilde{\sigma}$ must be small.  To see this, note that~\eqref{subeq:shiftedliouville} becomes
\begin{multline}
\label{eq:Liouvilledeltasigma}
-(\partial_\xi^2 + \partial_\phi^2)\delta\tilde{\sigma} + \frac{1}{\xi_0^2} \sec^2\left(\frac{\xi}{\xi_0}\right) \left\{\exp\left[2\sum_{k = -\infty}^\infty F(\xi,\phi-k\pi)\right] e^{2\delta\tilde{\sigma}} - 1 \right. \\
	\left. - \sum_{k = -\infty}^\infty \left(e^{2F(\xi,\phi - k\pi)} -1 \right)\right\} = 0,
\end{multline}
where we have defined
\be
F(\xi,\phi) \equiv \frac{1}{\nu} \ln \left|\tan\left(\frac{\xi + i\phi}{2\xi_0}\right) \right| + f(\xi,\phi).
\ee
Per our expectations on~$f$,~$F$ should also be sharply peaked around~$\phi = 0$ when~$\xi_0$ is small.  Though~\eqref{eq:Liouvilledeltasigma} is an exact rewriting of the Liouville equation, in order to estimate the magnitude of~$\delta \tilde{\sigma}$ we may linearize in terms that are small.  Without loss of generality we restrict our attention to the region~$\phi \in (-\pi/2,\pi/2]$, since we can extend the solution to the entire strip by symmetry about~$\phi = k\pi/2$.  In this region,~$F(\xi,\phi - k\pi)$ is small for all~$k \neq 0$, and hence~\eqref{eq:Liouvilledeltasigma} gives
\begin{multline}
\xi_0^2 \cos^2\left(\frac{\xi}{\xi_0}\right) \left(\partial_\xi^2 + \partial_\phi^2\right)\delta \tilde{\sigma} - e^{2F(\xi,\phi)} \left(e^{2\delta\tilde{\sigma}} - 1\right) \\ = 2 \left(e^{2F(\xi,\phi) + 2\delta\tilde{\sigma}} - 1\right) \sum_{k \neq 0} F(\xi,\phi-k\pi) + \cdots,
\end{multline}
where the dots denote subleading terms in~$F(\xi,\phi-k\pi)$ for~$k \neq 0$.  In the region~$\phi \in (-\pi/2,\pi/2]$, the right-hand side is small; if it were to vanish, then clearly so would~$\delta \tilde{\sigma}$.  Hence~$\delta \tilde{\sigma}$ must be small, as claimed; linearizing in it, we find
\begin{multline}
\label{eq:deltasigmalinearized}
\left[\xi_0^2 \cos^2\left(\frac{\xi}{\xi_0}\right) \left(\partial_\xi^2 + \partial_\phi^2\right) - 2 e^{2F(\xi,\phi)} \right]\delta \tilde{\sigma} \\ = 2 \left(e^{2F(\xi,\phi)} - 1\right) \sum_{k \neq 0} F(\xi,\phi-k\pi) + \cdots,
\end{multline}
where the dots now also include subleading terms in~$\delta \tilde{\sigma}$.  Consequently we expect the magnitude of~$\delta \tilde{\sigma}$ to scale like the magnitude of the source on the right-hand side.

\begin{figure}[t]
\centering
\includegraphics[width=0.275\textwidth]{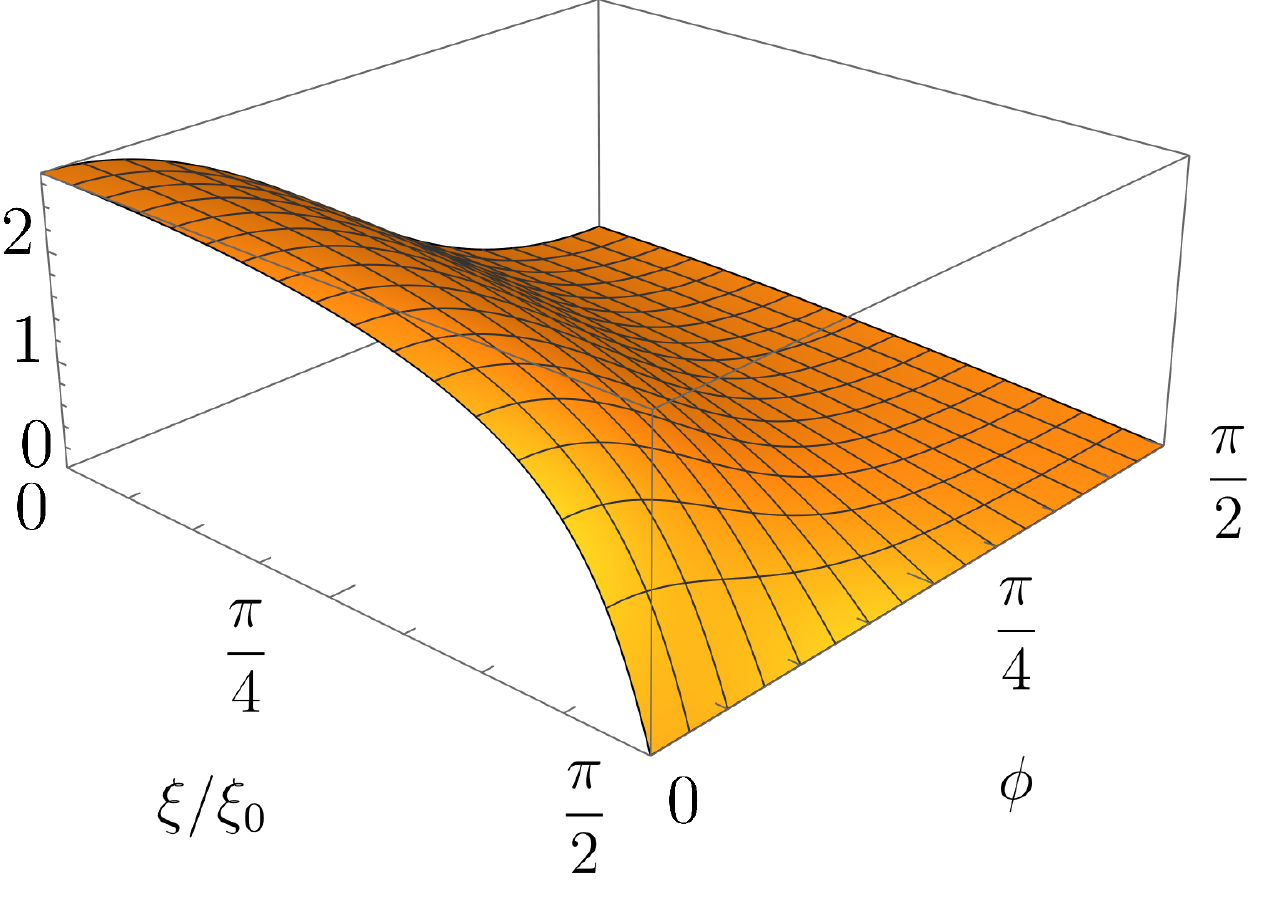}
\includegraphics[width=0.3\textwidth]{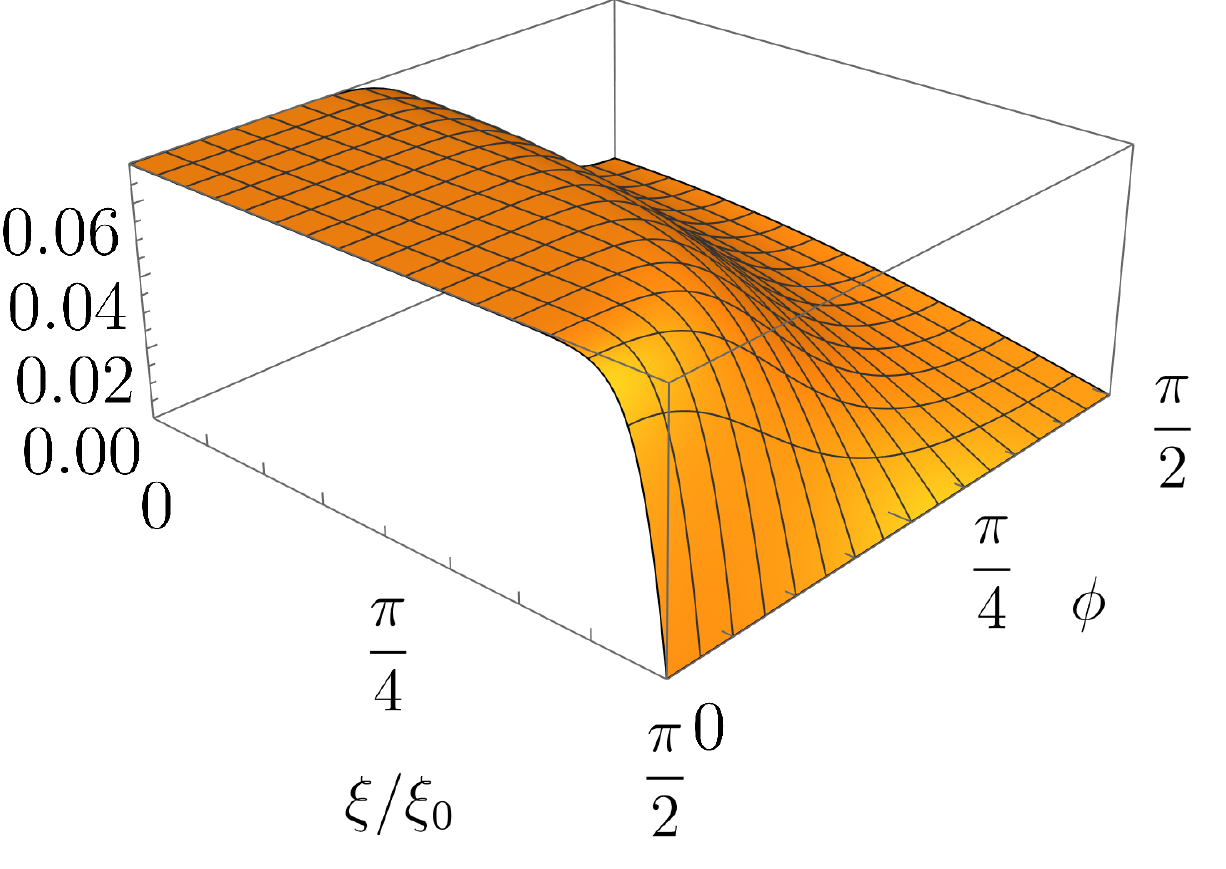}
\includegraphics[width=0.3\textwidth]{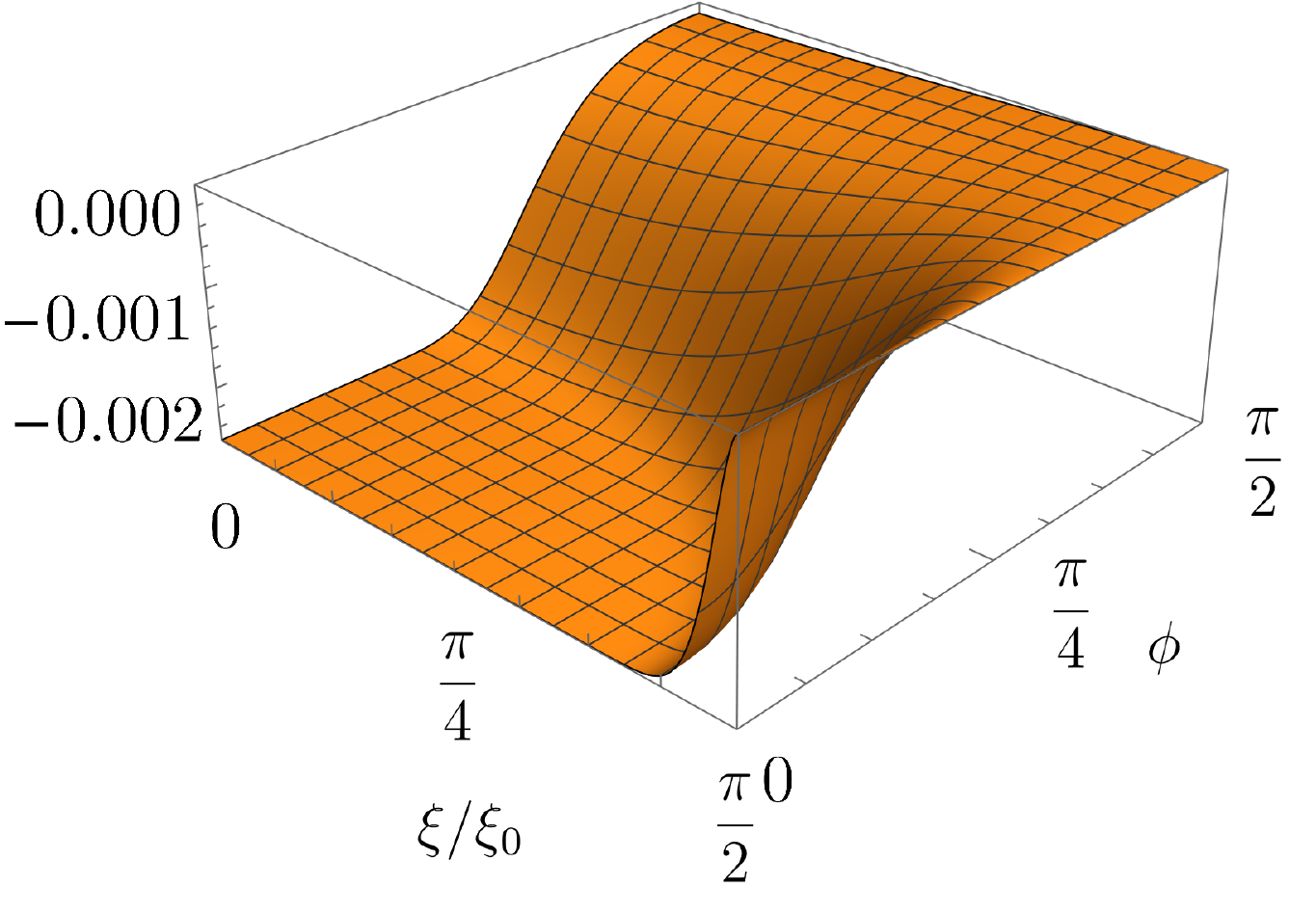}
\includegraphics[width=0.3\textwidth]{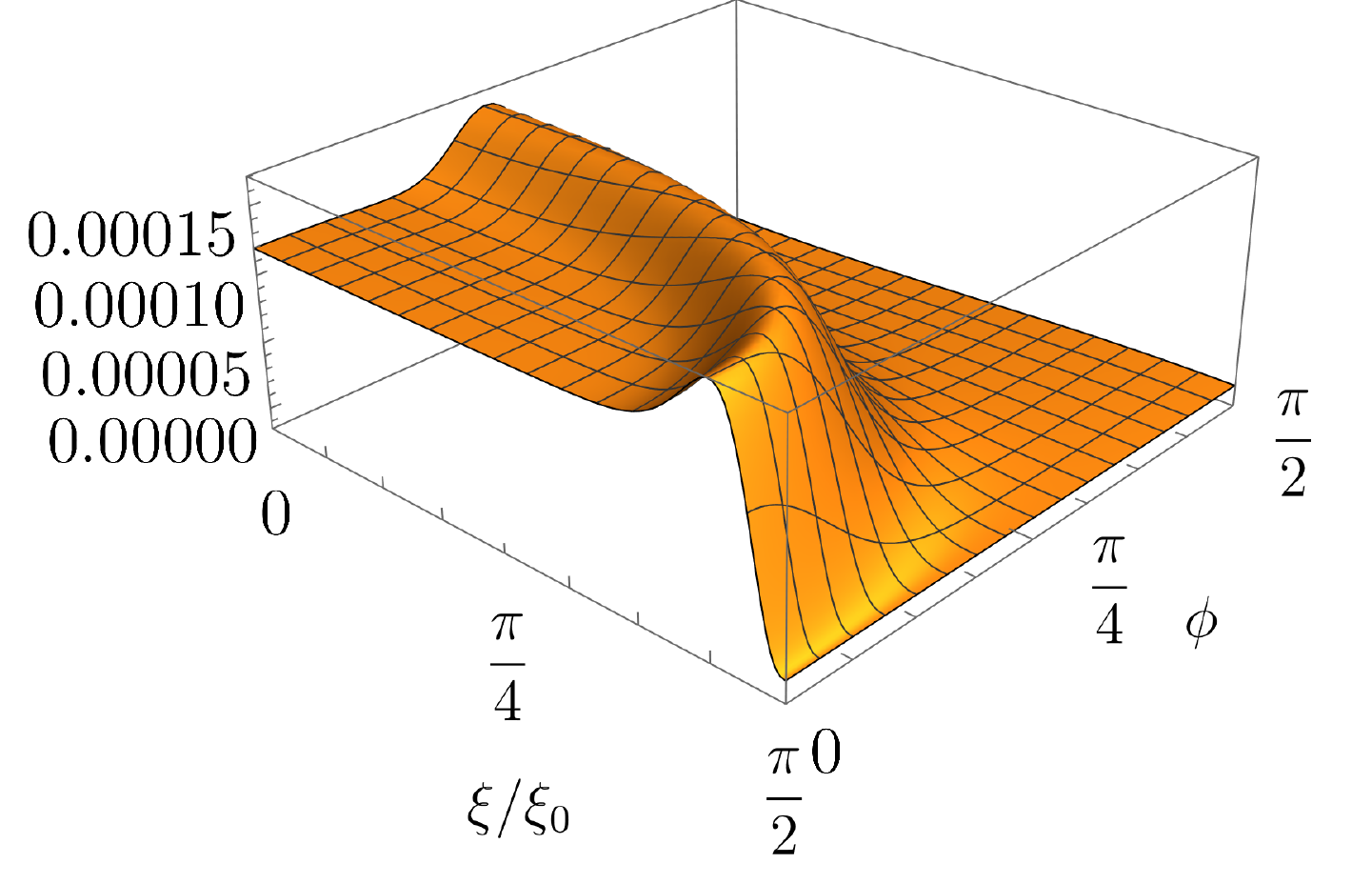}
\includegraphics[width=0.3\textwidth]{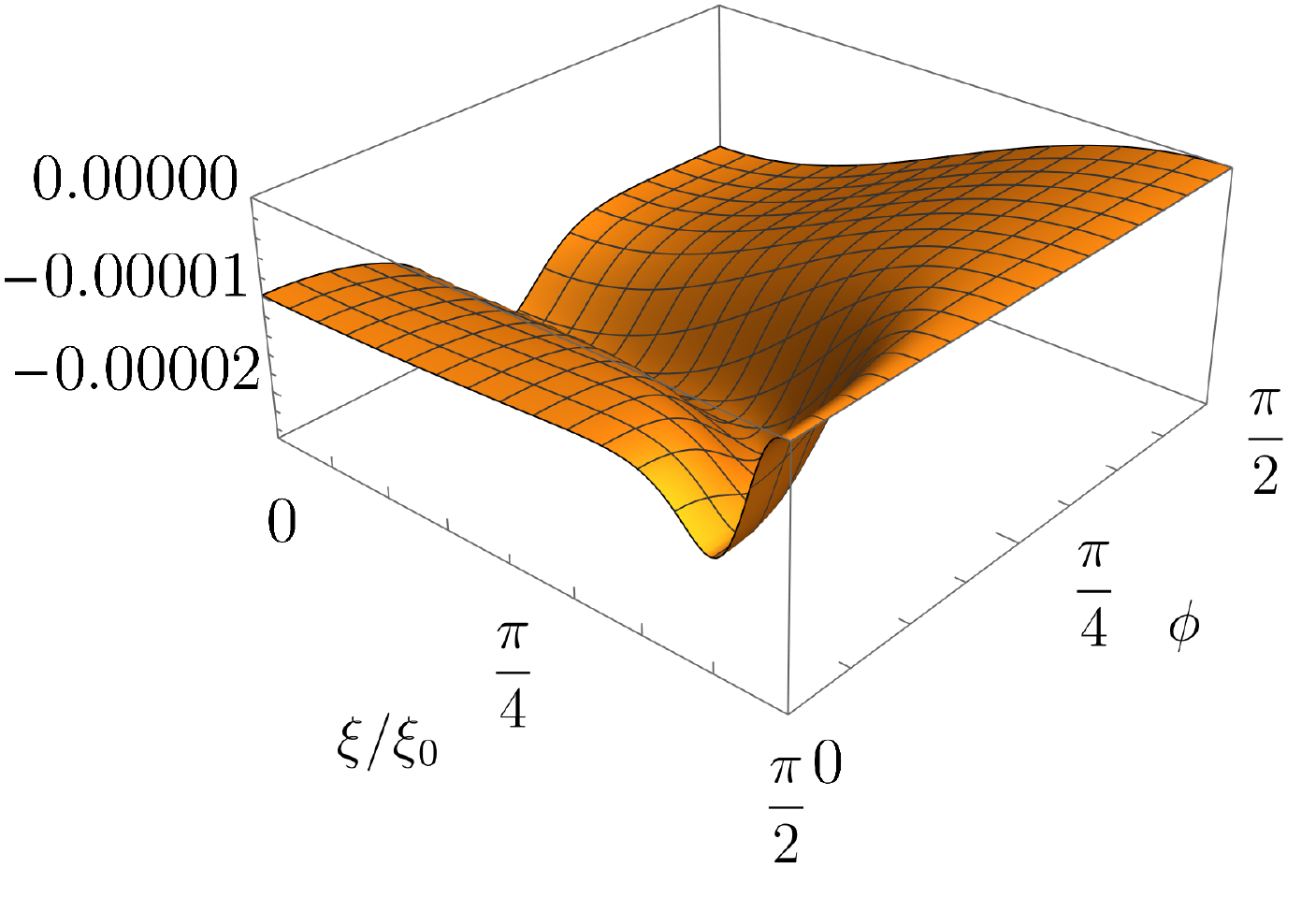}
\caption{A comparison between the approximation~$\tilde{\sigma}_\mathrm{approx}$ and~$\tilde{\sigma}_\mathrm{num}$.  The top left shows~$\tilde{\sigma}_\mathrm{num}$, while from left to right and top to bottom the subsequent plots show~$\tilde{\sigma}_\mathrm{num} - \tilde{\sigma}_\mathrm{approx}$ with successively higher-order terms in~$\nu$ included in~\eqref{eq:fexplicit}.  Here we take~$\xi_0 = 0.3$ and~$\nu = e^{-\pi/4\xi_0} \approx 0.07$, so that~$\tilde{\sigma}_\mathrm{approx}$ should agree with~$\tilde{\sigma}_\mathrm{num}$ up to corrections of order~$\Ocal(\nu^{-4} e^{-2\pi/\xi_0}) = \Ocal(\nu^4)$.}
\label{fig:Liouvillenumerics}
\end{figure}

All that remains is to solve~\eqref{eqs:fdef} for~$f$ to construct the solution~\eqref{eq:sigmaredef} and to quantify the size of~$\delta \tilde{\sigma}$ (as well as to verify the expectations that led to~\eqref{eq:deltasigmalinearized} in the first place).  At small~$\nu$,~$f$ can be constructed perturbatively in~$\nu$ using a matched asymptotic expansion, as shown in Appendix~\ref{subapp:Liovilleapprox}.  We ultimately find that
\begin{multline}
\label{eq:fexplicit}
f = -\ln \left[\left(\frac{1}{y} + \nu\right) \sinh y \right] + y + \nu\left(1 + y\left(1 -\coth y \right)\right) \\
	+ \frac{\nu^2}{6}\left(3y^2 \csch^2y + 2y^3 (1-\coth y) - 3\right) \\
	+\frac{\nu^3}{3} \left(1 + y^3 (1-\coth^3y) + y^4 \csch^2y\right) + \Ocal(\nu^4),
\end{multline}
where~$y \equiv \nu^{-1} \sech(\phi/\xi_0) \cos(\xi/\xi_0)$ and the neglected terms are~$\Ocal(\nu^4)$ for all~$y \in (0,1/\nu)$.  It then follows that if~$e^{-\pi/2\xi_0} \ll \nu$, the right-hand side of~\eqref{eq:deltasigmalinearized} -- and hence also~$\delta \tilde{\sigma}$ -- is~$\Ocal(\nu^{-4} e^{-2\pi/\xi_0})$.  We thus conclude that
\be
\label{eq:sigmaapprox}
\tilde{\sigma}_\mathrm{approx}(\xi,\phi) = \sum_{k = -\infty}^\infty f(\xi,\phi-k\pi)
\ee
with~$f$ given by~\eqref{eq:fexplicit}\footnote{In principle we can obtain~$f$ to any desired order in~$\nu$, but for constructing~\eqref{eq:sigmaapprox} there is no point computing~$f$ to higher order than the correction term~$\delta \tilde{\sigma}$, i.e.~to~$\Ocal(\nu^{-4} e^{-2\pi/\xi_0})$ or higher.} is a solution to the Liouville equation up to corrections of order~$\Ocal(\nu^{-4} e^{-2\pi/\xi_0})$.  To verify the validity of this approximation, in Figure~\ref{fig:Liouvillenumerics} we compare~$\tilde{\sigma}_\mathrm{approx}$ to the numerical solution~$\tilde{\sigma}_\mathrm{num}$ of~\eqref{subeq:shiftedliouville} for~$\nu = e^{-\pi/4\xi_0}$ and~$\xi_0 = 0.3$, in which case we should expect these solutions to agree to order~$\Ocal(\nu^3)$, as we find that they do.

\section{Pure JT for General \texorpdfstring{$m$}{m}}
\label{sec:pureJT}

Section~\ref{ssec:patchwork} described how to construct the two-dimensional quotient geometry of the replica trick by identifying two copies of the Poincar\'e disk with defects, yielding a geometry that depends on~$m$ and on the distance~$D$ between the defects.  This construction will allow us to compute the effective classical action of pure JT gravity as a function of~$D$ at arbitrary~$m$.  As noted above, this effective action should not exhibit any saddles for~$D$ except for~$m = 1$, and indeed as expected we do not find any classical wormhole solutions away from~$m = 1$.  However, many of the surprising nontrivial features of the continuation to small $m$ manifest already in pure JT, and will thus be illuminating for the subsequent more complicated model in Section~\ref{sec:JTCFT}. In particular, the equations of motion are critical for obtaining a single-valued analytic continuation to $m<1$.

\subsection{JT Gravity in the Boundary Formalism}
\label{subsec:JTboundary}

The tractability of pure JT stems largely from the fact that it can be reduced to the dynamics of a Schwarzian theory describing the boundary of nearly-AdS$_2$ space.  This same feature is what permits us to compute the action at general~$m$.  To do so, recall that the Euclidean JT action on a manifold~$M$ is
\be
I_\mathrm{JT} = -\frac{S_0}{4\pi}\left(\int_M R + 2\int_{\partial M} K\right) - \frac{1}{2} \int_M \varphi(R + 2) - \int_{\partial M} \varphi(K - 1),
\ee
where~$K$ is the extrinsic curvature of the boundary curve~$\partial M$ and we have set the AdS length to unity.  The boundary conditions at~$\partial M$ are that
\be
\label{eq:JTbndry}
\Length(\partial M) = \frac{\beta}{\delta}, \qquad \varphi|_{\partial M} = \frac{1}{\delta}, \qquad \delta \to 0.
\ee
These boundary conditions allow us to define a renormalized proper length coordinate~$u$ along~$\partial M$ through the condition that~$u/\delta$ be a proper distance on~$\partial M$.  Integrating out the dilaton~$\varphi$ as usual forces~$M$ to have constant negative curvature, reducing the partition function to an integral over the moduli space of~$M$ and an integral over the shape of~$\partial M$ -- hereafter referred to as the boundary ``wiggle''.

To evaluate the effective action for general~$m$, we now take $M$ to be the quotient geometry~$\widehat{M}_m$. The action will depend on the modulus~$D$ and on the wiggle, which is encoded in the embedding of~$\partial M$ in~$\widehat{M}_m$. To write this embedding explicitly, we proceed as follows.  Recall from Figure~\ref{fig:fundomainquotientgeometry} that~$\widehat{M}_m$ can be constructed from two identical pieces~$\widehat{M}_m^{(2)}$ joined along a geodesic~$\gamma$.  Since we expect an on-shell solution for the wiggle to be symmetric about~$\gamma$, we may restrict our attention to just a single copy of~$\widehat{M}_m^{(2)}$.  In terms of the complex coordinate~$z$ defined by~\eqref{eq:Poincaredisk}, we embed~$\partial M$ in~$\widehat{M}_m^{(2)}$ through the embedding~$z = R(u) e^{i \Theta(u)}$, where~$R(u)$ is related to~$\Theta(u)$ by the condition that~$u/\delta$ be a proper distance along~$\partial M$\footnote{
\label{fn:RTheta}
Explicitly,
\begin{equation*}
R = 1 - \Theta' \delta + \frac{\Theta'^2 \delta^2}{2} - \frac{\Theta''^2}{2\Theta'} \, \delta^3 + \Ocal(\delta^4).
\end{equation*}}.  Without loss of generality we also take~$\partial M$ to intersect~$\gamma$ at~$u = \pm \beta/4$.  Then because~$\widehat{M}_m^{(2)}$ is just a subregion of the Poincar\'e disk, the action on~$\widehat{M}_m$ is the usual Schwarzian action:
\bea
\widehat{I}_m &= \frac{m-2}{m} \, S_0 + \int_{-\beta/4}^{\beta/4} du \left[ \frac{\Theta''^2}{\Theta'^2} - \Theta'^2 - 2\left(\frac{\Theta''}{\Theta'}\right)'\right], \label{subeq:JTquotientaction} \\
	&= \frac{m-2}{m}\, S_0 - 2 \int_{-\beta/4}^{\beta/4} du \Sch\left(\tan (\Theta/2),u\right),
\eea
where~$\Sch(f,u) \equiv (f''/f')'-f''^2/(2f'^2)$ is the Schwarzian derivative.  However, our expectation that~$\partial M$ be symmetric about~$\gamma$ requires~$\gamma$ and~$\partial M$ to intersect orthogonally, which imposes nontrivial boundary conditions.  A straightforward computation finds these to be
\be
\label{eq:pureJTwigglebndryconditions}
\Theta\left(\pm \beta/4\right) = \pm\left(\frac{\pi}{m} - \alpha \right), \qquad \Theta''\left(\pm \beta/4\right) = \mp \cot\alpha \, \Theta'\left(\pm \beta/4\right)^2,
\ee
where we recall that~$\alpha$ is related to~$D$ via~\eqref{eq:alphaDrelationship}.  Importantly, recall that in order to properly accommodate the case~$m < 1$, we must \textit{not} identify~$\Theta(u)$ and~$\Theta(u) + 2\pi$.  

We now wish to put the wiggle~$\Theta(u)$ on shell to obtain an effective action for~$\alpha$ at general~$m$.  Taking a variation of~\eqref{subeq:JTquotientaction} yields the familiar equation of motion
\be
\label{eq:pureJTEOM}
\left(\frac{1}{\Theta'}\left(\frac{\Theta''}{\Theta'}\right)' + \Theta'\right)' = 0.
\ee
From the structure of the equation of motion and boundary conditions, we expect any solutions for~$\Theta(u)$ to be odd in~$u$.  The general such odd solution is
\be
\label{eq:pureJTgeneralTheta}
\tan\left(\frac{\Theta(u)}{2}\right) = a \tan\left(\frac{bu}{\beta}\right)
\ee
where~$a$ and~$b$ are arbitrary and will be fixed by the boundary conditions in \eqref{eq:pureJTwigglebndryconditions}.  There are two qualitatively different classes of solutions depending on whether~$a$ and~$b$ are both real or both imaginary, so we discuss them separately.

\subsubsection*{Exponential Solutions}

First consider the case that~$a$ and~$b$ are imaginary: take~$a = i a_i$ and~$b = -i b_i$ with~$a_i$ and~$b_i$ real, giving
\be
\label{eq:pureJTexponential}
\tan\left(\frac{\Theta(u)}{2}\right) = a_i \tanh\left(\frac{b_iu}{\beta}\right).
\ee
Since~$\Theta(u)$ must be monotonically increasing with~$u$,~$a_i$ and~$b_i$ must have the same sign; without loss of generality we take both to be positive.  Then imposing the boundary conditions~\eqref{eq:pureJTwigglebndryconditions}, we find that when a solution for~$a_i$ and~$b_i$ exists it is always unique and given by
\be
\label{eq:abexponential}
b_i = 2\arccosh\left(\frac{\sin(\pi/m)}{\sin\alpha}\right), \qquad a_i = \coth\left(\frac{b_i}{4}\right) \tan\left(\frac{\pi-m\alpha}{2m}\right).
\ee
Thus these solutions only exist when~$\sin(\pi/m)/\sin\alpha \geq 1$.  Moreover, note that the right-hand side of~\eqref{eq:pureJTexponential} is a regular function of~$u \in (-\infty,\infty)$, while the left-hand side is singular when~$\Theta = \pi$ (mod~$2\pi$).  Hence in order for solutions of this exponential type to be smooth in~$u$, we also require that~$\pi/m - \alpha < \pi$.  From these two constraints on~$m$ and~$\alpha$, we find that these classes of solutions always exist for~$m \geq 2$, never exist for~$m \leq 1$, and exist for certain values of~$\alpha$ but not others when~$1 < m < 2$.

\subsubsection*{Oscillatory Solutions}

Now take~$a = a_r$ and~$b = b_r$ with~$a_r$ and~$b_r$ real, giving the general solution
\be
\label{eq:pureJTocillatory}
\tan\left(\frac{\Theta(u)}{2}\right) = a_r \tan\left(\frac{b_ru}{\beta}\right).
\ee
Again, monotonicity of~$\Theta(u)$ allows us to restrict to positive~$a_r$ and~$b_r$.  Now, note that as~$\Theta$ runs from zero to~$\pi/m-\alpha$, the left-hand side of this expression goes through~$N$ poles, where
\be
N \equiv \left\lfloor \frac{1}{2m} + \frac{\pi - \alpha}{2\pi}\right\rfloor.
\ee
To obtain a monotonic and smooth~$\Theta(u)$, the right-hand side must go through the same number of poles as~$u$ goes from zero to~$\beta/4$, yielding a constraint on~$b_r$:
\be
\label{eq:pureJTbrconstraint}
\frac{b_r}{2\pi} \in (2N-1,2N+1].
\ee
With this constraint in mind, we impose the boundary conditions~\eqref{eq:pureJTwigglebndryconditions}; again we find that when a solution exists it is unique and given by
\begin{subequations}
\label{eqs:aboscillatory}
\begin{align}
b_r &= 2 (-1)^{\left\lfloor 1/m-\alpha/\pi\right\rfloor} \arccos\left(\frac{\sin(\pi/m)}{\sin\alpha}\right) + 4\pi N, \\
a_r &= \cot\left(\frac{b_r}{4}\right) \tan\left(\frac{\pi-m\alpha}{2m}\right),
\end{align}
\end{subequations}
where the principal branch of the inverse cosine is understood.  Consequently we conclude that there is always a unique solution of this form whenever~$|\sin(\pi/m)|/\sin\alpha < 1$.

\subsection{Effective Action}

Using the above solutions, it is straightforward to put the wiggle on shell and thereby obtain an effective action for the modulus~$\alpha$.  We find
\be
\label{eq:pureJTIeff}
\widehat{I}_m[\alpha] = \frac{m-2}{m}\, S_0 + \begin{dcases} \frac{2 b_i^2}{\beta}, & m > 1 \mbox{ and } \frac{\sin(\pi/m)}{\sin\alpha} \geq 1, \\ -\frac{2 b_r^2}{\beta}, & \frac{|\sin(\pi/m)|}{\sin\alpha} < 1, \end{dcases}
\ee
with~$b_i$ and~$b_r$ given by~\eqref{eq:abexponential} and~\eqref{eqs:aboscillatory}.  As a check, note that for~$m = 1$ we recover~$\widehat{I}_1 = S_0 - 2\pi^2/\beta$, which is the classical Schwarzian action of the disk~\cite{MalSta16b,HarJaf18}.  On the other hand, for~$m = 2$ and using~\eqref{eq:alphaDrelationship} we obtain~$\widehat{I}_2 = 2D^2/\beta$, which is half the classical action of the double-trumpet once we recognize~$D$ as half the length of the trumpet's throat~\cite{SSS}.

\begin{figure}[t]
\centering
\includegraphics[width=0.6\textwidth]{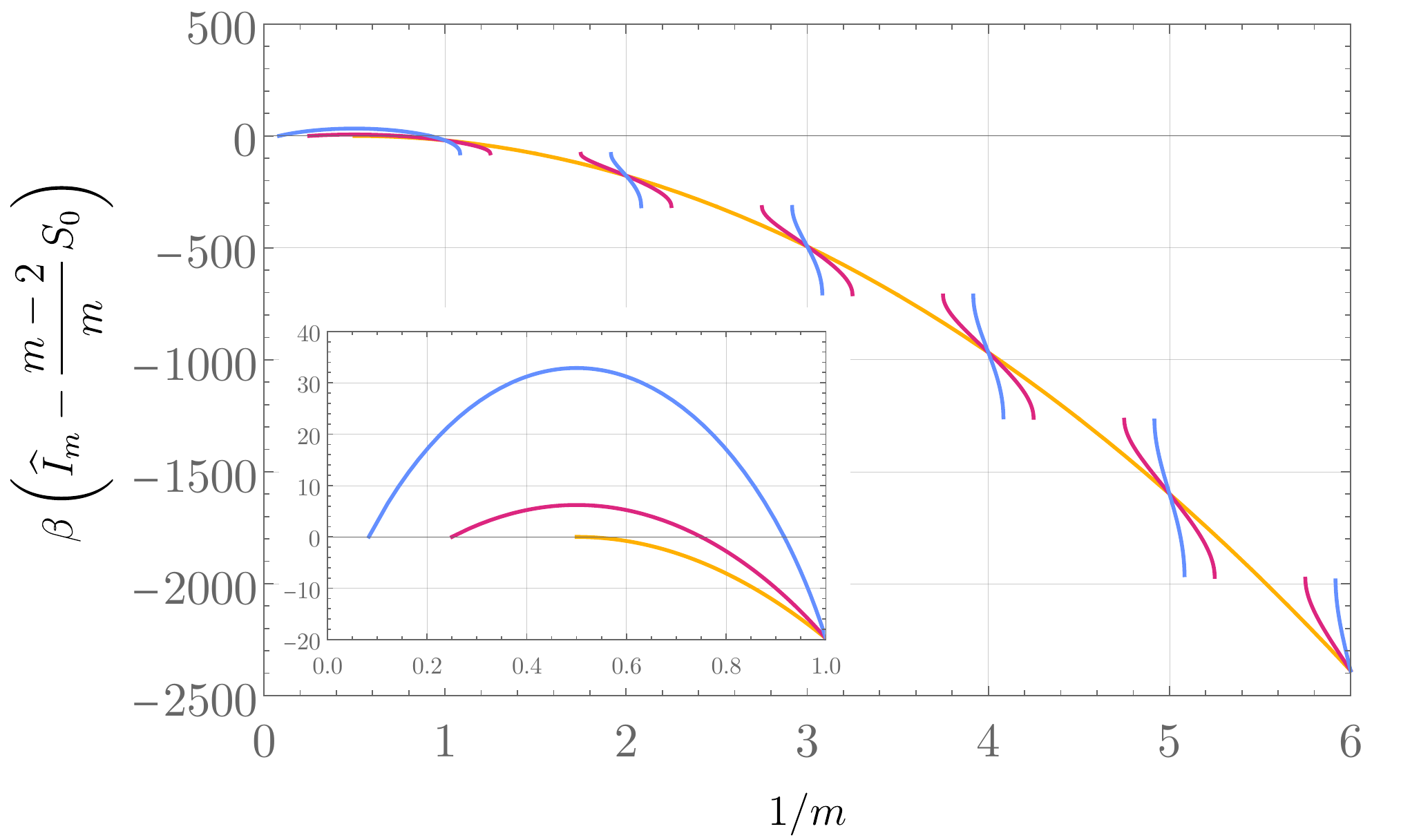}
\caption{The on-shell quotient space action~$\widehat{I}_m[\alpha]$ for pure JT, given in~\eqref{eq:pureJTIeff}, as a function of~$1/m$ for various values of~$\alpha$: the blue, red, and orange curves (steepest to least steep) correspond to~$\alpha = \pi/12$,~$\pi/4$, and~$\pi/2$; since~$\alpha$ is required to be strictly less than~$\pi/2$, the orange curve can be thought of as a limiting case.  Note that when~$m < 1$, classical solutions for the wiggle do not always exist, and hence there are ``gaps'' in which the on-shell action is not defined.  Solutions always exist whenever~$1/m$ is an integer, in which case they yield an action independent of~$\alpha$, while solutions never exist for any value of~$\alpha < \pi/2$ when~$1/m + 1/2$ is an integer.  The inset zooms in on the region~$m > 1$ and shows the action to the smallest value of~$1/m$ consistent with the constraint~$\alpha < \min(\pi/2,\pi/m)$.}
\label{fig:pureJTaction}
\end{figure}

Note that when~$m > 1$ we found saddles for the wiggle for all allowed values of~$\alpha$, but for~$m < 1$ we have found no solutions whenever~$|\sin(\pi/m)|/\sin\alpha > 1$.  Hence~$\widehat{I}_m[\alpha]$ is not defined for all~$m$ and~$\alpha$.  In Figure~\ref{fig:pureJTaction} we show~$\widehat{I}_m[\alpha]$ as a function of~$m$, giving some indication of why this is the case: at a given value of~$\alpha$, as we decrease~$m$ we eventually reach a branch point of the inverse cosine, after which a real solution ceases to exist.  Decreasing~$m$ further we reach new branch points at which solutions reappear.  The locations of these branch points depend on~$\alpha$, but it is clear from~\eqref{eq:pureJTIeff} that a solution exists for all~$\alpha$ whenever~$1/m$ is an integer, and no solution exists for \textit{any} values of~$\alpha$ whenever~$1/m + 1/2$ is an integer.  It is also clear that at a fixed value of~$m$, the action~$\widehat{I}_m[\alpha]$ is a monotonic function of~$\alpha$, and hence exhibits no saddles in~$\alpha$.

\begin{figure}[t]
\centering
\includegraphics[width=0.6\textwidth]{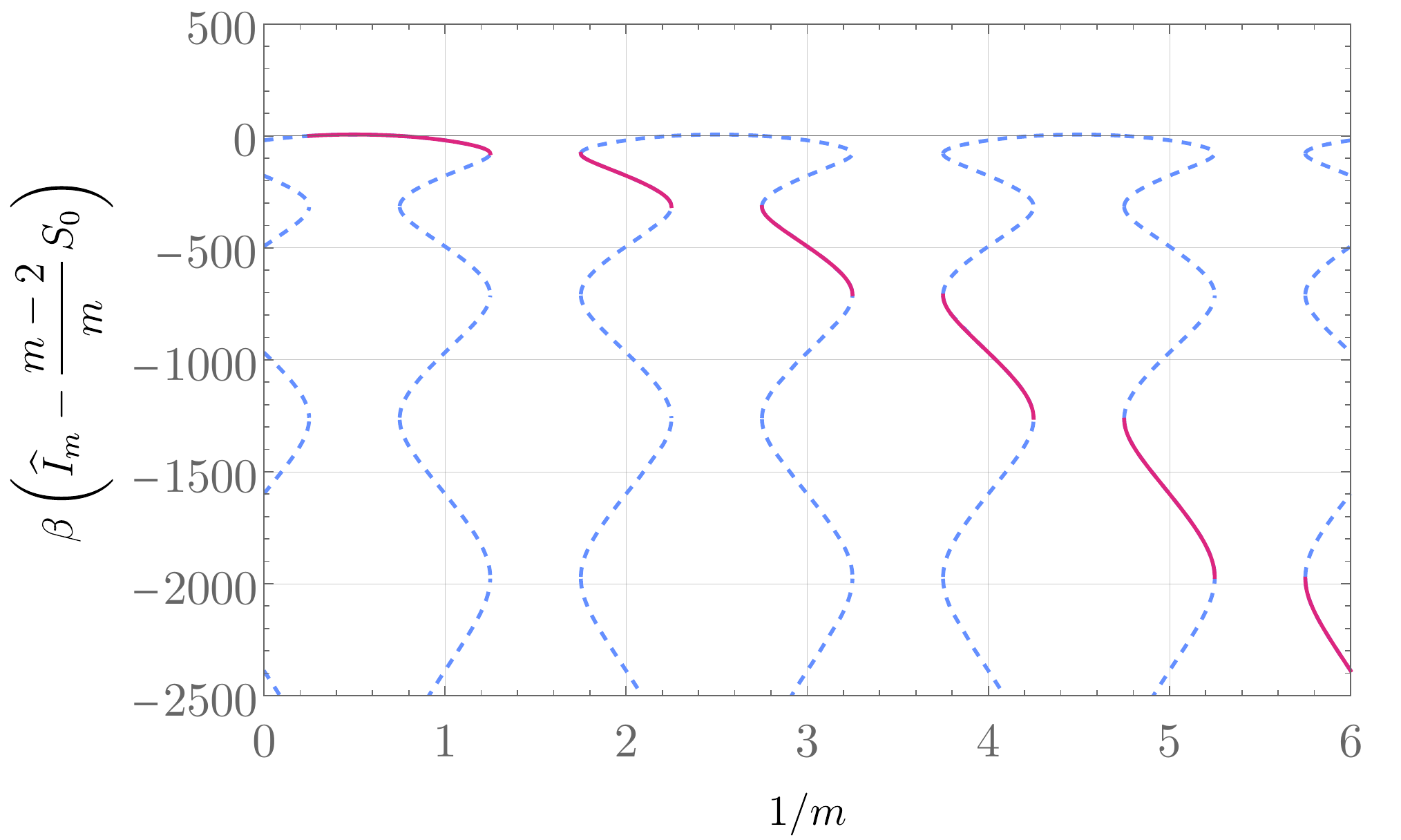}
\caption{The blue dashed lines show a cross-section through the Riemann surface obtained by analytically continuing the effective action~$\widehat{I}_m[\alpha]$ to all~$m$; here we take~$\alpha = \pi/4$.  We only plot this Riemann surface where it takes on real values: in the ``gaps'' it is complex.  This analytic continuation of~$\widehat{I}_m[\alpha]$ exhibits infinitely many branches of which only one, shown as a solid red line, computes the correct action.  The other branches correspond to violating the boundary conditions by having the wiggle~$\Theta(u)$ wrap around the circle too many or too few times.}
\label{fig:Riemannsurface}
\end{figure}

\begin{figure}[t]
\centering
\includegraphics[width=0.3\textwidth]{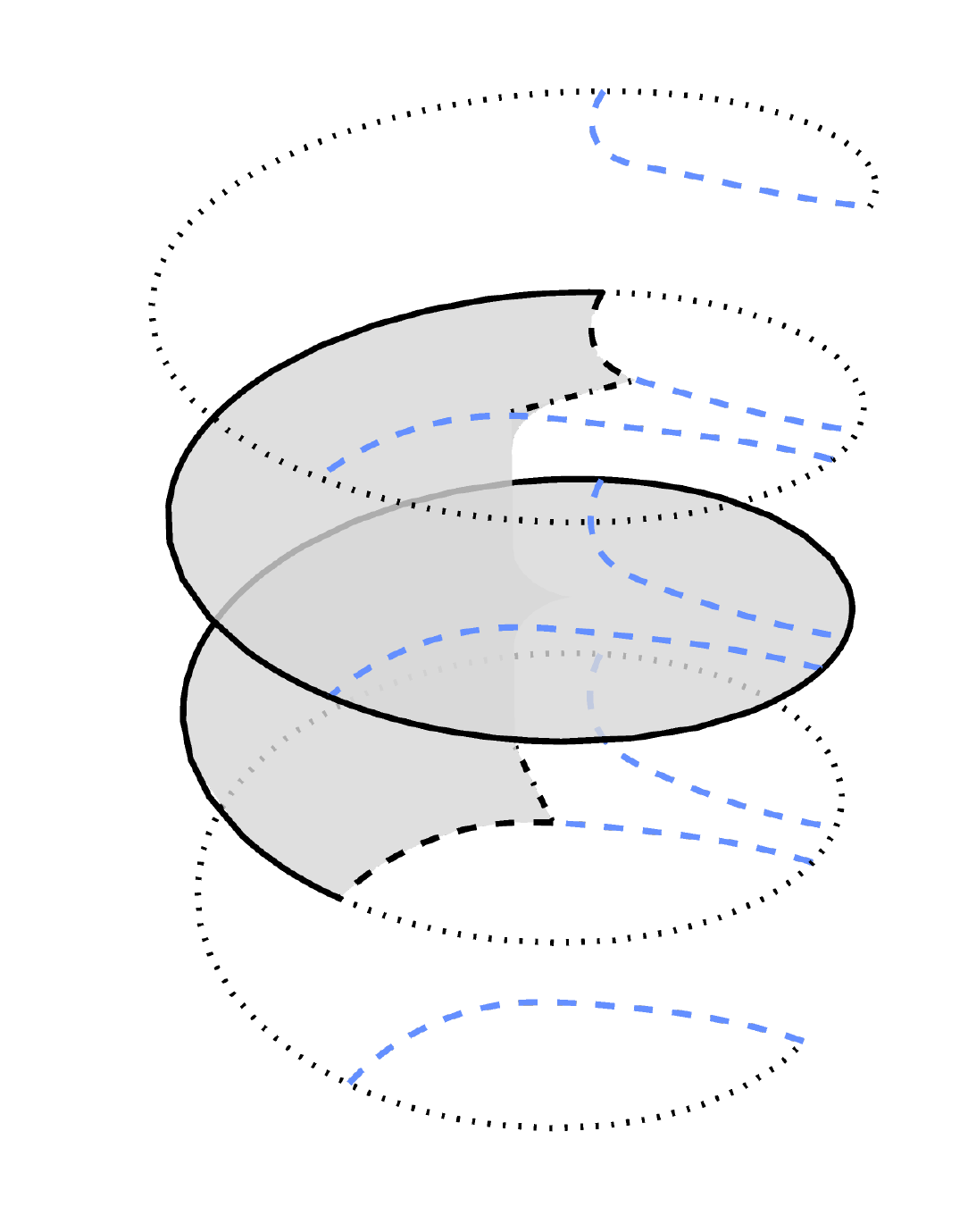}
\caption{On the covering of the disk, the geodesics~$\gamma$ that bound~$\widehat{M}^{(2)}_m$ have infinitely many images (blue dashed lines).  Anchoring the boundary wiggle~$\partial M$ to these images rather than to~$\gamma$ gives unphysical solutions that wrap the circle too many or too few times; these additional solutions correspond to the additional branches in Figure~\ref{fig:Riemannsurface}.}
\label{fig:helicoidsheets}
\end{figure}

Nevertheless, the analytic structure of~$\widehat{I}_m[\alpha]$ highlights an important lesson: in the spirit of the replica trick, one might have expected that knowledge of~$\widehat{I}_m[\alpha]$ for any arbitrarily small interval in~$m$ should have allowed us to analytically continue to all~$m$.  In a sense, this is indeed the case: for instance, knowing~$\widehat{I}_m[\alpha]$ for~$m > 1$ does allow us to analytically continue to~$m < 1$.  However, as shown in Figure~\ref{fig:Riemannsurface}, this analytic continuation does not give a single-valued function of~$m$.  Instead, it yields a Riemann surface with infinitely many branches.  In order to identify which, if any, of these branches correspond to the ``correct'' value of the action, we needed to make use of the equations of motion.  The ``other'' branches appearing in Figure~\ref{fig:Riemannsurface} are unphysical: they violate the constraint~\eqref{eq:pureJTbrconstraint}.  We may therefore interpret them as arising from having~$\Theta(u)$ wrap around the circle too many or too few times, corresponding to~$\partial M$ ending at different images of~$\gamma$ on the covering space of the disk, as shown in Figure~\ref{fig:helicoidsheets}.  The upshot is that even working in a saddle-point approximation, a mere analytic continuation from the on-shell action at~$m > 1$ is not sufficient to determine its correct behavior at~$m < 1$: one needs to analytically continue the \textit{equations of motion} themselves to~$m < 1$ in order to determine the correct analytic continuation\footnote{A noteworthy exception is~$\alpha = \pi/2$.  Although a regular wormhole geometry requires~$\alpha < \pi/2$ as a strict inequality, the case~$\alpha = \pi/2$ can be understood as a limit in which the conical defects merge together, analogous to the ``double-cone'' limit of the double-trumpet.  In this case, classical solutions for the wiggle exist for any~$m \leq 2$ and have the on-shell action~$\widehat{I}_m[\alpha = \pi/2] = (1-2/m)S_0 - 2\pi^2(1-2/m)^2/\beta$; clearly the analytic continuation of this function from any interval in~$m$ to all~$m$ is just itself, with no additional structure appearing.  As can be seen in Figure~\ref{fig:pureJTaction}, the correct branch of the Riemann surface for~$\widehat{I}_m[\alpha]$ that appears for~$\alpha < \pi/2$ is the one obtained via a smooth deformation away from~$\widehat{I}_m[\alpha = \pi/2]$.}.

The attentive reader will note a potential concern here: even though there are no saddles for the modulus~$\alpha$ (except when~$1/m$ is an integer),~$\widehat{I}_m[\alpha]$ becomes arbitrarily negative as~$m \to 0$ whenever solutions exist.  If we were to perform a saddle-point approximation for the path integral over the wiggle but perform a full integral over~$\alpha$, we might conclude that the partition function on the quotient manifold should be dominated by small-$m$ geometries and hence exhibit a divergence in the~$m \to 0$ limit of the replica trick.  This would indicate a complete failure of the replica trick (or a deep pathology of pure JT gravity).  However, it turns out that the saddles we have found for the wiggle are in fact unstable at sufficiently small~$m$, and hence they should not contribute to any saddle-point approximation, which resolves this tension.  We now discuss this stability analysis.

\subsection{Stability Analysis}
\label{subsec:JTstability}

To perform a stability analysis, we write~$\Theta(u) = \widetilde{\Theta}(u) + \vartheta(u)$ where~$\widetilde{\Theta}(u)$ is a solution to the equations of motion.  We restrict our analysis here to perturbations that preserve not only the~$\mathbb{Z}_2$ symmetry corresponding to reflecting across~$\gamma$, but also the additional~$\mathbb{Z}_2$ symmetry corresponding to reflections about the real~$z$-axis in the left diagram of Figure~\ref{fig:fundomainquotientgeometry}.  This latter symmetry amounts to taking~$u \to -u$, and requires that~$\Theta$ be odd in~$u$.  Then expanding the boundary conditions~\eqref{eq:pureJTwigglebndryconditions} to linear order in~$\vartheta$, these symmetries require that
\be
\label{eq:perturbbndryconds}
\vartheta(0) = 0 = \vartheta''(0), \qquad \vartheta(\beta/4) = 0, \qquad \vartheta''(\beta/4) = - 2\cot\alpha \, \widetilde{\Theta}'(\beta/4) \vartheta'(\beta/4).
\ee
Expanding the action to quadratic order in~$\vartheta$, we obtain
\be
\widehat{I}_m[\Theta] = \widehat{I}_m[\widetilde{\Theta}] + 2 \int_0^{\beta/4} du \, \vartheta L \vartheta + \Ocal(\vartheta^3),
\ee
where the linear operator~$L$ is defined by
\be
\label{eq:JTfluctuationL}
L \vartheta \equiv \left(\frac{\widetilde{\Theta}' \vartheta'' - 2 \widetilde{\Theta}'' \vartheta'}{\widetilde{\Theta}'^3}\right)'' + \left(\frac{2\widetilde{\Theta}' \widetilde{\Theta}'' \vartheta'' - 3 \widetilde{\Theta}''^2 \vartheta'}{\widetilde{\Theta}'^4} + \vartheta'\right)'.
\ee
Note that~$L$ depends implicitly on~$m$ and~$\alpha$ through its dependence on~$\widetilde{\Theta}$.

\begin{figure}[t]
\centering
\includegraphics[width=0.6\textwidth]{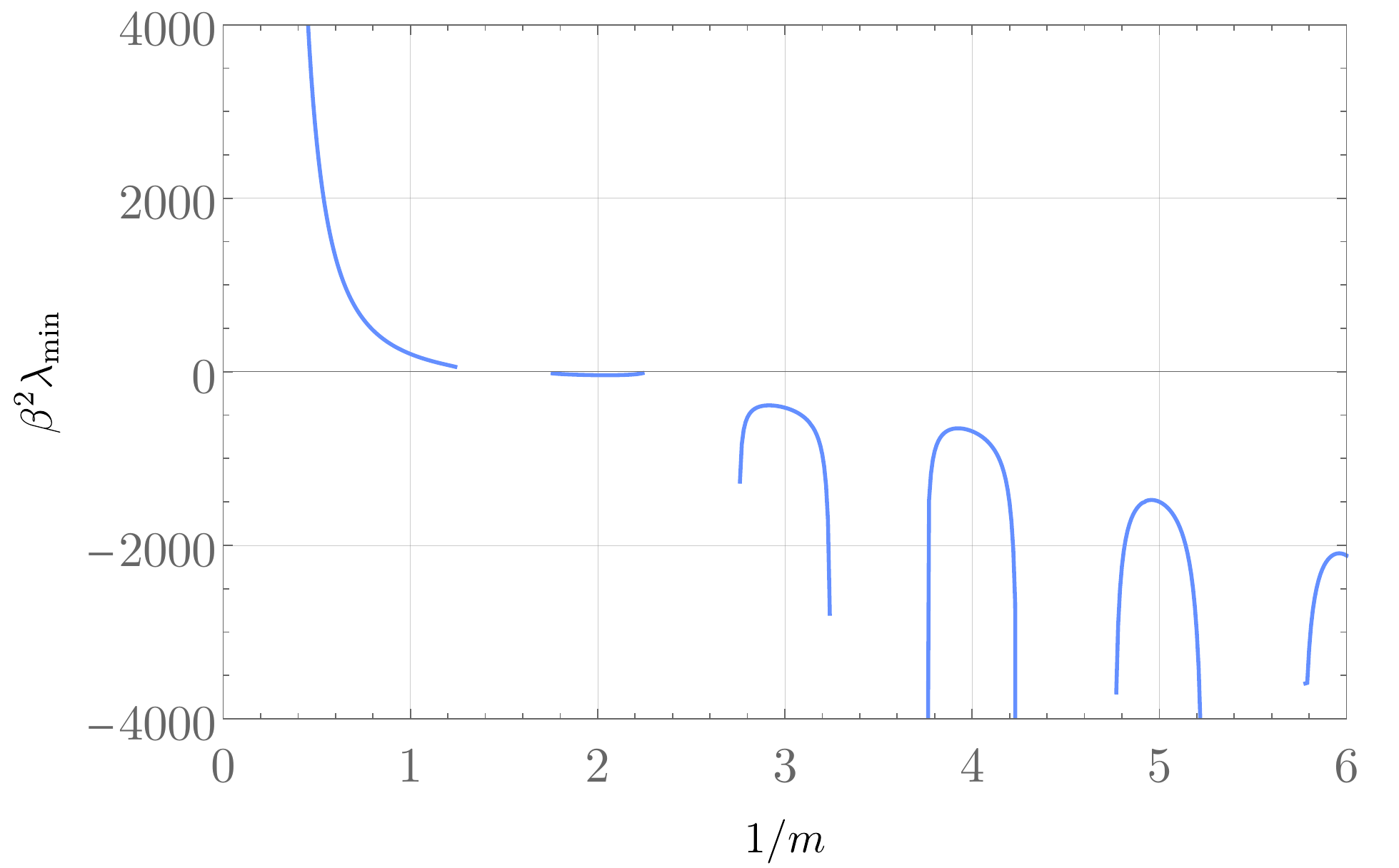}
\caption{The lowest eigenvalue~$\lambda_\mathrm{min}$ of~$L$ as a function of~$m$, with~$\alpha = \pi/4$.  Note that~$\lambda_\mathrm{min}$ is positive on the branch of solutions that continuously connects to~$m > 1$, but then becomes negative on the other branches at smaller~$m$, indicating that these branches are all unstable.}
\label{fig:lambdamin}
\end{figure}

It is straightforward to check that~$L$ is self-adjoint (with respect to the usual~$L_2$ norm on~$[0,\beta/4]$) on the space of functions obeying the boundary conditions~\eqref{eq:perturbbndryconds}.  Consequently a solution~$\widetilde{\Theta}$ to the equations of motion is a local minimum of the action if and only if the spectrum of~$L$ is nonnegative.  Because the background solutions~$\widetilde{\Theta}$ are known analytically (when they exist), it is straightforward to compute the spectrum of~$L$ numerically using standard pseudospectral collocation methods~\cite{Trefethen}.  In Figure~\ref{fig:lambdamin} we show the smallest eigenvalue~$\lambda_\mathrm{min}$ of~$L$ as a function of~$m$ for various values of~$\alpha$.  As~$m$ is decreased at fixed~$\alpha$,~$\lambda_\mathrm{min}$ remains positive until the first branch point of the action is reached and (real) classical solutions stop existing.  When solutions reappear at smaller values of~$m$,~$\lambda_\mathrm{min}$ is negative.  This indicates that the branch of solutions that connects continuously to~$m > 1$ is stable, but the solutions that appear at smaller~$m$ past the branch points are not.  So as advertised, we see that the saddles at small~$m$ that yield an arbitrarily large and negative action should not be picked up in a saddle-point approximation, and there is no worry of the replica trick giving a divergent~$m \to 0$ contribution.

It is worth pausing to note the role of the~$\mathbb{Z}_2 \times \mathbb{Z}_2$ symmetry that we imposed at the level of the stability analysis.  Indeed, a careful reader might notice that~$\lambda_\mathrm{min} > 0$ at~$m = 1$, despite the fact that for~$m = 1$ we expect there should be three zero modes associated with the~$SL(2,\mathbb{R})$ symmetry of the disk.  The point is that these zero modes break the~$\mathbb{Z}_2 \times \mathbb{Z}_2$ symmetry that we have imposed and hence do not appear in our analysis.  We can verify this claim by breaking, for instance, the~$\mathbb{Z}_2$ corresponding to reflection symmetry about the real~$z$-axis in the left diagram of Figure~\ref{fig:fundomainquotientgeometry}.  In repeating the stability analysis with this symmetry removed, we find that~$L$ does indeed exhibit a zero mode at~$m = 1$, and that a negative mode appears for all~$m < 1$.  Breaking the other~$\mathbb{Z}_2$ corresponding to reflection symmetry across~$\gamma$ should recover the other two zero modes at~$m = 1$.  Thus we see that the additional~$\mathbb{Z}_2 \times \mathbb{Z}_2$ symmetry we have enforced has a stabilizing effect on the wiggle, stabilizing some of the~$m < 1$ solutions that would have otherwise been unstable.  This stabilizing effect will be crucial in our next model, where we will find saddles for both the wiggle \textit{and the modulus} at~$m < 1$ that are stable only if we restrict to~$\mathbb{Z}_2 \times \mathbb{Z}_2$-symmetric configurations.

\section{JT with a Massless Scalar}
\label{sec:JTCFT}

We now turn to our main model of interest: JT gravity coupled to a massless scalar.  We will take the scalar $\psi$ to be real and minimally-coupled, so that the Euclidean action of the theory is~$I = I_\mathrm{JT} + I_\mathrm{mat}$ where
\be
I_\mathrm{mat} = \frac{1}{2} \int_M (\grad \psi)^2. \label{eq:matteract}
\ee
We can easily integrate out the scalar field by placing it on shell: since its equation of motion is~$\grad^2 \psi = 0$, the on-shell matter action just contributes a boundary term:
\be
\label{eq:psionshellaction}
I_\mathrm{mat} = \frac{1}{2} \int_{\partial M} \psi \, n^a \grad_a \psi,
\ee
where~$n^a$ is the unit outward-pointing normal to~$\partial M$.

In order to attempt to stabilize the replica wormholes at the classical level, we will need to turn on sources for the scalar:
\be
\psi|_{\partial M} = \psi_\partial(u),
\ee
where the profile~$\psi_\partial(u)$ can be specified arbitrarily.  Note that we will require this profile to have nontrivial dependence on~$u$: if it were a constant~$\psi_\partial = c$, then the equations of motion for~$\psi$ would be solved by the constant solution~$\psi = c$ everywhere.  Such a solution gives a vanishing on-shell action~\eqref{eq:psionshellaction} and reduces the total action to that of pure JT.  But if~$\psi_\partial(u)$ depends nontrivially on~$u$, then the fact that the proper length~$u$ is defined by the shape of the boundary wiggle means that the on-shell action~\eqref{eq:psionshellaction} induces a nontrivial coupling between the wiggle and the scalar.  Our goal is now to understand this coupling on the quotient geometry~$\widehat{I}_m$ and to show that it can provide the classical solutions for the wiggle and modulus.

\subsection{JT + Scalar in the Boundary Formalism}
\label{subsec:JTCFTboundary}

Why not proceed as in pure JT by constructing the quotient manifold~$\widehat{M}_m$ by taking two identical subregions of the Poincar\'e disk and stitching them together along a geodesic~$\gamma$?  The reason is that it is nontrivial to impose appropriate boundary conditions on~$\psi$ at~$\gamma$, which is required for the construction of a general solution to the equations of motion and the subsequent evaluation of the on-shell matter action~\eqref{eq:psionshellaction}.  Instead, it will be convenient to work with a single coordinate chart on~$\widehat{M}_m$ so that we may solve for~$\psi$ by imposing only a Dirichlet condition on~$\partial M$.  We will use the elliptical coordinate chart introduced in Section~\ref{subsec:liouville}, in terms of which the metric on~$\widehat{M}_m$ can be written as
\be
ds^2 = e^{2\sigma(\xi,\phi)} \left(d\xi^2 + d\phi^2\right),
\ee
where recall~$\xi \in [0,\pi \xi_0/2)$,~$\phi \in [0,2\pi)$, the conical defects lie at~$(\xi,\phi) = (0,0)$ and~$(0,\pi)$, and~$\partial M$ corresponds to~$\xi = \pi \xi_0/2$.  We must therefore express the JT and matter parts of the action in this coordinate chart.

\subsubsection*{JT Action}

To construct the JT part of the action in the elliptical coordinate chart, we note that the near-boundary expansion of the metric takes the form
\be
\label{eq:nearboundaryexpansion}
ds^2 = \left(\frac{1}{(\pi\xi_0/2-\xi)^2} + g_2(\phi) + \Ocal\left(\frac{\pi\xi_0}{2} - \xi\right)\right)\left(d\xi^2 + d\phi^2\right),
\ee
where~$g_2$ can be extracted from the solutions to the Liouville equation constructed in Section~\ref{subsec:liouville}:
\be
g_2(\phi) = \partial_\xi^2 \tilde{\sigma}|_{\xi = \pi \xi_0/2} + \frac{1}{3\xi_0^2},
\ee
where~$\tilde{\sigma}$ is defined by~\eqref{subeq:conformalshift}.
In particular, in the limit of small~$\xi_0$ and~$\nu$ with~$e^{-\pi/2\xi_0} \ll \nu$ (where~$\nu \equiv m/(2-m)$),~\eqref{eq:sigmaapprox} gives\footnote{More compactly, we note that
\begin{equation*}
    \sum_{k = -\infty}^\infty \sech^2\left(\frac{\phi - k \pi}{\xi_0}\right) = \xi_0^2 \, \partial_\phi^2 \, \log\left( \vartheta_2\!\left(i \phi/\xi_0,\,e^{\pi /\xi_0}\right)\right),
\end{equation*}
where~$\vartheta_2(z,q) = \sum_{n=-\infty}^\infty q^{(n+1/2)^2} e^{(2 n + 1) i z}$ is the Jacobi theta function of the second kind.}
\be
g_2(\phi) = -\frac{1+2\nu}{3\nu^2 \xi_0^2} \sum_{k = -\infty}^\infty \sech^2\left(\frac{\phi - k \pi}{\xi_0}\right) + \frac{1}{3\xi_0^2} + \Ocal(\nu^2,\nu^{-6} e^{-2\pi/\xi_0}).
\ee
Interestingly, comparison with the numerical solutions to the Liouville equation shows that this expression for~$g_2$ actually captures the~$\phi$-dependence of~$g_2$ \textit{exactly}: that is, we find that for all~$\nu$ and~$\xi_0$,
\be
\label{eq:g2smallxi0}
g_2(\phi) = -\frac{1+2\nu}{3\nu^2 \xi_0^2} \sum_{k = -\infty}^\infty \sech^2\left(\frac{\phi - k \pi}{\xi_0}\right) + C(\nu,\xi_0),
\ee
where~$C(\nu,\xi_0)$ is independent of~$\phi$ up to numerical resolution.  We do not know of an analytic argument for why this is the case, but in practice it means that we only need to numerically solve the Liouville equation to extract the constant~$C(\nu,\xi_0)$, rather than the entire functional form of~$g_2(\phi)$.  We discuss the computation of~$C(\nu,\xi_0)$ and illustrate its behavior in Appendix~\ref{subapp:Liouville}.

We may now construct the boundary JT action: we embed~$\partial M$ in~$\widehat{M}_m$ through the embeddings~$(\xi,\phi) = (X(u),\Phi(u))$, where as before the requirement that~$u/\delta$ be a proper length along~$\partial M$ relates the embeddings through
\be
X(u) = \frac{\pi \xi_0}{2} - \Phi'(u) \, \delta - \frac{g_2(\Phi(u)) \Phi'(u)^4 + \Phi''(u)^2}{2\Phi'(u)} \, \delta^3 + \Ocal(\delta^4).
\ee
In terms of these embeddings, the dynamical part of the JT action becomes
\bea
-\int_{\partial M} \varphi(K-1) &= \frac{1}{2} \int_0^\beta du \left(\frac{\Phi''(u)^2}{\Phi'(u)^2} + 3 g_2(\Phi(u)) \Phi'(u)^2 \right), \\
	&= \frac{1}{2} \int_0^{2\pi} d\phi \left(\frac{u''(\phi)^2}{u'(\phi)^3} + \frac{3 g_2(\phi)}{u'(\phi)} \right),
\eea
where we have written the second line in terms of the inverse function~$u(\phi)$ defined by~$\Phi(u(\phi)) = \phi$, the existence of which is guaranteed by the requirement that~$\Phi(u)$ be monotonic in~$u$.  Note that~$\Phi$ now wraps once around the ellipse:~$\Phi(u + \beta) = \Phi(u) + 2\pi$, or~$u(\phi + 2\pi) = u(\phi) + \beta$.

\subsubsection*{Matter Action}

To compute the contribution of the scalar to the action for the wiggle, we make use of the fact that the scalar field is a CFT, allowing us to compute the classical action~\eqref{eq:psionshellaction} in any choice of conformal frame.  A natural such choice is given by taking the conformal factor~$\sigma = 0$, thereby putting the scalar on a strip:
\be
ds^2 = d\xi^2 + d\phi^2, \qquad \xi \in [0,\pi\xi_0/2), \qquad \phi \in [0,2\pi).
\ee
Because this domain is conformal to half of the double-trumpet, in principle we could proceed as in e.g.~\cite{GarGot21,GarGot22} and express~$\psi$ as an integral against a bulk-to-boundary propagator on the double-trumpet with appropriate replica symmetry imposed to relate the boundary conditions on the two ends.  In practice, for implementing the numerical methods described in Appendix~\ref{app:numerics}, it is more convenient to directly solve the Laplace equation for~$\psi$ via Fourier series by constructing a solution subject to appropriate boundary conditions:
\begin{itemize}
	\item Periodicity in~$\phi$:~$\psi(\xi,\phi + 2\pi) = \psi(\xi,\phi)$.
	\item Smoothness at~$\xi = 0$ when conformally mapped back to the ellipse\footnote{The reader may be concerned that  the demand that~$\psi$ be regular on the ellipse excludes  solutions where~$\psi$ is regular on the full wormhole geometry~$M_m$ but singular at the conical defects in~$\widehat{M}_m$.  But this is not the case: because classical solutions for~$\psi$ are harmonic, smooth solutions on~$M_m$ must also be smooth on~$\widehat{M}_m$.  One way to see this is to consider an arbitrary~$\mathbb{Z}_m$-invariant holomorphic function~$f$ on the Poincar\'e disk~\eqref{eq:Poincaredisk}, which must have an expansion in powers of~$z^m$.  Taking a quotient~$z \to z^{1/m}$ to go to the Poincar\'e disk with a defect, one finds that~$f$ has a standard expansion in integer powers of~$z$, which is regular.  This analysis applies locally near any defect, and since the real and imaginary parts of~$f$ are harmonic, we conclude that any~$\mathbb{Z}_m$-invariant harmonic function on~$M_m$ must be regular on~$\widehat{M}_m$, including at the defects.}:~$$\partial_\xi^p \psi(0,\phi) = (-1)^p \partial_\xi^p\psi(0,-\phi) \qquad \forall p\in\mathbb{Z}_{\ge0}$$
	\item Scalar sources:~$\psi(\pi\xi_0/2,\phi) = \psi_\partial(u(\phi))$.
\end{itemize}
The general solution to the Laplace equation on this domain satisfying the first two boundary conditions is given by
\be
\label{eq:psigeneral}
\psi(\xi,\phi) = \sum_{k = -\infty}^\infty \left(a_k \cosh(k \xi) + i b_k \sinh(k \xi)\right) e^{ik\phi},
\ee
where the coefficients~$a_k$ and~$b_k$ are real and obey~$a_k = a_{-k}$,~$b_k = b_{-k}$.  These coefficients can be determined by imposing the Dirichlet condition at~$\xi = \pi\xi_0/2$:
\be
a_k \cosh(k \pi \xi_0/2) + i b_k \sinh(k \pi \xi_0/2) = \frac{1}{2\pi} \int_0^{2\pi} d\phi \, \psi_\partial(u(\phi)) e^{-ik\phi}.
\ee
Using these relations to express~\eqref{eq:psigeneral} in terms of~$\psi_\partial(u(\phi))$, we ultimately find that the on-shell action~\eqref{eq:psionshellaction} takes the form of a smearing of~$\psi_\partial(u(\phi))$:
\be
\widehat{I}_\mathrm{mat} = \frac{1}{2} \int_0^{2\pi} d\phi \, d\tilde{\phi} \,\, \psi_\partial(u(\phi)) \, \psi_\partial(u(\tilde{\phi}))\, G(\phi,\tilde{\phi}),
\ee
where
\begin{multline}
\label{eq:Gkerneldef}
G(\phi,\tilde{\phi}) \equiv \frac{1}{\pi} \sum_{k = 1}^\infty k \left[\tanh\left(\frac{k\pi\xi_0}{2}\right) \cos(k\phi)\cos(k\tilde{\phi}) \right. \\ \left. + \coth\left(\frac{k\pi\xi_0}{2}\right)\sin(k\phi)\sin(k\tilde{\phi}) \right].
\end{multline}
Note that~$G(\phi,\tilde{\phi})$ is essentially the boundary-to-boundary propagator for~$\psi$.  As written, the sum defining~$G(\phi,\tilde{\phi})$ is not convergent for all~$\phi$ and~$\tilde{\phi}$ due to contact terms; for the purposes of evaluating the action,~$G(\phi,\tilde{\phi})$ should be understood distributionally.  For computing the matter two-point function,~$G(\phi,\tilde{\phi})$ should instead be renormalized appropriately.

\subsubsection*{Boundary Action and Equations of Motion}

Putting these results together, we find that integrating out the scalar field~$\psi$ leaves us with the boundary action
\be
\label{eq:JTCFTaction}
\widehat{I}_m = -\frac{S_0}{\nu} + \frac{1}{2} \int_0^{2\pi} d\phi \left( \frac{u''(\phi)^2}{u'(\phi)^3} + \frac{3 g_2(\phi)}{u'(\phi)} + \int_0^{2\pi} d\tilde{\phi} \, \psi_\partial(u(\phi)) \psi_\partial(u(\tilde{\phi})) G(\phi,\tilde{\phi}) \right),
\ee
which we have expressed entirely in terms of the inverse wiggle~$u(\phi)$. Due to the nonlocality of the matter part of the action, the resulting equation of motion for~$u(\phi)$ is an integro-differential equation:
\be
\label{eq:JTCFTeom}
\left(\left(\frac{u''(\phi)}{u'(\phi)^3}\right)' + \frac{3 (u''(\phi)^2 + u'(\phi)^2 g_2(\phi))}{2u'(\phi)^4} \right)' + \dot{\psi}_\partial(u(\phi))\int_0^{2\pi} d\tilde{\phi} \, \psi_\partial(u(\tilde{\phi})) G(\phi,\tilde{\phi}) = 0,
\ee
where~$\dot{\psi}_\partial \equiv d\psi_\partial/du$.

We will also need to perform a stability analysis of the solutions to~\eqref{eq:JTCFTeom}, which is done as in pure JT: we write the wiggle as~$u(\phi) = \tilde{u}(\phi) + \kappa(\phi)$, where~$\tilde{u}(\phi)$ solves the equation of motion.  Expanding the action to quadratic order in~$\kappa$, we obtain
\be
\widehat{I}_m[u] = \widehat{I}_m[\tilde{u}] + \frac{1}{2} \int_0^{2\pi} d\phi \, \kappa L \kappa + \Ocal(\kappa^3),
\ee
where now the fluctuation operator~$L$ is
\begin{multline}
\label{eq:JTCFTfluctuationoperator}
L \kappa(\phi) \equiv \left(\frac{\kappa''(\phi)}{\tilde{u}'(\phi)^3} - \frac{3 \tilde{u}''(\phi) \kappa'(\phi)}{\tilde{u}'(\phi)^4}\right)'' + 3\left(\frac{\tilde{u}''(\phi) \kappa''(\phi)}{\tilde{u}'(\phi)^4} - \left(\frac{g_2(\phi)}{\tilde{u}'(\phi)^3} + \frac{2 \tilde{u}''(\phi)^2}{\tilde{u}'(\phi)^5}\right) \kappa'(\phi)\right)' \\
	+ \int_0^{2\pi} d\tilde{\phi} \left[\ddot{\psi}_\partial(\tilde{u}(\phi))  \psi_\partial(\tilde{u}(\tilde{\phi})) \kappa(\phi) + \dot{\psi}_\partial(\tilde{u}(\phi)) \dot{\psi}_\partial(\tilde{u}(\tilde{\phi})) \kappa(\tilde{\phi)} \right] G(\phi,\tilde{\phi}),
\end{multline}
where~$\ddot{\psi}_\partial \equiv d^2 \psi_\partial/du^2$.  It is straightforward to verify that~$L$ is self-adjoint (with respect to the usual~$L_2$ norm on~$[0,2\pi])$ on the space of functions periodic in~$\phi$ with period~$2\pi$, and hence a saddle~$\tilde{u}$ is a local minimum of the action if and only if the spectrum of~$L$ is nonnegative.

\subsection{Stabilizing the Double-Trumpet}
\label{subsec:analyticJTCFT}

For general~$\psi_\partial$,~\eqref{eq:JTCFTeom} cannot be solved analytically; we will require a numerical solution.  However, for~$m = 1$ and~$m = 2$, it is possible to obtain analytic solutions by looking for \textit{bulk} solutions for the dilaton~$\Phi$ and reconstructing the corresponding behavior of the boundary wiggle~$u(\phi)$.  We discuss the construction of these bulk solutions in Appendix~\ref{app:bulksolutions}; here we simply exhibit them in order to study the effect of turning on the CFT sources.  In particular, we will show that taking the amplitude of the boundary source~$\psi_\partial$ large enough allows us to stabilize the double-trumpet, and moreover that the double-trumpet dominates over the disk at sufficiently small temperatures.

The solutions constructed in Appendix~\ref{app:bulksolutions} correspond to the family of boundary profiles
\be
\label{eq:psipartialanalytic}
\psi_\partial(u) = J \cos\left(\frac{2\pi n u}{\beta}\right) \sqrt{\frac{1+A}{1+A\cos(4\pi n u/\beta)}},
\ee
with boundary wiggle given by
\be
\label{eq:m12generalu}
\tan\left(\frac{2\pi n u(\phi)}{\beta}\right) = \sqrt{\frac{1+A}{1-A}} \tan(n\phi),
\ee
where~$J$ and~$A \in (-1,1)$ are arbitrary constants and~$n$ is an arbitrary positive integer.  This class of profiles and form of the wiggle is tractable because~$\psi_\partial(u(\phi)) = J \cos(n\phi)$, so the smearing of~$\psi_\partial$ against~$G(\phi,\tilde{\phi})$ is straightforward to compute.  Saddles (for both the wiggle and the modulus~$\xi_0$) are then obtained through an appropriate choice of~$A$.  For instance, for~$m = 2$, we have~$g_2(\phi) = 1/(3\xi_0^2)$, and hence the action of the wiggle profile~\eqref{eq:m12generalu} is
\be
\label{eq:I2analytic}
\widehat{I}_2 = \frac{n^2}{\beta} \left[\frac{2\pi^2(4+1/(n\xi_0)^2)}{\sqrt{1-A^2}} - 8\pi^2 + \frac{\pi\beta J^2}{2n} \tanh\left(\frac{n\pi\xi_0}{2}\right)\right].
\ee
For a given choice of~$A$, the wiggle equation of motion is only satisfied for particular values~$\xi_*$ of the modulus~$\xi_0$: evaluating~\eqref{eq:JTCFTeom} on~\eqref{eq:psipartialanalytic} and~\eqref{eq:m12generalu}, we obtain the constraint
\be
\label{eq:m2wigglesaddle}
\frac{1}{\sqrt{1-A^2}} = \sqrt{1 + \left(\frac{\beta J^2/n}{8\pi(4+1/(n\xi_*)^2)\coth(n\pi\xi_*/2)}\right)^2}.
\ee
However, if we wish to find a saddle to the \textit{full} path integral rather than just for the integral over the boundary wiggle, we must further require that~$\xi_*$ be a stationary point with respect to~$\xi_0$ of the action~\eqref{eq:I2analytic}.  This requirement gives
\be
\label{eq:m2xi0saddle}
\frac{1}{\sqrt{1-A^2}} = \frac{n^2\xi_*^3 \beta J^2}{16 \cosh^2(n\pi\xi_*/2)}.
\ee
Hence by simultaneously solving~\eqref{eq:m2wigglesaddle} and~\eqref{eq:m2xi0saddle} for~$A$ and~$\xi_*$, we can obtain a boundary profile that gives rise to a classical saddle for both the wiggle~$u(\phi)$ and the modulus~$\xi_0$.  It is straightforward to see that such a saddle can only exist when~$\beta J^2/n$ is sufficiently large: any~$\xi_*$ that simultaneously satisfies~\eqref{eq:m2wigglesaddle} and~\eqref{eq:m2xi0saddle} obeys~$f(n\xi_*) = n^2/(\beta^2 J^4)$, where
\be
f(x) \equiv 
\left(\frac{x^3}{16\cosh^2(\pi x/2)}\right)^2 - \frac{1}{(8\pi(4+ 1/x^2) \coth(\pi x/2))^2}.
\ee
$f(x)$ has a global maximum at~$x_\mathrm{max} \approx 0.93$, where it attains the value~$f_\mathrm{max} \approx 4.7 \times 10^{-5}$, so saddles for~$\xi_0$ exist if and only if~$\sqrt{\beta} J \geq f_\mathrm{max}^{-1/4} \sqrt{n} \approx 12 \sqrt{n}$.  So turning on matter sources with sufficiently large amplitude, or taking the temperature sufficiently small, can give rise to classical saddles for the double-trumpet.

While the approach just described allows us to find simultaneous saddles for both the wiggle and the modulus, for the purposes of a stability analysis it is illuminating to construct the effective action~$\widehat{I}_2[\xi_0]$ for the modulus obtained by only putting the wiggle on shell.  To do so, we consider a family of boundary sources~$\psi_\partial(u)$ of the form~\eqref{eq:psipartialanalytic} with~$A = A_*(\beta J^2/n)$, where
\begin{subequations}
\label{eq:Afamily}
\begin{align}
A_*(y) &\equiv \left[1+\left(\frac{8\pi \left(4+\frac{1}{x_*(y)^2}\right)\coth\left(\frac{\pi x_*(y)}{2}\right)}{y}\right)^2\right]^{-1/2}, \\
x_*(y) &\equiv \begin{dcases} x_\mathrm{max}, & y < f_\mathrm{max}^{-1/2}, \\
		\mbox{smallest positive solution of } y^2 f(x) = 1, & y \geq f_\mathrm{max}^{-1/2}. \end{dcases}
\end{align}
\end{subequations}
This form of~$A$ ensures that~\eqref{eq:m2wigglesaddle} is satisfied (and hence~\eqref{eq:m12generalu} is a solution to the equation of motion) when~$n \xi_0 = x_*(\beta J^2/n)$, which for~$\sqrt{\beta} J \geq f_\mathrm{max}^{-1/4} \sqrt{n}$ corresponds to the location~$\xi_*$ of a saddle in~$\xi_0$.  We will also take~$n = 2$ in order to ensure that~$\psi_\partial(u)$ is symmetric about~$u = 0$ and~$u = \beta/4$.

\begin{figure}[t]
\centering
\includegraphics[width=0.32\textwidth]{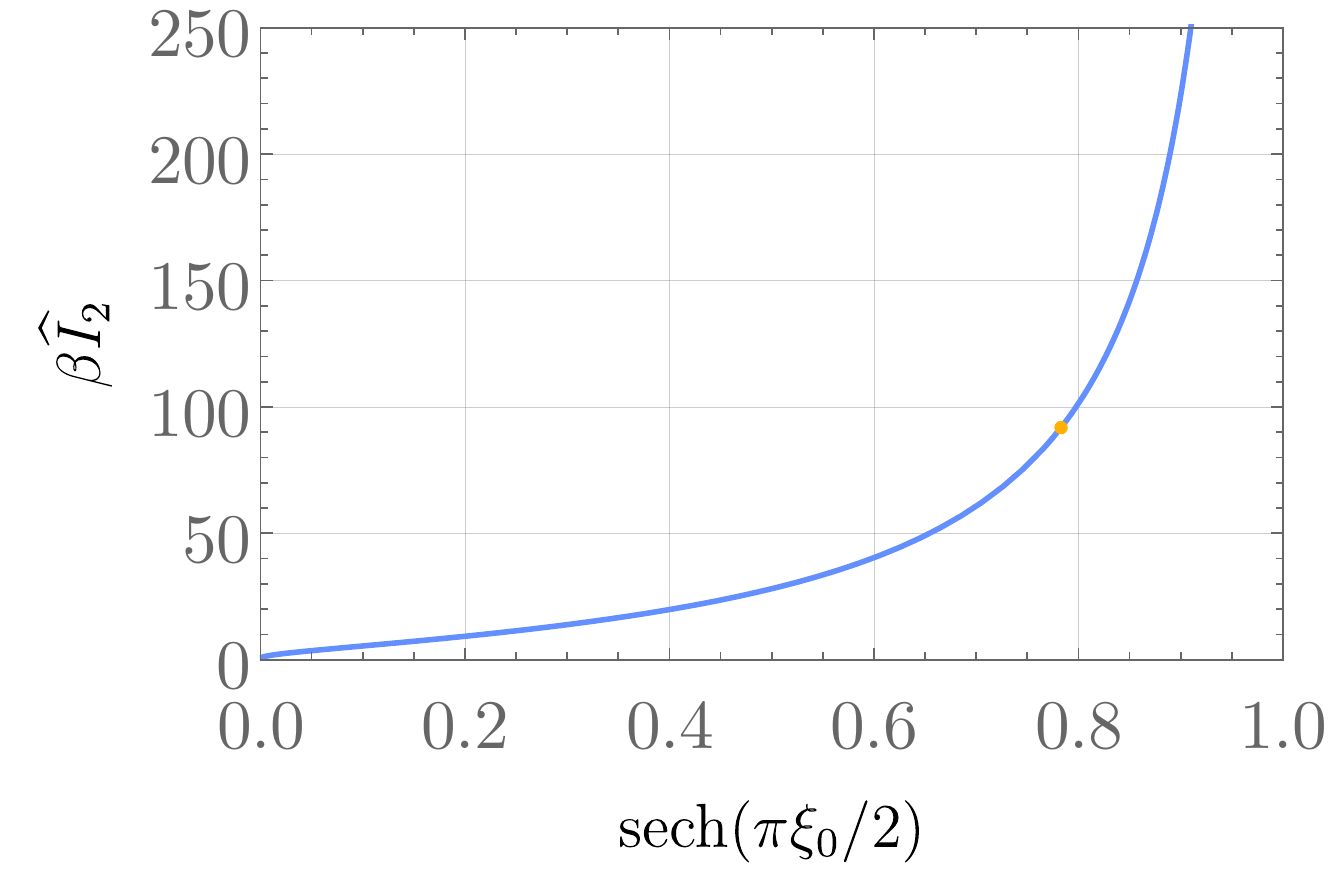}
\includegraphics[width=0.32\textwidth]{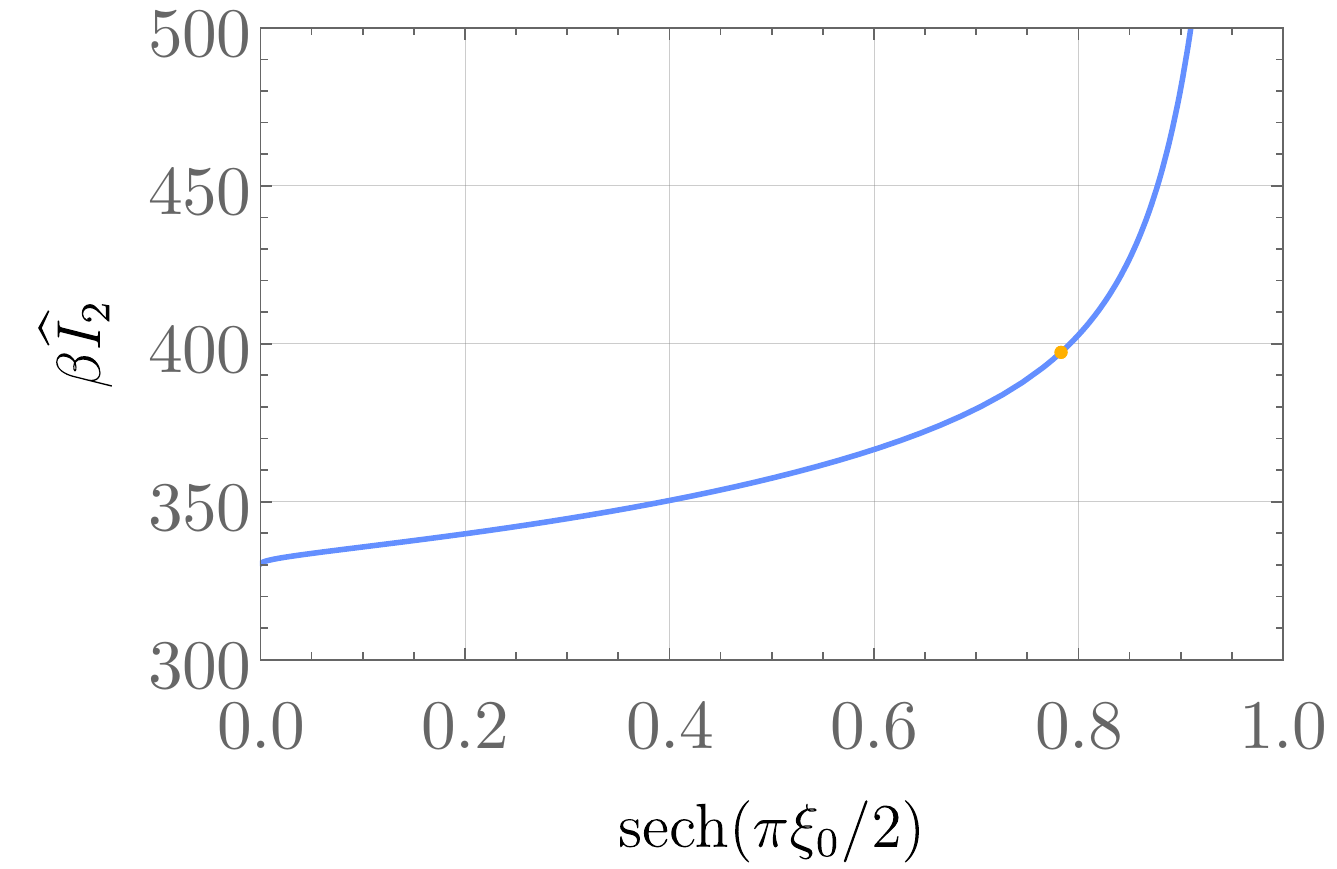}
\includegraphics[width=0.32\textwidth]{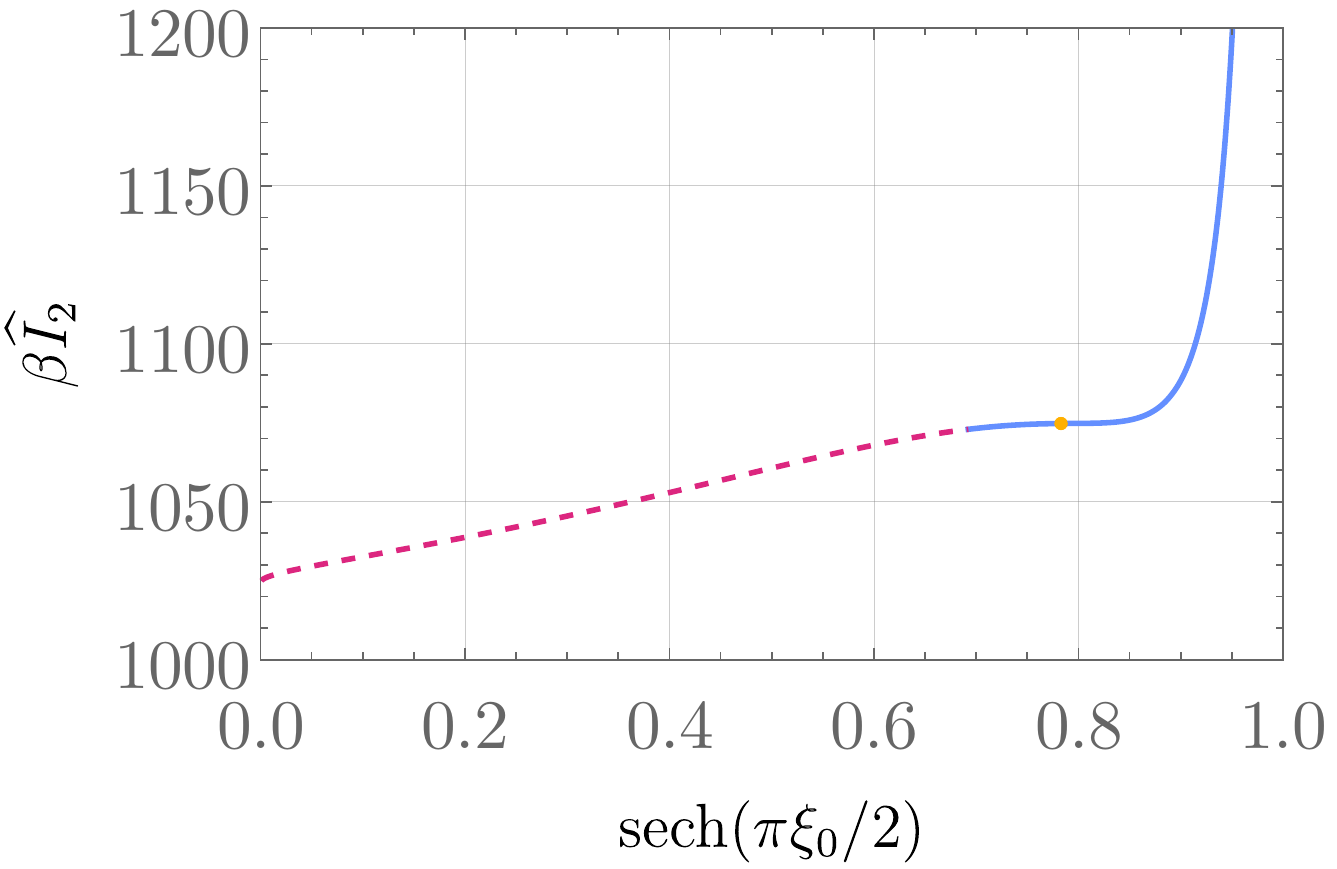}
\includegraphics[width=0.32\textwidth]{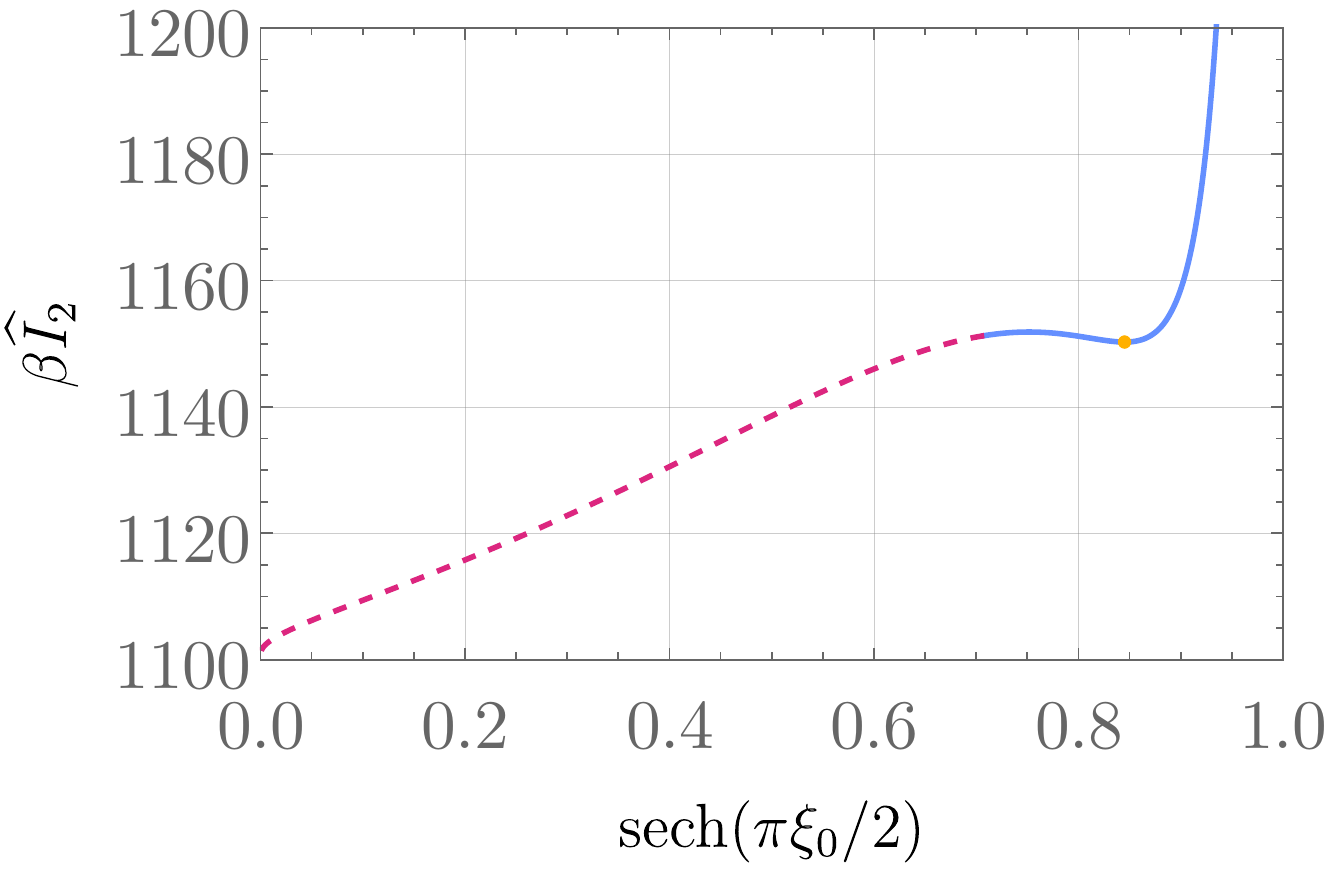}
\includegraphics[width=0.32\textwidth]{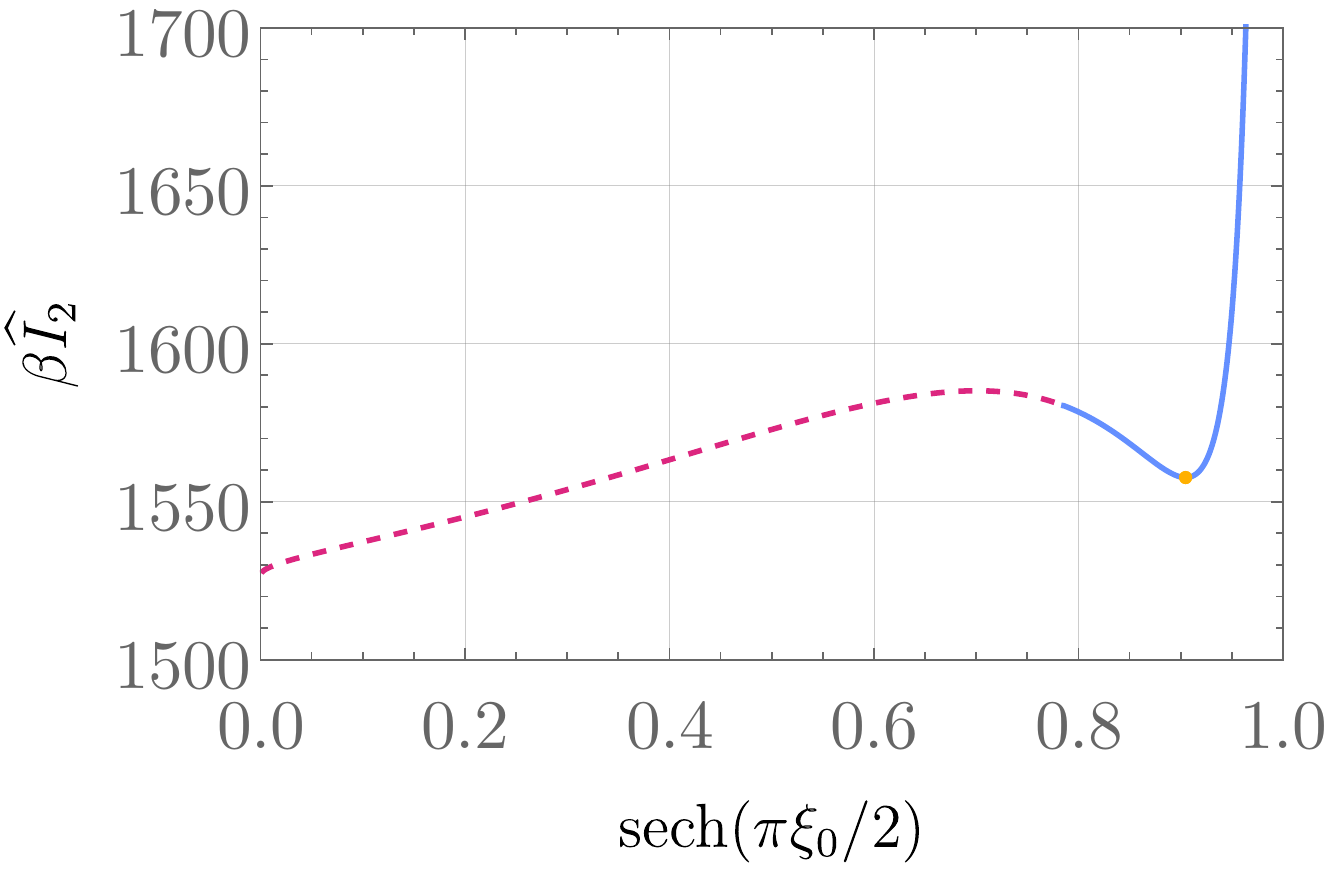}
\includegraphics[width=0.32\textwidth]{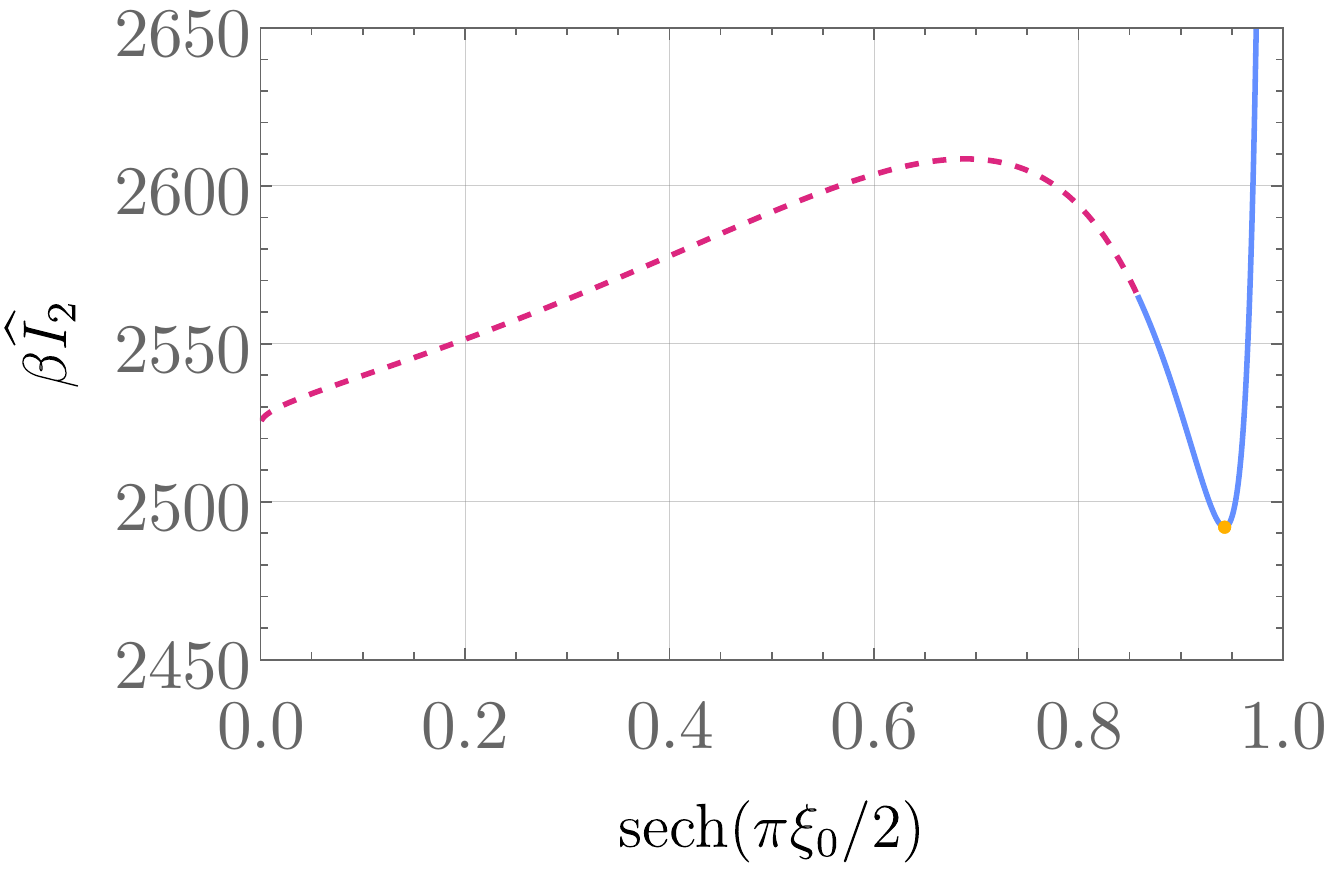}
\caption{The effective action~$\widehat{I}_2[\xi_0]$ obtained by placing the wiggle on shell, with boundary matter profile~\eqref{eq:psipartialanalytic} with~$n = 2$ and~$A$ given by~\eqref{eq:Afamily}; note that we plot the action as a function of the eccentricity~$\sech(\pi\xi_0/2)$ of the elliptical domain.  From left to right and top to bottom, we show~$\sqrt{\beta} J = 0$,~10,~17,~18,~22, and~30.  We plot the action as a solid blue curve where the spectrum of~$L$ (on the space of perturbations that preserve the~$\mathbb{Z}_2 \times \mathbb{Z}_2$ symmetry) is nonnegative, and as a dashed red curve when~$L$ has a negative eigenvalue.  The orange dots show the action~\eqref{eq:I2analytic} when~$\xi_0 = x_*(\beta J^2/2)/2$, for which the wiggle~\eqref{eq:m12generalu} with~$A$ given by~\eqref{eq:Afamily} satisfies the equation of motion~\eqref{eq:JTCFTeom}.}
\label{fig:m2action}
\end{figure}

With such a profile, the equation of motion~\eqref{eq:JTCFTeom} cannot be solved analytically for general~$\xi_0$ (except, by construction, for the special case~$n \xi_0 = x_*(\beta J^2/n)$), so we must proceed numerically.  The details of the numerical computation are presented in Appendix~\ref{app:numerics}, and the resulting action~$\widehat{I}_2[\xi_0]$ is shown in Figure~\ref{fig:m2action}.  When~$J = 0$, we recover the pure JT trumpet action~$\widehat{I}_2[\xi_0] = 2\pi^2/(\xi_0^2\beta) = b^2/(2\beta)$, where~$b = 2\pi/\xi_0$ is the circumference of the trumpet throat.  At sufficiently small values of~$\sqrt{\beta} J$, the action remains a monotonic function of~$\xi_0$, exhibiting no saddles in~$\xi_0$.  When~$\sqrt{\beta} J$ becomes sufficiently large, two saddles in~$\xi_0$ appear, with one stable and the other unstable with respect to perturbations in~$\xi_0$.  At these intermediate values of~$\sqrt{\beta} J$, the stable saddle does not \textit{globally} minimize the action: it corresponds to a metastable solution.  Further increasing~$\sqrt{\beta} J$, however, decreases the action of the stable saddle until it becomes a global minimum in~$\xi_0$.

We might expect that this global minimum of~$\widehat{I}_2[\xi_0]$ should dominate the path integral.  This well may be the case, but a stability analysis indicates a subtlety: not all of the saddles obtained for the wiggle are stable.  Restricting our consideration to perturbations that preserve the~$\mathbb{Z}_2 \times \mathbb{Z}_2$ symmetry, the solid blue curves in Figure~\ref{fig:m2action} correspond to solutions for which the fluctuation operator~$L$ defined in~\eqref{eq:JTCFTfluctuationoperator} has a nonnegative spectrum, while the dashed red curves correspond to solutions for which~$L$ exhibits a negative eigenvalue.  So although for large~$\sqrt{\beta} J$ the stable saddles are global minima of the effective action~$\widehat{I}_2[\xi_0]$ obtained by keeping the wiggle on shell, it is conceivable that off-shell configurations of the wiggle could decrease the action below that of these putative global minima.  We have not investigated this possibility further, but for now we assume that the~$m = 2$ path integral can be approximated by this new global minimum of the effective action~$\widehat{I}_2[\xi_0]$, when the minimum exists.

\begin{figure}[t]
\centering
\includegraphics[width=0.5\textwidth]{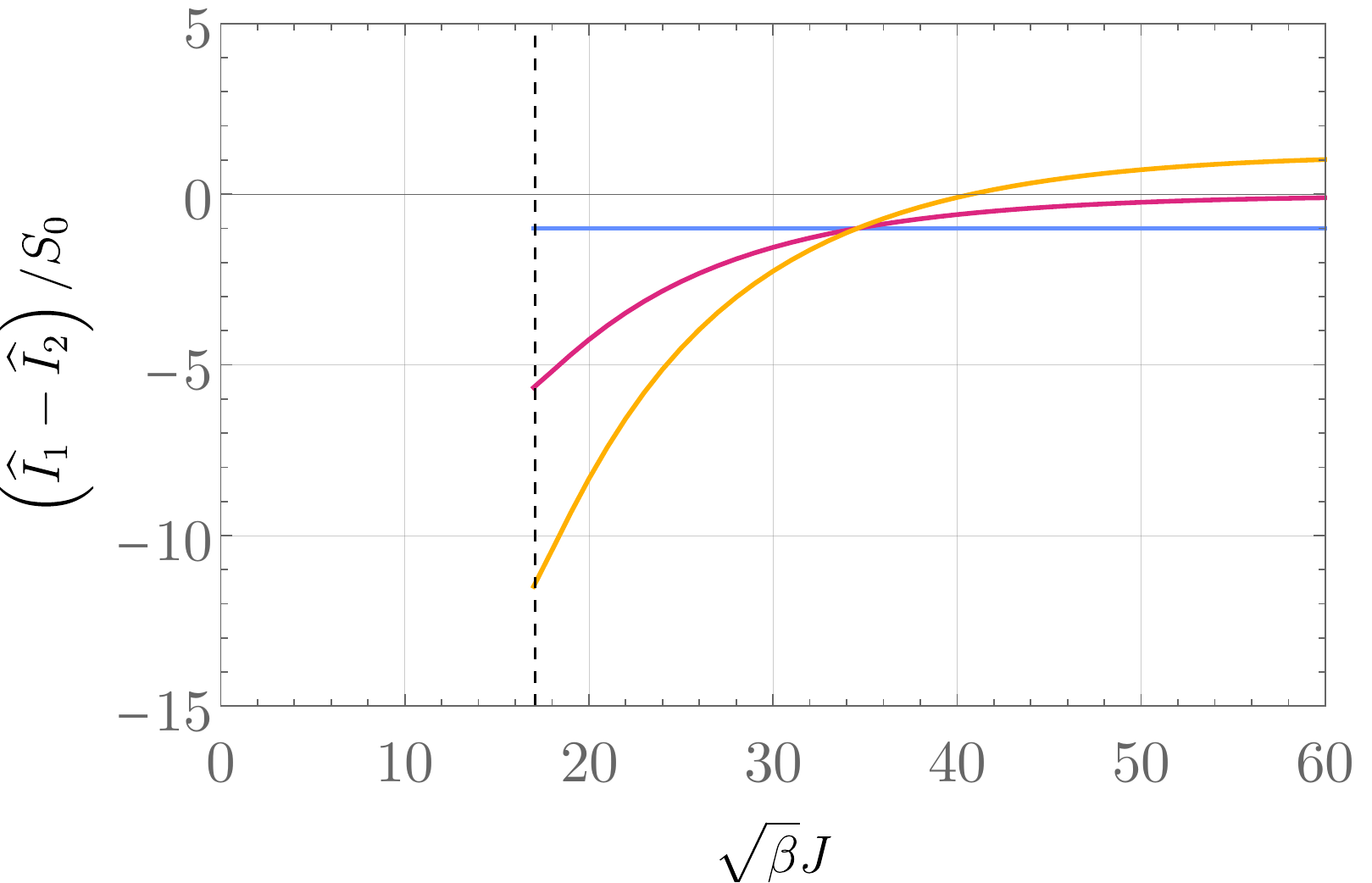}
\caption{The difference~$\widehat{I}_1 - \widehat{I}_2$ between the on-shell actions of the disk and the trumpet obtained by putting both the wiggle and the modulus~$\xi_0$ on shell.  The boundary matter profile is~\eqref{eq:psipartialanalytic} with~$n = 2$ and~$A$ given by~\eqref{eq:Afamily}; the corresponding trumpet action~$\widehat{I}_2$ is given by~\eqref{eq:I2analytic} with~$\xi_0 = x_*(\beta J^2/2)/2$, while the disk action~$\widehat{I}_1$ is computed numerically.  The blue, red, and orange curves correspond to~$J/\sqrt{S_0} = 0$,~4, and~6; the orange curve becomes positive at large enough~$\sqrt{\beta} J$, indicating dominance of the double-trumpet over the disk.  No saddles for~$\widehat{I}_2$ exist to the left of the dashed black line.}
\label{fig:doubletrumpet}
\end{figure}

If the double-trumpet can be stabilized, can it ever dominate over the disk in a computation of, say,~$\overline{Z^2}$?  Such dominance could only ever occur in a classical limit if~$J$ and~$\beta$ scale appropriately with~$S_0$, since otherwise the topological part of the action will trivially cause the disk to dominate.  This is analogous to the need for the matter partition function to be of order~$e^{S_0}$ in models of black hole evaporation before wormholes can start to dominate after the Page time~\cite{AlmHar19,PenShe19}.  Indeed, we find that with an appropriate scaling of~$J$ and~$\beta$ with~$S_0$, an exchange of dominance between the disk and the double-trumpet can occur: in Figure~\ref{fig:doubletrumpet} we show the difference~$\widehat{I}_1 - \widehat{I}_2$ between the actions of the disk and the trumpet for various values of~$J/\sqrt{S_0}$.  At values of~$J/\sqrt{S_0}$ around order unity or smaller, this difference is everywhere-negative, so the disk always dominates.  But at larger values of~$J/\sqrt{S_0}$, this difference becomes positive at sufficiently large values of~$\sqrt{\beta} J$, indicating that the double-trumpet dominates at sufficiently low temperatures.  This transition requires the amplitude~$J$ to scale with~$S_0$ like~$J \gtrsim \sqrt{S_0}$, so at fixed~$J/\sqrt{S_0}$ the temperatures at which the double-trumpet dominates (when it does at all) scale with~$S_0$ like~$T \lesssim S_0$ .  This behavior is analogous to the results of~\cite{GarGot20,GarGot22}, where a phase transition between the disk and the double-trumpet was induced by turning on constant but complex sources for a massless scalar.  Here we see that such a transition can be supported with real, replica-symmetric sources with nontrivial Euclidean time dependence.  It is this nontrivial Euclidean time dependence that induces the stress tensor necessary to stabilize the wormhole (non-constant boundary sources are also key to the higher-dimensional constructions of~\cite{MarSan21}).

\subsection{Wormholes for \texorpdfstring{$m < 1$}{m<1}}
\label{subsec:JTCFTmleq1}

Having discussed the special cases~$m = 1$ and~$m = 2$, we now turn our investigation to the saddles at~$m < 1$ that appear in the replica trick.  The first distinction to note with the~$m > 1$ case is that the topological part of the action is monotonically decreasing with~$m$.  Consequently, taking~$S_0$ large at fixed~$\beta$ and~$\psi_\partial$ will always cause the disk to win out over the~$m > 1$ wormholes; this was the reason that we needed to scale~$J$ and~$\beta$ with~$S_0$ in the previous section to get the double-trumpet to dominate over the disk.  However, for this same reason, taking~$S_0$ large will always cause the disk to be \textit{subdominant} to any saddles at~$m < 1$.  Hence if there are any saddles at~$m < 1$ at all, they will always dominate over the disk in a classical limit~$S_0 \to \infty$ with~$\beta$ and~$\psi_\partial$ kept fixed.  Thus we only need to look for stable saddles at~$m < 1$, without needing to worry about their dominance over the disk.

\begin{figure}[t]
\centering
\includegraphics[width=0.32\textwidth]{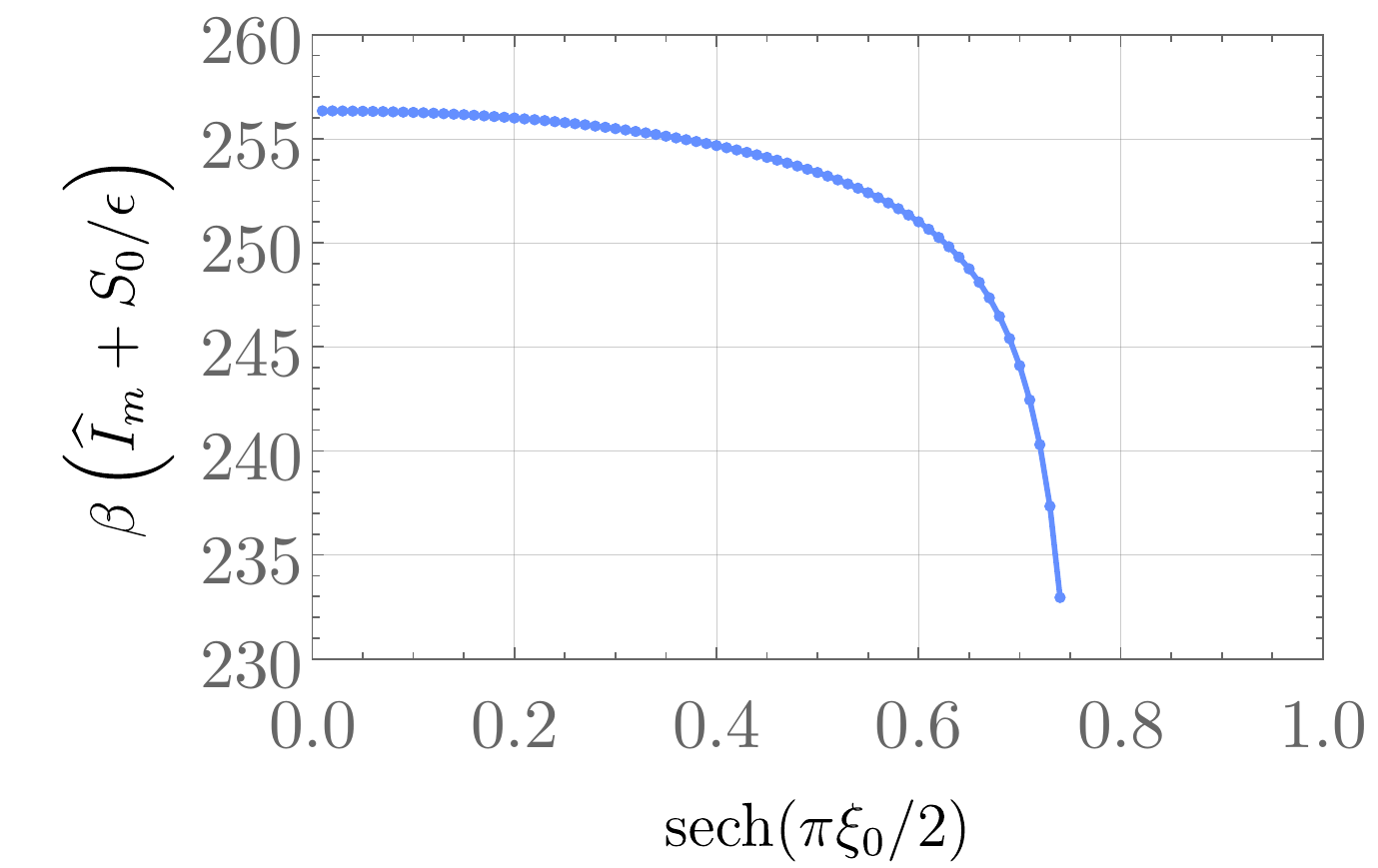}
\includegraphics[width=0.32\textwidth]{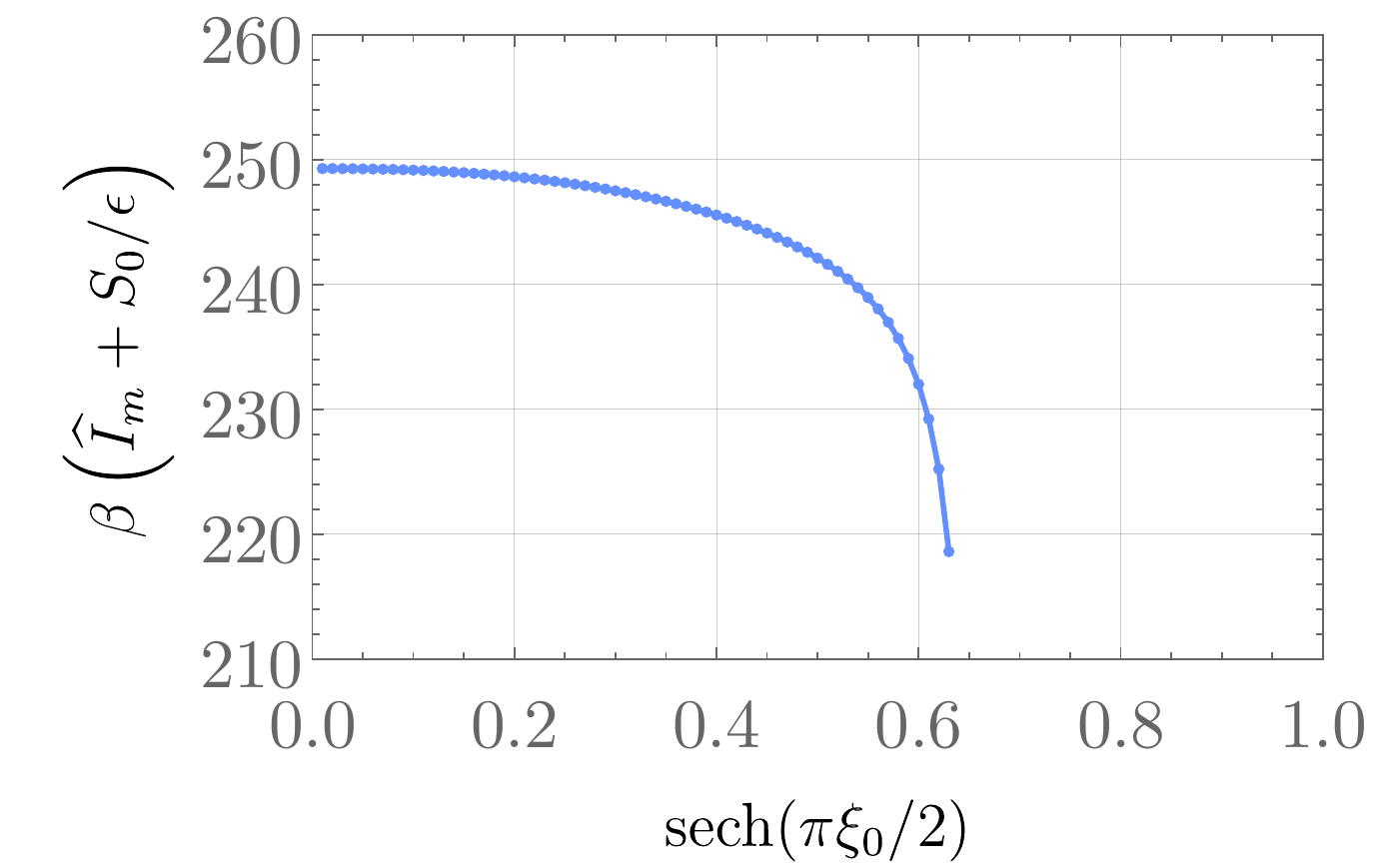}
\includegraphics[width=0.32\textwidth]{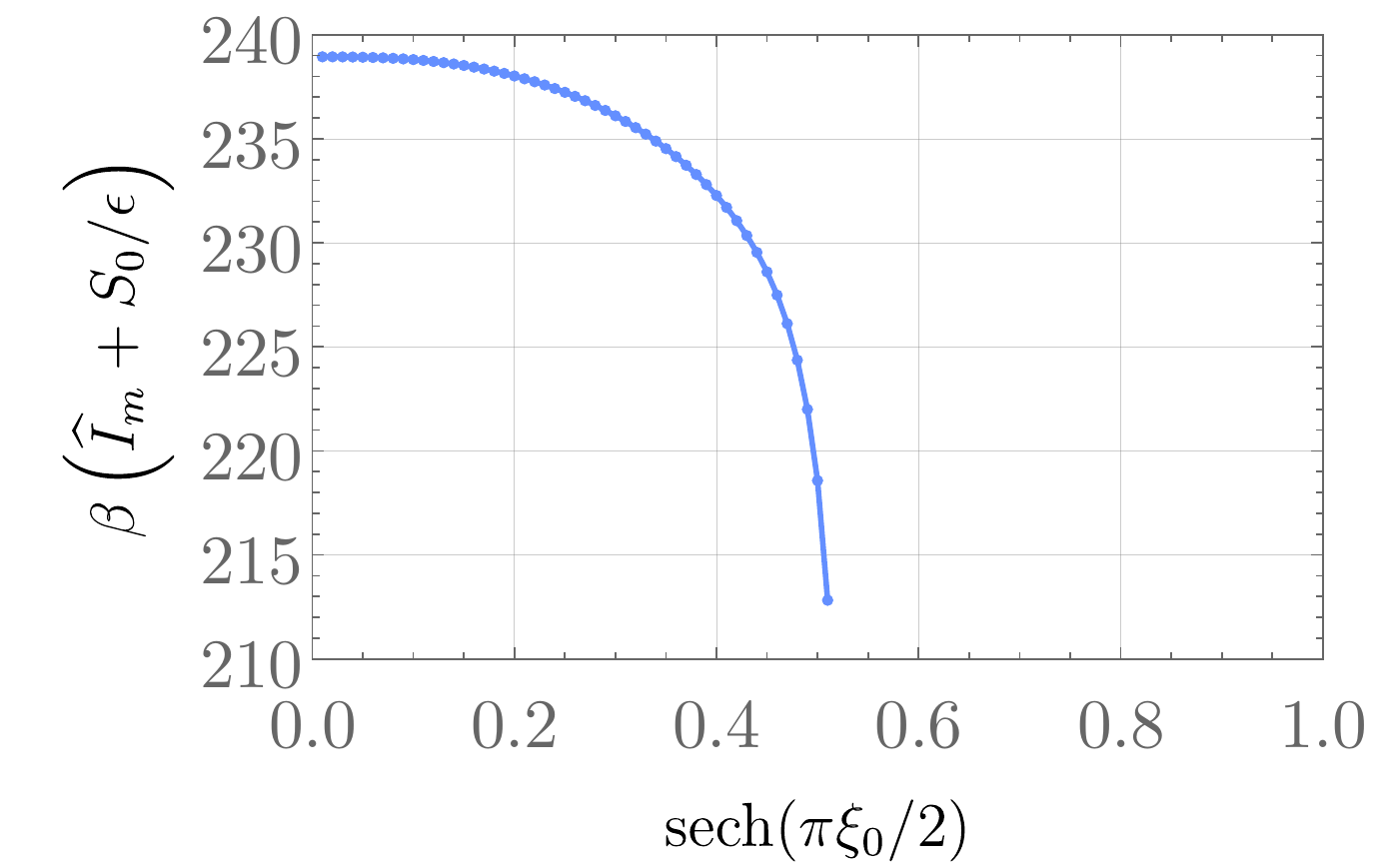}
\includegraphics[width=0.32\textwidth]{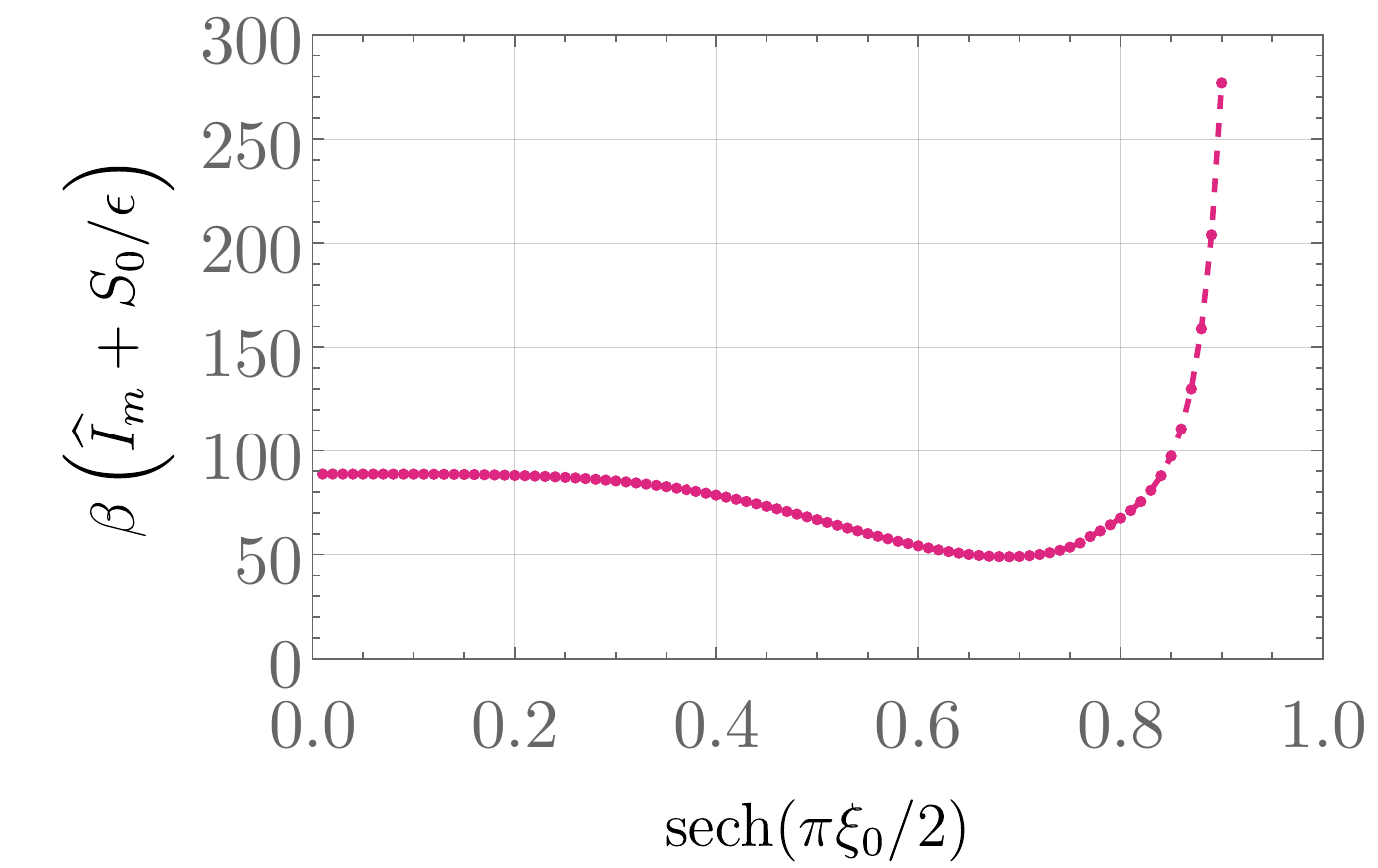}
\includegraphics[width=0.32\textwidth]{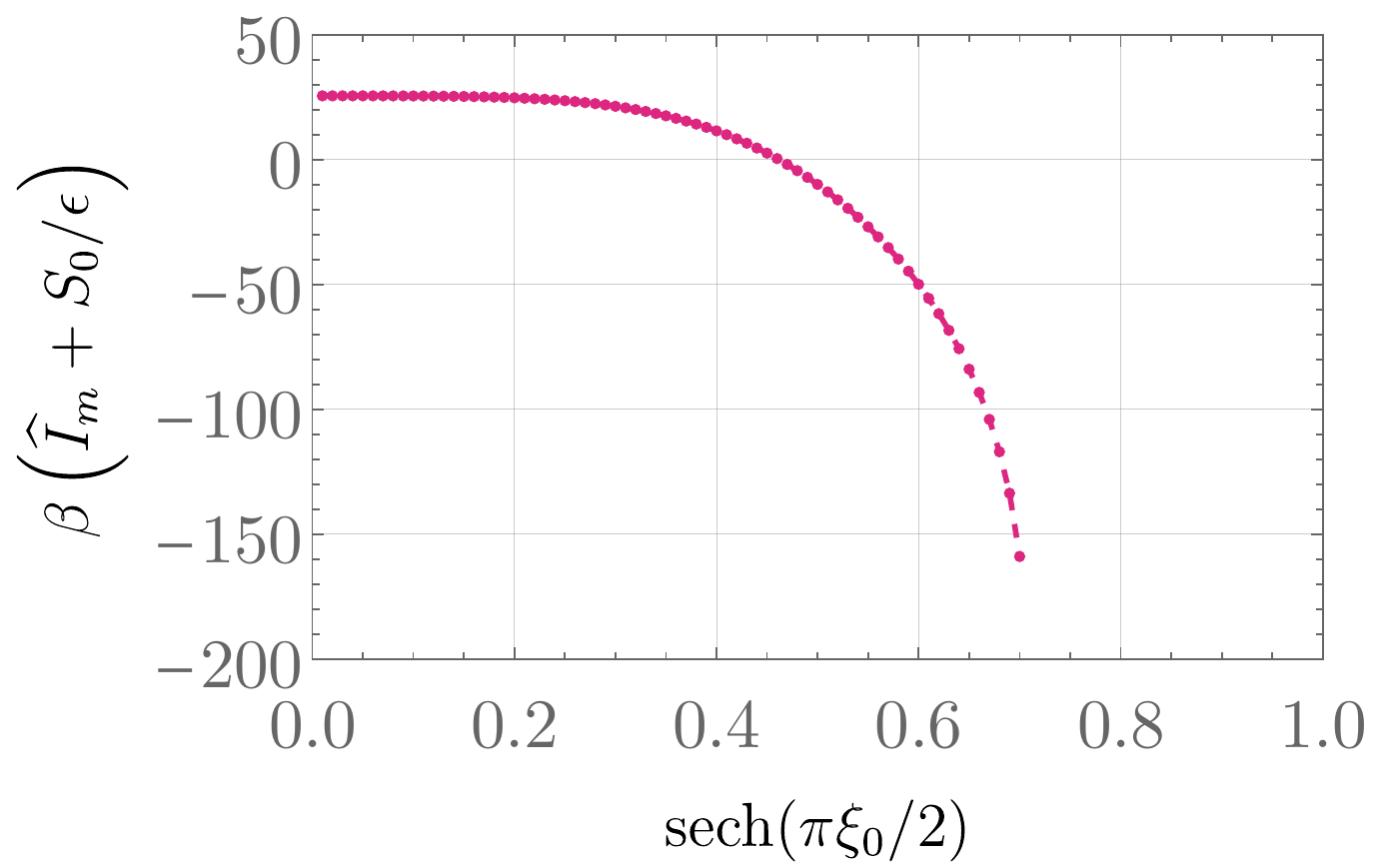}
\includegraphics[width=0.32\textwidth]{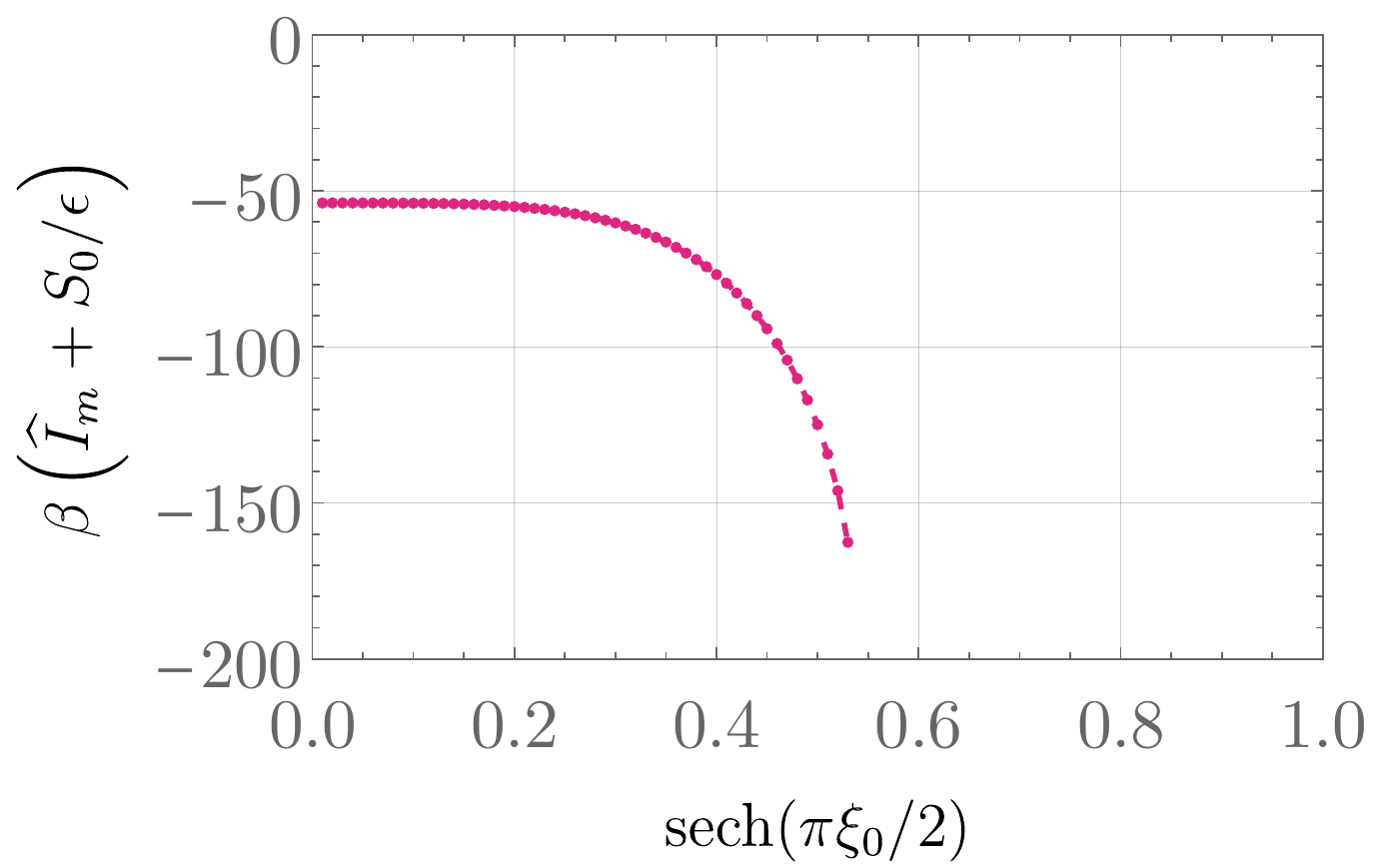}
\caption{The effective action~$\widehat{I}_m[\xi_0]$ of JT + CFT with boundary source given by~\eqref{eq:JTCFTbndrysource} with $\sqrt{\beta}J = 10$.  From left to right and top to bottom we show~$\nu = 0.9$, 0.8, 0.7, 0.34, 0.3, 0.27, corresponding to~$m \approx 0.95$, 0.89, 0.82, 0.51, 0.46, 0.43.  Points correspond to numerical data; curves are drawn to guide the eye.  Blue points connected by solid lines indicate that the spectrum of~$L$ is nonnegative; red points connected by dashed lines indicate that~$L$ has a negative eigenvalue.  No stable saddles for both~$\xi_0$ and the wiggle are present.}
\label{fig:JTCFTJ10}
\end{figure}

Our analysis will be entirely numerical, so for simplicity we now fix the matter sources to be the lowest nontrivial Fourier mode on the thermal circle compatible with our assumed~$\mathbb{Z}_2 \times \mathbb{Z}_2$ symmetry:
\be
\label{eq:JTCFTbndrysource}
\psi_\partial(u) = J \cos\left(\frac{4\pi u}{\beta}\right).
\ee
Again we leave the numerical details to Appendix~\ref{app:numerics}.  At relatively small values of~$\sqrt{\beta} J$, we do not find any stable saddles at any~$m$.  In Figure~\ref{fig:JTCFTJ10} we show the effective action~$\widehat{I}_m[\xi_0]$ with~$\sqrt{\beta}J = 10$ for several values of~$m$.  This effective action is either monotonic in~$\xi_0$, exhibiting no saddles for the modulus, or may exhibit a stable saddle for~$\xi_0$ which is however unstable to perturbations of the wiggle, as in the fourth plot in the figure.  Note that solutions do not exist for all~$\xi_0$: the wiggle becomes singular at sufficiently small~$\xi_0 = \xi_\mathrm{end}$ below which we found no more solutions.  This behavior is qualitatively analogous to what we observed in pure JT in Section~\ref{sec:pureJT}: on-shell solutions for the wiggle did not exist for all~$\alpha$.  In that case, the end points at which solutions stopped existing corresponded to branch points of the Riemann surface for the analytic continuation of~$\widehat{I}_m[\alpha]$ to complex~$m$ (and~$\alpha$).  The endpoints~$\xi_\mathrm{end}$ shown in Figure~\ref{fig:JTCFTJ10} may play the same role: they may be branch points of the analytic continuation of~$\widehat{I}_m[\xi_0]$ to complex~$m$ and~$\xi_0$.

\begin{figure}[t]
\centering
\includegraphics[width=0.32\textwidth]{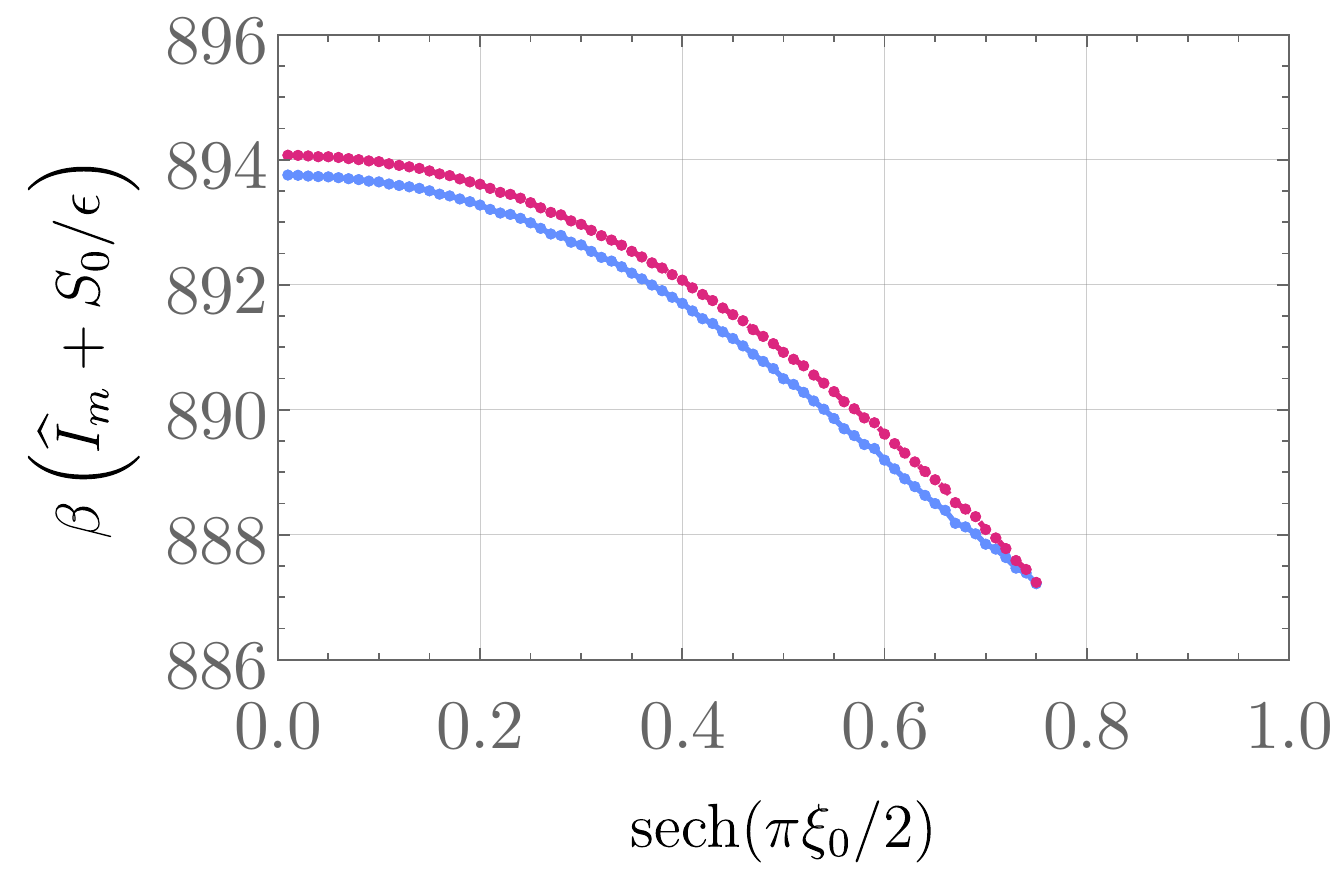}
\includegraphics[width=0.32\textwidth]{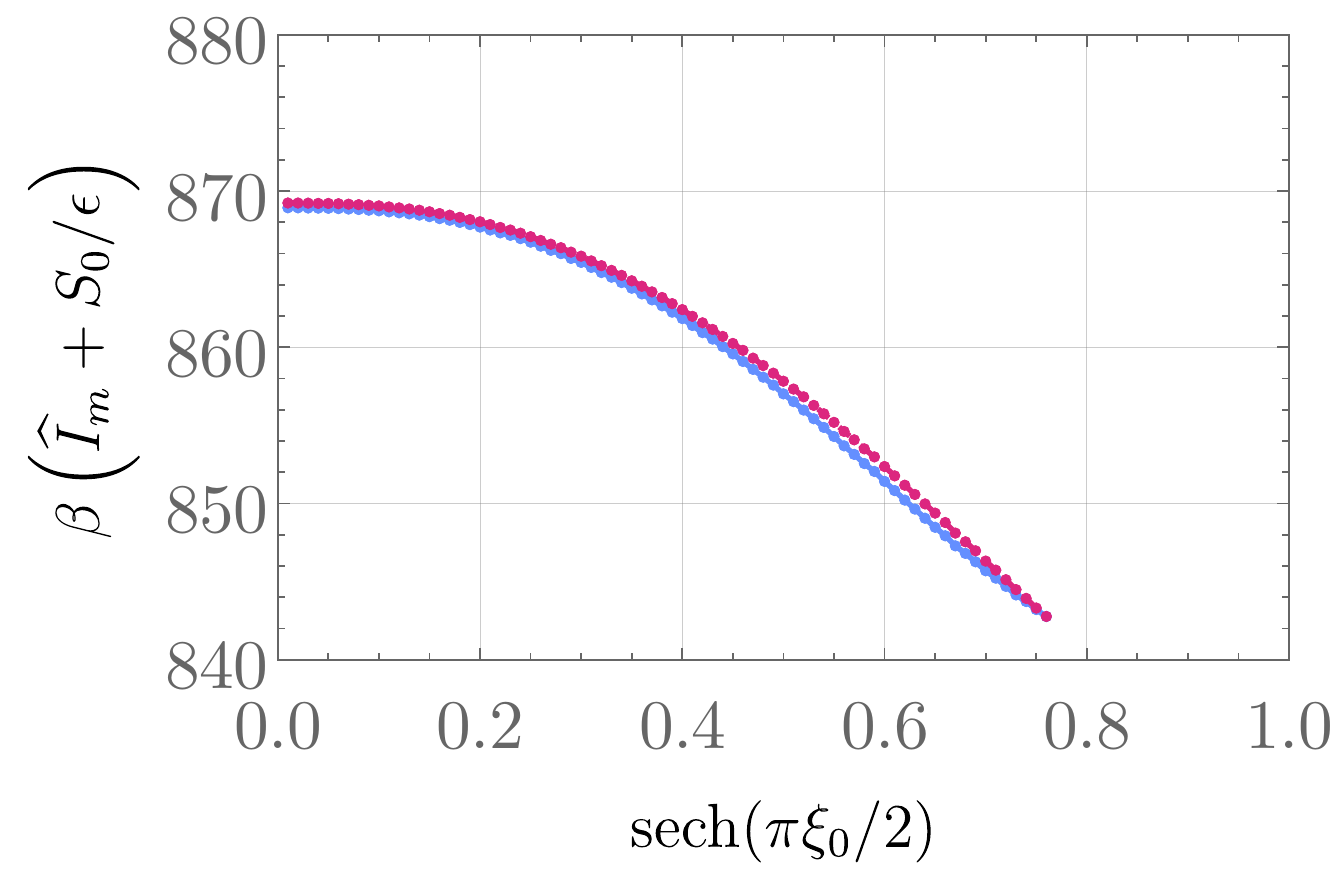}
\includegraphics[width=0.32\textwidth]{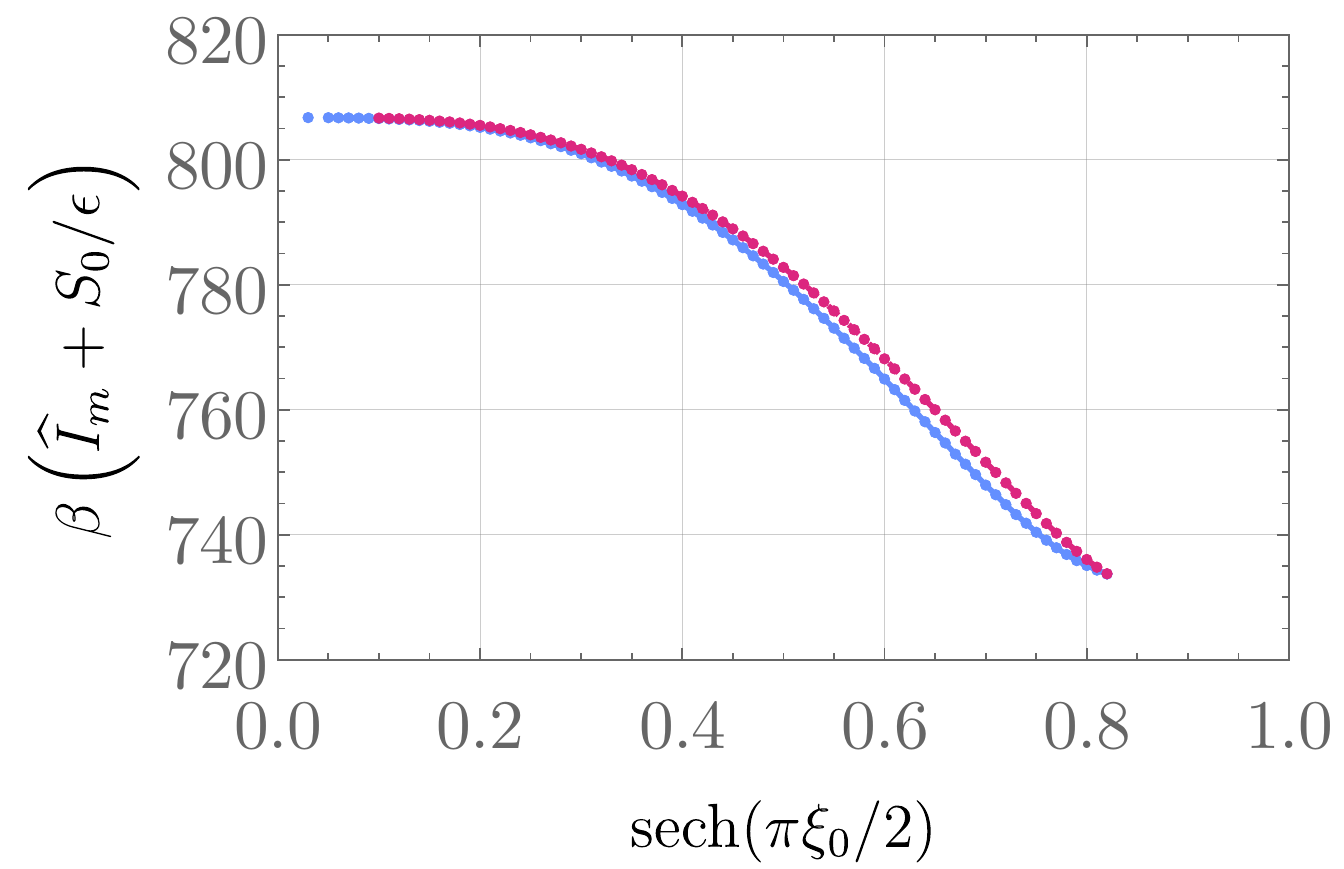}
\includegraphics[width=0.32\textwidth]{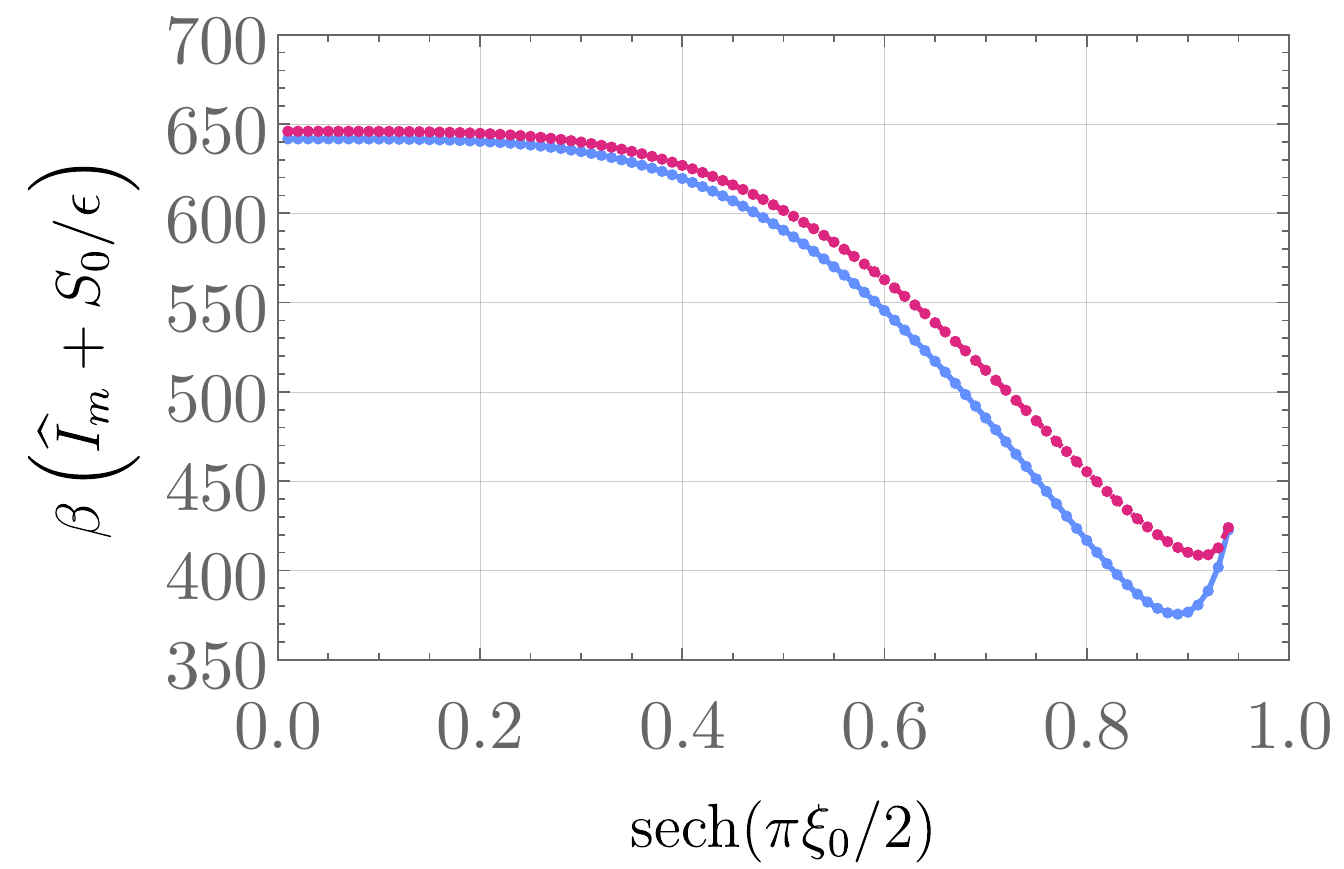}
\includegraphics[width=0.32\textwidth]{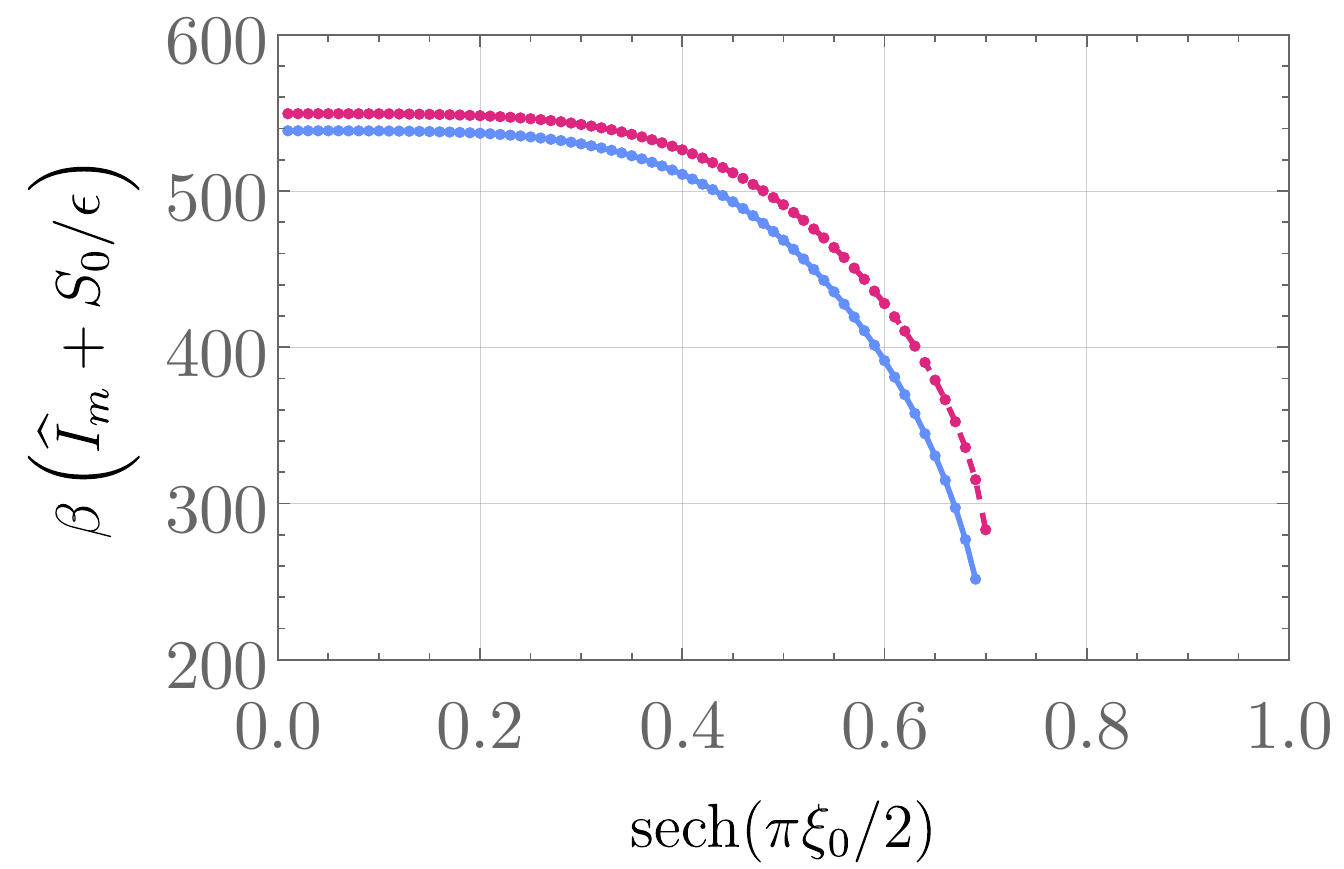}
\includegraphics[width=0.32\textwidth]{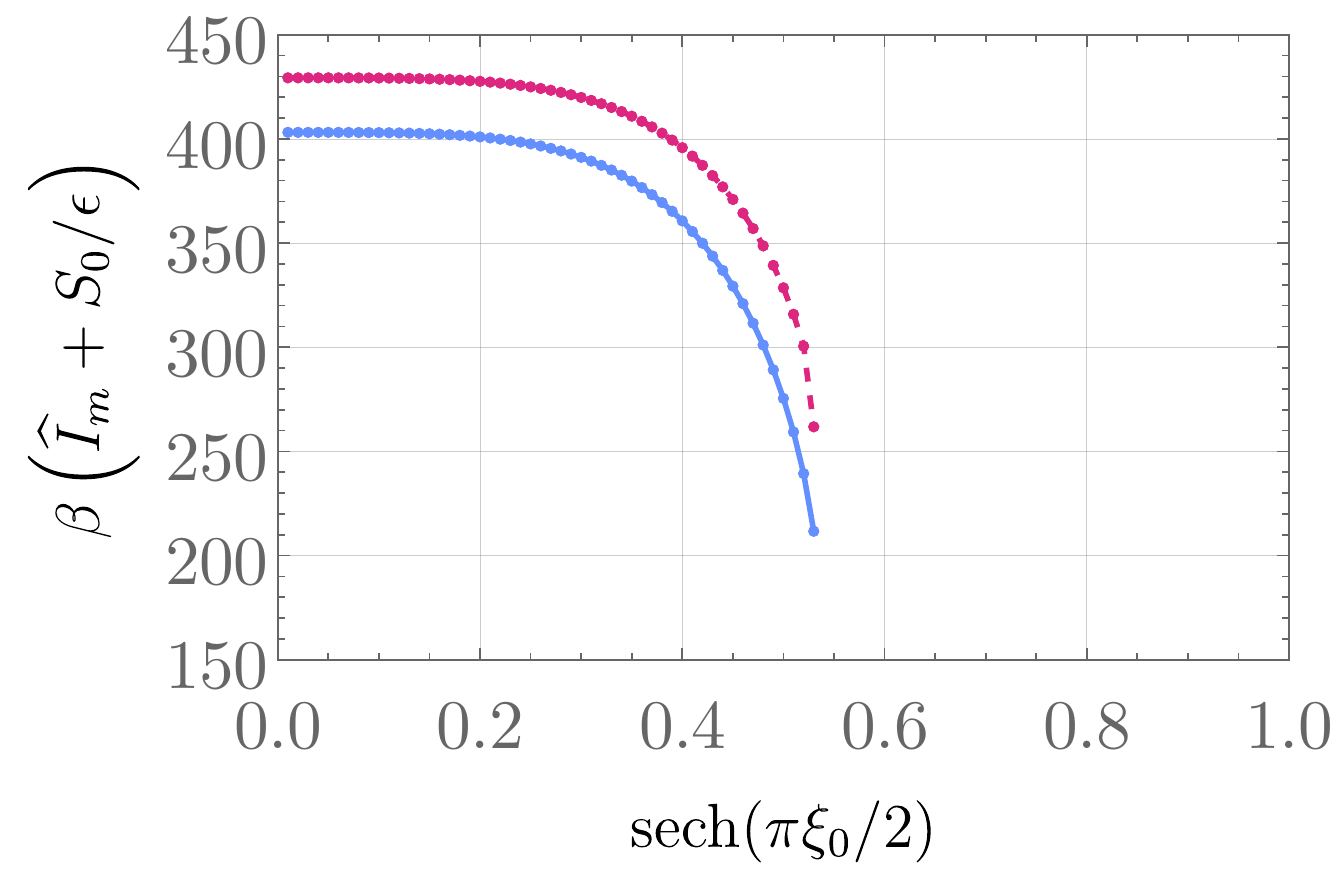}
\caption{The effective action~$\widehat{I}_m[\xi_0]$ of JT + CFT with boundary source given by~\eqref{eq:JTCFTbndrysource} with $\sqrt{\beta}J = 20$.  From left to right and top to bottom we show~$\nu = 0.9$, 0.7, 0.5, 0.34, 0.3, 0.27, corresponding to~$m \approx 0.95$, 0.82, 0.67, 0.51, 0.46, 0.43.  Points correspond to numerical data; curves are drawn to guide the eye.  Blue points connected by solid lines (lowermost curves) indicate that the spectrum of~$L$ is nonnegative; red points connected by dashed lines (uppermost curves) indicate that~$L$ has a negative eigenvalue.  For all~$\nu$ shown here there are two branches of solutions, with one stable and the other unstable; for~$\nu > 1/3$ ($m > 1/2$) these branches meet at a zero mode, but for~$\nu < 1/3$ these branches cease to join.  Note that there is a stable saddle in fourth plot.}
\label{fig:JTCFTJ20}
\end{figure}

Increasing~$\sqrt{\beta}J$ turns out to change the story substantially, however: in Figure~\ref{fig:JTCFTJ20} we show~$\widehat{I}_m[\xi_0]$ with~$\sqrt{\beta}J = 20$.  We are still unable to find on-shell solutions for the wiggle for all~$\xi_0$, but we also find two independent branches of solutions, with one unstable and the other stable.  For~$\nu \gtrsim 1/3$ (or~$m \gtrsim 1/2$), these branches meet at a zero mode at which the wiggle is regular.  As~$\nu$ is decreased below~$1/3$, the zero mode becomes singular and the two branches separate, with each one terminating at a singular endpoint analogous to those in Figure~\ref{fig:JTCFTJ10}.  The key feature of these two branches is that because one is stable, we need only find a stable saddle in~$\xi_0$ to deduce the existence of a stable wormhole.  And indeed, such a saddle exists, as seen in the fourth plot in the figure.  Interestingly, this saddle does not persist to smaller~$m$: as can be seen in the last two plots of the figure, when~$\nu$ is decreased below~$1/3$ and the branches separate, the saddle vanishes.  We are unable to find additional stable saddles by further decreasing~$\nu$.  This qualitative behavior is independent of the value of~$\sqrt{\beta} J$ up to the largest value~$\sqrt{\beta} J = 25$ for which we have constructed solutions.  The upshot is that at sufficiently large~$\sqrt{\beta} J$, stable classical replica wormholes exist at~$m < 1$, but only down to around~$m \approx 1/2$.

\begin{figure}[t]
\centering
\includegraphics[width=0.4\textwidth]{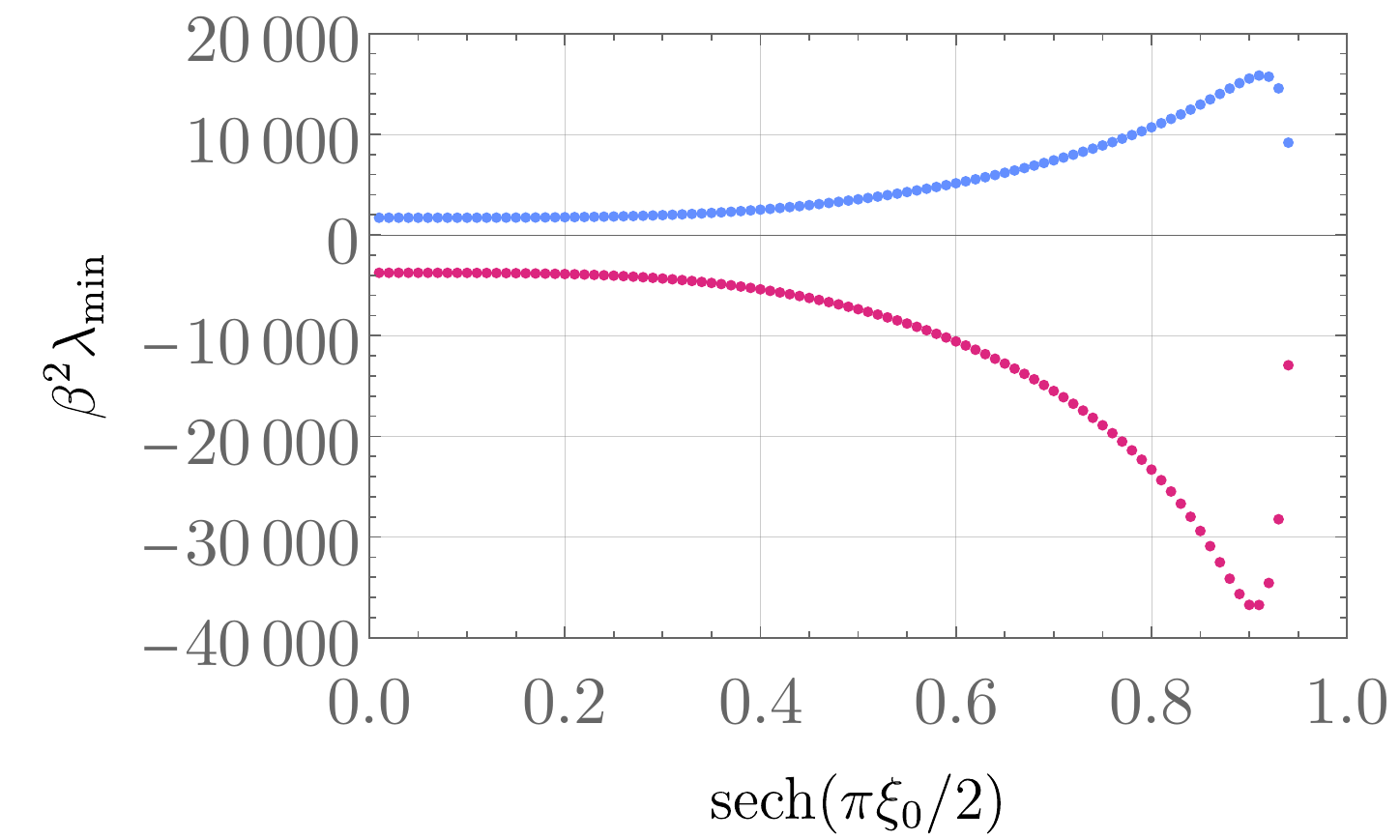}
\includegraphics[width=0.4\textwidth]{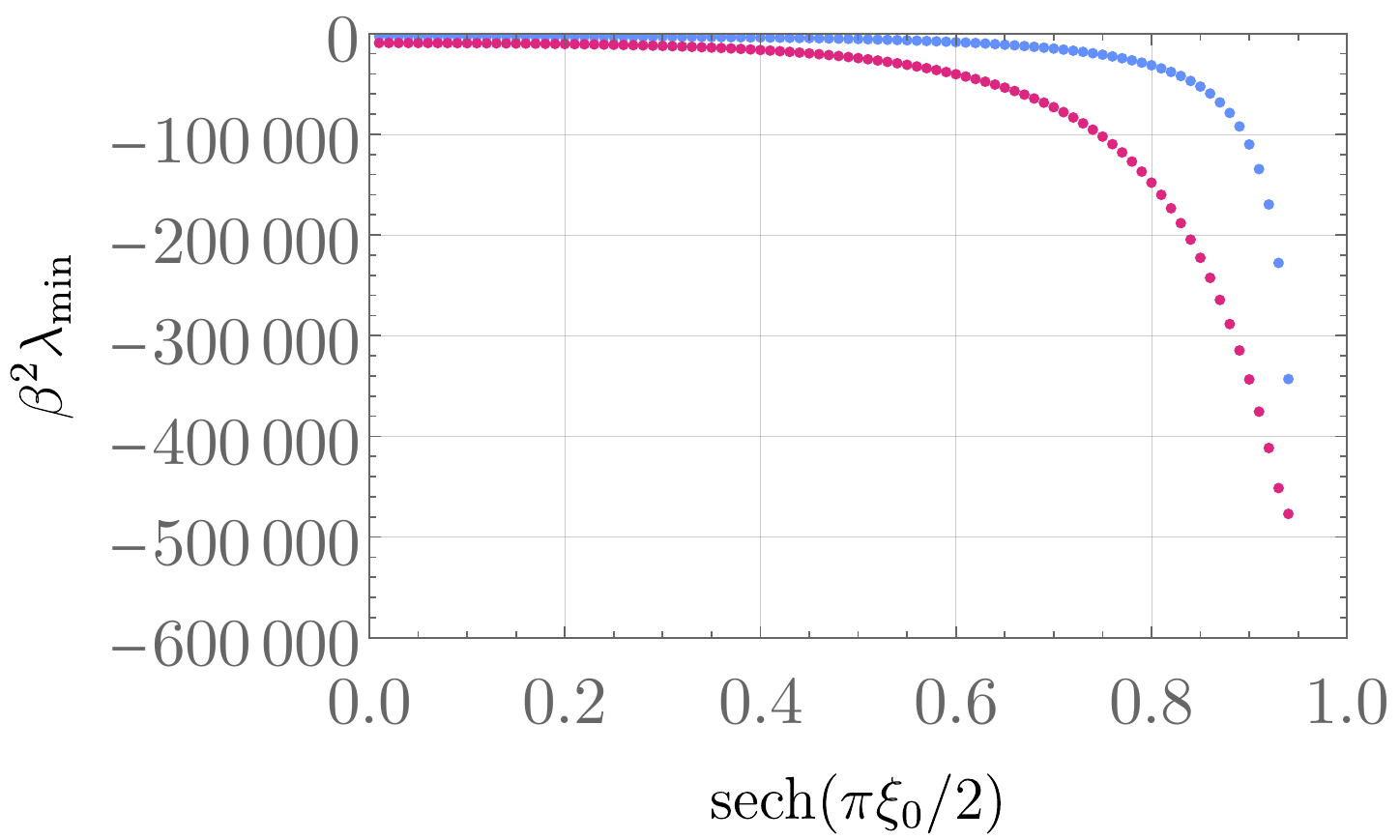}
\includegraphics[width=0.4\textwidth]{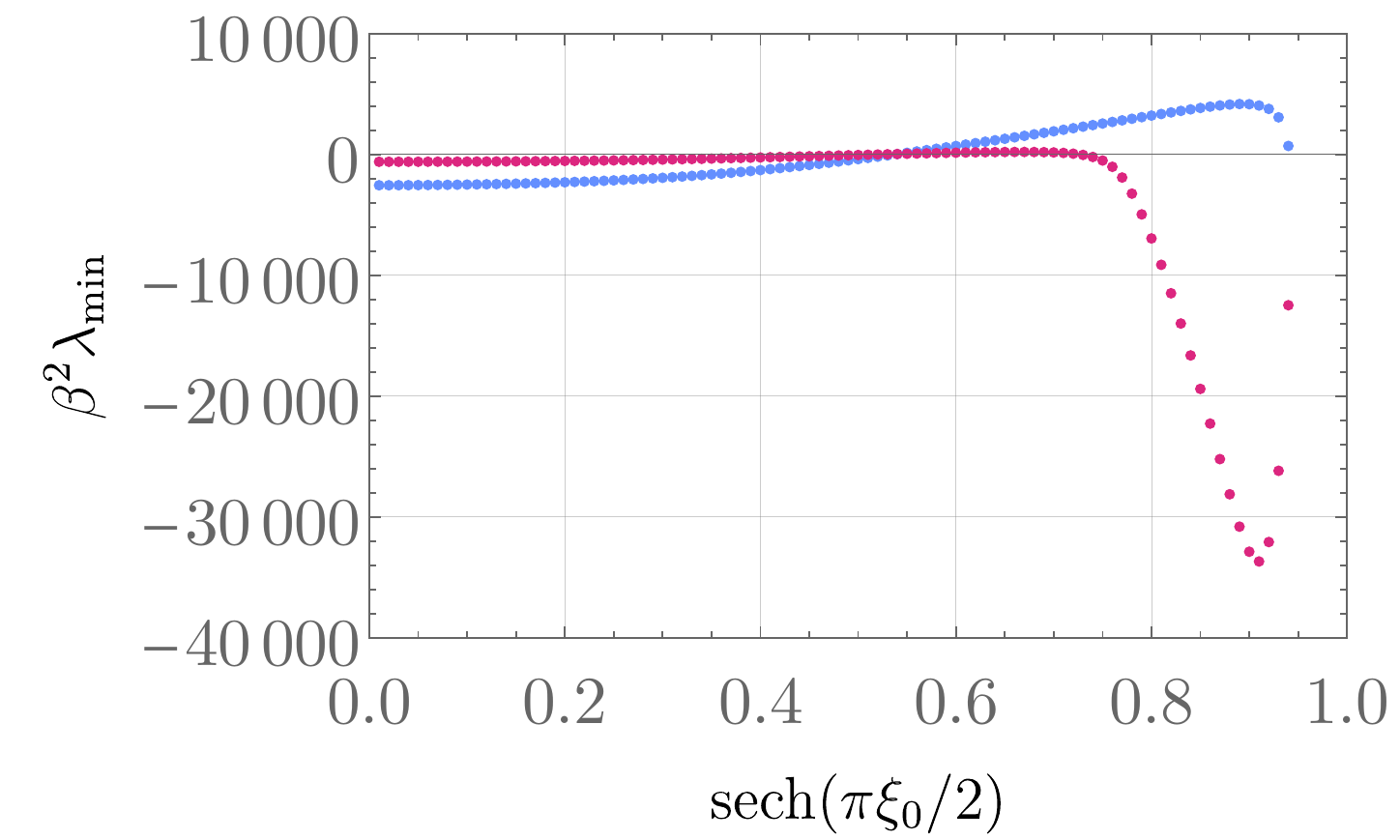}
\includegraphics[width=0.4\textwidth]{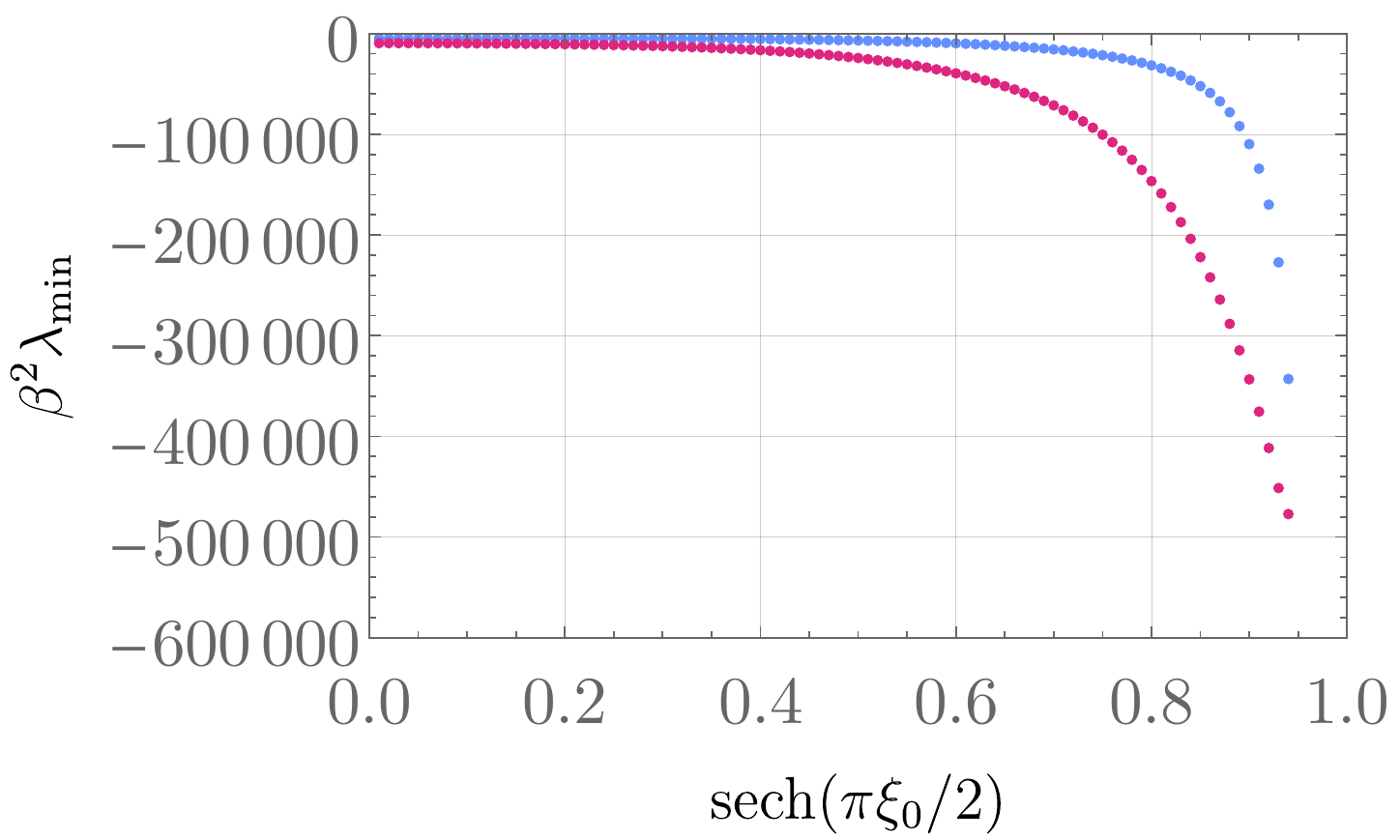}
\caption{The lowest eigenvalue~$\lambda_\mathrm{min}$ of the fluctuation operator~$L$ under different parities of the perturbation~$\nu(\phi)$ defined in~\eqref{eq:JTCFTfluctuationoperator}; here we show~$\sqrt{\beta} J = 20$ and~$\nu = 0.34$, i.e.~the fourth plot in Figure~\ref{fig:JTCFTJ20}.  The left (right) plots take~$\nu(\phi)$ to be odd (even) about~$\phi = 0$, while the top (bottom) plots take~$\nu(\phi)$ to be odd (even) about~$\phi = \pi/2$.  The colors label the two branches shown in Figure~\ref{fig:JTCFTJ20}; note that~$\nu(\phi)$ needs to be odd about~$\phi = 0$ for at least one of the branches to be stable at the location of the saddle in~$\xi_0$.  This corresponds to the boundary curve~$\partial M$ being even about the major axis of the ellipse.}
\label{fig:JTCFTeigvals}
\end{figure}

Before examining the on-shell action of these wormholes in more detail, let us pause to note that requiring the~$\mathbb{Z}_2 \times \mathbb{Z}_2$ symmetry was crucial to finding stable saddles.  For reference, in Figure~\ref{fig:JTCFTeigvals} we show the lowest eigenvalue of~$L$ when the parity of the perturbations about~$\phi = 0$ and~$\phi = \pi/2$ is modified.  Only for perturbations odd about~$\phi = 0$ (corresponding to the shape of~$\partial M$ being symmetric about the major axis of the ellipse) is one of the branches of solutions stable at the saddle for~$\xi_0$, giving a stable wormhole.  This symmetry about~$\phi = 0$ is necessary for a real Lorentzian continuation obtained by cutting the ellipse across its major axis.  So the stability of these wormholes -- and consequently whether the quenched and annealed generating functionals~$\Gamma_Q$ and~$\Gamma_A$ differ -- depends crucially on whether we demand that perturbations about the saddle admit a real Lorentzian section that contains the defects.  Note that it is crucial that the Lorentzian section contain the defects: a real Lorentzian geometry could also be generated by cutting the ellipse about its minor axis, but requiring reality of such a section is not sufficient to stabilize the wormholes.

How should the need for such a real Lorentzian section be interpreted?  If the path integral is to be understood as a purely Euclidean object completely removed from any Lorentzian underpinning, then there is no reason to impose any such condition.  In this case, our saddles are simply unstable and never contribute to the quenched generating functional.  But this interpretation is rather odd: after all, we are ultimately interested in theories with a Lorentzian counterpart, and moreover to even make the JT Euclidean path integral well-defined in the first place the dilaton~$\varphi$ needs to be Wick rotated to an imaginary contour: a \textit{strictly} Euclidean definition of the JT path integral is manifestly divergent.  In addition, there is the question of which~$\mathbb{Z}_2$ symmetry to preserve; that is, which of the principal axes of the ellipse should correspond to the~``$t = 0$'' slice of the Lorentzian section.  Comparison with other replica tricks make it natural to require the conical defects to live on this~$t = 0$ slice: for example, in the LM construction of von Neumann entropy, it is the~$m \to 1$ limit of the conical defects that turns into the minimal or QES surfaces in the RT or QES formulas.  In order for these surfaces to live in the Lorentzian section, the~$t = 0$ Lorentzian slice of the quotient geometry must therefore contain the conical defects.  Ultimately, whether or not the new saddles should genuinely contribute to~$\Gamma_Q$ will depends on the desired properties of the theory; we will revisit this question in Section~\ref{sec:disc}.

As a final note, the need to study the~$m < 1$ wormholes completely numerically somewhat obstructs the origin of the new branch of solutions and renders it difficult to completely scan the parameter space in a controlled way.  To shed some light into the qualitative features exhibited by Figures~\ref{fig:JTCFTJ10} and~\ref{fig:JTCFTJ20}, in Appendix~\ref{app:WCmodel} we study a simpler model of JT gravity coupled to end-of-the-world branes, similar to that considered in~\cite{PenShe19}.  This model can be studied analytically (up to a single transcendental equation), and we find that turning on a brane tension gives rise to stable wormholes for~$m > 1$ and to two branches of solutions for the wiggle when~$m < 1$, as we have seen in JT coupled to a massless scalar.  However, it does not exhibit the stable wormholes at~$m < 1$ we have found, so it is not sufficiently rich to exhibit our desired behavior.  Nevertheless, it is instructive in showing explicitly why two branches of solutions for the wiggle can exist when~$m < 1$.

\subsection{It was the Best of Saddles, it was the Worst of Saddles}
\label{subsec:quenched}

\begin{figure}[t]
\centering
\includegraphics[width=0.5\textwidth]{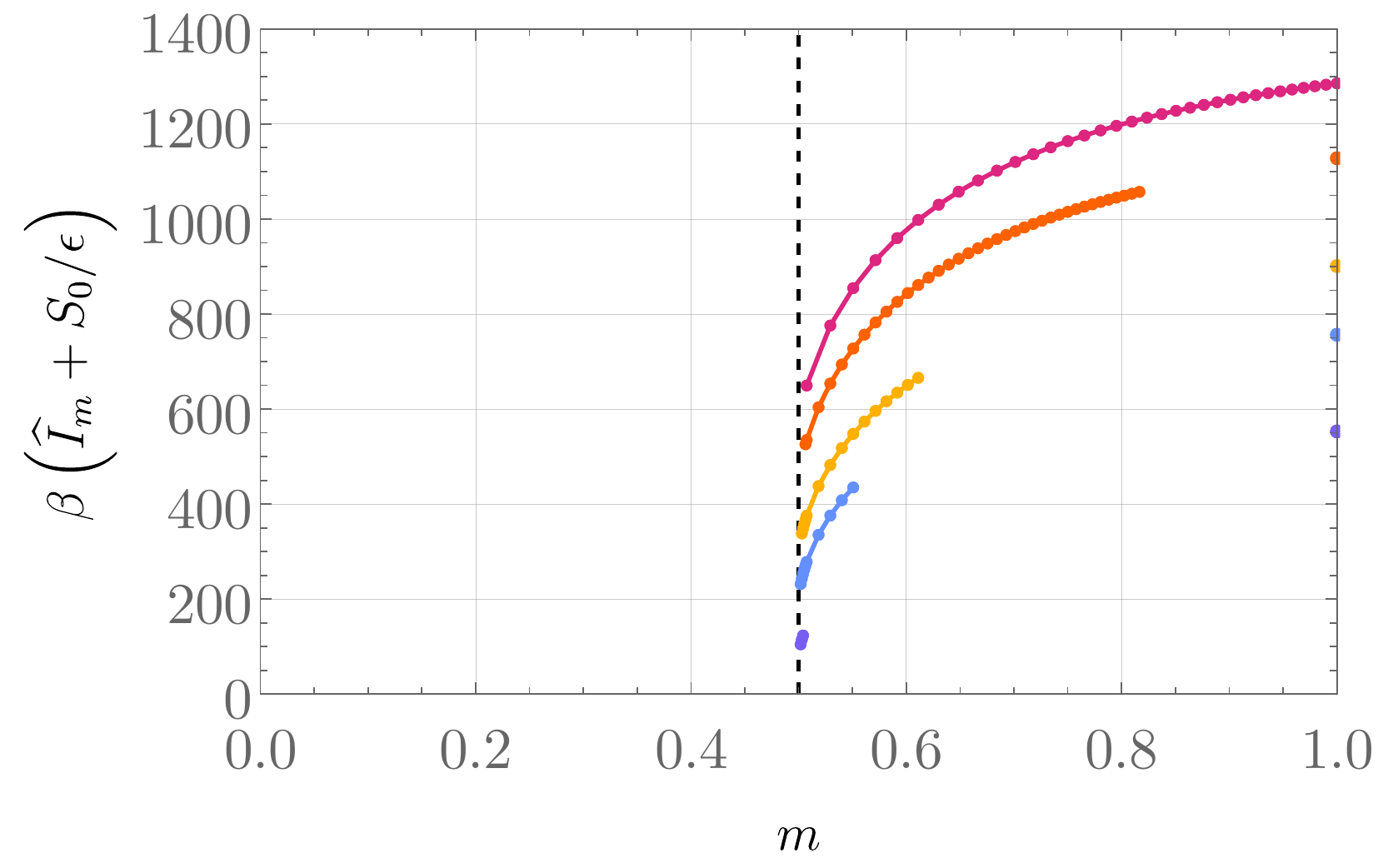}
\caption{The on-shell action~$\widehat{I}_m$ obtained by putting both the wiggle and the modulus on shell; from uppermost to lowermost, the curves correspond to~$\sqrt{\beta} J = 25$,~23,~20,~18, and~15.  Note that the stable saddles we have found do not persist below~$m = 1/2$, marked as a dashed black line, and we are unable to find further saddles at smaller values of~$m$.  For sufficiently large values of~$\sqrt{\beta} J$ saddles exist in the entire region~$m \in (1/2,1]$; decreasing~$\sqrt{\beta} J$ shrinks the region in~$m$ in which saddles exist, until for~$\sqrt{\beta} J \lesssim 15$ we find no saddles at all for~$m < 1$ (the isolated points on the right-hand side of the plot are the~$m = 1$ solutions, which exist for any value of~$\sqrt{\beta} J$).}
\label{fig:JTCFTaction}
\end{figure}

Assuming that we restrict to perturbations with a real Lorentzian section in the sense discussed above (in particular, with the defects contained on the moment of time symmetry of the Lorentzian section), we may now compute the saddle-point approximation of the effective action~$\widehat{I}_m$ by putting the modulus on-shell, and consequently obtain~$\Gamma_Q$.  The action~$\widehat{I}_m$ as a function of~$m$ is shown in Figure~\ref{fig:JTCFTaction}.  Besides the aforementioned fact that the saddles, when they exist, do not appear to persist below~$m \approx 1/2$ (at least in the parameter space we have been able to probe numerically), an additional noteworthy feature is that for intermediate values of~$\sqrt{\beta} J$ these saddles also do not extend all the way to~$m = 1$: there can be an isolated interval of saddles for~$1/2 < m < m_\mathrm{max} < 1$.  Regardless, the upshot is that in the classical~$S_0 \to \infty$ limit in which the dominance of geometries is controlled solely by the topological term in the action, the replica trick~\eqref{eq:lnZRSB} gives the quenched generating functional
\be
\label{eq:Ionehalf}
\Gamma_Q = -\widehat{I}_{1/2}.
\ee
We should of course be clear that our numerics are unable to determine that solutions stop existing \textit{precisely} at~$m = 1/2$, so the above equality should really be understood as taking the limit towards the leftmost endpoint of the curves shown in Figure~\ref{fig:JTCFTaction}.

An outstanding question worth considering is whether the minimization involved in the one-step RSB prescription is justified in the present context, given that the minimum of~$\widehat{I}_m$ at~$m = 1/2$ is not a local minimum, but rather a global extremum at the boundary of the space of solutions: that is,~$m_1 = 1/2$ is not a saddle of~$\widehat{I}_{m_1}$.  To some extent an answer to this question requires a more comprehensive understanding of the replica trick, but there is no obvious need for the minimization over~$m_1$ in the replica trick to be treated on the same footing as the other saddles.  Some intuition can be obtained from going back to the original wormhole geometries with integer~$m \geq 1$.  In these geometries,~$m_1$ is only allowed to take on a discrete set of values (namely, the divisors of~$m$), and a dominant solution is found by minimizing~$m_1$ over this discrete set; there is no sense in which we can look for ``saddles'' of~$m_1$.   When~$m$, and consequently~$m_1$, are continued away from the integers in the replica trick, it is reasonable to expect that the minimization over~$m_1$ should remain a mere global minimization with no requirement that~$m_1$ be a saddle.  (Put differently: we must look for saddles in the wiggle and the modulus~$\xi_0$ because these are degrees of freedom we integrate over in the path integral, and we approximate this integral by a saddle-point approximation.  On the other hand,~$m_1$ represents degrees of freedom that are \textit{summed} over in the path integral, i.e.~different topologies, so we only need to minimize with respect to~$m_1$ with no need for a saddle.)

\begin{figure}[t]
\centering
\includegraphics[height=0.2\textwidth]{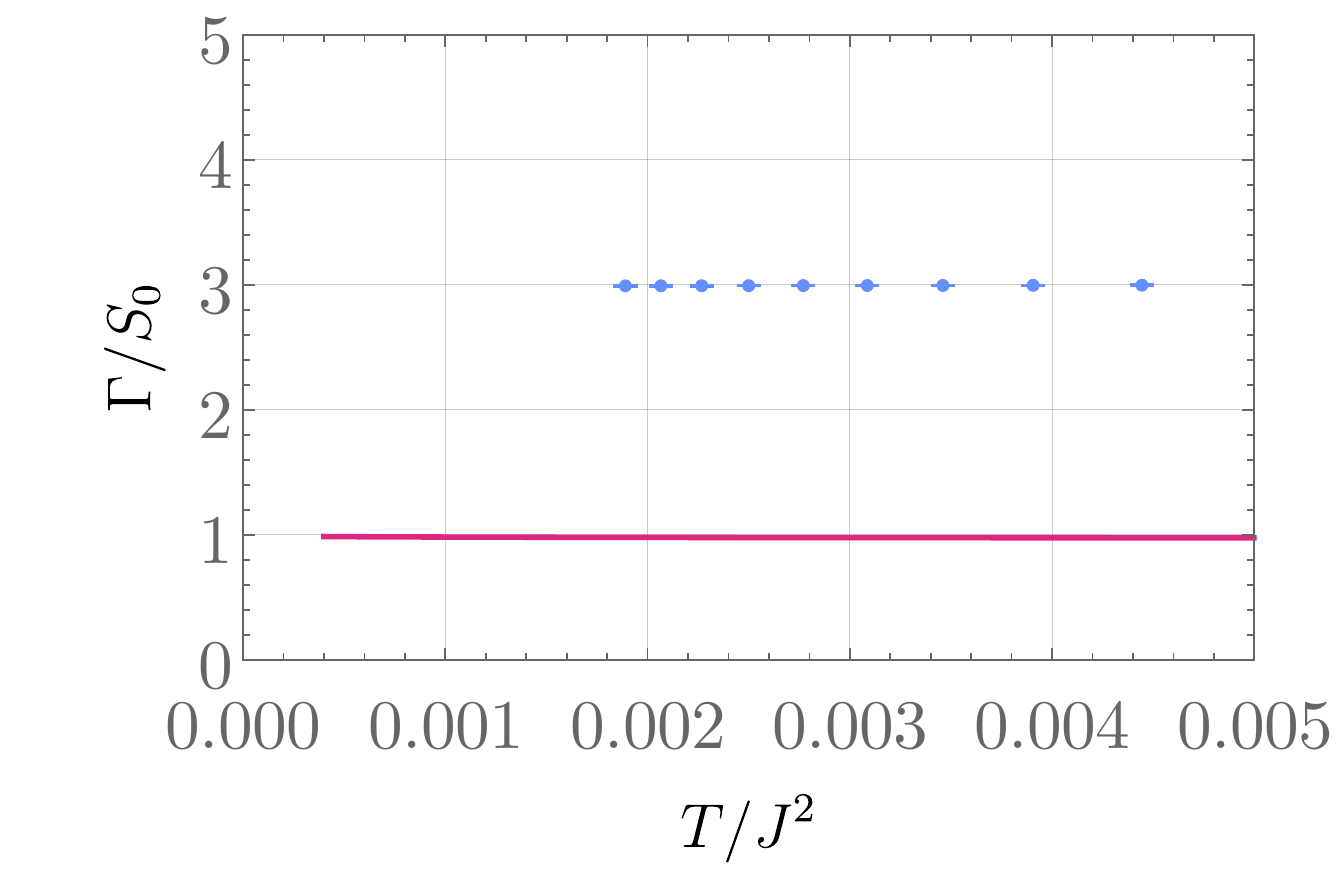}
\includegraphics[height=0.2\textwidth]{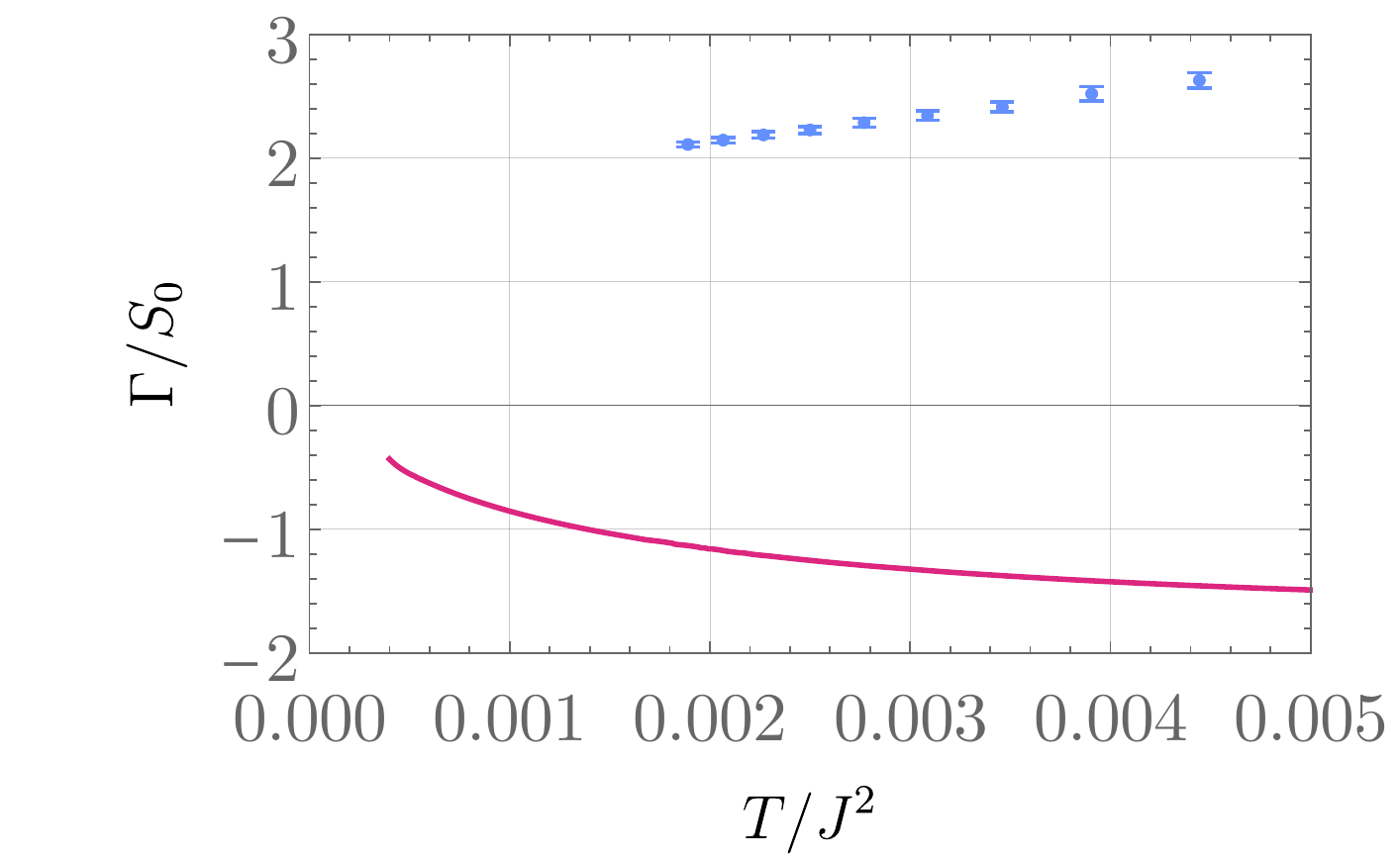}
\includegraphics[height=0.2\textwidth]{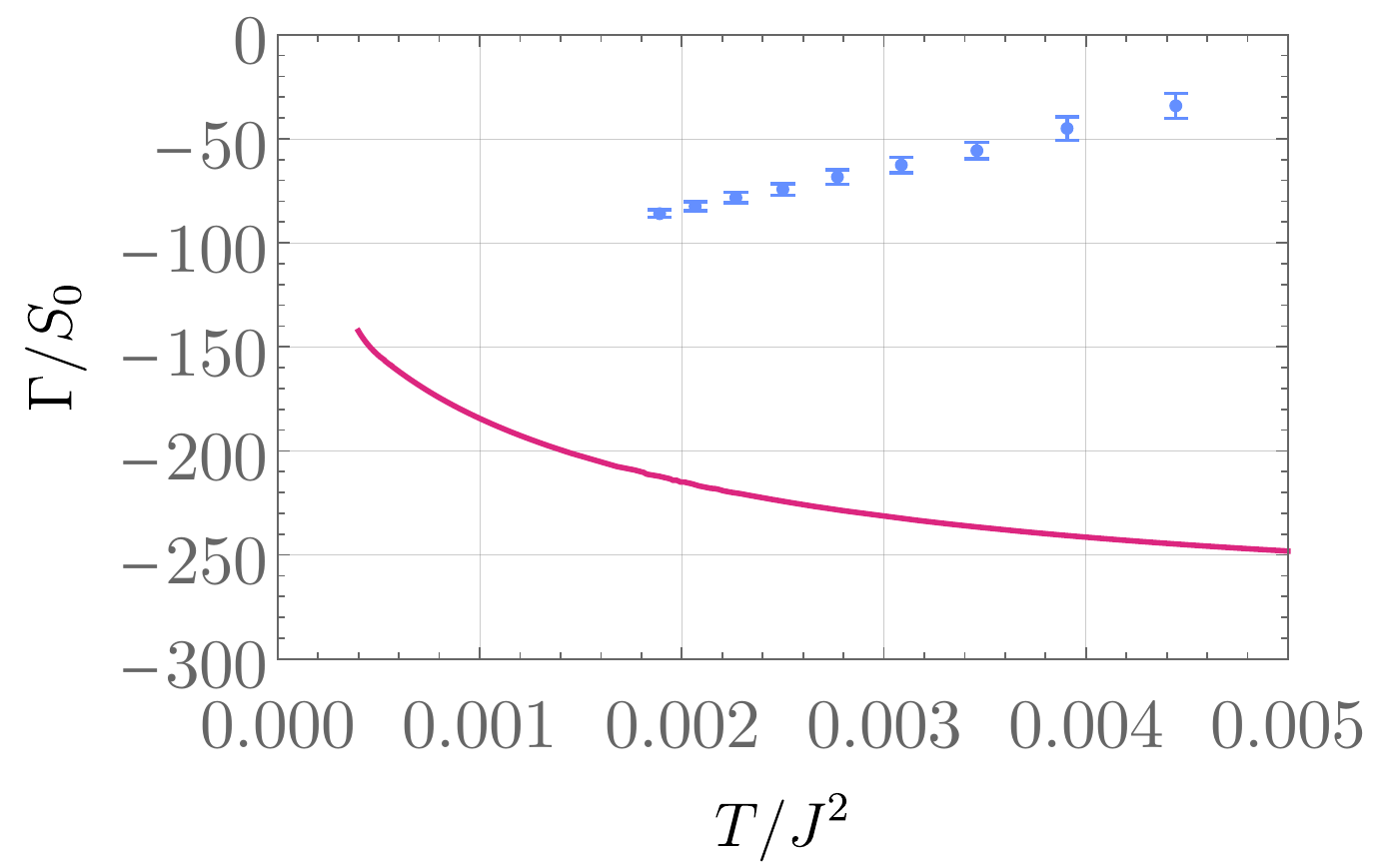}
\caption{The low-temperature quenched and annealed generating functionals~$\Gamma_Q$ (blue points) and~$\Gamma_A$ (red line).  From left to right, the plots correspond to~$J/\sqrt{S_0} = 0.1$,~$1$, and~$10$.  The error bars on the blue points are a rough estimate of the uncertainty introduced in extrapolating the data to~$m = 1/2$ as required in~\eqref{eq:Ionehalf}.}
\label{fig:JTCFTquenchedannealed}
\end{figure}

With this interpretation understood, we can now compare the quenched and annealed generating functionals using these new saddles at~$m = 1/2$: we show~$\Gamma_Q$ and~$\Gamma_A$ in Figure~\ref{fig:JTCFTquenchedannealed} as functions of the ``temperature''~$T \equiv 1/\beta$ (though recall that these are not thermal states).  Since our data is consistent with stable saddles for~$m < 1$ existing to arbitrarily low temperatures, we expect~$\Gamma_Q$ and~$\Gamma_A$ to continue to be distinct down to~$T = 0$; Figure~\ref{fig:JTCFTquenchedannealed} merely displays our results to the lowest temperatures for which we have generated data.  On the other hand, the apparent lack of stable saddles with~$m < 1$ at high temperatures implies that the only saddle that can contribute to~$\Gamma_Q$ is the disk, and so we would expect that at temperatures higher than those shown in Figure~\ref{fig:JTCFTquenchedannealed},~$\Gamma_Q = \Gamma_A$.  But if this were the case, it is clear from the figure that~$\Gamma_Q$ would be discontinuous at this transition temperature, which is pathological behavior!  This discontinuity stems from the fact that the~$m = 1/2$ saddles that contribute to~$\Gamma_Q$ do not smoothly exchange dominance with the~$m = 1$ saddle that defines~$\Gamma_A$, but rather they dominate \textit{immediately} as soon as they start existing.  What are we to make of this apparent discontinuity?

One possibility is to object to the~$m < 1$ saddles in the first place: after all, if we modify our stability criterion to require that saddles be stable under any real Euclidean perturbation, rather than just perturbations that admit a real Lorentzian section, then the~$m < 1$ saddles are unstable.  Without this stability condition, we would thus trivially find that~$\Gamma_Q = \Gamma_A$ at all temperatures.  However, we have discussed above our reasoning for taking  the requirement of stability under real Lorentzian perturbations seriously, so we do not find this objection compelling.  What we deem more likely is that the story so far is incomplete: as alluded to earlier, the structure of solutions to the equations of motion of the JT + CFT model at~$m < 1$ is richer than one might have otherwise expected, and unfortunately a numerical analysis cannot ensure that we have found all relevant saddles.  It may be that there are additional saddles we have missed at smaller values of~$\sqrt{\beta} J$ that allow~$F_Q$ to transition \textit{continuously} to~$F_A$ at high temperatures.  In fact, there could even be additional saddles that are not captured by our one-step RSB ansatz.  As we will see in Section~\ref{sec:quantum}, it may even be that quantum effects are needed to give rise to new semiclassical saddles, which may then yield a continuous~$\Gamma_Q$.  Or perhaps, since our states are far from thermal there is no reason to require~$\Gamma_Q$ to be continuous after all, and the discontinuity is simply an interesting quirk of the theory.

Because of this incompleteness, we remain agnostic on what the ``correct'' answer for the quenched generating functional is.  We claim that our main finding is not the particular functional form of~$\Gamma_Q$ shown in Figure~\ref{fig:JTCFTquenchedannealed}, but rather the existence of new saddles at~$m < 1$ whose stability properties depend crucially on whether or not perturbations about them are required to admit a real Lorentzian continuation.  Additional investigation, either into the space of saddles, the structure of the replica trick, or non-gravitational models of JT $+$ matter, is needed to determine what the right answer for $\Gamma_Q$ actually is.

\section{Quantum Corrections: The Adventures of Operator Twist}
\label{sec:quantum}

To assess whether our purely classical results are robust to semiclassical corrections, we now investigate the effect of turning on quantum corrections to the scalar field.  Because the scalar only couples to the wiggle through the boundary sources~$\psi_\partial$, and because these sources have already been incorporated into the classical analysis, excitations of the scalar about its classical background will obey homogeneous Dirichlet boundary conditions.  Such quantum excitations will not couple to the wiggle, so we need only concern ourselves with the coupling to the modulus.  To do so, we construct the effective matter action
\be
\label{eq:effact}
\widehat{I}^\mathrm{quant}[g_m] = -\frac{1}{m} \ln Z^\mathrm{quant}[g_m],
\ee
where~$Z^\mathrm{quant}[g_m]$ is the scalar partition function on the wormhole geometry~$g_m$, which will depend on the modulus~$D$.  Note that we have included the prefactor of~$1/m$ to match conventions with the classical quotient action~$\widehat{I}_m$.

After going to the quotient space, this partition function is computed in a standard way~\cite{CalCar09} by inserting appropriate twist operators at the locations of the conical defects:
\be
\label{eq:Sigmacorr}
Z^\mathrm{quant}[g_m] = \int \mathcal{D}\psi \, \Sigma(p_1) \Sigma^*(p_2) e^{-I_{\text{mat}}[\widehat{g}_m]} \equiv Z^\mathrm{quant}[\widehat{g}_m] \langle \Sigma(p_1) \Sigma^*(p_2)\rangle,
\ee
where~$p_1$ and~$p_2$ are the locations of the conical defects,~$\Sigma$ and~$\Sigma^*$ are the twist and anti-twist operators, which identify fields in the clockwise and anti-clockwise directions between the different Riemann sheets of the~$m$-copy replica manifold, and the normalization factor~$Z^\mathrm{quant}[\widehat{g}_m]$ is the matter partition function on the quotient geometry (with no twist operator insertions).  Because the massless scalar is a CFT, the partition function~$Z^\mathrm{quant}[\widehat{g}_m]$ can be computed straightforwardly from the conformal anomaly~$\int R[\widehat{g}_m]$, which is topological and hence independent of the modulus~$D$ (though it will depend on~$m$).  Thus the normalization factor only adds an~$m$-dependent additive constant to~$\widehat{I}^\mathrm{quant}$, so it can safely be ignored.  Instead, all nontrivial dependence on~$D$ is contained in the correlator~$\langle \Sigma(p_1) \Sigma^*(p_2)\rangle$.

In order to compute~$Z^\mathrm{quant}$, we will make two simplifications.  First, we will work perturbatively about~$m = 1$, since this is the regime in which the correlator~$\langle \Sigma \Sigma^*\rangle$ is most tractable.  Second, we will work in the limit of a CFT with large central charge.  This limit is compatible with the regime in which our classical~$m < 1$ saddles exist in the following sense.  In Section~\ref{sec:JTCFT} we worked with a single scalar field (central charge~$c = 1$ in appropriate conventions) with boundary sources with amplitude~$J$, giving an on-shell classical action proportional to~$J^2$.  But we could equally well have worked with, say, many independent scalar fields with total central charge~$c$, giving an on-shell classical action proportional to~$c J^2$.  Consequently, the results of Section~\ref{subsec:JTCFTmleq1} can be interpreted as showing that~$m < 1$ saddles can be supported by a large-$c$ CFT as long as~$c J^2$ is sufficiently large relative to~$1/\beta$.

With these assumptions understood, we turn to computing~$\langle \Sigma \Sigma^*\rangle$.  To do so, we first take the quotient geometry to be the unit disk with the defects located at~$z_{\pm} = \pm a$, as in the left diagram of Figure~\ref{fig:fundomainquotientgeometry}.  Here~$a$ is the modulus that sets the proper distance~$D$ between the defects.  The metric takes the form
\be
ds^2 = e^{2\sigma(z,\bar{z})} dz \, d\bar{z} = \frac{4}{(1-z\bar{z})^2} \left(1 + \Ocal(\vareps)\right) dz \,d\bar{z},
\ee
where we have defined~$\vareps \equiv 1-m$.  We then map the disk to the upper half-plane via the M\"obius transformation~$z = (1+iw)/(i+w)$, giving the metric
\be
ds^2 = \frac{1}{4} \, e^{2\sigma(z,\bar{z})} |z - i|^4 dw \, d\bar{w}.
\ee
Since the twist operators are primaries with scaling dimension~\cite{CalCar09}\footnote{Note that we are analytically continuing the scaling dimension from~$m > 1$ down to~$m < 1$, which should be valid since we are only interested in the behavior in a neighborhood of~$m = 1$ and we expect the scaling dimension to be analytic in~$m$ at $m = 1$.  Analogous arguments apply for the rest of this section.}
\be
\Delta_m = \frac{c}{12} \left(m - \frac{1}{m}\right) = - \frac{c}{6} \,\varepsilon + \mathcal{O}(\varepsilon^2),
\ee
the two-point function of twist operators on the quotient geometry~$\widehat{g}_m$ can be related to that on the flat upper half-plane by a standard scaling under Weyl and conformal transformations:
\bea
\langle \Sigma(a) \Sigma^*(-a) \rangle &= \left(\frac{1}{4} \, e^{\sigma(a,a) + \sigma(-a,-a)}(1+a^2)^2\right)^{-\Delta_m}  \langle \Sigma(w_+) \Sigma^*(w_-) \rangle_{\mathrm{UHP}}, \\
    &= \left[1 + \frac{c}{3} \ln \left(\frac{1+a^2}{1-a^2}\right) \vareps + \Ocal(\vareps^2)\right]\langle \Sigma(w_+) \Sigma^*(w_-) \rangle_{\mathrm{UHP}}, \label{subeq:Sigmacorrscaling}
\eea
where~$w_{\pm} = (1 \mp i a)/(\pm a - i)$ are the locations of the twist operator insertions in the upper half-plane.

We thus need only to compute the correlator~$\langle \Sigma(w_+) \Sigma^*(w_-) \rangle_{\mathrm{UHP}}$ on the (flat) upper half-plane.  This setting corresponds to working in a boundary CFT (BCFT); this will typically induce excitations on the boundary due to the breaking of full conformal symmetry, as realized by the presence of a bulk to boundary OPE.  To evaluate~$\langle \Sigma(w_+) \Sigma^*(w_-) \rangle_{\mathrm{UHP}}$, we can therefore follow~\cite{SulRaa20}.  The basic idea is that the product~$\Sigma(w_+) \Sigma^*(w_-)$ can be expanded in terms of an OPE.  Because we have a BCFT, there are two channels the OPE can be expanded in.  The first of these is the boundary channel, where we expand each bulk operator in terms of a bulk to boundary OPE, corresponding to the boundary operator spectrum of the bulk operator.  This would convert the twist two-point function into a sum over boundary two-point functions.  There is also a bulk channel in which we expand the product of bulk operators in terms of a bulk OPE; in this channel the two-point function becomes a sum over bulk one-point functions.  In general, these two channels are only valid in their respective OPE limits: the boundary channel is valid when the twist operators are brought closer to the boundary than they are to each other, while the bulk channel is valid when the twist operators are brought closer to each other than they are to the boundary. However, if we work in the~$c\rightarrow \infty$ limit then we can use the channels for all values of $\eta \in [0,1]$, with the two channels exchanging dominance at some $\eta_*$, where $\eta$ is the cross-ratio:
\be
\eta \equiv \frac{(w_+ - \overline{w}_+)(w_- - \overline{w}_-)}{(w_+ - \overline{w}_-)(w_- - \overline{w}_+)} = \left(\frac{1-a^2}{1+a^2}\right)^2.
\ee
Moreover, we can approximate the OPE expansion in a given channel by the identity block of that channel.

\begin{figure}[t]
\centering
\includegraphics[width=0.4\textwidth]{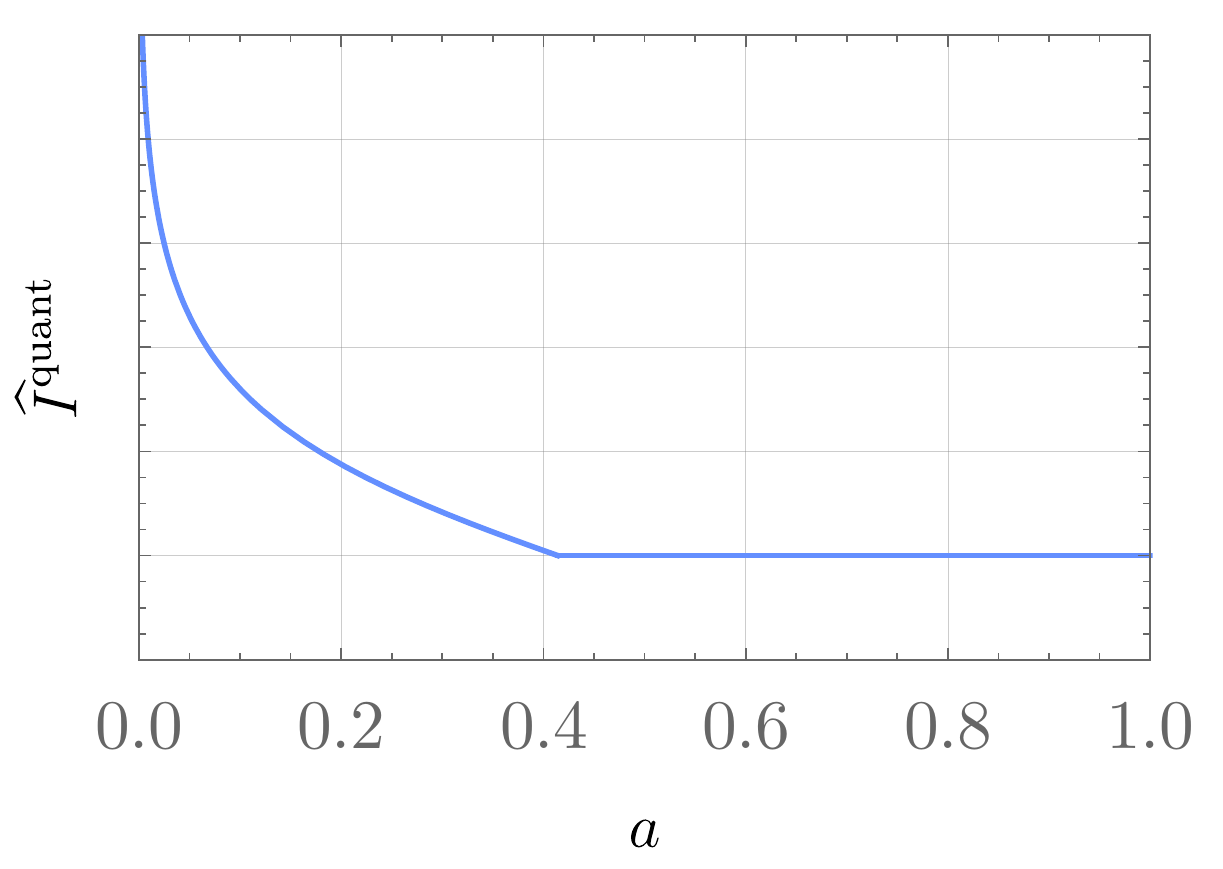}
\caption{The large-$c$, leading-order in~$m-1$ behavior of the quantum effective action~$I^\mathrm{quant}_m[a]$ for~$m < 1$ (up to various additive constants that we have ignored).  Note the divergence at~$a = 0$ due to a Casimir effect.}
\label{fig:effact}
\end{figure}

This calculation of~$\langle \Sigma(w_+) \Sigma^*(w_-) \rangle_{\mathrm{UHP}}$ in the $c\rightarrow \infty$ limit was done explicitly in~\cite{SulRaa20} to linear order in~$\varepsilon$, and yields
\be
\langle \Sigma(w_+) \Sigma^*(w_-) \rangle_{\text{UHP}} = 1 - \frac{\varepsilon c}{6}\left(2 \ln \epsilon_{\text{UV}} - \ln \min(\eta,1-\eta)  \right) + \mathcal{O}(\varepsilon^2),
\ee
where~$\eps_\mathrm{UV}$ is a cutoff that regulates the region around the operator insertions.  Consequently, using~\eqref{subeq:Sigmacorrscaling},~\eqref{eq:effact}, and~\eqref{eq:Sigmacorr} we obtain the matter effective action
\be
\label{eq:Iquant}
\widehat{I}_m^\mathrm{quant}[a] = -\frac{c (m-1)}{3} \max\left(0, \ln \left(\frac{1-a^2}{2a}\right)\right) + \Ocal(m-1)^2,
\ee
up to overall additive constants independent of~$a$.  This action is shown in Figure~\ref{fig:effact}.  Note that for~$a > \sqrt{2} -1$,~$\widehat{I}_m^\mathrm{quant}[a]$ is independent of~$a$: this is to be expected from the fact that~$a > \sqrt{2} - 1$ corresponds to the boundary OPE channel~$\eta < 1/2$ in which the two-point function is dominated by the proximity of the twist operators to the boundary rather than to each other.  When mapped back to hyperbolic space the distance from the twist operators to the boundary is renormalized to a constant, and hence we expect that correlators dominated by this channel should be independent of~$a$.

For~$m > 1$,~$\widehat{I}^\mathrm{quant}_m[a]$ is negatively divergent at~$a = 0$.  For the double-trumpet case~$m = 2$, this behavior (exhibited by bosonic matter) is well-understood and is due to a divergent Casimir energy as the throat of the wormhole pinches off.  This divergence is extremely destabilizing: the quantum effective action is unbounded below, suggesting that a classically-stable double-trumpet cannot remain globally stable under the inclusion of semiclassical effects; see e.g.~\cite{MoiSak19,MoiSak21}. The action~\eqref{eq:Iquant} suggests that this destabilizing Casimir energy persists for wormholes with~$m > 1$, and presumably means that any classical saddles with~$m > 1$ do not remain well-defined saddles of a semiclassical theory (though they may remain metastable).

However, for the~$m < 1$ wormholes relevant to the quenched generating functional~$\Gamma_Q$,~$\widehat{I}^\mathrm{quant}_m$ is \textit{positively} divergent at~$a = 0$.  Hence quantum effects do not render these wormholes pathologically unstable.  Whether or not a classical saddle remains stable under the inclusion of quantum corrections then depends on the details of the combined action~$\widehat{I}_m[a] + \widehat{I}_m^\mathrm{quant}[a]$.  For instance, a classical saddle at some~$a > \sqrt{2}-1$ will remain a semiclassical saddle, but a classical saddle at~$a < \sqrt{2}-1$ may or may not remain a semiclassical saddle.  Conversely, it is possible for semiclassical saddles to appear even when no classical ones existed: for example, if~$d\widehat{I}_m/da > 0$ for all~$a$, then no classical saddles exist, but the fact that~$d\widehat{I}^\mathrm{quant}_m/da \to -\infty$ at~$a = 0$ (and that~$d\widehat{I}^\mathrm{quant}_m/da = 0$ for~$a > \sqrt{2}-1$) is sufficient to ensure that a semiclassical saddle will exist.  Indeed, using the fact that~$a$ is roughly inversely related to the parameter~$\xi_0$ used in earlier sections, it is tantalizing to note that of the plots shown in Figures~\ref{fig:JTCFTJ10} and~\ref{fig:JTCFTJ20},~$d\widehat{I}_m/da > 0$ on all branches with no stable modulus but with a stable wiggle.  Hence it is entirely possible that semiclassical effects will stabilize some of the classically unstable wormholes we have looked at -- this may even be sufficient to remedy the apparent discontinuity of~$\Gamma_Q$ in Figure~\ref{fig:JTCFTquenchedannealed}.

Investigating these possibilities in more detail will require us to construct the quantum effective action away from a neighborhood of~$m = 1$, which we leave to future work.  The upshot is that the same quantum effects that manifest pathologies in~$m > 1$ wormholes seem to be benign, or even beneficial, in the~$m < 1$ wormholes necessary to construct the quenched generating functional.

\section{Discussion: Great Expectations}
\label{sec:disc}

We have investigated the potential contributions of connected saddles to the quenched generating functional $\Gamma_{Q}$ in the semiclassical approximation by developing an LM-inspired procedure for replica symmetry breaking that admits continuation to zero replicas. Using this technology, we have shown that in a model of JT gravity coupled to a massless scalar, a computation of~$\Gamma_Q$ reveals the existence of a new on-shell wormhole that dominates over the usual disconnected contribution that gives the annealed result~$\Gamma_A$.  This new wormhole is unstable to arbitrary Euclidean perturbations, but it is stabilized by restricting to perturbations that admit a real Lorentzian section with a moment of time symmetry on which the conical defects live.  Moreover, we have shown that quantum effects do not destabilize this wormhole in the same way that they destabilize conventional wormholes with~$m > 1$, at least when working perturbatively about~$m =1$.  In fact, they may even have a stabilizing influence, potentially making semiclassical saddles appear where no classical ones existed.  

It is not clear whether the (in)stability of the new saddle is a feature or a problem.  On the one hand, the Euclidean gravitational path integral is notorious for its superior intelligence compared with its Lorentzian counterpart, as manifested in e.g.~computations of the black hole entropy by Gibbons and Hawking \cite{GibHaw77a}.  Perhaps, then, we should take a purely Euclidean perspective, excluding saddles like the connected one we have found that are unstable under some Euclidean perturbations.  On the other hand, due to the conformal mode problem, a strictly Euclidean gravitational path integral is in fact divergent (and in JT gravity, inconsistent with canonical quantization): the contour of integration of the conformal mode (or, in JT gravity, the dilaton) needs to be Wick rotated to the imaginary axis to give sensible results. Perhaps this is a hint that a Lorentzian treatment is more fundamental after all, as suggested in~\cite{Mar22}.  If so, then only behavior under Lorentzian perturbations is relevant, and the quenched generating functional should be dominated by the connected saddle after all.

Ultimately, whether or not the new saddle should contribute to~$\Gamma_Q$ should be diagnosed by whether its inclusion yields the desired physics.  In prior discussions of replica wormholes, arguments against the inclusion of Euclidean wormholes were countered by appeal to unitarity.   Is there an analogous fundamental physical guiding principle that justifies (or excludes) these wormholes from contributing to the replica trick~\eqref{eq:lnZRSB} for~$\Gamma_Q$?  Unfortunately, any such principle is not immediately evident. Because the states that we consider are not thermal due to matter sources that break the~$U(1)$ Euclidean time-translation symmetry, the generating functional is no longer required to obey typical constraints of thermal states. Some other constraints do exist of course: for instance, correlation functions computed from the generating functional must satisfy a number of properties, from large-distance (or late-time) behavior to triangle inequalities. Some of these are manifest already semiclassically, while others (e.g.~very late-time recurrences) are expected to be inherited from the microscopic, nonperturbative description.  An investigation of whether or not these constraints are satisfied by certain saddles requires an analysis of observables computed from the generating functional. For instance,~\cite{Saa19} found that a full nonperturbative calculation of correlation functions in JT gravity reproduces the correct late time recurrences predicted in~\cite{Mal01}; the problem -- often dubbed the Maldacena information paradox -- is that the standard semiclassical gravity calculation predicts no late time recurrences. It would be interesting to investigate whether replica wormhole contributions to the generating functional can resolve this tension in a similar way to how connected topologies resolved the tension between the Page and Hawking calculations of the entropy of Hawking radiation. 

Alternatively, some guidance could be provided by appealing to dual models.  For example, pure JT gravity is dual to a double-scaling limit of a matrix model; such models give nonperturbative completions of JT that can be used to explicitly compute the quenched generating functional and confirm that it differs from its annealed counterpart at low temperatures~\cite{Joh19,Joh20a,Joh20b,Joh20c,Joh21a,Joh21b,Joh21c,Joh22a,Joh22b}.  Is there an analogous dual to JT gravity coupled to matter that would allow us to compute the quenched free energy directly -- without reference to replica tricks or wormholes -- in a semiclassical limit?  For example, are there sources we could turn on in the SYK model that would approximately reproduce, in an appropriate low-energy limit, the Schwarzian theory of JT coupled to matter?  If so, then an explicit computation of the quenched generating functional should reveal whether or not our prescription for computing~$\Gamma_Q$ via the replica trick should include contributions from the new saddles we have found.

Identifying the contributing saddle to the generating functional is part of a larger quest to understand the rigorous underpinning of the gravitational path integral in general. Without a guiding principle such as unitarity for the von Neumann entropy to help us in picking the correct saddle for  $\Gamma_Q$, for now we simply raise this question and bring up the possibility of future investigations into stability of saddles as a potential avenue for furthering our understanding of the gravitational path integral.

\acknowledgments

It is a pleasure to thank Ahmed Almheiri, Tarek Anous, Alejandra Castro, Ping Gao, Daniel Harlow, Matt Headrick, Kristan Jensen, Clifford Johnson, Arjun Kar, Adam Levine, Alex Maloney, Jason Pollack, Mykhaylo Usatyuk, Herman Verlinde, and Wayne Weng for stimulating discussions. VC is supported by a grant from the Simons Foundation (816048, VC). NE is supported in part by NSF grant no. PHY-2011905, by the U.S. Department of Energy Early Career Award DE-SC0021886, by the John Templeton Foundation and the Gordon and Betty Moore Foundation via the Black Hole Initiative, by DOE grant no. DE-SC0012647, and by funds from the MIT department of physics. Research of SF is supported in part by the Simons Foundation Grant No.~385602 and the Natural Sciences and Engineering Research Council of Canada
(NSERC), funding reference number SAPIN/00032-2015. 
SHC is supported by NSF grant PHY-2107939 and by funds from UCSB.

\appendix

\section{Liouville Equation}
\label{app:Liouville}

In this Appendix we fill in some of the details omitted from Section~\ref{subsec:liouville}.  On the rectangular domain shown in Figure~\ref{fig:ellipse}, the conformal factor~$\sigma$ obeys the Liouville equation~\eqref{eq:preliouville} subject to the requirements that~$\sigma$ diverge at the conformal boundary~$\xi = \pi \xi_0/2$; that~$\phi = 0$ and~$\phi = 2\pi$ be identified; and that~$\sigma$ be continuous in the interior of the ellipse in the~$z$ plane.  This last condition imposes boundary conditions at the~$\xi = 0$ edge of the coordinate rectangle, which is mapped to a double-cover of the line segment connecting~$z = \pm 1$: it requires that for all nonnegative integer~$p$,
\be
\label{eq:sigmasmooth}
\partial_\xi^p \sigma(\xi = 0,\phi) = (-1)^p \partial_\xi^p \sigma(\xi = 0,-\phi).
\ee
These various boundary conditions make it convenient to extend~$\sigma$ from the coordinate rectangle to the infinite strip~$\xi \in (-\pi \xi_0/2,\pi \xi_0/2)$,~$\phi \in (-\infty,\infty)$.  The point is that the~$\mathbb{Z}_2 \times \mathbb{Z}_2$ symmetry of the ellipse implies that on the strip~$\sigma$ is symmetric about the lines~$\xi = 0$ and~$\phi = k\pi/2$ for all integer~$k$, which is sufficient to ensure that~$\sigma$ is appropriately periodic in~$\phi$ and that it obeys~\eqref{eq:sigmasmooth}.  Since the strip contains infinitely many images of the foci at~$z = \pm 1$ under the map~\eqref{eq:ellipticcoords}, the Liouville equation takes the form\footnote{The infinite delta functions effectively ensure that the problem has the desired periodicity in~$\phi$.}
\be
\label{eq:liouvilleimages}
-4\partial_u \partial_{\bar{u}} \sigma + e^{2\sigma} = -\frac{2\pi}{\nu} \sum_{k = -\infty}^\infty \delta_{ik\pi}(u,\bar{u}),
\ee
where recall that~$\nu \equiv m/(2-m)$.  It is convenient to absorb the singularities, both from the delta functions and the boundary condition at the conformal boundary, into~$\sigma$ by defining
\be
\sigma \equiv \tilde{\sigma} + \ln\left(\frac{1}{\xi_0} \sec \left(\frac{u + \bar{u}}{2\xi_0} \right) \right) + \frac{1}{\nu} \sum_{k = -\infty}^\infty G_{ik\pi}(u,\bar{u}),
\ee
where~$G_{u_*}(u,\bar{u})$ is a Green's function of the Laplacian.  The Green's function term absorbs the delta functions in~\eqref{eq:liouvilleimages}, while the logarithmic term ensures that~$\sigma$ diverges at the conformal boundary if~$\tilde{\sigma}$ is regular there.  Its particular form is chosen to be the conformal factor of the double-trumpet, so that for~$m = 2$ (or~$\nu \to \infty$) the Liouville equation is solved by~$\tilde{\sigma} = 0$.

It is in fact convenient to take~$G_{u_*}$ to be the Dirichlet Green's function on the strip:
\be
4\partial_u \partial_{\bar{u}} G_{u_*}(u,\bar{u}) = 2\pi \delta_{u_*}(u,\bar{u}), \qquad G_{u_*}\left(\pm \frac{\pi \xi_0}{2} + i\phi,\pm \frac{\pi \xi_0}{2} - i\phi\right) = 0.
\ee
An explicit form of~$G_{u_*}$ can be obtained by starting with the Dirichlet Green's function on the upper half-plane,~$G_{w_*}(w,\overline{w}) = \ln \left|(w - w_*)/(w - \overline{w}_*)\right|$, and then mapping the upper half-plane to the strip via~$w = i e^{-iu/\xi_0}$, yielding
\be
G_{u_*}(u,\bar{u}) = \ln \left| \frac{\sin \left(\frac{u-u_*}{2\xi_0}\right)}{\cos \left(\frac{u+\bar{u}_*}{2\xi_0}\right)}\right|.
\ee
The Liouville equation consequently yields~\eqref{subeq:shiftedliouville}.  Moreover, since by construction~$H$ vanishes at the conformal boundary~$\xi = \pi\xi_0/2$, regularity of~$\tilde{\sigma}$ in fact requires that~$\tilde{\sigma}$ \textit{vanish} there, giving the boundary conditions~\eqref{subeq:sigmatbndry}.

\subsection{Approximate Solutions}
\label{subapp:Liovilleapprox}

To construct the function~$f$ defined by~\eqref{eqs:fdef}, first note that~\eqref{eqs:fdef} can be reduced to an ODE: exchanging~$\xi$ for a new variable~$x \equiv \sech(\phi/\xi_0)\cos(\xi/\xi_0) \in (0,1)$, we obtain
\begin{multline}
\left[x^2(x^2-1) \partial_x^2 + 2 x^3 \partial_x + \xi_0 x^2 \cosh^2\left(\frac{\phi}{\xi_0}\right) \left( 2x \tanh \left(\frac{\phi}{\xi_0}\right) \partial_x\partial_\phi - \partial_\phi^2 \right) \right] f \\ + \left(\frac{1-x}{1+x}\right)^{1/\nu} e^{2f} -1 = 0,
\end{multline}
while the boundary conditions require~$f(x = 0,\phi) = 0$ and that~$f$ be regular at~$x = 1$. Consequently it is consistent to take~$f$ to be a function of~$x$ only, satisfying
\be
\label{eq:Liouvillefy}
\left( x^2(x^2-1) \partial_x^2 + 2 x^3 \partial_x \right) f + \left(\frac{1-x}{1+x}\right)^{1/\nu} e^{2f} -1 = 0.
\ee
We now take~$\nu$ to be small to construct a solution to this equation.  In such a case, two scaling regimes emerge depending on the behavior of~$((1-x)/(1+x))^{1/\nu}$: when~$x/\nu$ is large,~$((1-x)/(1+x))^{1/\nu}$ is nonperturbatively small in~$\nu$, while when~$x/\nu$ is order unity or smaller,~$((1-x)/(1+x))^{1/\nu}$ can be expanded perturbatively in~$\nu$.  We can therefore obtain a solution for~$f$ perturbatively in~$\nu$ by performing a matched asymptotic expansion: we solve for~$f$ in these two regimes and match the solutions together.

Large~$x/\nu$ corresponds to the interior of the quotient geometry; in this regime, the penultimate term in~\eqref{eq:Liouvillefy} can be ignored, leaving simply
\be
\left( x^2(x^2-1) \partial_x^2 + 2 x^3 \partial_x \right) f_\mathrm{int} -1 = 0.
\ee
Requiring that~$f_\mathrm{int}$ be regular at~$x = 1$ gives the family of solutions
\be
f_\mathrm{int}(x) = C + \ln\left(\frac{x}{1+x}\right),
\ee
where~$C$ is an arbitrary constant (that may depend on~$\nu$).  On the other hand,~$x/\nu$ of order unity or smaller corresponds to a thin layer near the conformal boundary.  Here we define~$y \equiv x/\nu$ and expand~\eqref{eq:Liouvillefy} in~$\nu$ at fixed~$y$:
\be
\label{eq:Liouvillefbndry}
\left( -y^2 \partial_y^2 + \nu^2 y^3 \left(y \partial_y^2 + 2 \partial_y\right)\right) f_\mathrm{bndry} + \left(1-\frac{2y^3}{3}\,\nu^2 + \Ocal(\nu^4)\right) e^{2(f_\mathrm{bndry}-y)} -1 = 0.
\ee
We then look for solutions order-by-order in~$\nu$ by writing
\be
f_\mathrm{bndry}(y) = \sum_{n = 0}^\infty \nu^n f_n(y).
\ee
Plugging this expansion into~\eqref{eq:Liouvillefbndry} and imposing the boundary condition~$f_\mathrm{bndry}(0) = 0$ allows us to solve for the~$f_n$ order-by-order.  For example, to~$\Ocal(\nu)$ we have
\be
f_0 = y -\ln\left(\frac{\sinh(\alpha_0 y)}{\alpha_0 y}\right), \qquad f_1 = \alpha_1\left(\alpha_0 y \coth(\alpha_0 y) - 1\right),
\ee
where~$\alpha_0 > 0$ and~$\alpha_1$ are arbitrary constants (independent of~$\nu$).  One obtains a new constant~$\alpha_n$ at each order in~$\nu$.

The constants of integration~$C$ and~$\alpha_n$ are fixed by matching the solutions~$f_\mathrm{int}$ and~$f_\mathrm{bndry}$ in the overlap region: that is, we require that the expansion of~$f_\mathrm{int}(x)$ at small~$x = \nu y$ agree with the expansion of~$f_\mathrm{bndry}(y)$ at large~$y$.  To~$\Ocal(\nu)$, the relevant expansions are
\bea
f_\mathrm{int} &= C + \ln \nu + \ln y - \nu y + \Ocal(\nu^2), \\
f_\mathrm{bndry} &= y (1-\alpha_0) + \ln y + \ln (2\alpha_0) + \nu \alpha_1 \left(\alpha_0 y - 1\right) + \Ocal(\nu^2,e^{-\alpha_0y}),
\eea
and hence matching order-by-order in~$y$ and~$\nu$ gives~$\alpha_0 = 1$,~$\alpha_1 = -1$, and~$C = \ln(2/\nu) + \nu + \Ocal(\nu^2)$.  We can then form a composite solution by superimposing the interior and near-boundary solutions:
\be
f = f_\mathrm{int} + f_\mathrm{bndry} - f_\mathrm{overlap},
\ee
where~$f_\mathrm{overlap}$ is the common behavior shared by~$f_\mathrm{int}$ and~$f_\mathrm{bndry}$ in the matching region.  Proceeding in this manner up to order~$\nu^3$ ultimately yields~\eqref{eq:fexplicit} in the main text.  To then compute the magnitude of~$\delta \tilde{\sigma}$, we use~\eqref{eq:fexplicit} to quantify the falloff of~$F$ outside the interval~$\phi \in (-\pi/2,\pi/2)$: for~$\phi$ outside this interval,~$y$ is~$\Ocal(\nu^{-1} e^{-|\phi|/\xi_0})$, which is small if~$e^{-\pi/2\xi_0} \ll \nu$.  Expanding~$F$ at small~$y$ gives~$F = \Ocal(y^2)$, which leads to the right-hand side of~\eqref{eq:deltasigmalinearized} being~$\Ocal(\nu^{-4} e^{-2\pi/\xi_0})$, as claimed.

\section{Numerical Details}
\label{app:numerics}

In this Appendix we briefly go into some details on the numerical approaches we use both in solving the Liouville equation for the conformal factor~$\sigma$ as in solving the equation of motion~\eqref{eq:JTCFTeom} for the boundary wiggle in JT coupled to a massless scalar field.  For a more detailed description of some of the pseudospectral discretization methods we implement, as well as of the Newton-Raphson method for solving nonlinear problems, see e.g.~\cite{Trefethen,DiaSan15}.

\subsection{Liouville Equation}
\label{subapp:Liouville}

We are interested in solving the Liouville equation~\eqref{subeq:shiftedliouville}, which we reproduce here for convenience:
\be
-\left(\partial_\xi^2 + \partial_\phi^2\right) \tilde{\sigma} + \frac{1}{\xi_0^2} \sec^2 \left(\frac{\xi}{\xi_0}\right)\left(e^{2H(\xi,\phi)/\nu} e^{2 \tilde{\sigma}} - 1 \right) = 0,
\ee
where~$H(\xi,\phi)$ is defined in~\eqref{subeq:Hdef}.  The coordinate domain corresponding to the ellipse is given by~$\xi \in [0,\pi\xi_0/2)$,~$\phi \in [0,2\pi)$, and we require~$\tilde{\sigma}|_{\xi = \pi\xi_0/2} = 0$ and for~$\tilde{\sigma}$ to be regular everywhere in the interior of the ellipse except potentially at the foci~$(\xi,\phi) = (0,0)$ and~$(0,\pi)$.  In fact, since we expect~$\tilde{\sigma}$ to be symmetric about the major and minor axes of the ellipse, it suffices to instead solve the equation on just the quarter-ellipse~$\xi \in [0,\pi\xi_0/2)$,~$\phi \in [0,\pi/2]$.  Symmetry about the major axis of the ellipse is imposed by Neumann boundary conditions at~$\phi = 0$ and~$\xi = 0$, while symmetry about the minor axis is impose by a Neumann condition at~$\phi = \pi/2$, so the boundary conditions on the computational domain are
\be
\partial_\xi \tilde{\sigma}|_{\xi = 0} = 0, \quad \tilde{\sigma}|_{\xi = \pi\xi_0/2} = 0, \quad \partial_\phi \tilde{\sigma}|_{\phi = 0} = 0, \quad \partial_\phi \tilde{\sigma}|_{\phi = \pi/2} = 0.
\ee
We are thus left with a nonlinear elliptic boundary-value problem which can be solved by standard methods; we implement a Newton-Raphson nonlinear problem solver after discretization using pseudospectral methods with Chebyshev grids in both the~$\xi$ and~$\phi$ directions.  For the data discussed below, we used a grid size of~$100$ points in~$\xi$ and~$101$ points in~$\phi$, which is sufficient to reach machine precision except at very small values of~$\nu$ and~$\xi_0$, where large gradients make the numerics poorly-behaved.

\begin{figure}[t]
\centering
\includegraphics[width=0.5\textwidth]{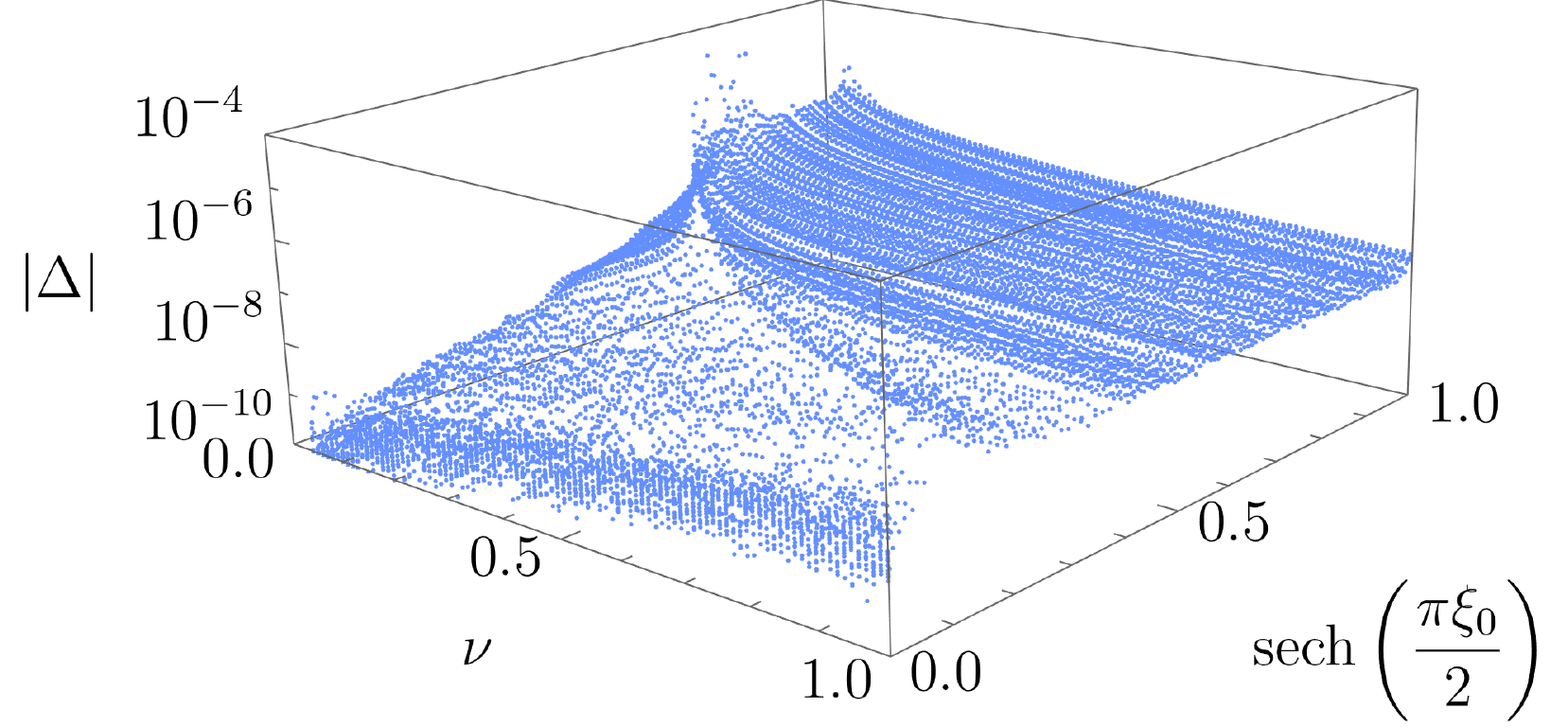}
\caption{The quantity~$\Delta$ illustrating the~$\phi$-independence of~$C(\nu,\xi_0)$; $\Delta \lesssim 10^{-7}$ in almost the entire parameter range, consistent with~$\Delta = 0$ within the resolution of our numerics.  The apparent spike in~$\Delta$ appears around a region where~$C(\nu,\xi_0)$ vanishes, and where~$\Delta$ is consequently a poor diagnostic of the~$\phi$-independence of~$C(\nu,\xi_0)$.  Nevertheless, even in that region~$\Delta$ never exceeds~$10^{-2}$ or so.}
\label{fig:CDelta}
\end{figure}

Once a solution for~$\tilde{\sigma}$ has been obtained, we can extract the near-boundary metric function~$g_2(\phi)$ appearing in~\eqref{eq:nearboundaryexpansion}:
\be
\label{eq:g2numerical}
g_2(\phi) = \partial_\xi^2 \tilde{\sigma}|_{\xi = \pi \xi_0/2} + \frac{1}{3\xi_0^2}.
\ee
As discussed in Section~\ref{subsec:JTCFTboundary}, comparison between our numerical results and analytic approximations reveals that
\be
g_2(\phi) = -\frac{1+2\nu}{3\nu^2 \xi_0^2} \sum_{k = -\infty}^\infty \sech^2\left(\frac{\phi - k \pi}{\xi_0}\right) + C(\nu,\xi_0),
\ee
where~$C(\nu,\xi_0)$ is independent of~$\phi$ to within the accuracy of our numerics.  In particular, the difference between the analytic approximation~\eqref{eq:sigmaapprox} and the numerical solution of~\eqref{subeq:sigmatbndry} \textit{does} exhibit nontrivial dependence on~$\phi$, as can be seen in the final plot of Figure~\ref{fig:Liouvillenumerics}; it is only in~$g_2(\phi)$ that the difference becomes independent of~$\phi$.  In Figure~\ref{fig:CDelta} we show the~$\phi$-independence of~$C(\nu,\xi_0)$ by plotting the relative difference
\be
\Delta \equiv 1 - \frac{\displaystyle \min_\phi \left[g^\mathrm{num}_2(\phi) + \frac{1+2\nu}{3\nu^2 \xi_0^2} \sum_{k = -\infty}^\infty \sech^2\left(\frac{\phi - k \pi}{\xi_0}\right)\right]}{\displaystyle \max_\phi \left[g^\mathrm{num}_2(\phi) + \frac{1+2\nu}{3\nu^2 \xi_0^2} \sum_{k = -\infty}^\infty \sech^2\left(\frac{\phi - k \pi}{\xi_0}\right)\right]},
\ee
where~$g^\mathrm{num}_2(\phi)$ is extracted from the numerical solutions of the Liouville equation via~\eqref{eq:g2numerical}.  $\Delta$ is order~$10^{-7}$ or smaller in the entire parameter range (except for a narrow region around where~$C(\nu,\xi_0)$ vanishes, where~$\Delta$ is consequently a poor diagnostic of the~$\phi$-independence of~$C(\nu,\xi_0)$).  This is a consistent with a value of~$\Delta = 0$ to the resolution of our numerics, and hence with~$C(\nu,\xi_0)$ being independent of~$\phi$.  Because numerical errors in our computation of~$\tilde{\sigma}$ (and hence~$g_2^\mathrm{num}$) are largest near~$\phi = 0$, in practice we determine~$C(\nu,\xi_0)$ by evaluating at~$\phi = \pi/2$:
\be
C(\nu,\xi_0) = g^\mathrm{num}_2(\pi/2) + \frac{1+2\nu}{3\nu^2 \xi_0^2} \sum_{k = -\infty}^\infty \sech^2\left(\frac{(2k-1)\pi}{2\xi_0}\right).
\ee

\begin{figure}[t]
\centering
\includegraphics[width=0.6\textwidth]{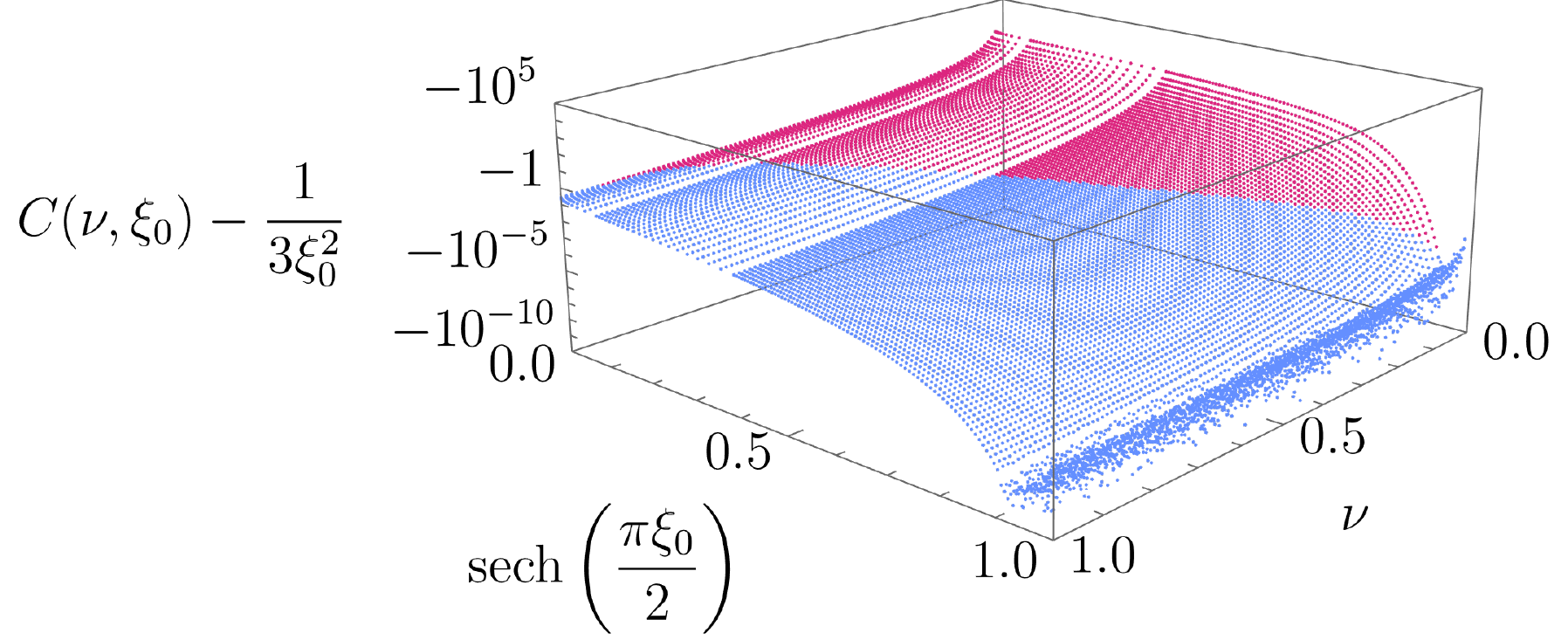}
\caption{The difference~$C(\nu,\xi_0) - 1/(3\xi_0^2)$, expected to be small when~$\nu$ is large, and also when~$\xi_0$ and~$\nu$ are small with~$e^{-\pi/4\xi_0} \lesssim \nu$.  This difference becomes very small at small~$\xi_0$, where it is eventually limited by machine precision.  Points colored in blue (red) lie within (outside) the region~$e^{-\pi/4\xi_0} < \nu$.}
\label{fig:Cdiff}
\end{figure}

As a check, we may compare the behavior of~$C(\nu,\xi_0)$ extracted from the numerics with the analytic approximations in various limiting cases.  For example, when~$\xi_0$ and~$\nu$ are small with~$e^{-\pi/4\xi_0} \lesssim \nu$, we have that~$C(\nu,\xi_0) = 1/(3\xi_0^2) + \Ocal(\nu^2)$; likewise we have~$C(\nu,\xi_0) = 1/(3\xi_0^2)$ exactly when~$m = 2$ (corresponding to~$\nu = \infty$); and in the limit~$\xi_0 \to \infty$ we have~$C(\nu,\xi_0) \to -1/(3\nu^2)$.  Figure~\ref{fig:Cdiff} shows the difference~$C(\nu,\xi_0) - 1/(3\xi_0^2)$; as expected, it is very small (in fact, limited by machine precision) at small~$\xi_0$.  In fact, the quality of the approximation is much better than would be expected from the analytic arguments: even at values of~$\nu$ of order unity,~$C(\nu,\xi_0) - 1/(3\xi_0^2)$ becomes arbitrarily small at small~$\xi_0$.  Evidently the analytic approximation for~$g_2$ is much better than that for~$\bar{\sigma}$.

\subsection{Boundary Wiggle}
\label{subapp:wiggle}

In our study of JT gravity coupled to a CFT, we must solve the equation of motion~\eqref{eq:JTCFTeom} and find the lowest eigenvalue of the fluctuation operator~$L$ defined in~\eqref{eq:JTCFTfluctuationoperator}.  The equation of motion for~$u(\phi)$ is nonlinear, so we solve it using a standard Newton-Raphson method.  We will not discuss this method further, except to note that the basin of attraction for solutions of~\eqref{eq:JTCFTeom} is relatively small, so in practice we need to implement a \textit{damped} Newton-Raphson iterative scheme in order to achieve convergence.  Likewise, the discretization of the ``local'' parts of the equation of motion and fluctuation operator can be performed using standard methods, which we will also avoid discussing in any detail.  Instead, here we will focus on discretization of the more uncommon ``nonlocal'' parts of~\eqref{eq:JTCFTeom} and~\eqref{eq:JTCFTfluctuationoperator} defined by a smearing against the kernel~$G(\phi,\tilde{\phi})$: that is, the discretization of the linear operator~$\Gcal$ defined by
\be
(\Gcal f)(\phi) = \int_0^{2\pi} d\tilde{\phi} \, G(\phi,\tilde{\phi}) f(\tilde{\phi}),
\ee
where~$G(\phi,\tilde{\phi})$ is defined in~\eqref{eq:Gkerneldef} and~$f(\phi)$ is a periodic function of~$\phi$.  We implement two approaches using pseudospectral methods on either a Fourier or a Chebyshev grid.

\subsubsection*{Fourier Discretization}

Because the functions in which we are interested are periodic in~$\phi$, it is perhaps most natural to work on an evenly-spaced grid: that is, we discretize the domain into~$N$ grid points~$\phi_j = 2\pi j/N$ for~$j = 1, \ldots, N$ and work with the discrete values~$f_j = f(\phi_j)$.  We take~$N$ to be even.  We then consider the discrete Fourier and inverse Fourier transforms of~$f_j$, constructed by restricting wave numbers to just a single copy of the Brillouin zone:
\be
\hat{f}_k = \frac{1}{N} \sum_{j = 1}^N f_j \, e^{-i k \phi_j}, \qquad f_j = \sideset{}{'}\sum_{k = -N/2}^{N/2} \hat{f}_k \, e^{i k \phi_j}, \qquad \forall j, k \in \mathbb{Z},
\ee
where the prime on the second sum indicates that the terms at~$k = \pm N/2$ are to be multiplied by one half to compensate for the fact that the wave numbers~$k = -N/2$ and~$k = N/2$ should be identified.  To compute the discretization of~$\Gcal$, we first consider the approximant
\be
p(\phi) = \sideset{}{'}\sum_{k = -N/2}^{N/2} \hat{f}_k \, e^{i k \phi},
\ee
which by construction satisfies~$p(\phi_j) = f_j$.  We then compute the action of~$\Gcal$ on~$p(\phi)$, evaluate the result at the grid points~$\phi = \phi_j$, and express the results in terms of~$f_j$ to obtain a matrix representation of~$\Gcal$.  This procedure automatically truncates the Fourier series that defines~$G(\phi,\tilde{\phi})$, and we find~$(\Gcal f)(\phi_i) = \sum_{j = 1}^N \Gcal_{ij} f_j$ where
\be
\Gcal_{ij} \equiv \frac{2}{N} \sideset{}{'}\sum_{k = 1}^{N/2} k \left[\tanh\left(\frac{k \pi \xi_0}{2} \right) \cos(k \phi_i) \cos(k \phi_j) + \coth\left(\frac{k \pi \xi_0}{2} \right) \sin(k \phi_i) \sin(k \phi_j)\right],
\ee
with the prime still denoting that the term~$k = N/2$ is multiplied by a factor of one half.  By construction, this discretization of~$\Gcal$ (as well as the analogous discretizations of the derivative operators~$d/d\phi$,~$d^2/d\phi^2$, etc.) is \textit{exact} when acting on any Fourier mode with wave number up to~$N/2$.

In practice, we expect solutions for the boundary wiggle to be symmetric about the axes of the ellipse, which in terms of~$u(\phi)$ corresponds to~$u'(\phi)$ being even in~$\phi$ at~$\phi = 0$ and~$\pi/2$, assuming~$u(\phi = 0) = 0$.  Likewise, we expect eigenfunctions of the fluctuation operator~$L$ to have definite parity about~$\phi = 0$ and~$\pi/2$.  It is numerically economical to exploit these symmetries by working on only the quarter-ellipse~$\phi \in [0,\pi/2]$, which effectively improves our numerical resolution at fixed grid size by a factor of four: a grid size of~$n_\phi$ (including~$\phi = 0$) on the quarter-ellipse corresponds to a grid size of~$N = 4(n_\phi-1)$ around the full ellipse.  Doing so is straightforward: for example, if a function~$f(\phi)$ is symmetric about~$\phi = 0$ and~$\pi/2$, we have that~$f_{N-i} = f_{N/2-i} = f_i$.  Under the assumption of such a symmetry, any~$N \times N$ matrix acting on~$f_i$ for~$i = 1, \ldots, N$ can be expressed as an~$n_\phi \times n_\phi$ matrix acting on~$f_i$ for~$i = 0,\ldots,n_\phi-1$ (where~$f_0 \equiv f_N$).  Consequently we restrict our attention to the quarter-ellipse.

We work with grid sizes up to~$n_\phi = 500$ (corresponding to keeping Fourier modes up to wave number~$k = 998$).  However, in certain regions of parameter space, the wiggle function~$u(\phi)$ exhibits large gradients near the computational boundaries~$\phi = 0$ and~$\phi = \pi/2$ which even such a large grid has difficulty resolving.  Rather than continuing to increase the grid size, in such cases we implement discretization based on a Chebyshev grid, which naturally clusters more grid points near the computational boundaries and is much more effective at resolving large gradients there with smaller grid sizes.

\subsubsection*{Chebyshev Discretization}

To implement pseudospectral methods on the quarter-ellipse with a Chebyshev grid, we could proceed directly by truncating the sum defining~$G(\phi,\tilde{\phi})$ and then using Clenshaw-Curtis quadrature to compute the integral defining the operator~$\Gcal$.  However, unlike the case of Fourier discretization just described, with a Chebyshev grid there is no natural rule for how to truncate the sum defining~$G(\phi,\tilde{\phi})$, so one would need to introduce an additional parameter to capture this order of truncation.  Consequently, we find it more convenient to rewrite~$G(\phi,\tilde{\phi})$ as a much more rapidly-converging sum using Poisson summation.

To do so, we write~$G(\phi,\tilde{\phi})$ in the form
\begin{multline}
G(\phi,\tilde{\phi}) \equiv \frac{1}{2\pi} \sum_{k = -\infty}^\infty \left[\left(k \tanh\left(\frac{k\pi\xi_0}{2}\right) - |k|\right) \cos(k\phi)\cos(k\tilde{\phi}) \right. \\ \left. + \left(k \coth\left(\frac{k\pi\xi_0}{2}\right)- |k|\right) \sin(k\phi)\sin(k\tilde{\phi}) \right] + \frac{1}{2\pi} \sum_{k = -\infty}^\infty |k| e^{ik(\phi - \tilde{\phi})}.
\end{multline}
The first sum converges exponentially in~$k$, so we may freely use the Poisson summation formula on it.  The second sum is not convergent but is still a well-defined distribution: noting that the Fourier transform of~$|x|$ is~$\Dcal_\omega^2 \ln|\omega|/(2\pi^2)$ with~$\Dcal_\omega$ a distributional derivative\footnote{That is,~$\Dcal_\omega^n \ln |\omega|$ is a homogeneous distribution of degree~$-n$.  Essentially, when integrated against a test function one integrates by parts ``ignoring the singularity'': e.g.~for~$a,b \neq 0$,
\begin{align*}
\int_a^b d\omega \, f(\omega) \Dcal_\omega \ln|\omega| &\equiv \left[f(\omega) \ln|\omega| \right]_a^b - \int_a^b d\omega \, f'(\omega) \ln|\omega| = \mathrm{PV} \int_a^b d\omega \, \frac{f(\omega)}{\omega}, \\
\int_a^b d\omega \, f(\omega) \Dcal_\omega^2 \ln|\omega| &\equiv \left[\frac{f(\omega)}{\omega}\right]_a^b - \left[f'(\omega) \ln|\omega| \right]_a^b + \int_a^b d\omega \, f''(\omega) \ln|\omega|,
\end{align*}
where~PV denotes the Cauchy principal value.}, the Poisson summation formula gives the distributional relation
\be
\frac{1}{2\pi} \sum_{k = -\infty}^\infty |k| e^{ik(\phi - \tilde{\phi})} = \frac{1}{\pi} \sum_{k = -\infty}^\infty \Dcal_{\tilde{\phi}}^2 \ln\left|\tilde{\phi} - \phi + 2k\pi\right|;
\ee
the right-hand sum is convergent (it is~$(1/\pi) \Dcal^2_{\tilde{\phi}} \ln |\sin((\tilde{\phi}-\phi)/2)|$).  Consequently we find
\be
G(\phi,\tilde{\phi}) = \frac{1}{\pi} \sum_{k = -\infty}^\infty \Dcal_{\tilde{\phi}}^2 \ln \left|\frac{1- e^{-(\tilde{\phi}-\phi+2k\pi)/\xi_0}}{1+ e^{-(\tilde{\phi}+\phi+2k\pi)/\xi_0}}\right|,
\ee
so the action of~$\Gcal$ on a test function is
\be
\label{eq:Gcalaction}
(\Gcal f)(\phi) = \frac{2\phi}{\pi \xi_0}  f'(0) + \frac{1}{\pi} \sum_{k = -\infty}^\infty \int_0^{2\pi} d\tilde{\phi} \, f''(\tilde{\phi}) \ln \left|\frac{1- e^{-(\tilde{\phi}-\phi + 2k\pi)/\xi_0}}{1+ e^{-(\tilde{\phi}+\phi+2k\pi)/\xi_0}}\right|.
\ee
The terms in the sum are~$\Ocal(e^{-2\pi|k|/\xi_0})$ at large~$|k|$, so the sum is rapidly convergent.  However, the integrand of the~$k = 0$ term is singular at~$\tilde{\phi} = \phi$, so we cannot yet straightforwardly discretize this expression for~$\Gcal$.  Instead we integrate by parts to obtain
\begin{multline}
\label{eq:GcalPoisson}
(\Gcal f)(\phi) = \frac{2\phi}{\pi \xi_0} f'(0) - \frac{\pi \xi_0}{2} f''(0) \\ 
	+ \sum_{k = -\infty}^\infty \left[\frac{4\phi}{\xi_0} f''(0) H(-k) - \int_0^{2\pi} d\tilde{\phi} \, f'''(\tilde{\phi}) \widehat{G}_3(\phi,\tilde{\phi}+2k\pi)\right],
\end{multline}
where~$H(x)$ is the Heaviside step function (with the convention~$H(0) = 0$) and we have defined
\be
\widehat{G}_3(\phi,\tilde{\phi}) \equiv \frac{\xi_0}{\pi} \, \Re\left[\Li_2\left(e^{-(\tilde{\phi}-\phi)/\xi_0}\right) - \Li_2\left(-e^{-(\tilde{\phi}+\phi)/\xi_0}\right)\right],
\ee
with~$\Li_2(z)$ the dilogarithm.  The dilogarithm is a continuous function on the real line, and hence the integrand is everywhere-finite, as desired.  Moreover, note that in integrating by parts, we included a nonzero constant of integration to ensure that the sum is rapidly convergent: as can be verified using the asymptotics of the dilogarithm, the terms of this sum also decay like~$e^{-2\pi|k|/\xi_0}$ at large~$|k|$.  In practice we find that truncating to~$|k| \leq 15$ is more than sufficient for obtaining accurate results for all the values of~$\xi_0$ we consider (i.e.~up to~$\xi_0 = 10$).  To discretize~$\Gcal$ on the quarter-ellipse, we impose appropriate parity of~$f(\phi)$ across~$\phi = 0$ and~$\phi = \pi/2$, which allows us to evaluate all integrals on the reduced domain~$\phi \in [0,\pi/2]$.  We then discretize the integrals in~\eqref{eq:GcalPoisson} via Clenshaw-Curtis quadrature on a Chebyshev grid on this domain.

It should be noted that the price we pay for reexpressing~$\Gcal$ in the form~\eqref{eq:GcalPoisson} is the loss of spectral accuracy due to the need to integrate the dilogarithm~$\Li_2(z)$ through its non-analytic behavior at~$z = 1$.  Nevertheless, in practice we find that in this new formulation, a Chebyshev grid of size~201 on the quarter-ellipse is sufficient to comfortably resolve the large gradients in~$u(\phi)$ near~$\phi = 0$ and~$\phi = \pi/2$.

\section{Bulk Solutions to JT + Scalar for \texorpdfstring{$m = 1$}{m=1} and \texorpdfstring{$2$}{2}}
\label{app:bulksolutions}

In this Appendix we explain how to obtain the analytic solutions~\eqref{eq:m12generalu} for the wiggle which satisfy the equations of motion with the matter sources given by~\eqref{eq:psipartialanalytic} (under an appropriate choice of~$A$) on the Poincar\'e disk an on the double-trumpet.  The idea is to solve the \textit{bulk} problem by constructing a solution for the dilaton~$\Phi$ and the scalar~$\psi$ on the disk or the double-trumpet, and from this bulk solution to then extract the boundary quantities~$u(\phi)$ and~$\psi_\partial(u)$.

\subsection{General Dilaton Solution}
\label{subapp:generaldilaton}

We begin by constructing a general solution to the JT + matter equations of motion on any portion of a Riemann surface of constant negative curvature that can be covered with a single coordinate chart, and with arbitrary conserved matter stress tensor.  With bulk action~$I = I_\mathrm{JT} + I_\mathrm{mat}$, the equation of motion for the metric gives
\be
\label{eq:dilatonEOM}
-\grad_a \grad_b \Phi + \Phi g_{ab} = T_{ab} - T g_{ab},
\ee
where~$T_{ab}$ is the stress tensor obtained from the matter action~$I_\mathrm{mat}$ (and~$T$ is its trace).

We solve~\eqref{eq:dilatonEOM} in conformal gauge, in which the metric takes the form
\be
ds^2 = e^{2\sigma} dz \, d\bar{z}
\ee
where~$\sigma(z,\bar{z})$ solves the Liouville equation (except for potentially at isolated conical defects).  Conservation of the stress tensor yields
\be
\label{eq:Tconservation}
\partial_{\bar{z}} T_{zz} + e^{2\sigma} \partial_z \left(e^{-2\sigma} T_{z\bar{z}}\right) = 0,
\ee
and likewise with~$z \leftrightarrow \bar{z}$.  In this conformal gauge,~\eqref{eq:dilatonEOM} gives
\begin{subequations}
\label{eqs:dilatoncomplex}
\begin{align}
-\partial_z \partial_z \Phi + 2 \partial_z \sigma \partial_z \Phi &= T_{zz}, \label{subeq:Phizz} \\
-\partial_z \partial_{\bar{z}} \Phi + \frac{1}{2} e^{2\sigma} \Phi &= -T_{z\bar{z}}, \label{subeq:Phizzbar} \\
-\partial_{\bar{z}} \partial_{\bar{z}} \Phi + 2 \partial_{\bar{z}} \sigma \partial_{\bar{z}} \Phi &= T_{\bar{z}\bar{z}}, \label{subeq:Phizbarzbar}
\end{align}
\end{subequations}
which can be integrated explicitly.  To do so, we first integrate~\eqref{subeq:Phizz} to get
\be
\label{eq:Phigeneral}
\Phi(z,\bar{z}) = A(\bar{z}) + \int_{z_0}^z dw \, e^{2\sigma(w,\bar{z})} \left[B(\bar{z}) - \int_{z_0}^w du \, e^{-2\sigma(u,\bar{z})} T_{zz}(u,\bar{z})\right],
\ee
where~$A$ and~$B$ are arbitrary antiholomorphic functions and the integrals are contour integrals that start at an arbitrary point~$z_0$ and end at~$z$ (avoiding conical defects if there are any).  We can relate~$A$ and~$B$ by inserting this solution into~\eqref{subeq:Phizzbar}, which gives
\be
\label{eq:PhizzbarAB}
A(\bar{z}) = 2 B'(\bar{z}) - s_1(\bar{z}) B(\bar{z}) - \Tcal_1(\bar{z}),
\ee
where we have defined the antiholomorphic functions
\begin{subequations}
\be
s_1(\bar{z}) \equiv -4\partial_{\bar{z}} \sigma + \int_{z_0}^z dw \, e^{2\sigma(w,\bar{z})},
\ee
\begin{multline}
\Tcal_1(\bar{z}) \equiv 2e^{-2\sigma} T_{z\bar{z}} + 4 \partial_{\bar{z}} \sigma \int_{z_0}^z dw \, e^{-2\sigma(w,\bar{z})} T_{zz}(w,\bar{z}) \\ 
	+ \int_{z_0}^z dw \left[2\partial_{\bar{z}}\left(e^{-2\sigma(w,\bar{z})} T_{zz}(w,\bar{z})\right) - e^{2\sigma(w,\bar{z})} \int_{z_0}^w du \, e^{-2\sigma(u,\bar{z})} T_{zz}(u,\bar{z})\right];
\end{multline}
\end{subequations}
the fact that these are antiholomorphic follows from the Liouville equation and the conservation of the stress tensor.  Consequently we may evaluate them at any value of~$z$; choosing~$z = z_0$ gives the simpler expressions
\be
\label{eq:s1T1}
s_1(\bar{z}) = -4\partial_{\bar{z}} \sigma(z_0, \bar{z}), \qquad \Tcal_1(\bar{z}) = 2e^{-2\sigma(z_0,\bar{z})} T_{z\bar{z}}(z_0,\bar{z}).
\ee
Finally, using~\eqref{eq:PhizzbarAB} in~\eqref{subeq:Phizbarzbar} yields a third-order differential equation for~$B$:
\be
\label{eq:dilatonBEOM}
-2B'''(\bar{z}) + s_2(\bar{z}) B'(\bar{z}) + \frac{1}{2} s_2'(\bar{z}) B(\bar{z}) + \Tcal_2(\bar{z}) = 0,
\ee
where we have introduced the additional antiholomorphic functions
\bea
s_2(\bar{z}) &\equiv 8\left((\partial_{\bar{z}}\sigma)^2 - \partial_{\bar{z}}^2 \sigma\right), \\
\Tcal_2(\bar{z}) &\equiv e^{2\sigma} \partial_{\bar{z}}\left[e^{-2\sigma}\left(\Tcal_1' + \int_{z_0}^z dw\, \partial_{\bar{z}} \left(e^{2\sigma(w,\bar{z})} \int_{z_0}^w du\, e^{-2\sigma(u,\bar{z})} T_{zz}(u,\bar{z})\right)\right)\right] - T_{\bar{z}\bar{z}};
\eea
as with~$s_1$ and~$\Tcal_1$, one can verify that~$s_2$ and~$\Tcal_2$ are antiholomorphic using the Liouville equation and the conservation of the stress tensor.  So we may again evaluate these functions at~$z = z_0$, which gives the relations
\be
s_2 = 2 s_1' + \frac{s_1^2}{2}, \qquad \Tcal_2 = \Tcal_1'' + \frac{1}{2} \, s_1 \Tcal_1' - T_{\bar{z}\bar{z}}(z_0,\bar{z}).
\ee
Using these and~\eqref{eq:PhizzbarAB} we may express~\eqref{eq:dilatonBEOM} entirely in terms of~$A$:
\be
-A'' - \frac{s_1}{2} \, A' - T_{\bar{z}\bar{z}}(z_0,\bar{z}) = 0.
\ee
Using~\eqref{eq:s1T1} this can be integrated to give
\be
\label{eq:dilatonA}
A(\bar{z}) = c_1 + \int_{\bar{z}_0}^{\bar{z}} d\bar{w} \, e^{2\sigma(z_0,\bar{w})} \left[c_2 - \int_{\bar{z}_0}^{\bar{w}} d\bar{u} \, e^{-2\sigma(z_0,\bar{u})} T_{\bar{z}\bar{z}}(z_0,\bar{u})\right],
\ee
where~$c_1$,~$c_2$, and~$\bar{z}_0$ are arbitrary (complex) constants.  Finally, we then integrate~\eqref{eq:PhizzbarAB} to obtain
\be
\label{eq:dilatonB}
B(\bar{z}) = e^{-2\sigma(z_0,\bar{z})} \left[c_3 + \frac{1}{2} \int_{\bar{z}_0}^{\bar{z}} d\bar{w} \, e^{2\sigma(z_0,\bar{w})} \left(A(\bar{w}) + \Tcal_1(\bar{w})\right)\right],
\ee
where~$c_3$ is another arbitrary complex constant.  Consequently, equation~\eqref{eq:Phigeneral} with~$A$ and~$B$ given by~\eqref{eq:dilatonA} and~\eqref{eq:dilatonB} is the general solution for the dilaton for arbitrary~$\sigma$ and matter stress tensor.  Note that any changes to~$z_0$ and~$\bar{z}_0$ can be reabsorbed into the~$c_i$, so there are only three independent degrees of freedom in this solution.

When the matter is classical and conformal,~$T_{z\bar{z}} = 0$, and hence~$T_{zz}$ and~$T_{\bar{z}\bar{z}}$ are holomorphic and antiholomorphic, respectively.  This simplification, along with use of the Liouville equation, allows us to integrate~\eqref{eq:Phigeneral} to obtain
\begin{multline}
\label{eq:dilatonconformalT}
\Phi = 4e^{-2\sigma(z_0,\bar{z})} \left.\left[\tilde{c}_3 + 2\tilde{c}_1 \partial_z \sigma + 4\tilde{c}_2 \partial_z^2 \sigma - 2\partial_z (e^{-2\sigma} \partial_z(e^{2\sigma} \overline{F}))\right]\right|_{z = z_0}\left(\partial_{\bar{z}}\sigma(z,\bar{z}) - \partial_{\bar{z}}\sigma(z_0,\bar{z})\right) \\
	+ \tilde{c}_1 + 4\tilde{c}_2 \partial_z \sigma(z_0,\bar{z}) - 2 e^{-2\sigma(z_0,\bar{z})} \left.\partial_z\left(e^{2\sigma} \overline{F}\right)\right|_{z = z_0} - 2 e^{-2\sigma(z,\bar{z})} \partial_{\bar{z}}\left(e^{2\sigma} F\right),
\end{multline}
where the~$\tilde{c}_i$ are constants and we have defined
\be
F(z,\bar{z}) \equiv \int_{z_0}^z dw \, e^{-2\sigma(w,\bar{z})} T_{zz}(w), \qquad \overline{F}(z,\bar{z}) \equiv \int_{\bar{z}_0}^{\bar{z}} d\bar{w} \, e^{-2\sigma(z,\bar{w})} T_{\bar{z}\bar{z}}(\bar{w}).
\ee
We'd now like to use this expression to solve for the dilaton in the presence of conformal matter on the Poincar\'e disk and the double-trumpet.  For the real massless scalar field~$\psi$, the equation of motion~$\grad^2 \psi = 0$ is solved by
\be
\label{eq:scalarf}
\psi(z,\bar{z}) = f(z) + \bar{f}(\bar{z})
\ee
for an arbitrary holomorphic function~$f$, and the corresponding components of the stress tensor are~$T_{zz} = (f')^2$ and~$T_{\bar{z}\bar{z}} = (\bar{f}')^2$.  In principle we should determine~$f$ by imposing Dirichlet boundary conditions at~$\partial M$, but for our purposes in this section we will instead choose~$f$ and then determine the corresponding boundary conditions from it.

\subsection{Poincar\'e Disk}
\label{subapp:Poincaredisk}

The Poincar\'e disk is given by conformal factor~$\sigma = \ln(2/(1-z \bar{z}))$, with~$z$ covering the unit disk.  We take the scalar field to be given by~\eqref{eq:scalarf} with~$f(z) = J z^n/2$ for positive integer~$n$.  Then~\eqref{eq:dilatonconformalT} with~$z_0 = \bar{z}_0 = 0$ gives the general solution
\be
\label{eq:Phim1onebn}
\Phi = \frac{\alpha_1(1+r^2) + 2r(\alpha_2 \cos\theta + \alpha_3 \sin\theta) + \frac{n J^2 r^{2n}}{4}\left(\frac{r^2}{2n+1} - \frac{1}{2n-1}\right) \cos(2n\theta)}{1-r^2},
\ee
where the~$\alpha_i$ are real and we have converted to the usual polar coordinates using~$z = re^{i\theta}$.  We expect the boundary conditions~\eqref{eq:JTbndry} to constrain the~$\alpha_i$, but we also expect there to be residual freedom in these constants due to the~$SL(2,\mathbb{R})$ symmetry of the Poincar\'e disk.  So for the purposes of constructing a particular solution, we set~$\alpha_2 = 0 = \alpha_3$.  Then the boundary condition~$\Phi|_{\partial M} = 1/\delta$ gives an embedding of the boundary~$\partial M$:
\be
r = 1 - \Phi_0(\theta) \, \delta + \Ocal(\delta)^2, \mbox{ where } \Phi_0(\theta) = \alpha_1 - \frac{n J^2}{4(4n^2-1)} \cos(2n\theta).
\ee
With this embedding we may fix~$\alpha_1$ by imposing the requirement that the length of~$\partial M$ be~$\beta/\delta$:
\be
\beta = \int_0^{2\pi} \frac{d\theta}{\Phi_0(\theta)} \quad \Rightarrow \quad \alpha_1 = \frac{2\pi}{\beta} \sqrt{1 + \left(\frac{n \beta J^2}{8\pi(4n^2-1)}\right)^2}.
\ee
Finally, we may define the proper length coordinate~$u$ along~$\partial M$ using~$du = d\theta/\Phi_0(\theta)$, which gives the wiggle function~$u(\theta)$~\eqref{eq:m12generalu} with
\be
\label{eq:Am1bulk}
A = \left[1+\left(\frac{8\pi(4n^2-1)}{n\beta J^2}\right)^2\right]^{-1/2}.
\ee
Likewise, the boundary profile of the scalar is given simply by~$\psi|_{\partial M} = J \cos(n \theta)$, which written in terms of~$u$ takes the form~\eqref{eq:psipartialanalytic}.

These solutions solve the equation of motion~\eqref{eq:JTCFTeom} in the boundary formalism.  This can be verified most easily by noting that when~$m = 1$, the conical defects vanish and hence the on-shell action should be independent of~$\xi_0$.  For simplicity we can therefore work purely in the~$\xi_0 \to \infty$ limit of the action~\eqref{eq:JTCFTaction} in which the ellipse degenerates into the Poincar\'e disk (with no defects).  In this~$\xi_0 \to \infty$ limit with~$m = 1$, we have
\be
g_2(\phi) \to -\frac{1}{3}, \; G(\phi,\tilde{\phi}) \to \frac{1}{2\pi} \sum_{k = -\infty}^\infty |k| e^{ik(\phi-\tilde{\phi})}.
\ee
We then find that the equation of motion~\eqref{eq:JTCFTeom} is satisfied by~\eqref{eq:m12generalu} if
\be
\frac{1}{\sqrt{1-A^2}} = \sqrt{1 + \left(\frac{n \beta J^2}{8\pi(4n^2-1)}\right)^2};
\ee
this agrees precisely with the expression~\eqref{eq:Am1bulk} obtained from the bulk solutions.  The action of the wiggle profile~\eqref{eq:m12generalu} is
\be
\widehat{I}_1 = -S_0 + \frac{1}{\beta} \left[\frac{2\pi^2(4n^2-1)}{\sqrt{1-A^2}} - 8\pi^2 n^2 + \frac{\pi n \beta J^2}{2}\right],
\ee
which recovers the pure JT Schwarzian result~$\widehat{I}_1 = -S_0 - 2\pi^2/\beta$ when~$J = 0$.

\subsection{Double-Trumpet}
\label{subapp:doubletrumpet}

To treat the double-trumpet, we will work on the quotient space (i.e.~the single trumpet) with appropriate regularity conditions imposed to ensure smoothness of the scalar and the dilaton on the quotient geometry.  The conformal factor on the trumpet is
\be
\sigma = \ln\left[\frac{1}{\xi_0} \sec\left(\frac{z + \bar{z}}{2\xi_0}\right)\right],
\ee
with~$z \equiv \xi + i \phi$ covering the strip~$\Re(z) \in [0,\pi \xi_0/2)$,~$\Im(z) \in [0,2\pi)$.  A wormhole solution will involve fixing not just the dilaton~$\Phi$, but also~$\xi_0$, which sets the size of the wormhole.  We take the scalar field to be given by~\eqref{eq:scalarf} with
\be
f(z) = \frac{J \cosh(nz)}{2\cosh(n \pi \xi_0/2)}
\ee
for positive integer~$n$; this choice ensures that~$\psi$ obeys all the regularity conditions discussed in Section~\ref{subsec:JTCFTboundary}.  Using~\eqref{eq:dilatonconformalT} with~$z_0 = \bar{z}_0 = 0$, we find that imposing these same regularity conditions on~$\Phi$ fixes the three constants~$\tilde{c}_i$ uniquely, leaving the solution
\begin{multline}
\label{eq:Phim2onebn}
\Phi = \frac{n J^2 \xi_0}{4 \cosh^2(n\pi\xi_0/2)} \left[n\left(\xi_0 + \xi\tan\left(\frac{\xi}{\xi_0}\right)\right) \right. \\ \left. - \frac{\cos(2n\phi)}{1+4n^2\xi_0^2}\left(n \xi_0 \cosh(2n\xi) + \frac{1}{2} \sinh(2n\xi) \tan\left(\frac{\xi}{\xi_0}\right)\right)\right].
\end{multline}
Again we impose the boundary condition~$\Phi|_{\partial M} = 1/\delta$ to obtain an embedding of~$\partial M$:
\be
\xi = \frac{\pi \xi_0}{2} - \Phi_0(\phi) \, \delta + \Ocal(\delta^2),
\ee
where
\be
\Phi_0(\phi) = \frac{n J^2 \xi_0^2}{8\cosh^2(n\pi\xi_0/2)} \left(n \pi \xi_0 - \frac{\cos(2n \phi) \sinh(n\pi \xi_0)}{1+4n^2 \xi_0^2}\right).
\ee
We must have~$\Phi_0(\phi) > 0$ for all~$\phi$, but~$\Phi_0(0)$ becomes negative at large~$\xi_0$; hence solutions only exist when~$\xi_0$ is sufficiently small\footnote{Specifically, when~$n \xi_0 < x_*$, where~$x_* \approx 1.36$ is the positive solution of~$\pi x(1+4 x^2) = \sinh(\pi x)$.}.  Requiring that the length of~$\partial M$ be~$\beta/\delta$ then fixes the allowed values~$\xi_*$ of~$\xi_0$ in terms of~$J$ and~$n$:
\be
\label{eq:m2Jxi0}
\beta = \int_0^{2\pi} \frac{d\phi}{\Phi_0(\phi)} \quad \Rightarrow \quad \frac{\beta J^2}{n} = \frac{16 \cosh^2(n \pi \xi_*/2)}{n^3 \xi_*^3\sqrt{1 - \left(\frac{\sinh(n\pi\xi_*)}{n\pi \xi_*(1+4n^2 \xi_*^2)}\right)^2}}.
\ee
Again we may define the proper length coordinate~$u$ along~$\partial M$ using~$du = d\phi/\Phi_0(\phi)$, which gives the wiggle function~$u(\phi)$~\eqref{eq:m12generalu} with
\be
\label{eq:Am2bulk}
A = \frac{\sinh(n\pi \xi_*)}{n\pi \xi_*(1+4n^2\xi_*^2)}.
\ee
The boundary profile of the scalar is given simply by~$\psi|_{\partial M} = J \cos(n \phi)$, which written in terms of~$u$ takes the form~\eqref{eq:psipartialanalytic}.

It is straightforward to check that~\eqref{eq:m2Jxi0} and~\eqref{eq:Am2bulk} satisfy~\eqref{eq:m2wigglesaddle} and~\eqref{eq:m2xi0saddle}, so again we confirm that the solutions obtained using the bulk and boundary formalisms coincide.  In particular, the need for~$\sqrt{\beta} J$ to be sufficiently large in order for solutions to exist can be inferred from~\eqref{eq:m2Jxi0}: in the region where the right-hand side is real, it has a global minimum at~$n \xi_* \approx 0.93$ where it attains the value~$\approx 146$.  Hence solutions for~$\xi_0$ only exist when~$\sqrt{\beta}J \geq \sqrt{\beta}J_\mathrm{min} \approx 12 \sqrt{n}$, just as we found in Section~\ref{subsec:analyticJTCFT}.

\section{JT + Branes}
\label{app:WCmodel}

In this Appendix we discuss a model of JT gravity coupled to end-of-the-world (EOW) branes.  This model is effectively a classical version of that considered in~\cite{PenShe19}, except that we do not give the branes any internal degrees of freedom.  Our purpose is illustrative: though we do not find stable wormholes at~$m < 1$, we will see very clearly that for~$m < 1$ multiple branches of solutions for the wiggle can appear in a way analogous to the more involved model studied in Section~\ref{sec:JTCFT}.

\subsection{Boundary Action}
\label{subapp:JTbraneaction}

\begin{figure}[t]
\centering
\includegraphics[width=0.5\textwidth,page=3]{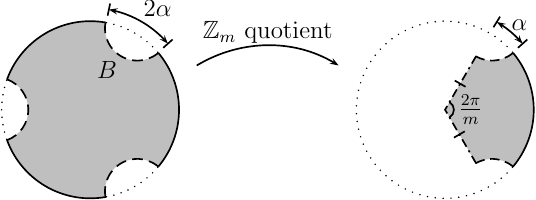}
\caption{For positive integer~$m$, the genus-zero wormhole geometry corresponding to the JT + brane model consists of the Poincar\'e disk with~$m$ geodesic ``bites''~$B$ removed, shown at left.  The quotient geometry is shown at right; the dot-dashed radial lines are identified, so the two half-geodesics correspond to a single copy of one of the branes~$B$.  This quotient geometry coincides with the geometry~$\widehat{M}_m^{(2)}$ shown in Figure~\ref{fig:fundomainquotientgeometry}.}
\label{fig:JTbranewormhole}
\end{figure}

The JT + brane model has the Euclidean action
\be
I_\mathrm{brane} = I_\mathrm{JT} + \mu \int_B ds,
\ee
where~$B$ is an EOW brane anchored to the boundary~$\partial M$ and~$\mu > 0$ is its tension.  With~$m$ boundaries, the geometry~$M$ consists of the Poincar\'e disk with~$m$ geodesic ``bites'' removed corresponding to the location of~$m$ disconnected portions of~$B$, as shown in Figure~\ref{fig:JTbranewormhole}.  The boundary~$\partial M$ consists of~$m$ disconnected pieces anchored to~$B$, and we take the length of each of these pieces of~$\partial M$ to be~$\beta/\delta$.  After quotienting by the~$\mathbb{Z}_m$ replica symmetry, it is clear that the quotient space geometry~$\widehat{M}_m$ is in fact identical to (a single copy of) the geometry~$\widehat{M}_m^{(2)}$ discussed in Section~\ref{subsec:JTboundary}.  Consequently, the JT part of the quotient action takes half its value in pure JT after the replacement~$\beta \to 2\beta$:
\be
\widehat{I}_m = \frac{m-2}{2m} \, S_0 - \int_{-\beta/2}^{\beta/2} du \Sch\left(\tan\left(\frac{\Theta}{2}\right),u\right) + \mu \Length(B).
\ee
(Hence the case~$\mu = 0$ can be thought of as a quotient of pure JT by a~$\mathbb{Z}_2$ symmetry about the geodesic~$B$.)   To put the brane on-shell, we must compute its length up to the cutoff boundary~$\partial M$.  From Figure~\ref{fig:JTbranewormhole}, it is clear that~$B$ is diffeomorphic to a geodesic on the Poincar\'e disk sweeping out an angle~$2\alpha$ on the boundary.  Moreover, footnote~\ref{fn:RTheta} relating the embedding functions~$\Theta(u)$ and~$R(u)$ indicates that this geodesic is cut off by~$\partial M$ at the radial cutoffs~$R_\pm = 1 - \Theta'(\pm \beta/2) \, \delta + \Ocal(\delta^2)$.  The length of~$B$ up to these cutoffs is then
\be
\mathrm{Length}(B) = \ln\left(\frac{\beta^2}{\delta^2}\right) -\ln\left(\frac{4\beta^2\Theta'(\beta/2) \Theta'(-\beta/2)}{\sin^2\alpha}\right) + \Ocal(\delta).
\ee
The first term is the expected UV divergent piece, and it can be cancelled out by adding an appropriate counterterm such as~$-2\mu\ln (\beta \Phi)|_{\partial M}$ to the action.  After this cancellation, the renormalized action for the wiggle is
\be
\label{eq:JTbraneaction}
\widehat{I}_m = \frac{m-2}{2m} \, S_0 - \int_{-\beta/2}^{\beta/2} du \Sch\left(\tan\left(\frac{\Theta}{2}\right),u\right) - \mu \ln\left(\frac{4\beta^2\Theta'(\beta/2) \Theta'(-\beta/2)}{\sin^2\alpha}\right).
\ee

We must next determine the boundary conditions to impose on the wiggle~$\Theta(u)$.  One condition is fixed as in pure JT by requiring that~$\Theta(u)$ wrap around the entire boundary of~$\widehat{M}_m^{(2)}$, while another will stem from fixing the angle at which~$\partial M$ intersects~$B$.  One way to infer this latter boundary condition is to treat the brane as a particle of mass~$\mu$ which scatters with the boundary trajectory~$\partial M$ and impose conservation of the~$SL(2,\mathbb{R})$ charges in this scattering process, as discussed in~\cite{MalSta17,KouMal17}.  Alternatively, we may impose the boundary condition~$n^a \grad_a \Phi|_B = \mu$ (with~$n^a$ a unit outward-pointing normal to~$B$) on the dilaton in the bulk~\cite{PenShe19} and convert it to the desired constraint on the intersection of~$\partial M$ and~$B$.  Ultimately we find that~$n \cdot u = \mu \delta + \Ocal(\delta^2)$, where~$n^a$ and~$u^a$ are outward-pointing unit normals to~$B$ and~$\partial M$\footnote{One way of determining this condition explicitly is to work on the Poincar\'e disk, placing the brane on the axis~$\theta = \pm \pi/2$, and considering the family of dilaton solutions
\begin{equation*}
\Phi = \frac{(1+r_0^2)(1+r^2) - 4r_0 r\cos\theta}{(1-r_0^2)(1-r^2)}
\end{equation*}
which are obtained from the ``standard'' solution~$\Phi = (1+r^2)/(1-r^2)$ by translating the origin to the right a coordinate distance~$r_0$ using an~$SL(2,\mathbb{R})$ transformation.  $r_0$ and~$\mu$ can be related using the bulk boundary condition~$n^a \grad_a \Phi|_B = 2r_0/(1-r_0^2) = \mu$, while it is easy to verify that where the level sets~$\Phi = 1/\delta$ intersect the brane, they satisfy~$n \cdot u = 2r_0 \delta/(1-r_0^2) + \Ocal(\delta^2) = \mu \delta + \Ocal(\delta^2)$.  This is a local condition on the intersection of~$\partial M$ and~$B$, and hence it must hold in any other geometry as well.}.  This constraint ultimately leads to the wiggle boundary conditions
\be
\label{eq:wigglebranebndryconditions}
\Theta\left(\pm \beta/2\right) = \pm\left(\frac{\pi}{m} - \alpha \right), \quad \Theta''\left(\pm \beta/2\right) = \mp \left(\cot\alpha \, \Theta'\left(\pm \beta/2\right)^2 - \mu \Theta'\left(\pm \beta/2\right)\right).
\ee

Finally, we will ultimately need to investigate the stability of the wiggle.  As usual, we restrict to perturbations that exhibit a~$\mathbb{Z}_2$ symmetry corresponding to reflection about~$u = 0$.  The stability analysis proceeds just as in the pure JT case discussed in Section~\ref{subsec:JTstability}: we write~$\Theta = \widetilde{\Theta} + \vartheta$ with~$\widetilde{\Theta}$ a solution to the equations of motion and expand the action to second order in~$\vartheta$.  The resulting fluctuation operator~$L$ is identical to the one for pure JT~\eqref{eq:JTfluctuationL}, except that it acts on the space of functions obeying the boundary conditions
\be
\vartheta(0) = 0 = \vartheta''(0), \quad \vartheta(\beta/2) = 0, \quad \vartheta''(\beta/2) = - \left(2\cot\alpha \, \widetilde{\Theta}'(\beta/2) - \mu\right) \vartheta'(\beta/2).
\ee
Hence a solution~$\widetilde{\Theta}$ is stable if and only if the spectrum of~$L$ is nonnegative on the space of functions obeying these boundary conditions.  We compute the spectrum of~$L$ numerically as described in Appendix~\ref{app:numerics}.

\subsection{Saddles}
\label{subapp:JTbranesaddles}

We now look for saddles of the action~\eqref{eq:JTbraneaction}.  As in the main text, we proceed by first looking for saddles for the wiggle~$\Theta(u)$ at fixed modulus~$\alpha$ and then evaluate the action on these saddles to obtain an effective action~$\widehat{I}_m[\alpha]$ for~$\alpha$, which we then examine to look for saddles for~$\alpha$.

\subsubsection*{Saddles for the Wiggle}

With the boundary conditions~\eqref{eq:wigglebranebndryconditions}, the variational problem for the action~\eqref{eq:JTbraneaction} is well-posed and leads to the same equation of motion~\eqref{eq:pureJTEOM} we obtained in pure JT.  Since the boundary conditions~\eqref{eq:wigglebranebndryconditions} are odd in~$u$, we may again consider the most general odd solution:
\be
\tan\left(\frac{\Theta(u)}{2}\right) = a \tan\left(\frac{bu}{2\beta}\right)
\ee
(the extra factor of~2 on the right-hand side relative to~\eqref{eq:pureJTgeneralTheta} is inserted for convenience to account for the relative factor of~2 in~$\beta$ between pure JT and the JT + brane model).  The constants~$a$ and~$b$ are determined by imposing the boundary conditions~\eqref{eq:wigglebranebndryconditions} just as in pure JT.  In short, we first take~$a = i a_i$ and~$b = -i b_i$ with~$a_i$ and~$b_i$ both real and positive.  Then solutions can only exist when~$\pi/m - \alpha < \pi$ in which case~$a_i$ and~$b_i$ must satisfy
\begin{subequations}
\label{eqs:JTbraneexponentialab}
\begin{align}
&\cosh\left(\frac{b_i}{2}\right) + \frac{\beta\mu}{b_i} \sinh\left(\frac{b_i}{2}\right) = \frac{\sin(\pi/m)}{\sin\alpha}, \label{subeq:JTbraneexponentialb} \\
&a_i = \coth\left(\frac{b_i}{4}\right) \tan\left(\frac{\pi-m\alpha}{2m}\right).
\end{align}
\end{subequations}
On the other hand, taking~$a = a_r$ and~$b = b_r$ with~$a_r$ and~$b_r$ both real and positive, we find that solutions must satisfy
\begin{subequations}
\label{eqs:JTbraneoscillatoryab}
\begin{align}
&\cos\left(\frac{b_r}{2}\right) + \frac{\beta\mu}{b_r} \sin\left(\frac{b_r}{2}\right) = \frac{\sin(\pi/m)}{\sin\alpha}, \label{subeq:JTbraneoscillatoryb} \\
&\frac{b_r}{2\pi} \in \begin{cases} (2N, 2N+1] &\mbox{if } \tan\left(\frac{\pi-m\alpha}{2m}\right) > 0, \\ (2N-1, 2N) & \mbox{if } \tan\left(\frac{\pi-m\alpha}{2m}\right) < 0, \end{cases} \quad N \equiv \left\lfloor \frac{1}{2m} + \frac{\pi - \alpha}{2\pi}\right\rfloor, \label{subeq:bconstraintbrane} \\
&a_r = \cot\left(\frac{b_r}{4}\right) \tan\left(\frac{\pi-m\alpha}{2m}\right). \label{subeq:JTbraneoscillatorya}
\end{align}
\end{subequations}

\begin{figure}[t]
\centering
\includegraphics[height=4cm]{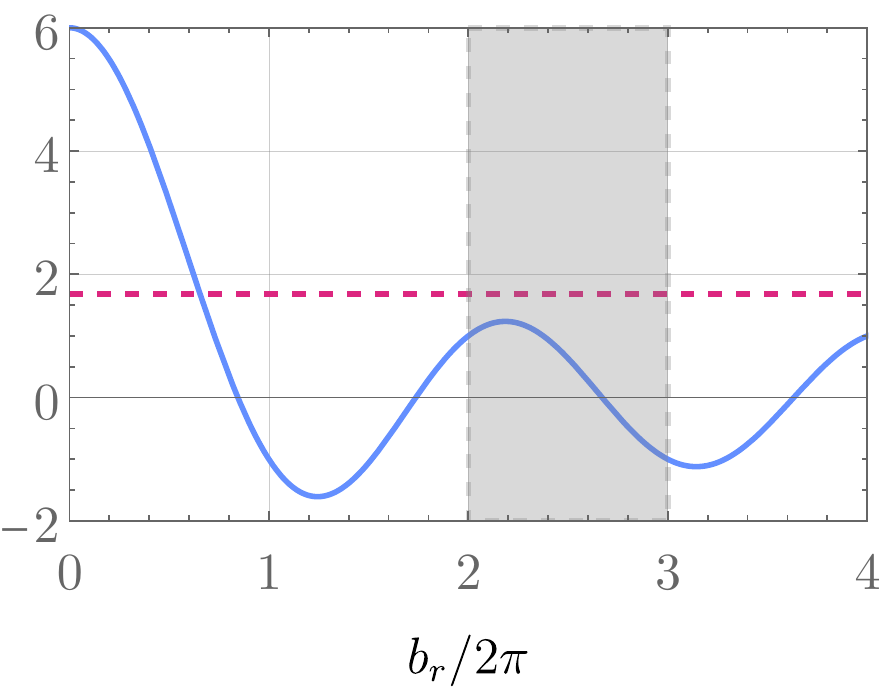}
\hspace{0.1cm}
\includegraphics[height=4cm]{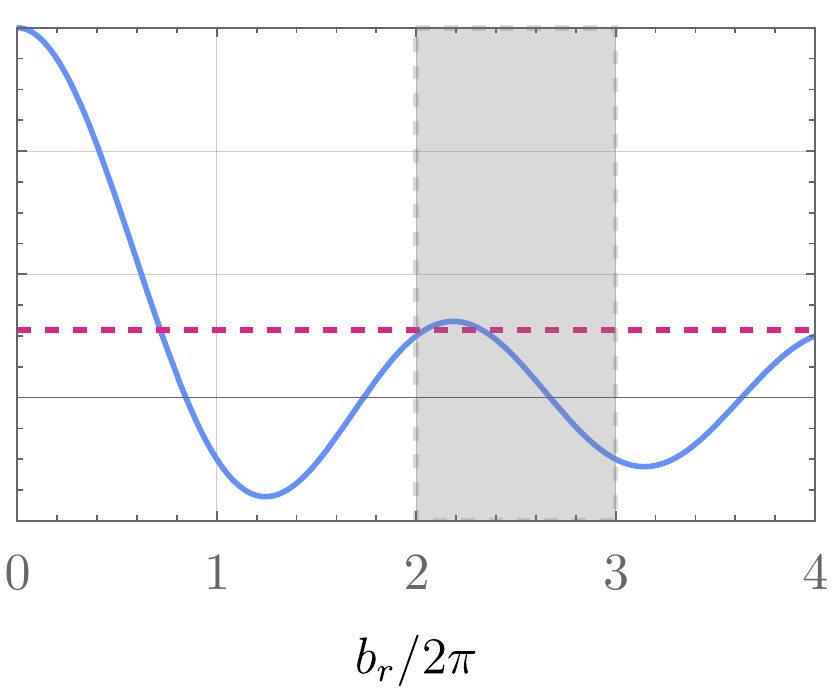}
\hspace{0.1cm}
\includegraphics[height=4cm]{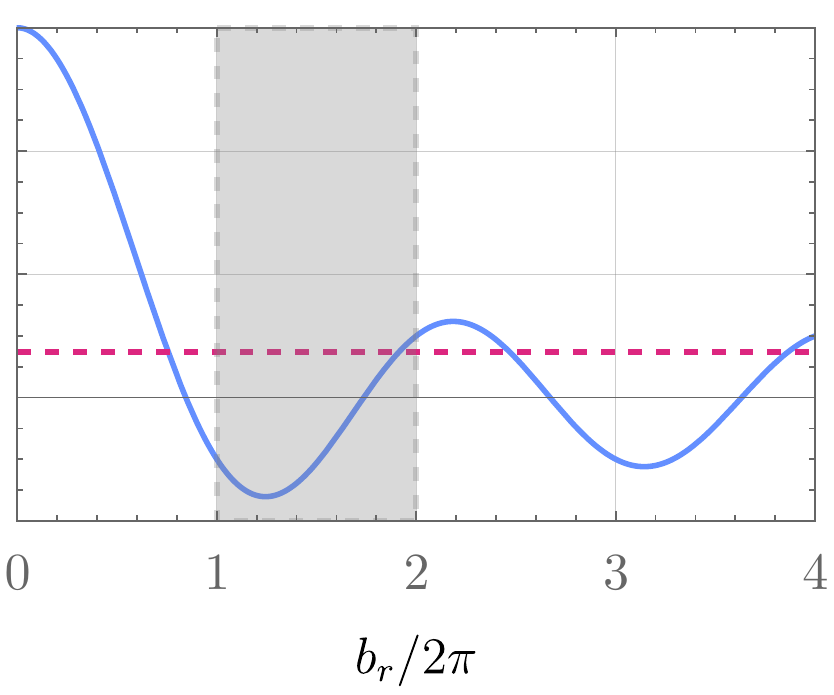}
\caption{Graphical solutions of~\eqref{eqs:JTbraneoscillatoryab} for~$b_r$ when~$m < 1$.  The solid blue curves and the dashed red lines show the left- and right-hand sides of~\eqref{subeq:JTbraneoscillatoryb}, respectively, while the shaded region is the interval satisfying~\eqref{subeq:bconstraintbrane}; hence solutions of~\eqref{eqs:JTbraneoscillatoryab} correspond to intersections of the blue and red lines in the shaded regions.  All three plots take~$\beta\mu = 10$ and~$m = 0.45$, while from left to right we show~$\alpha = \pi/8$,~$\pi/5$, and~$\pi/3$.  At small~$\alpha$ there are no solutions; as~$\alpha$ grows, two solutions appear at a common value of~$b_r$ and branch apart; as~$\alpha$ grows further, the larger of the two solutions disappears abruptly as the region satisfying~\eqref{subeq:bconstraintbrane} changes.}
\label{fig:oscillatorysolutions}
\end{figure}

It is now straightforward to see how turning on a brane can give rise to new branches of solutions.  For any value of~$\mu$, the left-hand side of~\eqref{subeq:JTbraneexponentialb} is monotonic in~$b_i > 0$, so when solutions of exponential type exist (i.e.~when~$\pi/m - \alpha < \pi$ and~$\sin(\pi/m)/\sin\alpha > 1 + \mu\beta/2$), then precisely one solution exists.  Similarly, for~$\mu = 0$ the left-hand side of~\eqref{subeq:JTbraneoscillatoryb} is monotonic in~$b_r$ in the interval allowed by the constraint~\eqref{subeq:bconstraintbrane}, so again at most one solution of oscillatory type can exist (and it is given by the pure JT solution~\eqref{eqs:aboscillatory}).  However, for~$\mu \neq 0$ the left-hand side of~\eqref{subeq:JTbraneoscillatoryb} is \textit{not} in general monotonic in the interval allowed by~\eqref{subeq:bconstraintbrane}, and consequently may admit an additional solution.  In Figure~\ref{fig:oscillatorysolutions} we graphically illustrate the structure of solutions for~$b_r$ for various values of~$\alpha$, showing how zero, one, or two solutions may exist.

\subsubsection*{Saddles for the Modulus}

Putting the wiggle on-shell, we are left with an effective action~$\widehat{I}_m[\alpha]$ for~$\alpha$:
\be
\label{eq:JTbraneonshellaction}
\widehat{I}_m[\alpha] = \frac{m-2}{2m} \, S_0 + \begin{dcases} \frac{b_i^2}{2\beta} - 2\mu \ln \left(\frac{2b_i}{\sinh(b_i/2)} \frac{\sin(\pi/m-\alpha)}{\sin\alpha} \right), & \mbox{exponential}, \\
-\frac{b_r^2}{2\beta} - 2\mu \ln \left(\frac{2b_r}{\sin(b_r/2)} \frac{\sin(\pi/m-\alpha)}{\sin\alpha} \right), & \mbox{oscillatory}, \end{dcases}
\ee
with~$b_i$ and~$b_r$ implicit functions of~$\alpha$ (as well as~$m$ and~$\beta\mu$) through~\eqref{eqs:JTbraneexponentialab} and~\eqref{eqs:JTbraneoscillatoryab}.

\begin{figure}[t]
\centering
\includegraphics[height=3.5cm]{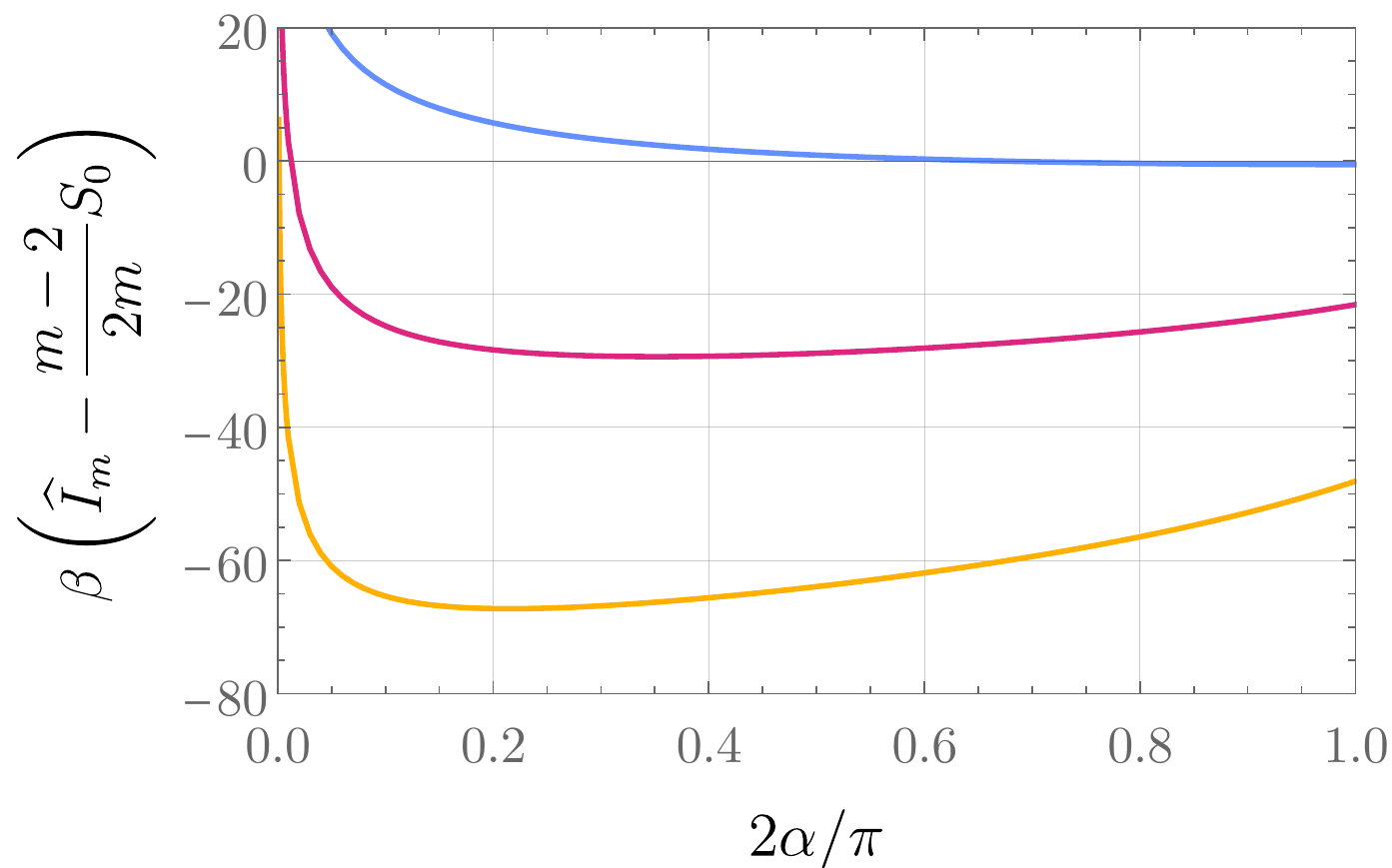}
\includegraphics[height=3.5cm]{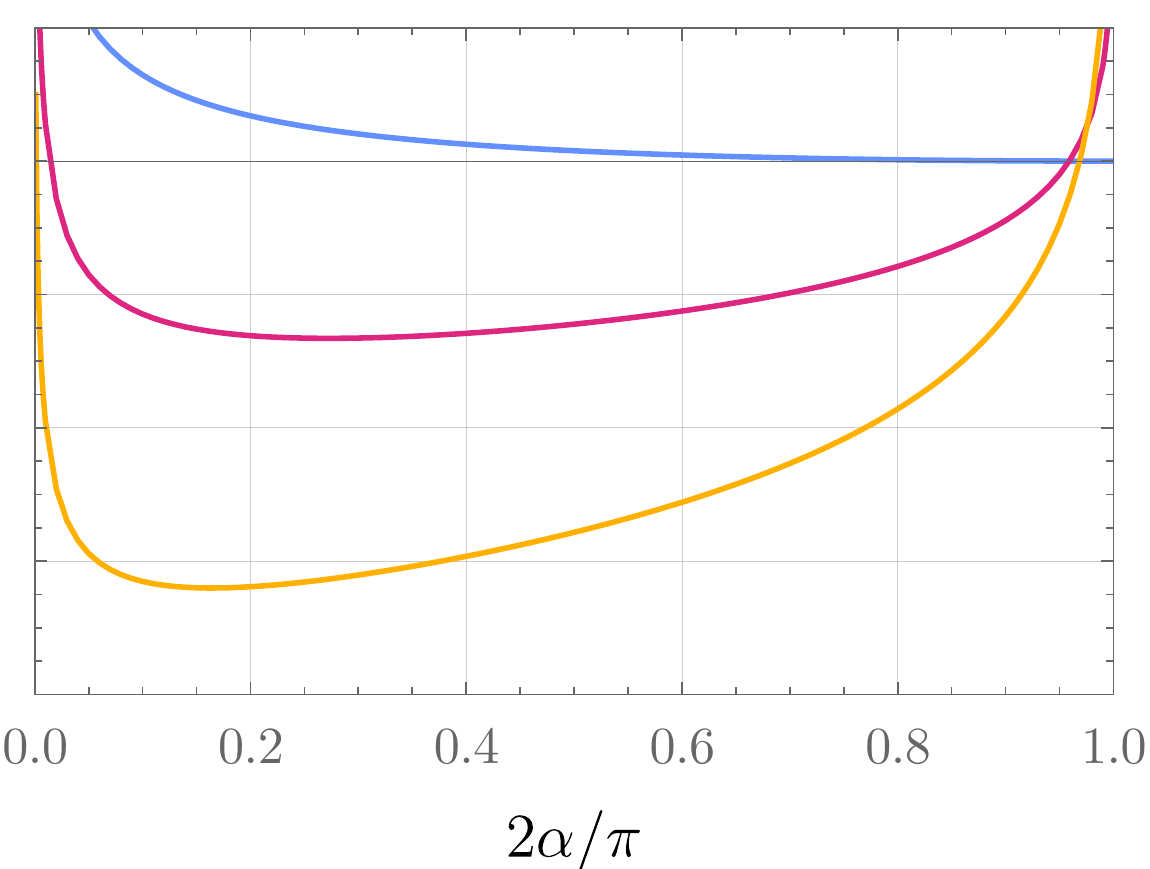}
\includegraphics[height=3.5cm]{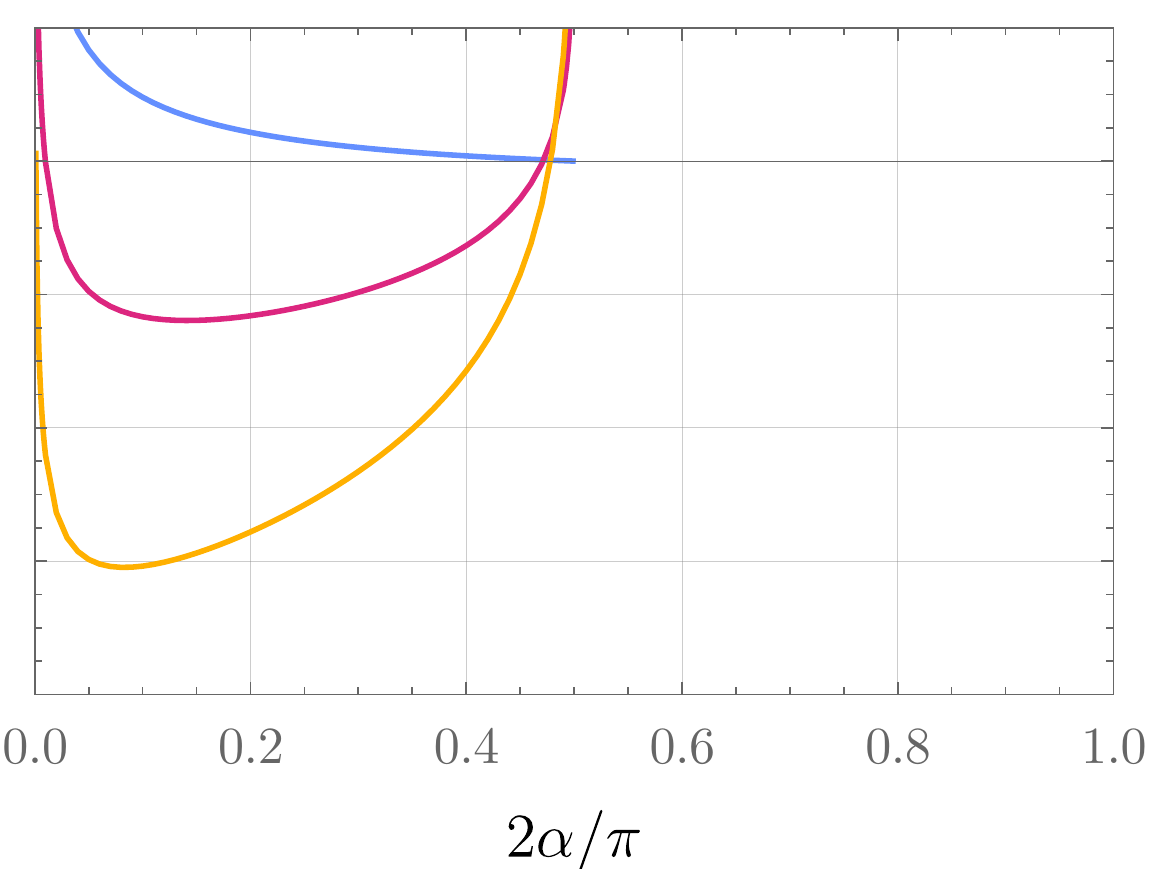}
\caption{The effective action~$\widehat{I}_m[\alpha]$ for various values of~$m > 1$ and~$\beta\mu$.  From left to right we show~$m = 3/2$,~2, and~4, while the blue, red, and orange curves (uppermost to lowermost within each plot) correspond to~$\beta\mu = 0$,~5, and~10, respectively.  Note that for any~$m > 1$ and~$\beta\mu \neq 0$,~$\widehat{I}_m[\alpha]$ exhibits a local minimum in~$\alpha$.  The spectrum of~$L$ is also nonnegative on all of these solutions, so these minima in~$\alpha$ correspond to stable wormholes.}
\label{fig:mg1actions}
\end{figure}

\begin{figure}[t]
\centering
\includegraphics[height=3.5cm]{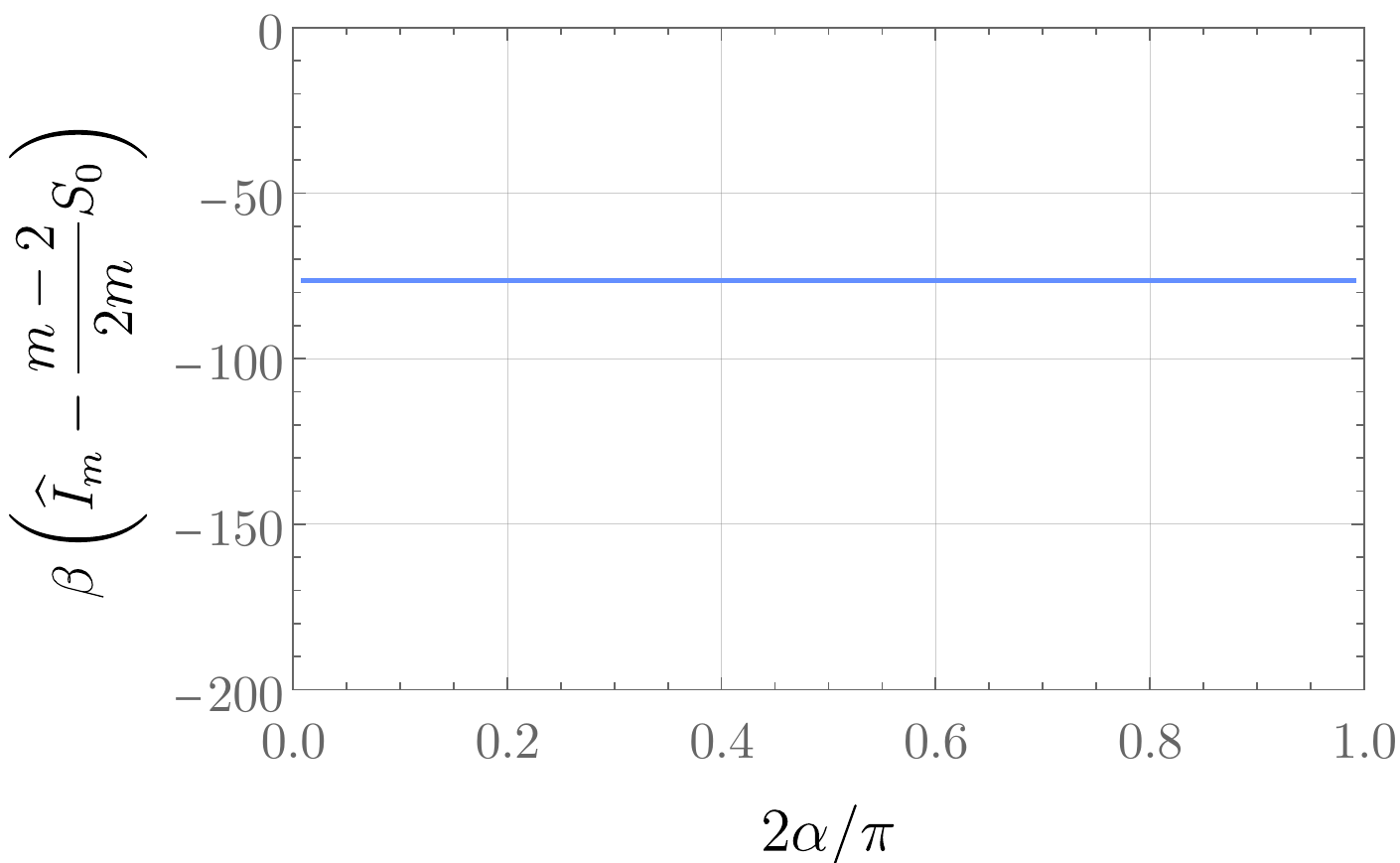}
\includegraphics[height=3.5cm]{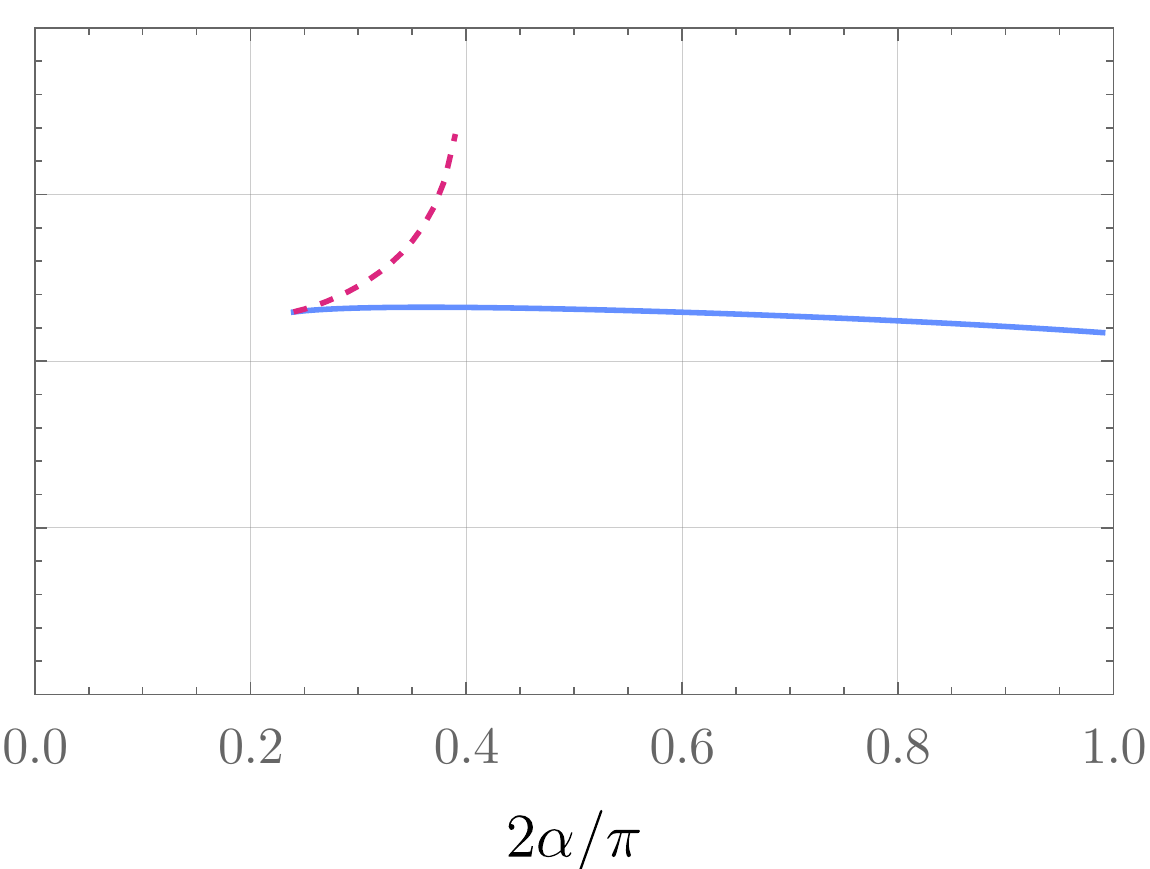}
\includegraphics[height=3.5cm]{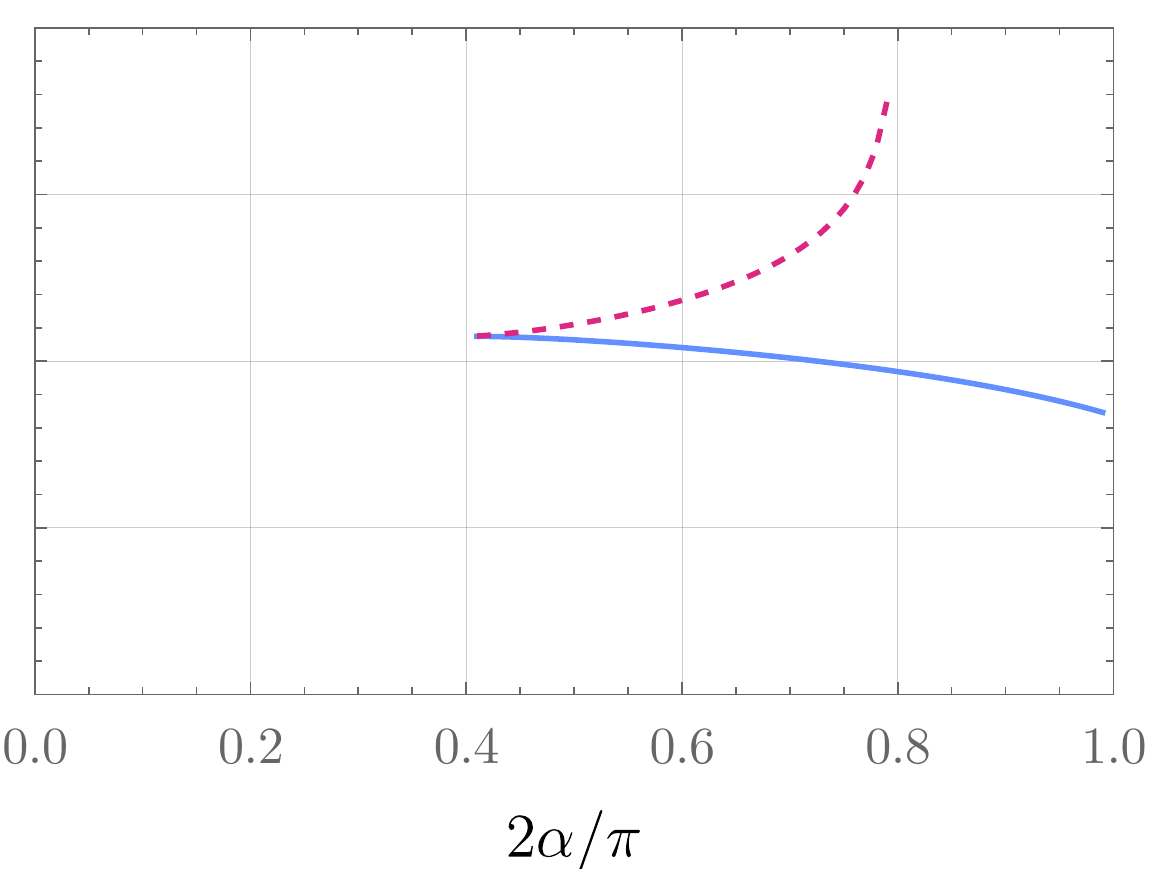} \\
\includegraphics[height=3.5cm]{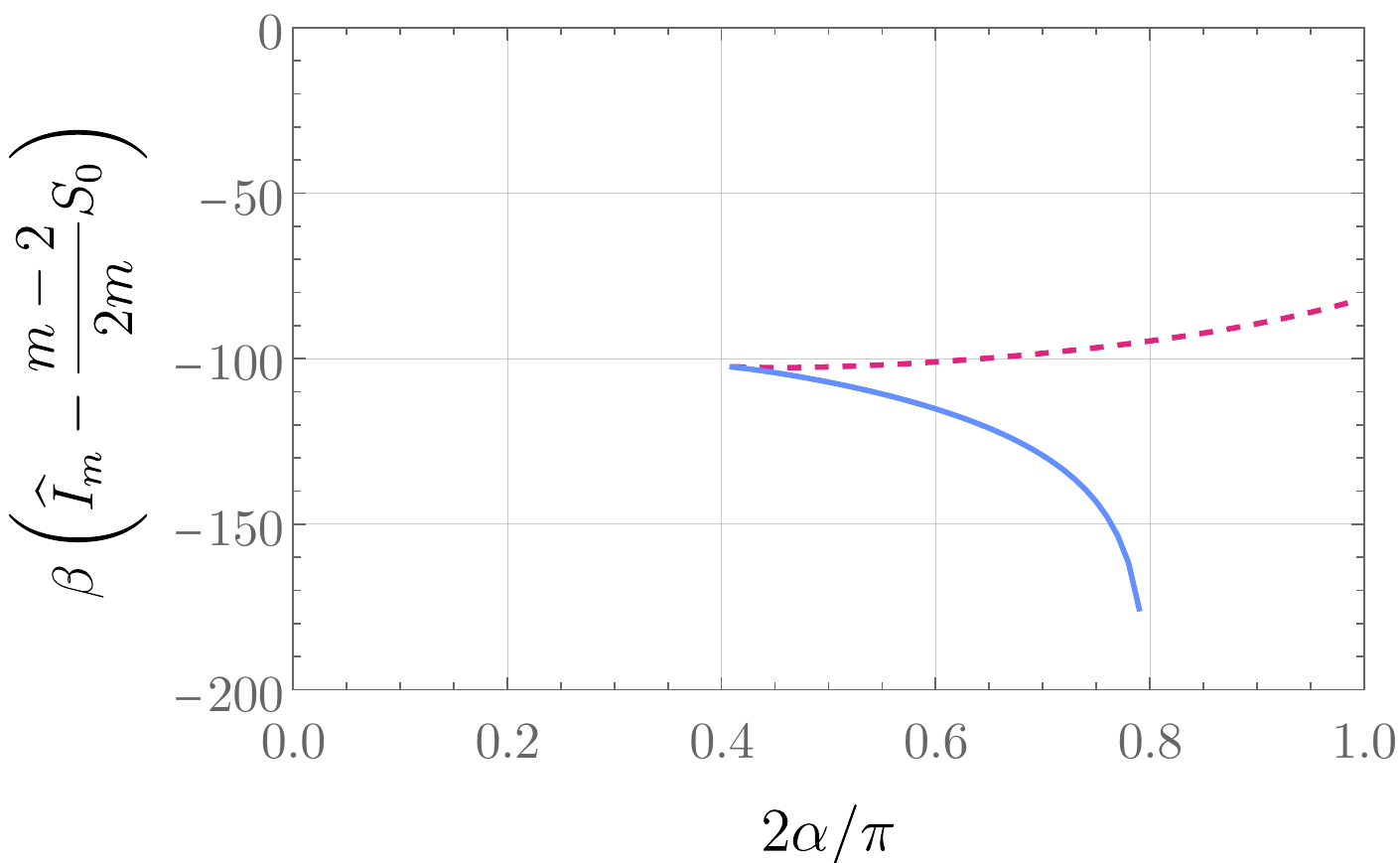}
\includegraphics[height=3.5cm]{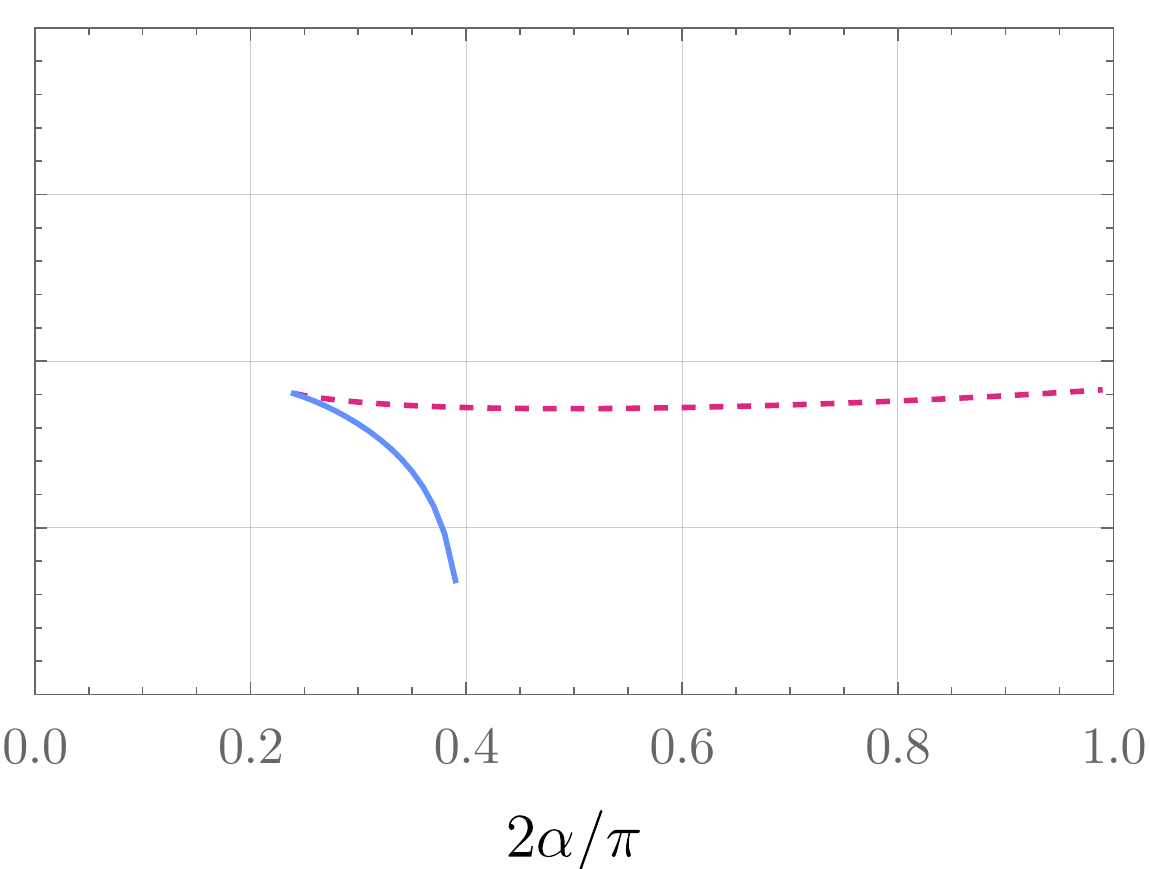}
\includegraphics[height=3.5cm]{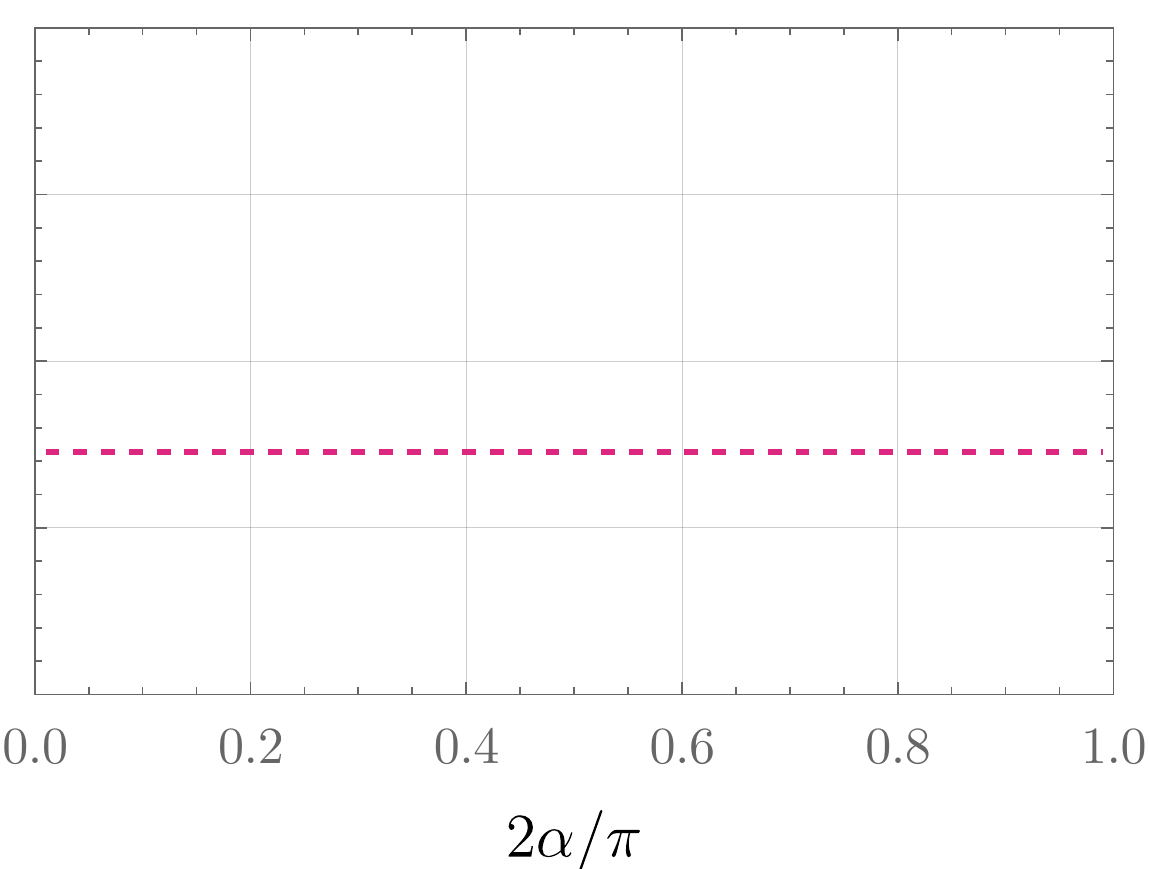}
\caption{The effective action~$\widehat{I}_m[\alpha]$ for various values of~$m \leq 1$ with~$\beta\mu = 10$.  From left to right and top to bottom, we show~$1/m = 1$,~1.2,~1.4,~1.6,~1.8, and~2.  Except for when~$1/m$ is an integer, there are two branches: on the ones drawn as solid blue curves, the spectrum of~$L$ is nonnegative, while on the dashed red curves~$L$ has a negative eigenvalue.  Note that although there are some saddles for~$\alpha$, these saddles are all unstable, either to perturbations of~$\alpha$ (as on the solid blue branch in the second plot) or of the wiggle (as in the dashed red branch in the fifth plot).  The behavior for larger~$1/m$ is analogous, except that all of the solutions for the wiggle are unstable.}
\label{fig:ml1actionsm}
\end{figure}

\begin{figure}[t]
\centering
\includegraphics[height=3.5cm]{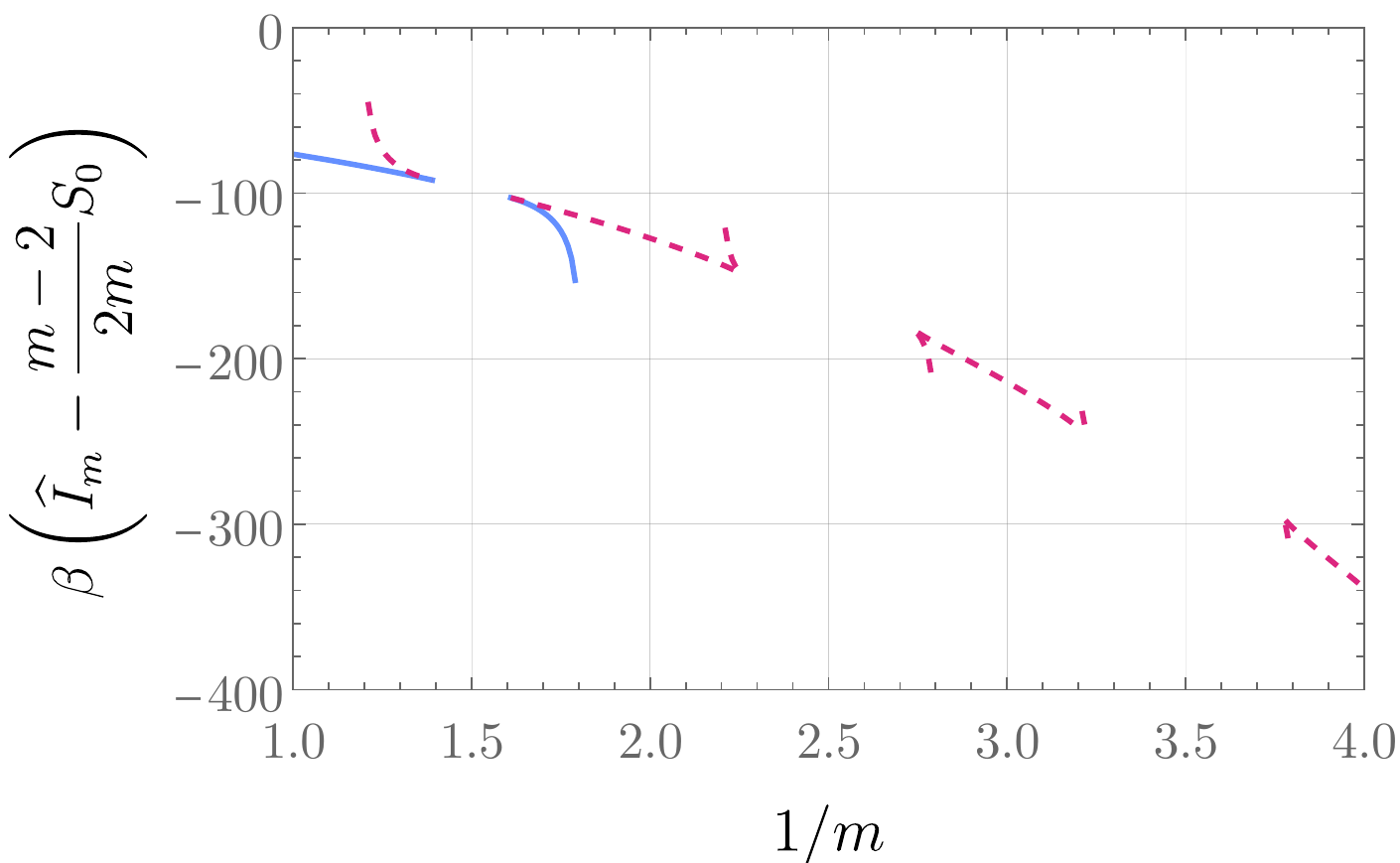}
\includegraphics[height=3.5cm]{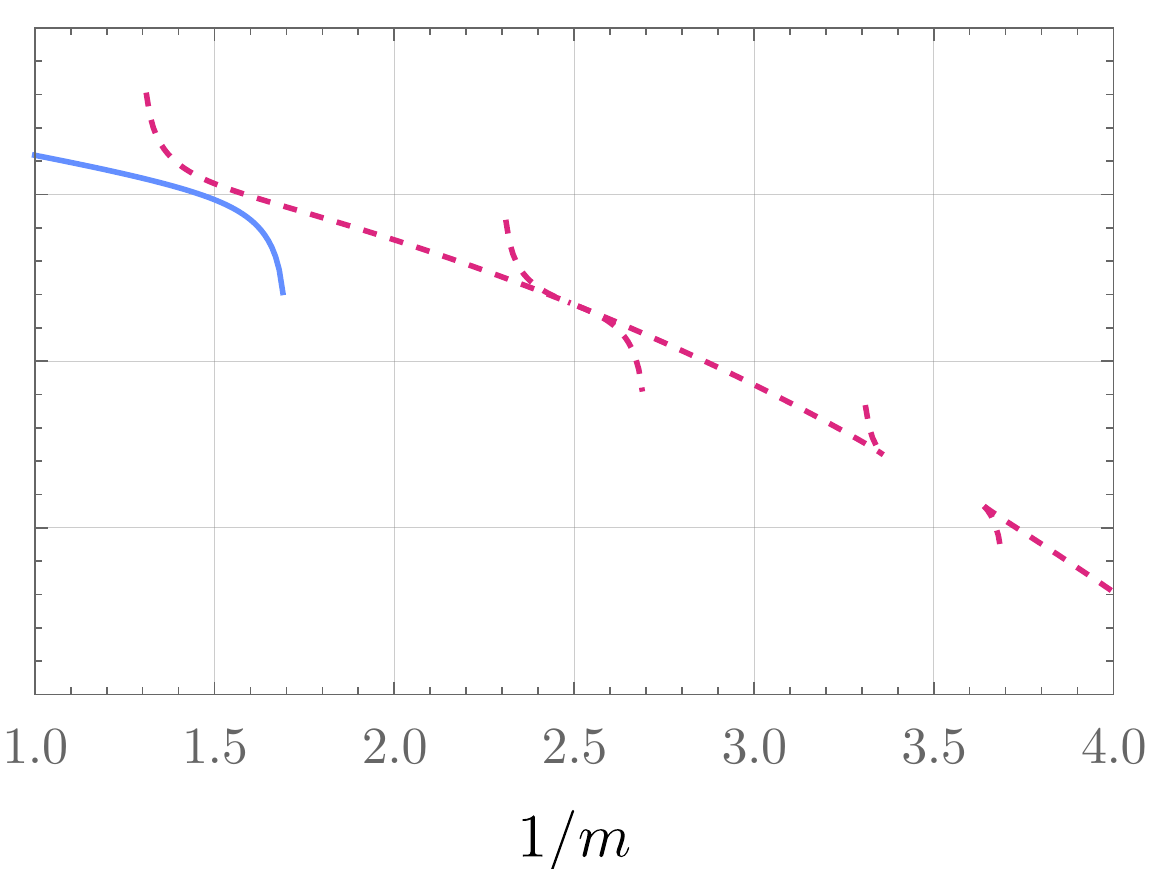}
\includegraphics[height=3.5cm]{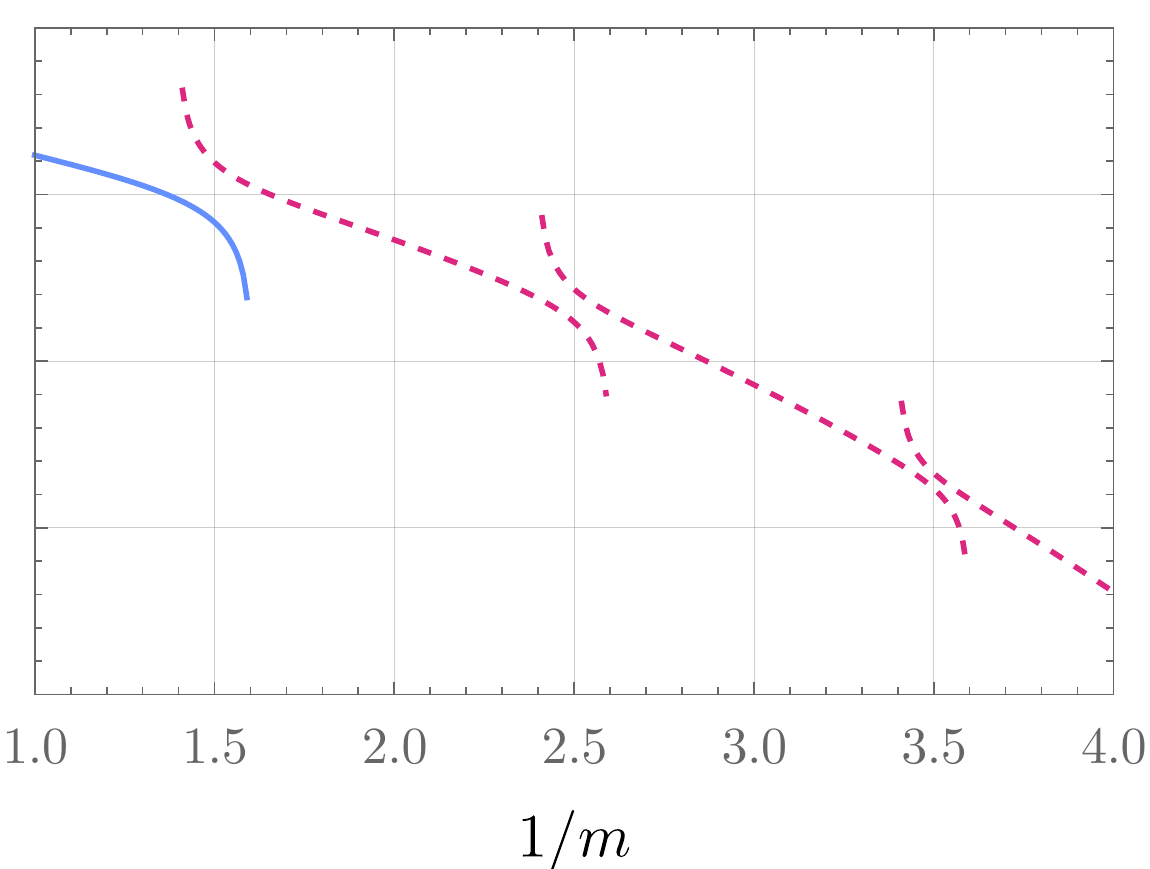}
\caption{The effective action~$\widehat{I}_m[\alpha]$ as a function of~$m$ for~$\beta\mu = 10$ and various values of~$\alpha$; from left to right we show~$\alpha = \pi/5$,~$3\pi/10$, and~$2\pi/5$.  On the solid blue branch(es) the spectrum of~$L$ is nonnegative; on the dashed red branches~$L$ has a negative eigenvalue.}
\label{fig:ml1actionsalpha}
\end{figure}

When~$m \geq 1$, we always have~$N = 0$ and~$\sin(\pi/m)/\sin\alpha \geq 0$ for any allowed~$\alpha$, from which it follows that precisely one solution will exist for any allowed value of~$\alpha$ or~$\mu$.  Figure~\ref{fig:mg1actions} shows~$\widehat{I}_m[\alpha]$ for various values of~$\beta\mu$ and~$m$.  Note that for any nonzero~$\beta\mu$ there is a local minimum, meaning that the modulus is stabilized.  Moreover, we have verified that the spectrum of~$L$ is nonnegative for all of these solutions, so we conclude that turning on the EOW branes stabilizes the wormholes when~$m > 1$ (though the wormholes do not appear to dominate over the disk: it is clear from Figure~\ref{fig:mg1actions} that the on-shell action~$\widehat{I}_m[\alpha_\mathrm{min}]$ evaluated at the saddle~$\alpha_\mathrm{min}$ grows with~$m$).  On the other hand, the behavior of~$\widehat{I}_m[\alpha]$ when~$m < 1$ is quite different.  It is clear from~\eqref{eqs:JTbraneoscillatoryab} that~$\widehat{I}_m[\alpha]$ is single-valued and independent of~$\alpha$ whenever~$1/m$ is an integer, just as in pure JT.  For intermediate values of~$1/m$, there are instead always two branches of~$\widehat{I}_m[\alpha]$, as shown in Figure~\ref{fig:ml1actionsm} for~$\beta\mu = 10$; see also Figure~\ref{fig:ml1actionsalpha} (the behavior for other nonzero values of~$\beta\mu$ is analogous).  These branches can exhibit either stable, unstable, or no saddles in~$\alpha$.  However, any stable saddles for~$\alpha$ coincide with unstable saddles for the wiggle; conversely, the branches of solutions on which the wiggle is stable (which only exist for~$m \geq 1/2$) only exhibit unstable saddles for the modulus.  Hence this classical JT+brane model does not admit any stable wormholes for~$m < 1$.

Just as in pure JT, the branches in Figures~\ref{fig:ml1actionsm} and~\ref{fig:ml1actionsalpha} that appear to simply end are indicative of additional sheets of the analytic continuation of~$\widehat{I}_m[\alpha]$ to complex~$m$ (and complex~$\alpha$, if desired).  This analytic continuation is obtained by inverting~\eqref{subeq:JTbraneoscillatoryb} for~$b_r$ ignoring the constraint~\eqref{subeq:bconstraintbrane}; the inverse function~$b_r(\sin(\pi/m)/\sin\alpha)$ is a meromorphic function of~$m$ whose Riemann surface contains infinitely many sheets, and the action~\eqref{eq:JTbraneonshellaction} is infinitely-sheeted as well.  Hence we can attribute the ends of the ``branches to nowhere'' in Figures~\ref{fig:ml1actionsm} and~\ref{fig:ml1actionsalpha} as stemming from the constraint~\eqref{subeq:bconstraintbrane} that fixes the allowed branches of this Riemann surface.  The upshot is that as in pure JT, imposing the equations of motions is crucial to constraining the allowed behavior of the analytic continuation of~$\widehat{I}_m$: merely continuing a portion of the action (from, say,~$m > 1$) to all~$m$ without invoking the equations of motion is insufficient to uniquely fix the allowed behavior of the action at~$m < 1$.

(It is also worth noting that the endpoints of the ``branches to nowhere'' in Figures~\ref{fig:ml1actionsm} and~\ref{fig:ml1actionsalpha} correspond to singular solutions for the wiggle, which can be seen as follows.  As the graphical analysis in Figure~\ref{fig:oscillatorysolutions} indicates, as~$\alpha$ and/or~$m$ are varied a single branch of solutions can end when the allowed interval for~$b_r$~\eqref{subeq:bconstraintbrane} changes due to a change in the sign of~$\tan((\pi-m\alpha)/2m)$.  On this branch,~$b_r/2\pi$ is not an integer as this transition is reached, so from~\eqref{subeq:JTbraneoscillatorya} and~\eqref{eq:JTbraneonshellaction} we conclude that~$a_r$ either vanishes or diverges there while the action remains finite.  Hence these endpoints correspond to singular configurations of the wiggle with finite action.)

\bibliographystyle{jhep}
\bibliography{all}

\providecommand{\href}[2]{#2}\begingroup\raggedright\begin{thebibliography}{10}

\bibitem{Col88}
S.~R. Coleman, \emph{{Black Holes as Red Herrings: Topological Fluctuations and
  the Loss of Quantum Coherence}},
  \href{https://doi.org/10.1016/0550-3213(88)90110-1}{\emph{Nucl. Phys. B}
  {\bfseries 307} (1988) 867}.

\bibitem{GidStr88}
S.~B. Giddings and A.~Strominger, \emph{{Loss of Incoherence and Determination
  of Coupling Constants in Quantum Gravity}},
  \href{https://doi.org/10.1016/0550-3213(88)90109-5}{\emph{Nucl. Phys. B}
  {\bfseries 307} (1988) 854}.

\bibitem{MalMao04}
J.~M. Maldacena and L.~Maoz, \emph{{Wormholes in AdS}},
  \href{https://doi.org/10.1088/1126-6708/2004/02/053}{\emph{JHEP} {\bfseries
  02} (2004) 053} [\href{https://arxiv.org/abs/hep-th/0401024}{{\ttfamily
  hep-th/0401024}}].

\bibitem{ArkOrg07}
N.~Arkani-Hamed, J.~Orgera and J.~Polchinski, \emph{{Euclidean wormholes in
  string theory}},
  \href{https://doi.org/10.1088/1126-6708/2007/12/018}{\emph{JHEP} {\bfseries
  12} (2007) 018} [\href{https://arxiv.org/abs/0705.2768}{{\ttfamily
  0705.2768}}].

\bibitem{HarJaf18}
D.~Harlow and D.~Jafferis, \emph{{The Factorization Problem in
  Jackiw-Teitelboim Gravity}},
  \href{https://doi.org/10.1007/JHEP02(2020)177}{\emph{JHEP} {\bfseries 02}
  (2020) 177} [\href{https://arxiv.org/abs/1804.01081}{{\ttfamily
  1804.01081}}].

\bibitem{SSS}
P.~Saad, S.~H. Shenker and D.~Stanford, \emph{{JT gravity as a matrix
  integral}},  \href{https://arxiv.org/abs/1903.11115}{{\ttfamily 1903.11115}}.

\bibitem{AlmHar19}
A.~Almheiri, T.~Hartman, J.~Maldacena, E.~Shaghoulian and A.~Tajdini,
  \emph{{Replica Wormholes and the Entropy of Hawking Radiation}},
  \href{https://doi.org/10.1007/JHEP05(2020)013}{\emph{JHEP} {\bfseries 05}
  (2020) 013} [\href{https://arxiv.org/abs/1911.12333}{{\ttfamily
  1911.12333}}].

\bibitem{PenShe19}
G.~Penington, S.~H. Shenker, D.~Stanford and Z.~Yang, \emph{{Replica wormholes
  and the black hole interior}},
  \href{https://doi.org/10.1007/JHEP03(2022)205}{\emph{JHEP} {\bfseries 03}
  (2022) 205} [\href{https://arxiv.org/abs/1911.11977}{{\ttfamily
  1911.11977}}].

\bibitem{StaWit19}
D.~Stanford and E.~Witten, \emph{{JT gravity and the ensembles of random matrix
  theory}}, \href{https://doi.org/10.4310/ATMP.2020.v24.n6.a4}{\emph{Adv.
  Theor. Math. Phys.} {\bfseries 24} (2020) 1475}
  [\href{https://arxiv.org/abs/1907.03363}{{\ttfamily 1907.03363}}].

\bibitem{Ili19}
L.~V. Iliesiu, \emph{{On 2D gauge theories in Jackiw-Teitelboim gravity}},
  \href{https://arxiv.org/abs/1909.05253}{{\ttfamily 1909.05253}}.

\bibitem{KapMah19}
D.~Kapec, R.~Mahajan and D.~Stanford, \emph{{Matrix ensembles with global
  symmetries and 't Hooft anomalies from 2d gauge theory}},
  \href{https://doi.org/10.1007/JHEP04(2020)186}{\emph{JHEP} {\bfseries 04}
  (2020) 186} [\href{https://arxiv.org/abs/1912.12285}{{\ttfamily
  1912.12285}}].

\bibitem{Saa19}
P.~Saad, \emph{{Late Time Correlation Functions, Baby Universes, and ETH in JT
  Gravity}},  \href{https://arxiv.org/abs/1910.10311}{{\ttfamily 1910.10311}}.

\bibitem{MarMax20}
D.~Marolf and H.~Maxfield, \emph{{Transcending the ensemble: baby universes,
  spacetime wormholes, and the order and disorder of black hole information}},
  \href{https://doi.org/10.1007/JHEP08(2020)044}{\emph{JHEP} {\bfseries 08}
  (2020) 044} [\href{https://arxiv.org/abs/2002.08950}{{\ttfamily
  2002.08950}}].

\bibitem{MarMax20b}
D.~Marolf and H.~Maxfield, \emph{{Observations of Hawking radiation: the Page
  curve and baby universes}},
  \href{https://doi.org/10.1007/JHEP04(2021)272}{\emph{JHEP} {\bfseries 04}
  (2021) 272} [\href{https://arxiv.org/abs/2010.06602}{{\ttfamily
  2010.06602}}].

\bibitem{GidTur20}
S.~B. Giddings and G.~J. Turiaci, \emph{{Wormhole calculus, replicas, and
  entropies}}, \href{https://doi.org/10.1007/JHEP09(2020)194}{\emph{JHEP}
  {\bfseries 09} (2020) 194}
  [\href{https://arxiv.org/abs/2004.02900}{{\ttfamily 2004.02900}}].

\bibitem{LewMal13}
A.~Lewkowycz and J.~Maldacena, \emph{{Generalized gravitational entropy}},
  \href{https://doi.org/10.1007/JHEP08(2013)090}{\emph{JHEP} {\bfseries 1308}
  (2013) 090} [\href{https://arxiv.org/abs/1304.4926}{{\ttfamily 1304.4926}}].

\bibitem{FauLew13}
T.~Faulkner, A.~Lewkowycz and J.~Maldacena, \emph{{Quantum corrections to
  holographic entanglement entropy}},
  \href{https://doi.org/10.1007/JHEP11(2013)074}{\emph{JHEP} {\bfseries 1311}
  (2013) 074} [\href{https://arxiv.org/abs/1307.2892}{{\ttfamily 1307.2892}}].

\bibitem{EngWal14}
N.~Engelhardt and A.~C. Wall, \emph{{Quantum Extremal Surfaces: Holographic
  Entanglement Entropy beyond the Classical Regime}},
  \href{https://doi.org/10.1007/JHEP01(2015)073}{\emph{JHEP} {\bfseries 01}
  (2015) 073} [\href{https://arxiv.org/abs/1408.3203}{{\ttfamily 1408.3203}}].

\bibitem{AEMM}
A.~Almheiri, N.~Engelhardt, D.~Marolf and H.~Maxfield, \emph{{The entropy of
  bulk quantum fields and the entanglement wedge of an evaporating black
  hole}}, \href{https://doi.org/10.1007/JHEP12(2019)063}{\emph{JHEP} {\bfseries
  12} (2019) 063} [\href{https://arxiv.org/abs/1905.08762}{{\ttfamily
  1905.08762}}].

\bibitem{Pen19}
G.~Penington, \emph{{Entanglement Wedge Reconstruction and the Information
  Paradox}}, \href{https://doi.org/10.1007/JHEP09(2020)002}{\emph{JHEP}
  {\bfseries 09} (2020) 002}
  [\href{https://arxiv.org/abs/1905.08255}{{\ttfamily 1905.08255}}].

\bibitem{AreVol19}
I.~Aref'eva and I.~Volovich, \emph{{Gas of baby universes in JT gravity and
  matrix models}}, \href{https://doi.org/10.3390/sym12060975}{\emph{Symmetry}
  {\bfseries 12} (2020) 975}
  [\href{https://arxiv.org/abs/1905.08207}{{\ttfamily 1905.08207}}].

\bibitem{BouWil20}
R.~Bousso and E.~Wildenhain, \emph{{Gravity/ensemble duality}},
  \href{https://doi.org/10.1103/PhysRevD.102.066005}{\emph{Phys. Rev. D}
  {\bfseries 102} (2020) 066005}
  [\href{https://arxiv.org/abs/2006.16289}{{\ttfamily 2006.16289}}].

\bibitem{Wit20b}
E.~Witten, \emph{{Matrix Models and Deformations of JT Gravity}},
  \href{https://doi.org/10.1098/rspa.2020.0582}{\emph{Proc. Roy. Soc. Lond. A}
  {\bfseries 476} (2020) 20200582}
  [\href{https://arxiv.org/abs/2006.13414}{{\ttfamily 2006.13414}}].

\bibitem{AfkCoh20}
N.~Afkhami-Jeddi, H.~Cohn, T.~Hartman and A.~Tajdini, \emph{{Free partition
  functions and an averaged holographic duality}},
  \href{https://doi.org/10.1007/JHEP01(2021)130}{\emph{JHEP} {\bfseries 01}
  (2021) 130} [\href{https://arxiv.org/abs/2006.04839}{{\ttfamily
  2006.04839}}].

\bibitem{BeldeB20b}
A.~Belin, J.~De~Boer, P.~Nayak and J.~Sonner, \emph{{Charged eigenstate
  thermalization, Euclidean wormholes and global symmetries in quantum
  gravity}}, \href{https://doi.org/10.21468/SciPostPhys.12.2.059}{\emph{SciPost
  Phys.} {\bfseries 12} (2022) 059}
  [\href{https://arxiv.org/abs/2012.07875}{{\ttfamily 2012.07875}}].

\bibitem{Van20}
M.~Van~Raamsdonk, \emph{{Comments on wormholes, ensembles, and cosmology}},
  \href{https://arxiv.org/abs/2008.02259}{{\ttfamily 2008.02259}}.

\bibitem{Blo20}
A.~Blommaert, \emph{{Dissecting the ensemble in JT gravity}},
  \href{https://arxiv.org/abs/2006.13971}{{\ttfamily 2006.13971}}.

\bibitem{MalWit20}
A.~Maloney and E.~Witten, \emph{{Averaging over Narain moduli space}},
  \href{https://doi.org/10.1007/JHEP10(2020)187}{\emph{JHEP} {\bfseries 10}
  (2020) 187} [\href{https://arxiv.org/abs/2006.04855}{{\ttfamily
  2006.04855}}].

\bibitem{CotJen20}
J.~Cotler and K.~Jensen, \emph{{AdS$_{3}$ gravity and random CFT}},
  \href{https://doi.org/10.1007/JHEP04(2021)033}{\emph{JHEP} {\bfseries 04}
  (2021) 033} [\href{https://arxiv.org/abs/2006.08648}{{\ttfamily
  2006.08648}}].

\bibitem{PerTro20}
A.~P\'erez and R.~Troncoso, \emph{{Gravitational dual of averaged free
  CFT\textquoteright{}s over the Narain lattice}},
  \href{https://doi.org/10.1007/JHEP11(2020)015}{\emph{JHEP} {\bfseries 11}
  (2020) 015} [\href{https://arxiv.org/abs/2006.08216}{{\ttfamily
  2006.08216}}].

\bibitem{JanMir21}
O.~Janssen, M.~Mirbabayi and P.~Zograf, \emph{{Gravity as an ensemble and the
  moment problem}}, \href{https://doi.org/10.1007/JHEP06(2021)184}{\emph{JHEP}
  {\bfseries 06} (2021) 184}
  [\href{https://arxiv.org/abs/2103.12078}{{\ttfamily 2103.12078}}].

\bibitem{TurUsa21}
G.~J. Turiaci, M.~Usatyuk and W.~W. Weng, \emph{{2D dilaton-gravity,
  deformations of the minimal string, and matrix models}},
  \href{https://doi.org/10.1088/1361-6382/ac25df}{\emph{Class. Quant. Grav.}
  {\bfseries 38} (2021) 204001}
  [\href{https://arxiv.org/abs/2011.06038}{{\ttfamily 2011.06038}}].

\bibitem{MaxTur21}
H.~Maxfield and G.~J. Turiaci, \emph{{The path integral of 3D gravity near
  extremality; or, JT gravity with defects as a matrix integral}},
  \href{https://doi.org/10.1007/JHEP01(2021)118}{\emph{JHEP} {\bfseries 01}
  (2021) 118} [\href{https://arxiv.org/abs/2006.11317}{{\ttfamily
  2006.11317}}].

\bibitem{BenKel21}
N.~Benjamin, C.~A. Keller, H.~Ooguri and I.~G. Zadeh, \emph{{Narain to
  Narnia}}, \href{https://doi.org/10.1007/s00220-021-04211-x}{\emph{Commun.
  Math. Phys.} {\bfseries 390} (2022) 425}
  [\href{https://arxiv.org/abs/2103.15826}{{\ttfamily 2103.15826}}].

\bibitem{PenTia22}
C.~Peng, J.~Tian and Y.~Yang, \emph{{Half-Wormholes and Ensemble Averages}},
  \href{https://arxiv.org/abs/2205.01288}{{\ttfamily 2205.01288}}.

\bibitem{EngFis20}
N.~Engelhardt, S.~Fischetti and A.~Maloney, \emph{{Free energy from replica
  wormholes}}, \href{https://doi.org/10.1103/PhysRevD.103.046021}{\emph{Phys.
  Rev. D} {\bfseries 103} (2021) 046021}
  [\href{https://arxiv.org/abs/2007.07444}{{\ttfamily 2007.07444}}].

\bibitem{Joh19}
C.~V. Johnson, \emph{{Nonperturbative Jackiw-Teitelboim gravity}},
  \href{https://doi.org/10.1103/PhysRevD.101.106023}{\emph{Phys. Rev. D}
  {\bfseries 101} (2020) 106023}
  [\href{https://arxiv.org/abs/1912.03637}{{\ttfamily 1912.03637}}].

\bibitem{Joh20a}
C.~V. Johnson, \emph{{Jackiw-Teitelboim supergravity, minimal strings, and
  matrix models}},
  \href{https://doi.org/10.1103/PhysRevD.103.046012}{\emph{Phys. Rev. D}
  {\bfseries 103} (2021) 046012}
  [\href{https://arxiv.org/abs/2005.01893}{{\ttfamily 2005.01893}}].

\bibitem{Joh20b}
C.~V. Johnson, \emph{{Explorations of nonperturbative Jackiw-Teitelboim gravity
  and supergravity}},
  \href{https://doi.org/10.1103/PhysRevD.103.046013}{\emph{Phys. Rev. D}
  {\bfseries 103} (2021) 046013}
  [\href{https://arxiv.org/abs/2006.10959}{{\ttfamily 2006.10959}}].

\bibitem{Joh20c}
C.~V. Johnson, \emph{{Low Energy Thermodynamics of JT Gravity and
  Supergravity}},  \href{https://arxiv.org/abs/2008.13120}{{\ttfamily
  2008.13120}}.

\bibitem{Joh21a}
C.~V. Johnson, \emph{{On the Quenched Free Energy of JT Gravity and
  Supergravity}},  \href{https://arxiv.org/abs/2104.02733}{{\ttfamily
  2104.02733}}.

\bibitem{Joh21b}
C.~V. Johnson, \emph{{Quantum Gravity Microstates from Fredholm Determinants}},
  \href{https://doi.org/10.1103/PhysRevLett.127.181602}{\emph{Phys. Rev. Lett.}
  {\bfseries 127} (2021) 181602}
  [\href{https://arxiv.org/abs/2106.09048}{{\ttfamily 2106.09048}}].

\bibitem{Joh21c}
C.~V. Johnson, \emph{{Consistency Conditions for Non-Perturbative Completions
  of JT Gravity}},  \href{https://arxiv.org/abs/2112.00766}{{\ttfamily
  2112.00766}}.

\bibitem{Joh22a}
C.~V. Johnson, \emph{{The Microstate Physics of JT Gravity and Supergravity}},
  \href{https://arxiv.org/abs/2201.11942}{{\ttfamily 2201.11942}}.

\bibitem{Joh22b}
C.~V. Johnson, \emph{{The Distribution of Ground State Energies in JT
  Gravity}},  \href{https://arxiv.org/abs/2206.00692}{{\ttfamily 2206.00692}}.

\bibitem{BeldeB20}
A.~Belin and J.~de~Boer, \emph{{Random statistics of OPE coefficients and
  Euclidean wormholes}},
  \href{https://doi.org/10.1088/1361-6382/ac1082}{\emph{Class. Quant. Grav.}
  {\bfseries 38} (2021) 164001}
  [\href{https://arxiv.org/abs/2006.05499}{{\ttfamily 2006.05499}}].

\bibitem{PolRoz20}
J.~Pollack, M.~Rozali, J.~Sully and D.~Wakeham, \emph{{Eigenstate
  Thermalization and Disorder Averaging in Gravity}},
  \href{https://doi.org/10.1103/PhysRevLett.125.021601}{\emph{Phys. Rev. Lett.}
  {\bfseries 125} (2020) 021601}
  [\href{https://arxiv.org/abs/2002.02971}{{\ttfamily 2002.02971}}].

\bibitem{Ebe21}
L.~Eberhardt, \emph{{Summing over Geometries in String Theory}},
  \href{https://doi.org/10.1007/JHEP05(2021)233}{\emph{JHEP} {\bfseries 05}
  (2021) 233} [\href{https://arxiv.org/abs/2102.12355}{{\ttfamily
  2102.12355}}].

\bibitem{SaaShe21}
P.~Saad, S.~H. Shenker, D.~Stanford and S.~Yao, \emph{{Wormholes without
  averaging}},  \href{https://arxiv.org/abs/2103.16754}{{\ttfamily
  2103.16754}}.

\bibitem{GarGot21}
A.~M. Garc\'\i{}a-Garc\'\i{}a and V.~Godet, \emph{{Half-wormholes in nearly
  AdS$_2$ holography}},
  \href{https://doi.org/10.21468/SciPostPhys.12.4.135}{\emph{SciPost Phys.}
  {\bfseries 12} (2022) 135}
  [\href{https://arxiv.org/abs/2107.07720}{{\ttfamily 2107.07720}}].

\bibitem{BloIli21}
A.~Blommaert, L.~V. Iliesiu and J.~Kruthoff, \emph{{Gravity factorized}},
  \href{https://arxiv.org/abs/2111.07863}{{\ttfamily 2111.07863}}.

\bibitem{Don16}
X.~Dong, \emph{{The Gravity Dual of Renyi Entropy}},
  \href{https://doi.org/10.1038/ncomms12472}{\emph{Nature Commun.} {\bfseries
  7} (2016) 12472} [\href{https://arxiv.org/abs/1601.06788}{{\ttfamily
  1601.06788}}].

\bibitem{MarSan21}
D.~Marolf and J.~E. Santos, \emph{{AdS Euclidean wormholes}},
  \href{https://doi.org/10.1088/1361-6382/ac2cb7}{\emph{Class. Quant. Grav.}
  {\bfseries 38} (2021) 224002}
  [\href{https://arxiv.org/abs/2101.08875}{{\ttfamily 2101.08875}}].

\bibitem{GarGot20}
A.~M. Garc\'\i{}a-Garc\'\i{}a and V.~Godet, \emph{{Euclidean wormhole in the
  Sachdev-Ye-Kitaev model}},
  \href{https://doi.org/10.1103/PhysRevD.103.046014}{\emph{Phys. Rev. D}
  {\bfseries 103} (2021) 046014}
  [\href{https://arxiv.org/abs/2010.11633}{{\ttfamily 2010.11633}}].

\bibitem{MoiSak21}
U.~Moitra, S.~K. Sake and S.~P. Trivedi, \emph{{Jackiw-Teitelboim gravity in
  the second order formalism}},
  \href{https://doi.org/10.1007/JHEP10(2021)204}{\emph{JHEP} {\bfseries 10}
  (2021) 204} [\href{https://arxiv.org/abs/2101.00596}{{\ttfamily
  2101.00596}}].

\bibitem{MoiSak22}
U.~Moitra, S.~K. Sake and S.~P. Trivedi, \emph{{Aspects of Jackiw-Teitelboim
  gravity in Anti-de Sitter and de Sitter spacetime}},
  \href{https://doi.org/10.1007/JHEP06(2022)138}{\emph{JHEP} {\bfseries 06}
  (2022) 138} [\href{https://arxiv.org/abs/2202.03130}{{\ttfamily
  2202.03130}}].

\bibitem{SpinGlassBook}
M.~Mezard, G.~Parisi and M.~Virasoro, \emph{Spin Glass Theory and Beyond}.
  World Scientific, 1986, \href{https://doi.org/10.1142/0271}{10.1142/0271}.

\bibitem{RyuTak06}
S.~Ryu and T.~Takayanagi, \emph{{Holographic derivation of entanglement entropy
  from AdS/CFT}},
  \href{https://doi.org/10.1103/PhysRevLett.96.181602}{\emph{Phys.Rev.Lett.}
  {\bfseries 96} (2006) 181602}
  [\href{https://arxiv.org/abs/hep-th/0603001}{{\ttfamily hep-th/0603001}}].

\bibitem{HubRan07}
V.~E. Hubeny, M.~Rangamani and T.~Takayanagi, \emph{{A Covariant holographic
  entanglement entropy proposal}},
  \href{https://doi.org/10.1088/1126-6708/2007/07/062}{\emph{JHEP} {\bfseries
  0707} (2007) 062} [\href{https://arxiv.org/abs/0705.0016}{{\ttfamily
  0705.0016}}].

\bibitem{MalSta16b}
J.~Maldacena, D.~Stanford and Z.~Yang, \emph{{Conformal symmetry and its
  breaking in two dimensional Nearly Anti-de-Sitter space}},
  \href{https://doi.org/10.1093/ptep/ptw124}{\emph{PTEP} {\bfseries 2016}
  (2016) 12C104} [\href{https://arxiv.org/abs/1606.01857}{{\ttfamily
  1606.01857}}].

\bibitem{Trefethen}
L.~N. Trefethen, \emph{Spectral Methods in MATLAB}. SIAM, 2000,
  \href{https://doi.org/10.1137/1.9780898719598}{10.1137/1.9780898719598}.

\bibitem{GarGot22}
A.~M. Garc\'ia-Garc\'ia, V.~Godet, C.~Yin and J.~P. Zheng,
  \emph{{Euclidean-to-Lorentzian wormhole transition and gravitational symmetry
  breaking in the Sachdev-Ye-Kitaev model}},
  \href{https://arxiv.org/abs/2204.08558}{{\ttfamily 2204.08558}}.

\bibitem{CalCar09}
P.~Calabrese and J.~Cardy, \emph{{Entanglement entropy and conformal field
  theory}},
  \href{https://doi.org/10.1088/1751-8113/42/50/504005}{\emph{J.Phys.}
  {\bfseries A42} (2009) 504005}
  [\href{https://arxiv.org/abs/0905.4013}{{\ttfamily 0905.4013}}].

\bibitem{SulRaa20}
J.~Sully, M.~V. Raamsdonk and D.~Wakeham, \emph{{BCFT entanglement entropy at
  large central charge and the black hole interior}},
  \href{https://doi.org/10.1007/JHEP03(2021)167}{\emph{JHEP} {\bfseries 03}
  (2021) 167} [\href{https://arxiv.org/abs/2004.13088}{{\ttfamily
  2004.13088}}].

\bibitem{MoiSak19}
U.~Moitra, S.~K. Sake, S.~P. Trivedi and V.~Vishal, \emph{{Jackiw-Teitelboim
  Model Coupled to Conformal Matter in the Semi-Classical Limit}},
  \href{https://doi.org/10.1007/JHEP04(2020)199}{\emph{JHEP} {\bfseries 04}
  (2020) 199} [\href{https://arxiv.org/abs/1908.08523}{{\ttfamily
  1908.08523}}].

\bibitem{GibHaw77a}
G.~W. Gibbons and S.~W. Hawking, \emph{{Cosmological Event Horizons,
  Thermodynamics, and Particle Creation}},
  \href{https://doi.org/10.1103/PhysRevD.15.2738}{\emph{Phys. Rev. D}
  {\bfseries 15} (1977) 2738}.

\bibitem{Mar22}
D.~Marolf, \emph{{Gravitational thermodynamics without the conformal factor
  problem: Partition functions and Euclidean saddles from Lorentzian Path
  Integrals}},  \href{https://arxiv.org/abs/2203.07421}{{\ttfamily
  2203.07421}}.

\bibitem{Mal01}
J.~M. Maldacena, \emph{{Eternal black holes in anti-de Sitter}},
  \href{https://doi.org/10.1088/1126-6708/2003/04/021}{\emph{JHEP} {\bfseries
  04} (2003) 021} [\href{https://arxiv.org/abs/hep-th/0106112}{{\ttfamily
  hep-th/0106112}}].

\bibitem{DiaSan15}
O.~J.~C. Dias, J.~E. Santos and B.~Way, \emph{{Numerical Methods for Finding
  Stationary Gravitational Solutions}},
  \href{https://doi.org/10.1088/0264-9381/33/13/133001}{\emph{Class. Quant.
  Grav.} {\bfseries 33} (2016) 133001}
  [\href{https://arxiv.org/abs/1510.02804}{{\ttfamily 1510.02804}}].

\bibitem{MalSta17}
J.~Maldacena, D.~Stanford and Z.~Yang, \emph{{Diving into traversable
  wormholes}}, \href{https://doi.org/10.1002/prop.201700034}{\emph{Fortsch.
  Phys.} {\bfseries 65} (2017) 1700034}
  [\href{https://arxiv.org/abs/1704.05333}{{\ttfamily 1704.05333}}].

\bibitem{KouMal17}
I.~Kourkoulou and J.~Maldacena, \emph{{Pure states in the SYK model and
  nearly-$AdS_2$ gravity}},  \href{https://arxiv.org/abs/1707.02325}{{\ttfamily
  1707.02325}}.

\end{thebibliography}\endgroup

\end{document}